\documentclass[12pt]{report}
\usepackage{graphicx,amssymb,makeidx,fancybox}
\usepackage[francais]{babel}
\usepackage[T1]{fontenc}
\usepackage[latin1]{inputenc}
\catcode`\«=\active
\catcode`\»=\active
\def«{\og\ignorespaces}%
\def»{{\fg}}%
\title{Bifurcation vers l'état d'Abrikosov et diagramme de phase}
\makeindex
\author{Mathieu Dutour}
\begin{document}

\bibliographystyle{plain}
\renewcommand{\baselinestretch}{1}
\renewcommand{\theequation}{\arabic{chapter}.\arabic{section}.\arabic{equation}}
\def\saction#1{\section{#1}\setcounter{equation}0}
\def\Tore#1{\leavevmode\kern-.0em\raise.2ex\hbox{$\R^2$}\kern-.1em/\kern-.1em\lower.25ex\hbox{$#1$}}

\newcommand{\D}{\displaystyle}
\newcommand{\R}{\ensuremath{\Bbb{R}}}
\newcommand{\N}{\ensuremath{\Bbb{N}}}
\newcommand{\Q}{\ensuremath{\Bbb{Q}}}
\newcommand{\C}{\ensuremath{\Bbb{C}}}
\newcommand{\Z}{\ensuremath{\Bbb{Z}}}
\newcommand{\T}{\ensuremath{\Bbb{T}}}
\newcommand{\rot}{\ensuremath{\mathop{\rm rot}\nolimits}}
\newcommand{\divergence}{\ensuremath{\mathop{\rm div}\nolimits}}
\newcommand{\determinant}{\ensuremath{\mathop{\rm det}\nolimits}}
\newcommand{\Rez}{\ensuremath{\mathop{\rm Re}\nolimits}}
\newcommand{\Imz}{\ensuremath{\mathop{\rm Im}\nolimits}}

\newtheorem{theorem}{Théorème}[section]
\newtheorem{lemme}[theorem]{Lemme}
\newtheorem{proposition}[theorem]{Proposition}
\newtheorem{definition}[theorem]{Définition}
\newtheorem{remarque}[theorem]{Remarque}
\newtheorem{corollaire}[theorem]{Corollaire}
\newtheorem{exemple}[theorem]{Exemple}
\newtheorem{probleme}[theorem]{Problème}
\newtheorem{hypoth}[theorem]{Hypothèse}

{
\begin{flushleft}
\large ORSAY \\
{\it \underline{$N^{\circ}$ D'ORDRE : $5961$}}
\end{flushleft}
\begin{center}
\doublebox{\parbox{11cm}{\begin{tabular}{c}
\Large \bf UNIVERSIT\'E de PARIS-SUD\\
\Large \bf U.F.R. SCIENTIFIQUE D'ORSAY
\end{tabular}
}}
\end{center}
\begin{center} \large
Thèse\\[2mm]
présentée\\[2mm]
Pour obtenir\\[3mm]
{ \Large Le GRADE de DOCTEUR EN SCIENCES\\
DE L'UNIVERSIT\'E PARIS XI}\\
\large PAR\\
\Large Mathieu Dutour
\end{center}
\large \underline{SUJET:} {\bf Bifurcation vers l'état d'Abrikosov et diagramme de phase}\\[2cm]
Soutenue le 10 décembre 1999 devant la commission d'examen\\[2cm]

\begin{tabular}{rll}
MM:&J. C. Saut&Président du Jury\\
&B. Helffer&Directeur de la thèse\\
&C. Bolley&Rapporteur de la thèse\\
&T. Rivière&Rapporteur de la thèse\\
&F. Bethuel&Examinateur
\end{tabular}}
\begin{abstract}
\noindent Nous \'etudions dans cette th\`ese la fonctionnelle de Ginzburg-Landau dans $\R^3$ sur des couples de fonctions $(\phi, \overrightarrow{A})$ qui v\'erifient des conditions de p\'eriodicit\'e de jauge en $x_3$ et selon un r\'eseau discret de $(x_1,x_2)$. Nous montrons que le probl\`eme variationnel est \'equivalent au probl\`eme de la minimisation d'une autre fonctionnelle sur un tore. Dans le cadre de la d\'emonstration, un fibr\'e vectoriel non trivial apparaît. On se limite alors pour la suite à une quantification de $1$. On montre ensuite que la fonctionnelle admet un minimum sur l'espace fonctionnel $H^{1}$ qui v\'erifie un syst\`eme d'\'equations aux d\'eriv\'ees partielles appel\'e syst\`eme de Ginzburg-Landau. Le minimum est $C^{\infty}$ par l'ellipticit\'e du syst\`eme d'\'equations de Ginzburg-Landau. On montre qu'il y a une bifurcation du couple $(0,0)$ pour le champ critique $H_{ext}=k$ o\`u $k$ est un param\`etre caract\'eristique du syst\`eme. On \'etudie alors la stabilit\'e de la solution bifurqu\'ee. On \'etudie la d\'ependance de l'\'energie minimale à l'\'egard de la g\'eom\'etrie du tore. Enfin nous d\'ecrivons toutes les solutions du syst\`eme d'\'equations de Ginzburg-Landau dans la limite $k$ tend vers l'infini. Dans le dernier chapitre, nous donnons pour notre mod\`ele la structure du diagramme des phases en pr\'ecisant quelles r\'egions sont normales, supraconductrices pure, mixte.
\end{abstract}

\begin{center}
\bf REMERCIEMENTS
\end{center}
\par Je remercie chaleureusement Bernard Helffer pour avoir dirigé avec patience ma thèse pendant ces trois ans et m'avoir orienté dans mes recherches. Travailler avec lui f\^ut pour moi un grand honneur et une occasion de profiter de ses très vastes connaissances.
\vspace{0.5cm}
\par Je remercie vivement Catherine Bolley et Tristan Rivière d'avoir eu la patience de lire ma thèse.
\vspace{0.5cm}
\par Je remercie également Jean Claude Saut et Fabrice Bethuel pour avoir accepté d'\^etre parmi le jury de ma thèse.
\vspace{0.5cm}
\par Toute mon admiration va au très grand professeur Vladimir Arnold. Il a décidé de ma carrière.
\vspace{0.5cm}
\par Je remercie également les excellents professeurs que sont Yann Brenier et Bernard Randé.

\setcounter{tocdepth}{3}
\tableofcontents

\chapter{Introduction}
%\label{sec:introduction}
\noindent
L'étude des supraconducteurs peut se faire de façon thermodynamique (voir par exemple \cite{Kittel}, p.~368) en minimisant une énergie libre.\\
Landau et Ginzburg, dans un article célèbre, ont introduit une fonctionnelle $E(\phi, \overrightarrow{A})$ comme expression possible pour l'énergie libre des supraconducteurs où $\phi$\index{$\phi$} est une fonction d'onde et $\overrightarrow{A}$\index{$\overrightarrow{A}$} est un potentiel magnétique.\\
Dans cette thèse, on envisage une modélisation particulière du problème dont l'idée appartient à Abrikosov (voir l'article \cite{Abrikosov}) qui cherchait à montrer l'apparition d'un état mixte entre la solution supraconductrice pure où $\phi=1$ et le champ magnétique est nul à l'intérieur du supraconducteur et la solution normale où $\phi=0$ et le champ magnétique interne est égal au champ extérieur qui est constant.\\
Le paramètre $k$\index{$k$} est le paramètre de Ginzburg Landau. Il est sans dimension et strictement positif. On définit pour tout ouvert ${\cal X}$ relativement compact de $\R^3$ une fonctionnelle\index{fonctionnelle}\index{$E_{\cal X}$} par
\begin{equation}\label{GL-Var-ABRIKOSOV}
\begin{array}{rcl}
H^{1}_{loc}(\R^3,\C)\times H^{1}_{loc}(\R^3,\R^3)&\mapsto& \R\\
(\phi,\overrightarrow{A})&\mapsto&E_{\cal X}(\phi,\overrightarrow{A})=\frac{1}{2}\int_{{\cal X}}\Vert \overrightarrow{\rot}\, {\overrightarrow{A}}-\overrightarrow{H}_{ext}\Vert ^2dx\\
&&+\frac{1}{4}\int_{{\cal X}}(1-|\phi|^2)^2dx\\
&&+\frac{1}{2}\int_{{\cal X}}\Vert ik^{-1}{\overrightarrow{\nabla}}\phi+{\overrightarrow{A}}\phi\Vert ^2dx .
\end{array}
\end{equation}
L'ouvert ${\cal X}$ correspond au domaine occupé par le matériau et sera choisi plus loin.\\
Le champ magnétique \index{$\overrightarrow{H}_{ext}$}$\overrightarrow{H}_{ext}$ est le champ extérieur que l'on applique sur le supraconducteur. On le suppose constant et dirigé suivant la direction $x_3$.\\
On écrira les changements de jauge de la même façon que dans le livre \cite{Nakahara} c'est à dire sous la forme
\begin{equation}\label{changement-de-jauge}
\begin{array}{c}
\left\lbrace\begin{array}{rcl}
\phi&\rightarrow&\phi e^{ikf}\\
\overrightarrow{A}&\rightarrow&\overrightarrow{A}+\overrightarrow{\nabla}\, f
\end{array}\right.
\mbox{avec~}f\in H^{2}_{loc}(\R^3,\R)
\end{array} .
\end{equation}
La fonctionnelle $E_{\cal X}$ est invariante par ces changements de jauge, c'est à dire que l'on a:
\begin{equation}\label{invarianec-Exi}
E_{\cal X}(\phi e^{ikf},\overrightarrow{A}+\overrightarrow{\nabla}\, f)=E_{\cal X}(\phi,\overrightarrow{A})\mbox{~avec~}f\in H^{2}_{loc}(\R^3,\R) .
\end{equation}
On considère dans cette thèse des états $(\phi, \overrightarrow{A})$ dont les translations selon la direction $x_3$ qui est celle du champ magnétique sont équivalentes de jauge à l'état original. On suppose aussi que les translatés selon un réseau discret sont équivalents à l'état original.\\
On associe à l'état $(\phi,\overrightarrow{A})$ le champ magnétique $\overrightarrow{B}$\index{$\overrightarrow{B}$}\index{champ magnétique}, la densité d'électrons supraconducteurs $n$\index{$n$}\index{densité d'électrons supraconducteurs} et le courant d'électrons\index{courant d'électron} $\overrightarrow{J}$\index{$\overrightarrow{J}$} sont donnés par:
\begin{equation}\label{expression-de-B-J}
\left\lbrace\begin{array}{rcl}
\overrightarrow{B}&=&\overrightarrow{\rot}\, \overrightarrow{A},\\
n&=&|\phi|^2,\\
\overrightarrow{J}&=&\Rez [\overline{\phi}(ik^{-1}\overrightarrow{\nabla}\phi+\overrightarrow{A}\phi)]
\end{array}\right.
\end{equation}
et qui sont les quantités physiques intéressantes. Elles sont en effet invariantes de jauge donc en particulier périodiques selon un réseau ${\cal L}$ fixé et constantes selon la direction $x_{3}$.\\
Dans la section~\ref{sec:reduction2-3}, on montre que le minimum de la fonctionnelle peut être pris sur les états $(\phi, \overrightarrow{A})$ constants selon $x_3$ et périodiques de jauge selon le réseau ${\cal L}$.\\
Dans la section~\ref{sec:reduction-tore}, on démontre que le problème peut se réduire à l'étude d'un problème défini sur l'espace quotient \index{$\Tore{\cal L}$}$\Tore{\cal L}$. Mais cette réduction est plus complexe et le résultat est que la fonction $\phi$ appartient à un fibré vectoriel non trivial sur le tore. Le problème est alors scindé en une infinité dénombrable de problèmes de minimisation sur un espace vectoriel ${\cal B}_d$. Notre démonstration rigoureuse semble \^etre originale, voir cependant  \cite{Barany-Golubitsky} et \cite{pirate-II} pour des démonstration partielles.\\
Ci-dessous on indique ce qui sera démontré.
%On démontrera un résultat de réduction à des espaces qui ne semble pas avoir été démontré nulle à notre connaissance, voir cependant \cite{Barany-Golubitsky} et \cite{pirate-II}.
\begin{theorem}.
Il existe une suite de fibrés en droites complexes $E_d$ indexée par $d\in\Z$ définis sur le tore $\Tore{\cal L}$, avec une fonctionnelle associée $E^{int,d}$ sur l'espace $H^{1}(E_{d})\oplus H^{1}_{\divergence,0}(\Tore{\cal L},\R^2)$ tel que, pour chaque état $(\phi,  \overrightarrow{A})$ périodique de jauge selon le réseau ${\cal L}$ et périodique de jauge selon la direction $x_3$, il existe un entier $d\in\Z$ et un état de la forme $(\phi', \overrightarrow{A}')$ avec $\phi'\in H^{1}(E_d)$ et $\overrightarrow{A}'\in H^{1}(\R^2,\R^2)$ de divergence nulle sur $\Tore{\cal L}$ tel que $E^{int,d}(\phi',\overrightarrow{A}')\leq E_{\cal X}(\phi, \overrightarrow{A})$.\\
De plus si $d\not=0$ alors on peut imposer que $\overrightarrow{A}'$ est d'intégrale nulle sur $\Tore{\cal L}$.\\
En particulier l'étude des minima de la fonctionnelle se ramène de ce fait à l'étude des minima de la famille de fonctionnelles réduites $(E^{int,d})_{d\in\Z}$.
\end{theorem}
L'entier $d$ définit une quantification de notre problème. Dans la section~\ref{sec:modelisation-finale} on fait l'hypothèse d'une quantification égale à $1$. On fait alors appara\^itre une nouvelle fonctionnelle $E^V_{k,H_{ext}}$.\\
Un premier problème variationnel appelé ``problème réduit'' est obtenu en éliminant une des variables (le flux magnétique); il consiste en la minimisation de la fonctionnelle\index{fonctionnelle réduite}
\begin{equation}
\begin{array}{l}
F_{\lambda,k}(\phi, \overrightarrow{a})=\int_{\Omega}\frac{1}{2}|i\overrightarrow{\nabla}\phi+(\overrightarrow{A}_{0}+\overrightarrow{a})\phi|^2+\frac{1}{4}(\lambda-|\phi|^2)^2+\frac{k^2}{2}|\rot\, \overrightarrow{a}|^2
\end{array}
\end{equation}
sur l'espace fonctionnel
\begin{equation}
{\cal A}=\left\lbrace 
\begin{array}{c}
(\phi,\overrightarrow{a})\in H^{1}(E_{1})\times H^{1}(\Tore{\cal L},\R^2)\\
\mbox{avec~}\phi(z+v_{i})=e^{i\pi \det (v_i,x)}\phi(z), \\
\overrightarrow{a} \mbox{~${\cal L}$-périodique,~} \divergence\,\overrightarrow{a}=0\mbox{~et~}\int_{\Omega}\overrightarrow{a}=0 .
\end{array}\right\rbrace , 
\end{equation}
o\`u les $v_i$ engendrent un réseau de $\R^2$ et $\Omega$ est un domaine fondamental pour ce réseau. Le minimum est appelé $m_F(\lambda,k)$.\\
On est ensuite amené à étudier le minimum sur $\R_+^*$ de
\begin{equation}\label{si-bolley-le-veut}
G^V_{k,H_{ext}}:H_{int}\mapsto (\frac{H_{int}}{2\pi k})^2m_F(\frac{2\pi k}{H_{int}},k)+\frac{1}{2}(H_{int}-H_{ext})^2 ,
\end{equation}
o\`u $H_{int}$ est la variable éliminée précédemment. Le minimum est appelé ${\cal E}^{V}_{k,H_{ext}}$.\\
La fonctionnelle $E^V_{k,H_{ext}}$ s'écrit:
\begin{equation}
\begin{array}{rcl}
E^V_{k,H_{ext}}:\R\times{\cal A}&\mapsto&\R\\
(H_{int},\phi,\overrightarrow{a})&\mapsto&\frac{F_{\lambda,k}(\phi,\overrightarrow{a})}{\lambda^2}+\frac{1}{2}(H_{int}-H_{ext})^2
\end{array}
\end{equation}
avec $\lambda=\frac{2\pi k}{H_{int}}$.\\
Dans cette opération appara\^it un champ intérieur $H_{int}$ qui est en fait la moyenne du champ magnétique à l'intérieur du supraconducteur (ou flux).\\
On appelle \index{couple}``couple'' tout élément de la forme $(\phi,\overrightarrow{a})$. On appelle \index{état}``état'' tout triplet $(H_{int},\phi,\overrightarrow{a})$. On n'exige pas que ces états et couples soient solutions des équations de Ginzburg-Landau. Les triplets $(H_{int},\phi,\overrightarrow{a})$ correspondent à des couples $(\phi,\overrightarrow{A})$ qui sont donc également appelés des états. \\
On dit que l'état du supraconducteur est pur si le minimum ${\cal E}^V_{k,H_{ext}}$ est égal à $\frac{H_{ext}^2}{2}$, cette énergie est atteinte par un état de quantification $0$; on dit que l'état du supraconducteur est normal si le minimum est atteint par l'état $(H_{ext},0,0)$. On dit qu'il est dans un état mixte dans les autres cas.\\
Dans la section~\ref{sec:struct-reseau-space} on étudie les symétries de l'espace des réseaux d'une façon géométrique et on montre ainsi l'annulation de certaines dérivées. Cette étude est réminiscente de celle des courbes elliptiques faite dans \cite{Siegel}, volume III, chapitre 6, section 5, bien que notre groupe de symétries soit plus grand que le groupe modulaire.\\
\\
{\it Le matériel du Chapitre \ref{chapitre-generalite-sur-Flambda-k} est le plus souvent classique. Il s'agit de l'étude variationnelle du minimum de la fonctionnelle $F_{\lambda,k}$.}\\
La section~\ref{sec:operateur-lineaire} étudie deux opérateurs différentiels linéaires classiques: le laplacien magnétique en champ constant et le laplacien simple, tous deux opérant sur des espaces de sections de fibrés vectoriels $C^{\infty}$ de base $\Tore{\cal L}$. Cette étude n'est pas originale, voir par exemple \cite{Barany-Golubitsky}.
\begin{definition}.
On dit qu'une fonctionnelle $E$ est coercive\index{coercive} pour $N$ o\`u $N$ est une norme sur l'espace de definition de la fonctionnelle si, pour tout réel $C$, il existe un réel $C'>0$, tel que $E(f)<C$ implique $N(f)<C'$.
\end{definition}
On montre dans la section~\ref{sec:functional-analysis} que la fonctionnelle réduite $F_{\lambda,k}$ est coercive sur l'espace fonctionnel $H^{1}(E_{1})\oplus H^{1}(\Tore{\cal L},\R^2)$, ce qui n'était pas le cas pour le problème initial. On peut alors prouver que la fonctionnelle réduite atteint son minimum.\\
%La fonctionnelle complète atteint son minimum sur $\R$ comme on le montre dans la section~\ref{sec:universelle}.\\
On montre dans la section~\ref{sec:regularite} que les couples réalisant le minimum en quantification $1$ vérifient le système d'équations aux dérivées partielles de Ginzburg-Landau et que ces couples sont $C^{\infty}$ en vertu de l'ellipticité de ce même système. On obtient en particulier le théorème suivant (voir par exemple \cite{pirate-II}).
\begin{theorem}.
Le problème variationnel réduit sur $H^{1}(E_{1})\bigoplus H^{1}(\Tore{\cal L},\R^2)$ est coercif, et le minimum de la fonctionnelle réduite $F_{\lambda,k}$ est atteint en au moins un point. Les solutions minimisantes sont $C^{\infty}$.
\end{theorem}
Dans la section~\ref{sec:universelle}, on montre que la fonction d'onde $\phi$ est bornée par $1$ par un principe du maximum. On montre ensuite que la fonctionnelle complète 
\begin{equation}
\begin{array}{rcl}
E^V_{k,H_{ext}}:\R\times{\cal A}&\mapsto&\R\\
(H_{int},\phi,\overrightarrow{a})&\mapsto&E^V_{k,H_{ext}}(H_{int},\phi,\overrightarrow{a})
\end{array}
\end{equation}
atteint son minimum sur un état $(H_{int},\phi,\overrightarrow{a})$.\\
On montre ensuite que le champ magnétique à l'intérieur d'un supraconducteur est inférieur au champ extérieur.\\
Dans la section~\ref{sec:Bochner-Kodaira-Nakano} on montre une formule de Bochner-Kodaira-Nakano et on l'utilise pour élucider le cas particulier $k\geq \frac{1}{\sqrt{2}}$ et $\lambda\leq 2\pi$. C'est la clé pour l'étude précise du diagramme de phase de la fonctionnelle entreprise dans le chapitre 5.\\
\\
{\it Le chapitre \ref{chapitre-bifurcation-abrikosov} est consacré à la bifurcation d'Abrikosov.}\\
On constate expérimentalement le fait suivant: Si on place un supraconducteur dans un champ magnétique alors ce champ dispara\^it à l'intérieur du supraconducteur (cas des supraconducteurs de type I avec $k\leq \frac{1}{\sqrt{2}}$) ou s'affaiblit (cas des supraconducteurs de type II avec $k\geq \frac{1}{\sqrt{2}}$). Cette propriété est appelé écrantement \index{écrantement}du champ magnétique. Dans notre modélisation, on considère le champ magnétique interne $H_{int}$ comme un paramètre.\\
La bifurcation d'Abrikosov consiste à effectuer des calculs au voisinage de l'état normal pour montrer l'existence de solutions des équations de Ginzburg-Landau.\\
Dans la section~\ref{sec:lyapunov-schmidt}, on fait varier le paramètre $H_{int}$ et on montre qu'il y a une bifurcation des solutions (du problème réduit pour le champ $H_{int}=k$) du couple $(0,0)$ vers un couple non trivial. Cette section est inspirée des travaux \cite{odehI}, \cite{odehII} et \cite{Barany-Golubitsky}, mais notre analyse est plus précise.\\
Cette bifurcation existe si $k\not= \sqrt{\frac{2K}{I}}$. Dans la section~\ref{sec:calcul-energie}, on effectue des calculs explicites sur l'énergie du couple bifurqué. On constate que l'état bifurqué est d'énergie plus faible que l'état normal si et seulement si la condition $k>\sqrt{\frac{2K}{I}}$ est vérifiée.\\
Enfin dans la section~\ref{sec:stabilite-etat-bifurque}, on discute la stabilité du couple bifurqué. On montre que l'état bifurqué est stable si $k>\sqrt{\frac{2K}{I}}$ et instable si $k<\sqrt{\frac{2K}{I}}$.
Dans la section~\ref{sec:analyse-asymptotique}, on décrit pour des $k$ assez grands le comportement de la fonctionnelle près de la bifurcation:
\begin{theorem}.
Il existe $\gamma>0$ et $k_{0}>0$ tel que si $k>k_{0}$ et $H_{int}>k-\gamma$ alors les points critiques de la fonctionnelle réduite sont les solutions déjà calculées dans la section~\ref{sec:lyapunov-schmidt}, c'est à dire les solutions bifurquées et la solution triviale. On montre également que la solution triviale est un point col pour la fonctionnelle et que la solution bifurquée est un minimum global.\\
\end{theorem}
{\it Le chapitre \ref{Diagramm-phase} étudie le diagramme des phases $(k,H_{ext})$ du supraconducteur.} Dans la section \ref{sec:minimum-et-monotonie}, on montre des théorèmes de monotonie. Si, en $(k,H_{ext})$, le supraconducteur est pur, alors il est aussi pur en $(k',H'_{ext})$ si $k'\leq k$ et $H'_{ext}\leq H_{ext}$ (cf~\ref{monotonie-etat-pur}). On a des théorèmes similaires pour l'état normal (cf~\ref{normal-k-inegalite} et \ref{normal-H-inegalite}). On obtient du m\^eme coup le diagramme si $k\leq\frac{1}{\sqrt{2}}$ (cf~\ref{usage-theoreme-monotonie}).\\
La section \ref{sec:salamon-merci} est consacrée à une étude approfondie du cas $k=\frac{1}{\sqrt{2}}$. La sous-section \ref{Recopie-Dietmar-Salamon} redémontre des résultats de Dietmar Salamon non publiés. Puis on détermine les couples minimisants $F_{\lambda,\frac{1}{\sqrt{2}}}$ (cf~\ref{piratage-salamon}); on démontre la relation remarquable $I-4K=1$ (cf~\ref{theorem-egalite-somme-clef}) et on montre que la condition $k>\sqrt{\frac{2K}{I}}$ est vérifiée dans la zone $k\geq\frac{1}{\sqrt{2}}$ (cf~\ref{corollaire-condition-hard}).\\
La description complète de l'état normal est effectuée dans la section \ref{sec:diagramme-des-phases}. Elle commence par un lemme très important de localisation (cf~\ref{mur-1-racine2-franchi}) et se poursuit sur le théorème de bifurcation d'Abrikosov (cf~\ref{Bifurcation-Abrikosov-real-nature}), o\`u l'on donne une description précise du minimum et des états minimisants autour de $H_{ext}=k$. On peut alors calculer explicitement le champ critique $H_{c2}(k)$ de notre modèle.\\
La section \ref{sec:diagramme-de-phase-2} décrit l'état pur. Le théorème \ref{apparition-intervalle} montre l'existence du champ critique $H_{c1}(k)$. On trouve ensuite une expression en $\inf$ pour $H_{c1}(k)$ (cf~\ref{expression-en-inf-assez-utile}), une estimée asymptotique (cf~\ref{estimee-asymptotique-weak-sylvia-saint-serfaty}) et la continuité de $H_{c1}$ (cf~\ref{proposition-merci-marie-monier}). On a alors une description presque complète des régions du plan $(k,H_{ext})$ o\`u l'état du supraconducteur est pur.\\
On a alors démontré que le diagramme de phase a pour allure (cf~le théorème \ref{usage-theoreme-monotonie} pour $k\leq \frac{1}{\sqrt{2}}$, le théorème \ref{apparition-intervalle2} pour l'état normal, le théorème \ref{proposition-merci-marie-monier} pour la décroissance de la région o\`u l'état est pur et le théorème \ref{Bifurcation-Abrikosov-real-nature} pour la bifurcation d'Abrikosov)\index{diagramme des phases}:\\
\resizebox{12cm}{!}{\rotatebox{0}{\includegraphics{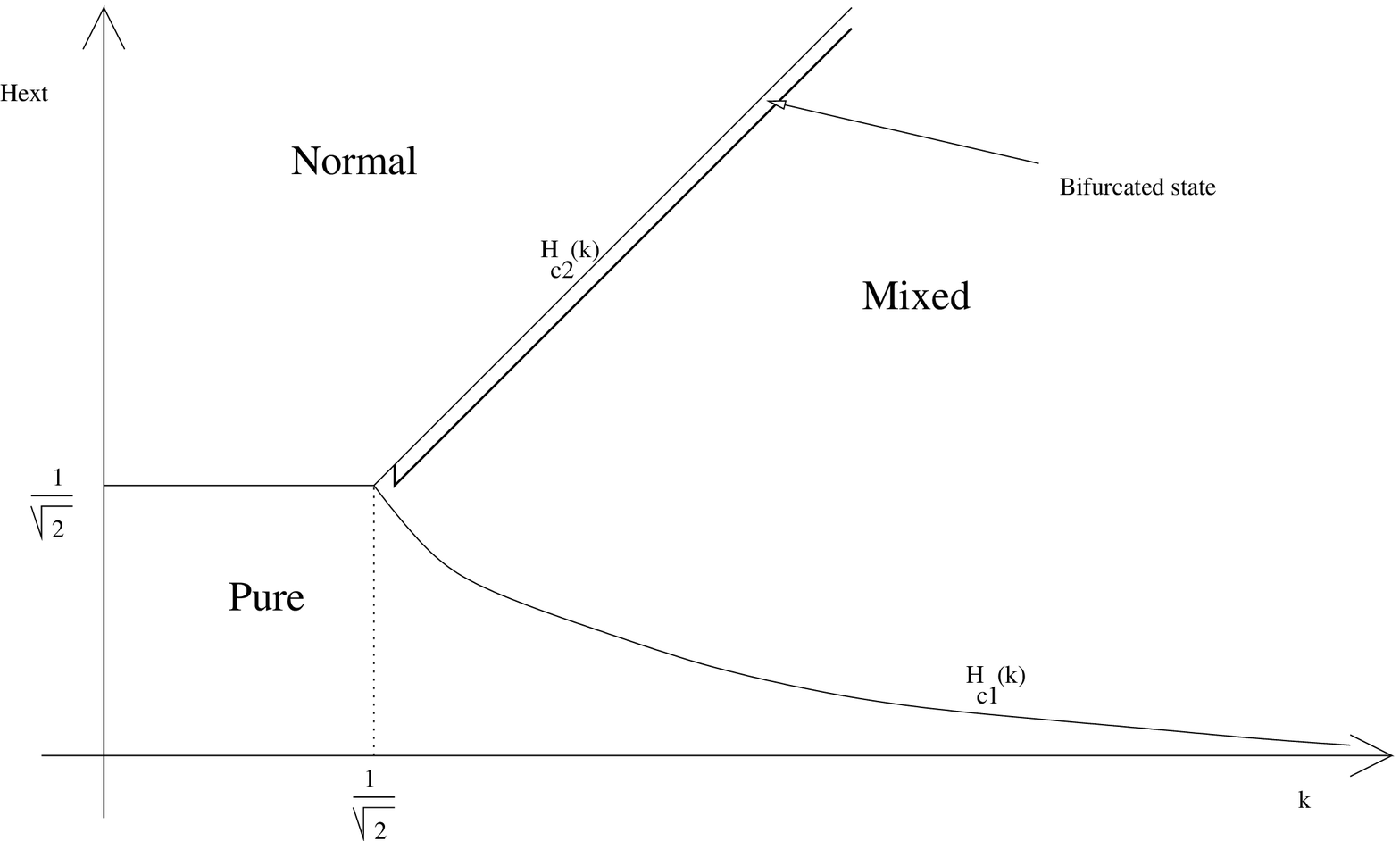}}}\\
o\`u
\begin{itemize}
\item $P$ désigne la région o\`u le minimum de la fonctionnelle est atteint par l'état pur,
\item $N$ désigne la région o\`u le minimum de la fonctionnelle est atteint par l'état normal,
\item $M$ désigne la région o\`u le minimum de la fonctionnelle est atteint par l'état mixte.
\end{itemize}

\chapter{Étude générale de la fonctionnelle}

\saction{Définition de la fonctionnelle et invariance de jauge}\label{sec:enonce}
\noindent On note \index{${\cal{E}}(\phi,\overrightarrow{A})$}${\cal{E}}(\phi,\overrightarrow{A})$ l'intégrande de la fonctionnelle \index{$E_{\cal X}(\phi, \overrightarrow{A})$}$E(\phi, \overrightarrow{A})$ définie en~(\ref{GL-Var-ABRIKOSOV}). C'est la fonction définie sur $\R^3$ par:
\begin{equation}\label{expression-de-la-fonctionnelle}
\begin{array}{rcccl}
{\cal{E}}(\phi,\overrightarrow{A})&:&\R^3&\mapsto&\R\\
&&x&\mapsto&
\frac{1}{2}\Vert \overrightarrow{\rot}\, {\overrightarrow{A}}(x)-\overrightarrow{H}_{ext}\Vert ^2+\frac{1}{4}(1-|\phi(x)|^2)^2\\
&&&&+\frac{1}{2}\Vert ik^{-1}{\overrightarrow{\nabla}}\phi(x)+{\overrightarrow{A}}(x)\phi(x)\Vert ^2 .
\end{array}
\end{equation}
On a alors pour tout ouvert ${\cal X}$ de $\R^3$ la relation
\begin{equation}
E_{\cal X}(\phi, \overrightarrow{A})=\int_{\cal X}{\cal{E}}(\phi,\overrightarrow{A}) dx
\end{equation}
qui montre bien que ${\cal{E}}$ est une densité d'énergie locale.\\
On verra plus loin que, si $\phi\in H^{1}_{loc}(\R^3,\C)$ et $\overrightarrow{A}\in H^{1}_{loc}(\R^3,\R^3)$, alors ${\cal E}(\phi,\overrightarrow{A})\in L^{1}_{loc}$.
\begin{hypoth}.
On supposera toujours que le champ extérieur $\overrightarrow{H}_{ext}$ est constant.
\end{hypoth}
On prend les coordonnées $x_1,x_2,x_3$ orthonormées de façon que \mbox{$\overrightarrow{H}_{ext}=(0,0,H_{ext})$}.\\
On note ${\cal L}$\index{${\cal L}$} le réseau de $\R^2$ défini par les deux vecteurs \index{$v_i$}\index{$\xi_i$}\index{$\eta_i$} suivants
\begin{equation}\label{definition-vecteur-du-reseau}
\left\lbrace\begin{array}{l}
v_{1}=(\xi_{1},\eta_{1}),\\
v_{2}=(\xi_{2},\eta_{2}).
\end{array}\right.
\end{equation}
Nous plongerons ce réseau dans $\R^3$ en posant que la composante selon $x_3$ est nulle.

\begin{remarque}.
Dans cette thèse nous aurons besoin de produits hermitiens. Par convention nous noterons
\begin{equation}
\langle \phi,\psi\rangle=\int_{\Omega}\overline{\phi}\psi
\end{equation}
Il y a donc antilinéarité par rapport à la première variable.
\end{remarque}

\begin{definition}.
Un ensemble $E\subset\R^2$ est un domaine fondamental\index{domaine fondamental} pour le réseau ${\cal L}$ si les deux conditions suivantes sont vérifiées
\begin{equation}
\begin{array}{l}
\R^2=\cup_{l\in{\cal L}}(E+l),\\
\mbox{si~}l\not=l'\mbox{~alors~}\{E+l\}\cap\{ E+l'\}=\emptyset .
\end{array}
\end{equation}
\end{definition}
On appelle $\Omega$\index{$\Omega$} le domaine fondamental de $\R^2$ associé à ${\cal L}$ défini par
\begin{equation}\label{definition-domaine}
\Omega=\lbrace w=t_1 v_1+t_2v_2\mbox{~tel~que~}t_1\in[0,1[\mbox{~et~}t_2\in[0,1[\rbrace
\end{equation}
que l'on dessine ci-dessous dans le plan $(x,y)$\\
\resizebox{8cm}{!}{\rotatebox{0}{\includegraphics{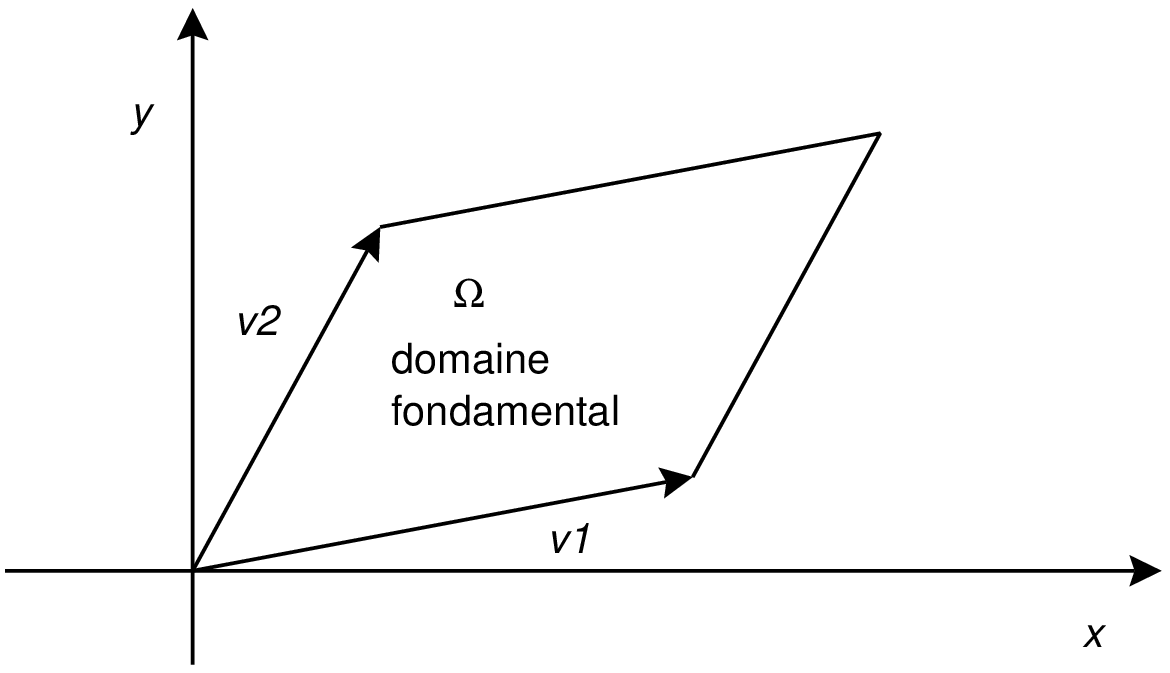}}}\\
L'aire du domaine est \index{$\vert\Omega\vert$}
\begin{equation}
|\Omega|=\xi_{1}\eta_{2}-\xi_{2}\eta_{1} .
\end{equation}
L'invariance selon $x_3$ du système signifie que le système translaté est équivalent de jauge avec le système initial; on dira que $(\phi,\overrightarrow{A})$ est << invariant de jauge selon $x_3$ >>\index{invariant de jauge selon $x_3$} si
\begin{equation}\label{invariance-en-z}
\left\lbrace\begin{array}{l}
\forall l\in \R,\,\,\exists f_{l}\in H^{2}_{loc}(\R^3,\R)\\
\phi(.,x_3+l)=e^{ikf_{l}(.,x_3)}\phi(.,x_3)\\
{\overrightarrow{A}}(.,x_3+l)={\overrightarrow{A}}(.,x_3)+{\overrightarrow{\nabla}}\,f_{l}(.,x_3)
\end{array}\right.
\end{equation}
et on dira que $(\phi,\overrightarrow{A})$ est << périodique de jauge selon le réseau ${\cal L}$ >>\index{périodique de jauge selon le réseau ${\cal L}$} si
\begin{equation}\label{jauge-equivalence-sur-L}
\left\lbrace\begin{array}{l}
\forall t\in {\cal{L}},\,\,\, \exists g^{t}\in H^{2}_{loc}(\R^3,\R)\mbox{~tel ~que~}\forall x\in \R^3\\
\phi(x+t)=e^{ikg^{t}(x)}\phi(x)\\
\overrightarrow{A}(x+t)=\overrightarrow{A}(x)+\overrightarrow{\nabla}\,g^{t}(x) .
\end{array}\right.
\end{equation}
Les fonctions \index{$f_l$}$f_{l}$ et $g^{t}$ des équations~(\ref{invariance-en-z}) et~(\ref{jauge-equivalence-sur-L}) sont définies à une constante près par le couple $(\phi, \overrightarrow{A})$.\\
On note ${\cal D}$ \index{${\cal D}$}l'espace des couples vérifiant ces conditions:
\begin{equation}
{\cal D}=\left\lbrace\begin{array}{c}
(\phi,\overrightarrow{A})\in H^{1}_{loc}(\R^3,\C)\times H^{1}_{loc}(\R^3,\R^3)\\
\mbox{~tel~que~} (\ref{invariance-en-z})\mbox{~et~}(\ref{jauge-equivalence-sur-L})\mbox{~sont~vérifiés.}
\end{array}
\right\rbrace
\end{equation}
La fonction $x\mapsto{\cal{E}}(\phi,\overrightarrow{A})(x)$ ne dépend que de $x_1,x_2$ et est périodique suivant le réseau ${\cal L}$ à cause de son invariance de jauge. Donc l'intégrale~(\ref{GL-Var-ABRIKOSOV}) diverge si ${\cal X}=\R^3$. Le choix de ${\cal X}$\index{${\cal X}$} suivant répond à notre problème
\begin{equation}\label{definition-ensemble}
{\cal X}=\Omega\times]a,b[\mbox{~avec~}]a,b[\mbox{~borné.}
%\lbrace(x_1,x_2,x_3)\in \R^3 \mbox{~tel~que~} (x_1,x_2)\in \Omega\mbox{~et~} x_3\in ]a,b[ . \rbrace
\end{equation}
Le domaine d'intégration ${\cal X}$ est relativement compact et la densité d'énergie ${\cal E}(\phi,\overrightarrow{A})$ est localement intégrable donc l'intégrale~(\ref{GL-Var-ABRIKOSOV}) est finie, si \mbox{$(\phi, \overrightarrow{A})\in {\cal D}$}.  Il est naturel de diviser par $b-a$ pour obtenir une fonctionnelle qui ne dépendent ni de $a$ ni de $b$. On cherche des énergies volumiques, on divise donc par $|\Omega|$. On appelle cette énergie \index{$E^3_n$}$E^3_{n}$
\begin{equation}\label{integrale-renormalise-de-facon-triviale}
E^3_{n}(\phi,\overrightarrow{A})=\frac{1}{(b-a)|\Omega|}\int_{a}^{b}\int_{\Omega}{\cal{E}}
(\phi,\overrightarrow{A})(x_1,x_2,l)dSdl .
\end{equation}
Cette intégrale ne dépend pas des valeurs précises de $a,b$ et du domaine fondamental sur lequel on intègre parce que le système est périodique de jauge. et que \mbox{${\cal{E}}(\phi,\overrightarrow{A})$} est invariant de jauge.\\
Le problème qui nous occupera dans cette thèse est le suivant:
\begin{probleme}.\label{probleme-de-la-these}
Trouver
\begin{equation}
\min_{(\phi,\overrightarrow{A})\in{\cal D}}E_n^3(\phi,\overrightarrow{A})
\end{equation}
et décrire les états $(H_{int},\phi,\overrightarrow{A})$ qui réalisent ce minimum.
\end{probleme}
La proposition qui suit précise le lien entre énergie volumique et notre fonctionnelle. Elle ne nous servira pas par la suite.
%\begin{equation}\label{energie-moyenne}
%\frac{1}{vol(V)}\int_{V}{\cal{E}}(\phi,\overrightarrow{A}) dx
%\end{equation}
%lorsque $V \nearrow \R^3$.

\begin{proposition}.
Soit $(\phi,\overrightarrow{A})\in{\cal D}$. On a l'égalité suivante
\begin{equation}
E_n^3(\phi,\overrightarrow{A})=\lim_{R\rightarrow\infty}\frac{1}{|B(0,R)|}\int_{B(0,R)}{\cal E}(\phi,\overrightarrow{A})(x)dx
\end{equation}
o\`u $B(0,R)=\{x=(x_1,x_2,x_3)\in\R^3,\,|x_i|\leq R\}$.
\end{proposition}
{\it Preuve.} On pose\index{${\cal I}$}
\begin{equation}
{\cal I}=\Omega\times[0,1[
%\{x\in\R^3\mbox{~tel~que~}(x_1,x_2)\in\Omega\mbox{~et~}0\leq x_3\leq 1\} .
\end{equation}
On définit un réseau \index{${\cal R}$}
\begin{equation}
{\cal R}={\cal L}\times\Z .
%\{t\in\R^3\mbox{~tel~que~}(t_1,t_2)\in{\cal L}\mbox{~et~}t_3\in\Z\} .
\end{equation}
L'ensemble ${\cal I}$ est un domaine fondamental pour le réseau ${\cal R}$ de $\R^3$.\\
Vu les propriétés d'invariance de jauge~(\ref{invariance-en-z}) et~(\ref{jauge-equivalence-sur-L}) et la propriété~(\ref{invarianec-Exi}), on a 
\begin{equation}\label{egalite-car-jauge-equiv}
\forall t\in{\cal R},\,\,\int_{{\cal I}+t}{\cal E}(\phi,\overrightarrow{A})(x)dx=\int_{\cal I}{\cal E}(\phi,\overrightarrow{A})(x)dx .
\end{equation}
On pose\index{$g(R)$}\index{$G(R)$}
\begin{equation}\label{ensemble-sur-reseau}
\begin{array}{l}
g(R)=\{t\in{\cal R},\,\mbox{~tel~que~}{\cal I}+t\subset B(0,R)\},\\
G(R)=\{t\in{\cal R},\,\mbox{~tel~que~}{\cal I}+t\cap B(0,R)\not=\emptyset \}
\end{array}
\end{equation}
et
\begin{equation}\label{encadrement-par-ensemble}
\begin{array}{l}
m(R)=\cup_{t\in g(R)}\{{\cal I}+t\}\\[2mm]
M(R)=\cup_{t\in G(R)}\{{\cal I}+t\} .
\end{array}
\end{equation}
Les définitions~(\ref{ensemble-sur-reseau}) et~(\ref{encadrement-par-ensemble}) correspondent à l'idée que l'on encadre l'ensemble $B(0,R)$ par deux ensembles sur lesquels on pourra calculer l'intégrale.\\
Puisque $m(R)\subset B(0,R)\subset M(R)$, on a l'inégalité
\begin{equation}\label{csq-pour-integrale}
\int_{m(R)}{\cal E}(\phi,\overrightarrow{A})(x)dx
\leq \int_{B(0,R)}{\cal E}(\phi,\overrightarrow{A})(x)dx
\leq \int_{M(R)}{\cal E}(\phi,\overrightarrow{A})(x)dx .
\end{equation}
On pose\index{$n(R)$}\index{$N(R)$}
\begin{equation}\label{cardinaux-de-ensemble-sur-reseau}
\begin{array}{l}
n(R)=\mbox{card}\,g(R),\\
N(R)=\mbox{card}\,G(R).
\end{array}
\end{equation}
Par la relation~(\ref{csq-pour-integrale}), on a l'inégalité
\begin{equation}\label{encadrement-pour-integrale}
n(R)|\Omega|E_n^3(\phi,\overrightarrow{A})
\leq \int_{B(0,R)}{\cal E}(\phi,\overrightarrow{A})(x)dx
\leq N(R)|\Omega|E_n^3(\phi,\overrightarrow{A}) .
\end{equation}
On montre par des arguments géométriques que 
\begin{equation}\label{equivalent-geometrique}
n(R)\sim N(R)\sim \frac{|B(0,R)|}{|\Omega|},
\end{equation}
ce qui signifie que le quotient des différentes quantités tend vers $1$ quand $R$ tend vers $\infty$.\\
On obtient alors gr\^ace à (\ref{equivalent-geometrique}) et (\ref{encadrement-pour-integrale})
\begin{equation}
\lim_{R\rightarrow\infty}\frac{1}{|B(0,R)|}\int_{B(0,R)}{\cal E}(\phi,\overrightarrow{A})(x)dx
= E_n^3(\phi,\overrightarrow{A}) .
\end{equation}
On a bien l'approximation annoncée. $\hfill{\bf CQFD}$

\saction{Cadre fonctionnel pour $E$ et réduction du problème à la dimension $2$}\label{sectio2}\label{sec:reduction2-3}
\noindent On se propose dans cette section de ramener le problème de la minimisation de la fonctionnelle $E_{n}^3$ à un problème bidimensionnel.\\
Rappelons tout d'abord quelques propriétés classiques des espaces de Sobolev que nous utiliserons.
\begin{definition}.\label{espace-W}
Soit $1\leq p\leq +\infty$, $m\in \Z$; on dit qu'une distribution  $f$ sur $\R^n$ appartient à $W^{p}_{m}(\R^n)$\index{$W^{p}_{m}(\R^n)$} si
\begin{equation}
\frac{\partial^{\alpha_1+\alpha_2+\dots+\alpha_n}f}{{\partial x_1^{\alpha_1}}\dots{\partial x_n^{\alpha_n}}}\in L^{p}(\R^n),\,\,\forall \alpha\in\N^n {~tel ~que~} |\alpha|\leq m \mbox{~~.}
\end{equation}
Si $O$ est un ouvert, on définit de la même manière les espaces $W^{p}_{m}(O)$\index{$W^{p}_{m}(O)$} (respectivement $W^{p}_{m, loc}(\R^n)$\index{$W^{p}_{m, loc}(\R^n)$}) en exigeant que les dérivées jusqu'à l'ordre $m$ appartiennent à $L^{p}(O)$ (respectivement $L^{p}_{loc}(\R^n)$).
\end{definition}
Dans la définition, les dérivées sont prises au sens des distributions.

\begin{definition}.(Espaces de Sobolev)\label{espace-de-sobolev}
On note \index{$H^{m}_{loc}(\R^n,\R)$}$H^{m}_{loc}(\R^n,\R)=W^{2}_{m, loc}(\R^n)$.
\end{definition}

\begin{theorem}. (Inclusions de Sobolev)\label{sobolev-imbedding}
Pour $p$ tel que $1<p<+\infty$, on a, avec $q=\frac{np}{n-p}$, les inclusions suivantes\\
\begin{tabular}{rcl}
(a)& Si $n>p$ & alors $W^{p}_{1, loc}(\R^n)\subset L^{q}_{loc}(\R^{n})$.\\
(b)& Si $n=p$ & alors $\forall r$ tel que $n<r<+\infty$ on a $W^{p}_{1, loc}(\R^n)\subset L^{r}_{loc}(\R^{n})$.\\
(c)& Si $n<p$ & alors $W^{p}_{1, loc}(\R^n)\subset C^{0}(\R^{n})$.
\end{tabular}\\
De plus si $1<r<q$ et si $O$ est un ouvert relativement compact de $\R^n$, alors $W^{p}_{1}(O)\subset L^{r}(O)$ et l'injection est compacte.
\end{theorem}
{\it Preuve.} Voir le traité d'Adams \cite{adams} sur les espaces de Sobolev.

\begin{theorem}.\label{long-time-to-get-it}
Si $f\in L^{2}_{loc}(\R^n)$ alors il existe $g\in H^2_{loc}(\R^n)$ tel que $\Delta g=f$.
\end{theorem}
{\it Preuve.} La fonction $f$ peut \^etre vue comme une distribution sur $\R^n$. Le corollaire 10.7.10 du tome II de \cite{Hormander-ALPDE-I} indique que l'on peut résoudre cette équation dans ${\cal D}'$ si $\R^n$ est $\Delta$-convexe pour les supports et les supports singuliers, ce qui est vrai par les théorèmes 10.7.2 et 10.6.2 du traité \cite{Hormander-ALPDE-I}. On a alors une distribution $g$ telle que $\Delta\,g=f$. L'ellipticité de $\Delta$ implique alors que $g\in H^2_{loc}(\R^n)$.  $\hfill{\bf CQFD}$
%p.~50 
%(p.~45)
%(p.~41)

\begin{theorem}.\label{espace-fonctionnel-pour-la-fonctionnelle}
La fonctionnelle~(\ref{integrale-renormalise-de-facon-triviale}) est bien définie si $(\phi, \overrightarrow{A})\in {\cal D}$.
\end{theorem}
{\it Preuve.} Le théorème~\ref{sobolev-imbedding} donne l'estimée
\begin{equation}
(\phi, \overrightarrow{A})\in L^{p}_{loc}(\R^3,\C\times\R^3)\mbox{~si~} p\leq 6 .
\end{equation}
Nous allons montrer que les différents termes de l'intégrande~(\ref{expression-de-la-fonctionnelle}) sont localement intégrables. On développe l'intégrande ci-dessous
\begin{tabbing}\label{termes-de-l-equation}
$2{\cal{E}}(\phi, \overrightarrow{A})$\= =$k^{-2}|i\overrightarrow{\nabla}\phi|^2+2k^{-1}\Rez[\overrightarrow{A}\overline{\phi}.i\overrightarrow{\nabla}\phi]+\overrightarrow{A}^2 |\phi|^2$\\
\> +$\frac{1}{2}(1-|\phi|^2)^2+\Vert \overrightarrow{\rot}\,{\overrightarrow{A}}-\overrightarrow{H}_{ext}\Vert ^2$ .
\end{tabbing}
On veut montrer que ${\cal{E}}$ appartient à $L^1_{loc}$; pour cela il faut montrer que les intégrales $\int_{O}|{\cal{E}}|dx$ sont finies quand on parcourt l'ensemble des ouverts relativement compacts de $\R^3$. On évalue chaque terme en utilisant l'inégalité de Hölder et le théorème~\ref{espace-fonctionnel-pour-la-fonctionnelle}.\\
Soit donc $O$ un ouvert relativement compact de $\R^{3}$; il existe une constante $C$ tel que $\forall (\phi,\overrightarrow{A})\in{\cal D}$
\begin{equation}\label{majorations brutales}
\left\lbrace\begin{array}{ccl}
\int_{O} \overrightarrow{A}^{2}|\phi|^2dx &\leq& \Vert \phi\Vert _{L^{4}(O)}^2 \Vert \overrightarrow{A}\Vert _{L^{4}(O)}^2\\
\int_{O} |\overrightarrow{A}\overline{\phi}.\overrightarrow{\nabla}\phi| dx&\leq& \Vert \overrightarrow{\nabla}\phi\Vert _{L^{2}(O)}\Vert \overrightarrow{A}\Vert _{L^{4}(O)}\Vert \phi\Vert _{L^{4}(O)}\\
\int_{O} \Vert\overrightarrow{\nabla}\phi\Vert^2 dx&=& \Vert \overrightarrow{\nabla}\phi\Vert _{L^{2}(O)}^2\\
\int_{O} (1-|\phi|^2)^2 dx&=&vol(O)+\Vert \phi\Vert _{L^{4}(O)}^4-2\Vert \phi\Vert _{L^{2}(O)}^2\\
\int_{O} \Vert \overrightarrow{\rot}\,{\overrightarrow{A}}-\overrightarrow{H}_{ext}\Vert ^2 dx&\leq& C(\Vert \overrightarrow{A}\Vert _{H^{1}(O)}^2+\Vert H_{ext}\Vert _{L^{2}(O)}^2) .\\
\end{array}\right.
\end{equation}
La constante $C$ qui apparaît à la dernière ligne de l'équation précédente est celle qui permet de contrôler la norme $L^2(O)$ de $\rot\,\overrightarrow{A}$ par la norme $H^1(O)$ de $\overrightarrow{A}$.\\
Tous les termes de ${\cal E}$ sont dans $L^1_{loc}$ donc $E^3_{n}$ est bien défini puisque l'intégrale~(\ref{integrale-renormalise-de-facon-triviale}) porte sur un ouvert relativement compact de $\R^3$. $\hfill{\bf CQFD}$

\begin{definition}.
L'ensemble des couples $(\phi,\overrightarrow{A})$ de ${\cal D}$ indépendant de $x_3$ et vérifiant $A_3=0$ est appelé \index{${\cal D}_{spec}$}${\cal D}_{spec}$.
%appartient à ${\cal D}_{spec}$ si ce couple appartient à ${\cal D}$ et si $(\phi,\overrightarrow{A})$ est indépendant de $x_3$ et vérifie $A_3=0$.
\end{definition}

\begin{theorem}.\label{reduction-dimension2-3}
Soit $(\phi,\overrightarrow{A})\in {\cal D}$ alors il existe $(\phi',\overrightarrow{A}')\in {\cal D}_{spec}$ vérifiant ${\cal E}(\phi',\overrightarrow{A}')\leq {\cal E}(\phi, \overrightarrow{A})$.
\end{theorem}
{\it Preuve.}\\
{\bf $1^{ere}$ étape :}\\
Soit $(\phi,\overrightarrow{A})\in {\cal D}$; la translation selon $x_3$ fait opérer un changement de jauge sur le système
\begin{equation}\label{invariance-en-z-deuxieme-apparition}
\begin{array}{c}
\forall l\in \R\mbox{,~il ~existe~} f_{l}\in H^{2}_{loc}(\R^3,\R)\mbox{~tel~que},\\
{\overrightarrow{A}}(.,x_3+l)={\overrightarrow{A}}(.,x_3)+{\overrightarrow{\nabla}}\,f_{l}(.,x_3) .
\end{array}
\end{equation}
Cela signifie que le champ magnétique $\overrightarrow{B}=\overrightarrow{\rot}\,{\overrightarrow{A}}$ est indépendant de $x_3$. Par ailleurs le champ $\overrightarrow{B}$ appartient à $(L^{2}_{loc}(\R^3))^3$; donc $\overrightarrow{B}(x_1,x_2, x_3)=\overrightarrow{\tilde{B}}(x_1,x_2)$ et $\overrightarrow{\tilde{B}}\in (L^{2}_{loc}(\R^2))^3$.\\
On cherche maintenant un potentiel vecteur $\overrightarrow{A'}$ qui ne dépende pas de $x_3$ et tel que $\overrightarrow{B}=\overrightarrow{\rot}\,{\overrightarrow{A'}}$. Pour cela on résout l'équation $B_{3}=\Delta\,f$, dans $H^{2}_{loc}(\R^2,\R)$ ce qui est possible (cf~\ref{long-time-to-get-it}). On a résolu en partie notre problème puisqu'on a la relation
\begin{equation}
\overrightarrow{\rot}\,
\left(\begin{array}{c}
-\frac{\partial f}{\partial x_2}\\
\frac{\partial f}{\partial x_1}\\
0
\end{array}\right)
=\left(\begin{array}{c}
0\\
0\\
B_{3}
\end{array}\right) .
\end{equation}
Puis on cherche une fonction \index{$A'_3$}$A'_{3}$ définie sur $\R^2$ telle que
\begin{equation}\label{equation-pour-Az}
\left(\begin{array}{c}
B_{1}\\
B_{2}
\end{array}\right)
=\left(\begin{array}{c}
\frac{\partial A'_{3}}{\partial x_2}\\
-\frac{\partial A'_{3}}{\partial x_1}
\end{array}\right) .
\end{equation}
L'équation~(\ref{equation-pour-Az}) est soluble dans $H^{2}_{loc}(\R^2,\R)$ parce que $\frac{\partial B_{1}}{\partial x_1}+\frac{\partial B_{2}}{\partial x_2}=0$.\\
Le potentiel suivant 
\begin{equation}
\overrightarrow{A'}(x_1,x_2,x_3)=\left(\begin{array}{c}
-\frac{\partial f}{\partial x_2}(x_1,x_2)\\
\frac{\partial f}{\partial x_1}(x_1,x_2)\\
A'_{3}(x_1,x_2)
\end{array}\right)
\end{equation}
vérifie l'équation $\overrightarrow{\rot}\,(\overrightarrow{A}-\overrightarrow{A'})=0$.\\
La différence $\overrightarrow{A}-\overrightarrow{A}'$ appartient à $H^{1}_{loc}(\R^3,\R^3)$. Donc $\exists g\in H^{2}_{loc}(\R^3,\R)$ tel que $\overrightarrow{A}-\overrightarrow{A'}=\overrightarrow{\nabla}\,g$. Pour une démonstration de ce résultat, voir \cite{Chabat}, p.~101-103 et le théorème~\ref{long-time-to-get-it}.\\
On utilise alors la fonction $g$ pour faire un changement de jauge sur le couple $(\phi, \overrightarrow{A})$
\begin{equation}\label{jauge-transformation-numero-un}
\left\lbrace\begin{array}{rcl}
\phi&\rightarrow&\phi e^{-ikg}\\
{\overrightarrow{A}}&\rightarrow&{\overrightarrow{A}}-{\overrightarrow{\nabla}}\,g .
\end{array}\right.
\end{equation}
Grâce à cette transformation, le potentiel vecteur devient indépendant de $x_3$.\\
Cela signifie par l'invariance de jauge selon $x_3$ que pour tout $l\in\R$, il existe un complexe de module $1$,  $g(l)$ tel que
\begin{equation}\label{fonction-g}
\phi(.,x_3+l)=\phi(.,x_3) g(l) .
\end{equation}
On souhaite montrer qu'il existe $c\in \C$ vérifiant $|c|=1$ tel que  $\forall l,\,\, g(l)=c^{l}$.\\
Si $\phi=0$ presque partout alors il suffit de prendre $c=1$ pour résoudre le problème.\\
Si $\phi(x_1, x_2, x_3)\not= 0$ alors $\forall l\in \R$, $\phi(x_1, x_2, l)\not= 0$ grâce à l'équation~(\ref{fonction-g}).\\
Si $mes\lbrace (x_1,x_2)\in \R^2\mbox{~tel~que~} \phi(x_1,x_2,x_3)\not= 0 \rbrace >0$ alors il existe un compact $K$ de mesure positive inclus dans l'ensemble précédent \cite{Malliavin}, p.~73. La fonction 
\begin{equation}
g(l)=\int_{K\times [0,1]}\frac{\phi(x_1, x_2, x_3+l)}{\phi(x_1, x_2, x_3)} dx
\end{equation}
est alors mesurable comme intégrale à paramètre de fonctions mesurables (cf~\cite{Malliavin}, p.~39).\\
On vérifie alors que $g(l+l')=g(l)g(l')$ par un calcul simple et en utilisant l'unicité de la fonction $g$.\\
On utilise alors le lemme suivant (cf~\cite{convex-plait-helffer}) dont nous donnons ici une démonstration courte utilisant la théorie des distributions.
\begin{lemme}.\label{mesurabilite-des-morphismes}
Soit $f$ un morphisme du groupe $\R$ dans le groupe $S^{1}$. On suppose qu'il est mesurable comme fonction de $\R$ dans $\C$ alors 
\begin{equation}
\exists a\in \R\mbox{~tel~que~} f(t)=e^{iat} .
\end{equation}
\end{lemme}
{\it Preuve.} La fonction $f$ est bornée et mesurable sur $\R$; donc elle peut être considérée comme une distribution tempérée. On note $\hat{f}$ sa transformée de Fourier. L'équation fonctionnelle pour la fonction $f$ implique alors, après changement de variable:
\begin{equation}\label{fourier-equation}
\begin{array}{rcl}
\hat{f}(k)&=&\int f(l)e^{-ikl}dl=\int f(l+l')e^{-ikl-ikl'}dl\\
&=&e^{-ikl'}f(l')\int f(l)e^{-ikl}dl=e^{-ikl'}f(l')\hat{f}(k),\,\forall l'\in \R\, .
\end{array}
\end{equation}
Cela s'écrit:
\begin{equation}
\forall l'\in\R,\,[1-e^{-ikl'}f(l')]\hat{f}(k)=0 .
\end{equation}
La distribution $\hat{f}$ est non nulle puisque la transformée de Fourier est injective. Cela implique qu'il existe $a\in \R$ tel que
\begin{equation}\label{construction-de-k}
\forall l'\in \R,~~1-e^{-ia l'}f(l')=0 .
\end{equation}
La dernière équation est équivalente à $f(l)=e^{ia l}$ . $\hfill{\bf CQFD}$\\
\\
Par le lemme~\ref{mesurabilite-des-morphismes}, la fonction $g$ est de la forme $g(l)=e^{ial}$ avec $a\in \R$; on fait le changement de jauge suivant:
\begin{equation}\label{changement-de-jauge-numero-deux}
\left\lbrace\begin{array}{rcl}
\phi&\rightarrow&\phi e^{-iax_3}\\
\overrightarrow{A}&\rightarrow&\overrightarrow{A}-\left(\begin{array}{c}
0\\
0\\
\frac{a}{k}
\end{array}\right) .
\end{array}\right.
\end{equation}
La fonction $\phi$ devient indépendante de $x_3$ et le potentiel vecteur reste indépendant de $x_3$.\\
Jusqu'à présent les opérations effectuées sur $(\phi, \overrightarrow{A})$ ont été des changements de jauge; elles n'ont donc pas changé $E^3_{n}(\phi, \overrightarrow{A})$. L'étape suivante ne suit pas ce schéma.\\
\\
{\bf $2^{eme}$ étape :}\\
On peut désormais supposer que $\phi$ et $\overrightarrow{A}$ sont indépendants de $x_3$. On a alors
\begin{equation}\label{champ magnetique}
{\overrightarrow{A}}=\left(\begin{array}{c}
A_{1}(x_1,x_2)\\
A_{2}(x_1,x_2)\\
A_{3}(x_1,x_2)
\end{array}\right) \mbox{~et~}
{\overrightarrow{B}}=\overrightarrow{\rot}\, {\overrightarrow{A}}=\left(\begin{array}{c}
\frac{\partial A_{3}}{\partial x_2}\\
-\frac{\partial A_{3}}{\partial x_1}\\
\frac{\partial A_{2}}{\partial x_1}-\frac{\partial A_{1}}{\partial x_2}
\end{array}\right) .
\end{equation}
On \index{${\overrightarrow{A}}^{plan}$}pose:
\begin{equation}\label{champ magnetique en z seulement}
{\overrightarrow{A}}^{plan}=\left(\begin{array}{c}
A_{1}(x_1,x_2)\\
A_{2}(x_1,x_2)\\
0
\end{array}\right)\mbox{~et~}\index{${\overrightarrow{B}}^{plan}$}
{\overrightarrow{B}}^{plan}=\overrightarrow{\rot}\, {\overrightarrow{A}}^{plan}=\left(\begin{array}{c}
0\\
0\\
\frac{\partial A_{2}}{\partial x_1}-\frac{\partial A_{1}}{\partial x_2}
\end{array}\right) .
\end{equation}
On calcule alors la différence des densités d'énergie entre les deux états ainsi définis:
\begin{equation}\label{majoration-simple-pour-virer-A3}
{\cal{E}}(\phi,\overrightarrow{A})-{\cal{E}}(\phi,\overrightarrow{A}^{plan})=\frac{(\frac{\partial A_{3}}{\partial x_2})^2+(\frac{\partial A_{3}}{\partial x_1})^2}{2}+\frac{1}{2}|A_{3}\phi|^2 \geq 0 .
\end{equation}
Par intégration on obtient l'inégalité $E^3_{n}(\phi,\overrightarrow{A})\geq E^3_{n}(\phi,\overrightarrow{A}^{plan})$; enfin le couple $(\phi,\overrightarrow{A}^{plan})\in {\cal D}_{spec}$.  $\hfill{\bf CQFD}$\\
\\
On définit le nouvel espace fonctionnel ${\cal C}$\index{${\cal C}$} sur $\R^2$
\begin{equation}\label{definition-de-espace-C}
{\cal C}=\left\lbrace\begin{array}{c}(\phi, \overrightarrow{A})\mbox{~tel ~que~} (\phi, \overrightarrow{A})\in H^{1}_{loc}(\R^2,\C)\times H^{1}_{loc}(\R^2,\R^2)\\
(\phi, \overrightarrow{A}) \mbox{~est ~périodique ~de ~jauge ~suivant ~le ~réseau ${\cal L}$.}
\end{array}
\right\rbrace
\end{equation}
Si $(\phi, \overrightarrow{A})$ appartient à ${\cal C}$, alors on peut construire un élément $(\phi', \overrightarrow{A}')$ de ${\cal D}_{spec}$ par la formule $(\phi', \overrightarrow{A}')(x_1,x_2,x_3)=(\phi, \overrightarrow{A})(x_1,x_2)$. Cela définit une application bijective de ${\cal C}$ dans ${\cal D}_{spec}$.\\
Maintenant tous les champs magnétiques utilisés seront de la forme $B\overrightarrow{e_{3}}$ tandis que les potentiels vecteurs vérifieront $\overrightarrow{A}.\overrightarrow{e_3}=0$. C'est pourquoi on note \index{$\rot$}maintenant
\begin{equation}
\rot\,\left(\begin{array}{c}
A_{1}\\
A_{2}
\end{array}\right)=\frac{\partial A_{2}}{\partial x_1}-\frac{\partial A_{1}}{\partial x_2} .
\end{equation}
\begin{definition}.\label{champ-C}
On introduit le potentiel vecteur $\overrightarrow{C}=\frac{1}{2}\left(\begin{array}{c}
-y\\
x
\end{array}\right)$. Ce potentiel \index{$\overrightarrow{C}$} vérifie $\divergence\,\overrightarrow{C}=0$ et $\rot\,\overrightarrow{C}=1$.
\end{definition}
On construit une fonctionnelle \index{$E^2_n$} sur ${\cal C}$
\begin{equation}\label{fonctionnelle-En2}
\begin{array}{rl}
E^2_{n}(\phi,\overrightarrow{A})&=\frac{1}{|\Omega|}\int_{\Omega}\frac{1}{2}\Vert i k^{-1}\overrightarrow{\nabla}\phi+\overrightarrow{A}\phi\Vert ^2+\frac{1}{4}(1-|\phi|^2)^2\\
&+\frac{1}{|\Omega|}\int_{\Omega}\frac{1}{2}(\rot\,\overrightarrow{A}-H_{ext})^2 .
\end{array}
\end{equation}
On rappelle que $\Omega$ est défini à l'équation~(\ref{definition-domaine}).\\
Si $(\phi', \overrightarrow{A}')$ appartient à ${\cal D}_{spec}$ alors le couple $(\phi,\overrightarrow{A})$ de ${\cal C}$ qui lui correspond vérifie $E^2_{n}(\phi,\overrightarrow{A})=E^3_{n}(\phi', \overrightarrow{A}')$.\\
Le théorème~\ref{espace-fonctionnel-pour-la-fonctionnelle} nous dit que $E^2_{n}$ est bien défini si le couple $(\phi, \overrightarrow{A})$ appartient à  ${\cal C}$. Le théorème suivant est évident d'après les calculs précédents et le théorème~\ref{reduction-dimension2-3}.
\begin{theorem}.
On a l'égalité suivante
\begin{equation}
\inf_{(\phi, \overrightarrow{A})\in {\cal C}}E^2_{n}(\phi, \overrightarrow{A})=\inf_{(\phi, \overrightarrow{A})\in {\cal D}}E^3_{n}(\phi, \overrightarrow{A}) .
\end{equation}
\end{theorem}
Cela implique que du point de vue énergétique la restriction de la fonctionnelle à l'espace ${\cal C}$ ne fait perdre aucune information sur la valeur du minimum de la fonctionnelle.\\
Notre problème est donc maintenant un problème en dimension $2$.

\saction{Réduction à un problème sur le tore}\label{sec:reduction-tore}
\noindent Dans cette section, on effectue un changement de jauge de telle sorte que les fonctions $g^{t}$ de l'équation~(\ref{jauge-equivalence-sur-L}) prennent une forme standard.\\
Conformément à l'article \cite{Barany-Golubitsky}, on pose \index{$x$}\index{$y$}\index{$z$}par la suite
\begin{equation}
\left\lbrace\begin{array}{c}
x=x_1\\
y=x_2\\
z=x_1+ix_2 .
\end{array}\right.
\end{equation}
Si $f$ est une fonction $C^{\infty}$ non constante sur le tore $\Tore{\cal L}$ alors la suite de couples $(\phi_{n}, \overrightarrow{A}_{n})=(\phi e^{in k f}, \overrightarrow{A}+n\overrightarrow{\nabla}\, f)$ n'admet aucune sous suite convergente et est d'énergie constante. La fonctionnelle $E_{n}^2$ n'est donc pas coercive sur ${\cal C}$. Cela signifie que le contr\^ole sur $E_n^2$ ne se transforme pas en contrôle sur les normes.\\
On cherche donc à lever cette dégénérescence en quotientant par  le groupe d'équivalence de jauge. Plus précisément, dans le théorème suivant, on cherche pour chaque classe un représentant naturel.\\
Le couple $(\phi, \overrightarrow{A})$ appartient à ${\cal C}$ et par conséquent \index{$g^t$}
\begin{equation}\label{jauge-equivalence-sur-L-dim2}
\left\lbrace\begin{array}{l}
\forall t\in {\cal{L}}\mbox{~il~existe~} g^{t}\in H^{2}_{loc}(\R^2,\R)\mbox{~tel ~que}\\
\phi(z+t)=e^{ikg^{t}(z)}\phi(z)\mbox{~et~}\\
\overrightarrow{A}(z+t)=\overrightarrow{A}(z)+\overrightarrow{\nabla}\,g^{t}(z).
\end{array}\right.
\end{equation}
On définit alors la fonction $g_{s,t}$\index{$g_{s,t}$} par 
\begin{equation}\label{definitions-de-gst}
g_{s,t}(z)=g^{t}(z+s)+g^{s}(z),
\end{equation}
et la fonction $K_{s,t}$\index{$K_{s,t}$} par
\begin{equation}\label{definitions-de-Kst}
K_{s,t}(z)=kg_{s,t}(z)-kg_{t,s}(z) .
\end{equation}

\begin{theorem}.
Pour tout $(\phi,\overrightarrow{A})\in{\cal C}$ et tout $(s,t)\in{\cal L}^2$ la fonction $K_{s,t}$ définie par~(\ref{jauge-equivalence-sur-L-dim2}), (\ref{definitions-de-gst}) et~(\ref{definitions-de-Kst}) est constante. Elle ne dépend pas de la fonction $g^t$ apparaissant dans~(\ref{jauge-equivalence-sur-L-dim2}).
%La fonction $K_{s,t}$ dépend seulement du couple $(\phi, \overrightarrow{A})$ et ne dépend pas de $z$. Elle est bilinéaire antisymétrique selon $s$ et $t$. Elle est donc uniquement définie par $K_{v_{1},v_{2}}$.\\
De plus si la fonction $\phi$ est non identiquement nulle alors $K_{v_{1},v_{2}}\in 2\pi\Z$ avec $v_i$ défini en (\ref{definition-vecteur-du-reseau}).
\end{theorem}
{\it Preuve.} En décomposant le potentiel $\overrightarrow{A}(z+t+s)$ de deux façons différentes, on obtient les équations suivantes
\begin{equation}\label{decomposition-cohomologie}
\begin{array}{rcl}
\overrightarrow{A}(z+t+s)&=&\overrightarrow{A}(z+s)+\overrightarrow{\nabla}\,g^{t}(z+s)\\
&=&\overrightarrow{A}(z)+\overrightarrow{\nabla}\,(g^{t}(z+s)+g^{s}(z))\\
&=&\overrightarrow{A}(z)+\overrightarrow{\nabla}\,g^{t+s}(z) .
\end{array}
\end{equation}
On obtient donc l'égalité
\begin{equation}
\overrightarrow{\nabla}\,(g^{t+s}(z)-g^{t}(z+s)-g^{s}(z))=0, 
\end{equation}
ce qui donne par intégration
\begin{equation}
g^{t+s}(z)-g^{t}(z+s)-g^{s}(z)=Cste .
\end{equation}
On appelle la constante d'intégration \index{$C_{s,t}$}$\frac{2\pi}{k}C_{s,t}$. La fonction $K_{s,t}$ est donc constante sur $\C$ puisque
\begin{equation}
K_{s,t}(z)=2\pi(C_{t,s}-C_{s,t}) .
\end{equation}
Le calcul suivant prouve que la fonction $K_{s,t}$ ne dépend que du couple $(\phi, \overrightarrow{A})$. On a:
\begin{eqnarray}
K_{s,t}(0) &=& kg_{s,t}(0)-kg_{t,s}(0) \nonumber \\
&=& [kg^{t}(s)+kg^{s}(0)]-[kg^{s}(t)+kg^{t}(0)]\nonumber \\
&=& [kg^{t}(s)-kg^{t}(0)]+[kg^{s}(0)-kg^{s}(t)] .
\end{eqnarray}
En effet les expressions $kg^{t}(s)-kg^{t}(0)$ ne changent pas, si on ajoute une constante à $g^{t}$.\\
On prouve maintenant que la fonction $K_{s,t}$ est bilinéaire sur le réseau ${\cal{L}}$ par le calcul suivant:
\begin{eqnarray}
K_{r+s,t}(0) &=& kg_{r+s,t}(0)-kg_{t,r+s}(0) \nonumber \\
&=& [kg^{t}(r+s)+kg^{r+s}(0)]-[kg^{r+s}(t)+kg^{t}(0)]\nonumber \\
&=& kg^{t}(r+s)+[kg^{s}(r)+kg^{r}(0)+\frac{2\pi}{k}C_{s,t}]\nonumber\\
&&-[kg^{s}(t+r)+kg^{r}(t)+\frac{2\pi}{k}C_{s,t}]-kg^{t}(0)\nonumber\\
&=& [kg^{t}(r+s)+kg^{s}(r)-kg^{t}(r)-kg^{s}(t+r)]\nonumber\\
&& +[kg^{t}(r)+kg^{r}(0)-kg^{r}(t)-kg^{t}(0)]\nonumber\\[2mm]
&=& K_{s,t}(r)+K_{r,t}(0) .
\end{eqnarray}
Le résultat est obtenu puisque la fonction $K_{s,t}$ est indépendante de $z$. L'antisymétrie de $K_{s,t}$ est évidente.\\
Le calcul~(\ref{decomposition-cohomologie}) pour $\overrightarrow{A}$ possède un analogue pour $\phi$ qui implique l'équation suivante. Il suffit de décomposer $\phi(z+s+t)$ et d'utiliser~(\ref{jauge-equivalence-sur-L-dim2}):
\begin{equation}\label{vrai-apparition}
\phi(z)(1-e^{i2\pi C_{s,t}})=0 .
\end{equation}
Si la fonction $\phi$ est non identiquement nulle alors l'égalité~(\ref{vrai-apparition}) implique que $C_{s,t}\in \Z$ pour tout $s,t\in {\cal L}^2$ et donc $K_{s,t}(0) \in 2\pi\Z$.\\
L'espace des formes bilinéaires antisymétriques sur ${\cal L}^2 \simeq \Z^2$ est de dimension~$1$. Cela implique l'existence d'un entier relatif $d$\index{$d$} tel que
\begin{equation}
\begin{array}{l}
%\forall v=n_{1}v_{1}+n_{2}v_{2}, v'=n'_{1}v_{1}+n'_{2}v_{2}\in {\cal{L}}\\
K_{n_{1}v_{1}+n_{2}v_{2},n'_{1}v_{1}+n'_{2}v_{2}}=2\pi d(n_{2}n'_{1}-n_{1}n'_{2}) .
\end{array}
\end{equation}
Le théorème est démontré puisque $2\pi d=-K_{v_{1},v_{2}}$. $\hfill{\bf CQFD}$

\begin{theorem}.\label{reduction-de-jauge}
Soit $(\phi, \overrightarrow{A})$ un élément de l'espace ${\cal C}$ (cf~(\ref{definition-de-espace-C})). Alors le couple $(\phi, \overrightarrow{A})$ est équivalent à un couple $(\psi, H_{int}\overrightarrow{C}+\overrightarrow{P})$ après un changement de jauge et une translation selon les variables $x$ et $y$.\\
Le couple $(\psi, H_{int}\overrightarrow{C}+\overrightarrow{P})$ vérifie les conditions:\\
\begin{tabular}{rl}
(a)&$\psi(z+t)=e^{i\frac{kH_{int}}{2}(t_{x}y-t_{y}x)}\psi(z)$,\\
(b)&$\overrightarrow{P}$ est ${\cal{L}}$-périodique,\\
(c)&$\divergence\, \overrightarrow{P}=0$.
\end{tabular}\\
Le potentiel $\overrightarrow{C}$ est défini à la définition \ref{champ-C} et le champ $H_{int}$ s'exprime à partir du couple $(\phi, \overrightarrow{A})$ comme
\begin{equation}
H_{int}|\Omega|=\int_{\Omega}\rot\,\overrightarrow{A} . 
\end{equation}
Si $H_{int}\not= 0$ alors on peut aussi imposer (d) $\int_{\Omega}\overrightarrow{P}=0$ .
\end{theorem}

\begin{remarque}.
Le théorème est énoncé dans le cadre $C^{\infty}$ dans \cite{Barany-Golubitsky}. Nous montrons ici une version $H^1$. La comparaison des techniques utilisées sera faite plus loin (Remarque~\ref{difference-deux-methode}).\\
Il n'y a pas a priori d'unicité pour le couple $(\psi, H_{int}\overrightarrow{C}+\overrightarrow{P})$.
\end{remarque}
{\it Preuve.}\\
{\bf $1^{ere}$ étape :}\\
Supposons que le résultat soit vrai, nous allons alors trouver une relation entre $H_{int}$, $d$ et $K_{v_1,v_2}$.\\
Le couple $(\phi, \overrightarrow{A})$ est équivalent de jauge au couple $(\psi, H_{int}\overrightarrow{C}+\overrightarrow{P})$ où $\overrightarrow{P}$ est ${\cal{L}}$-périodique. Cela signifie qu'il existe une fonction \index{$\eta(z)$}$\eta\in H^{2}_{loc}(\R^2,\R)$ tel que 
\begin{equation}\label{equation-de-changement-de-varaible}
\overrightarrow{A}(z)=\frac{H_{int}}{2}\left(\begin{array}{c}
-y\\
x
\end{array}\right)+\overrightarrow{P}(z)+\overrightarrow{\nabla}\,\eta(z) .
\end{equation}
On utilise alors les fonctions $g^{t}$ introduites auparavant
\begin{equation}\label{premiere-equation-pour-eta}
\begin{array}{ccl}
\overrightarrow{\nabla}\,g^{t}(z)&=&\overrightarrow{A}(z+t)-\overrightarrow{A}(z)\\
&=&\frac{\D H_{int}}{2}{\left(\begin{array}{c}
-t_{y}\\
t_{x}
\end{array}\right)}+{\overrightarrow{\nabla}}\,[\eta(z+t)-\eta(z)]
\end{array}
\end{equation}
On a utilisé l'expression~(\ref{equation-de-changement-de-varaible}). On intègre l'équation~(\ref{premiere-equation-pour-eta})
\begin{equation}\label{equation-de-reseau}
\left\lbrace\begin{array}{l}
g^{v_{1}}(z)=\frac{\D H_{int}}{2}(\xi_{1} y-\eta_{1} x)+\eta(z+v_{1})-\eta(z)+K_{v_{1}}\\
g^{v_{2}}(z)=\frac{\D H_{int}}{2}(\xi_{2} y-\eta_{2} x)+\eta(z+v_{2})-\eta(z)+K_{v_{2}} .
\end{array}\right.
\end{equation}
On rappelle (cf~(\ref{definition-vecteur-du-reseau})) que \index{$v_j$}$v_j=(\xi_j,\eta_j)\in\R^2$ (j=1,2).\\
Les réels \index{$K_{v_{1}}$}$K_{v_{1}}$ et \index{$K_{v_{2}}$}$K_{v_{2}}$ sont des constantes d'intégration.\\
Ensuite en faisant des combinaisons linéaires de ces expressions, on obtient l'équation suivante
\begin{eqnarray}\label{equation-de-flux}
2\pi d &=&-K_{v_{1},v_{2}}\nonumber\\
&=& kg_{v_{2},v_{1}}(0)-kg_{v_{1},v_{2}}(0)\nonumber\\
&=& [kg^{v_{1}}(v_{2})+kg^{v_{2}}(0)]-[kg^{v_{2}}(v_{1})+kg^{v_{1}}(0)]\nonumber\\
&=& [kg^{v_{1}}(v_{2})-kg^{v_{1}}(0)]-[kg^{v_{2}}(v_{1})-kg^{v_{2}}(0)]\nonumber\\
&=& kH_{int}|\Omega| .
\end{eqnarray}
Ce dernier calcul nous donne l'expression de $H_{int}$ en fonction de $d$, lien qui n'avait rien d'évident a priori. \\
On a donc obtenu:
\begin{equation}\label{relation-s-Hint}
H_{int}=\frac{2\pi d}{k|\Omega|} .
\end{equation}
La fonction $\overrightarrow{P}$ étant ${\cal L}$-périodique on a  $\int_{\Omega} \rot\,\overrightarrow{P}=0$. En intégrant le champ magnétique on obtient
\begin{equation}
\int_{\Omega}\rot\,\overrightarrow{A}=\int_{\Omega}H_{int}+\rot\,\overrightarrow{P}=H_{int}|\Omega|\, .
\end{equation}
Le flux magnétique est donc quantifié, $\int_{\Omega}\rot\,\overrightarrow{A}\in\frac{2\pi}{k}\Z$.
Pour la suite des calculs, on pose\index{$\overrightarrow{A}^{p}$}
\begin{equation}\label{petit-changement-de-variable}
\begin{array}{l}
\overrightarrow{A}^{p}(z)=
\overrightarrow{A}(z)-
\frac{\D H_{int}}{2}\left(\begin{array}{c}
-y\\
x
\end{array}\right),\\
\left\lbrace\begin{array}{l}
g_{p}^{v_{1}}(z)=g^{v_{1}}(z)-\frac{\D H_{int}}{2}(\xi_{1}y-\eta_{1}x),\\
g_{p}^{v_{2}}(z)=g^{v_{2}}(z)-\frac{\D H_{int}}{2}(\xi_{2}y-\eta_{2}x).
\end{array}\right.
\end{array}
\end{equation}
Ces nouvelles fonctions vérifient une équation identique à l'équation~(\ref{premiere-equation-pour-eta})
\begin{equation}\label{deuxieme-equation-pour-eta}
\overrightarrow{\nabla}\,g_{p}^{t}(z)=\overrightarrow{A}^{p}(z+t)-\overrightarrow{A}^{p}(z)\mbox{~avec~} t\in{\cal L} .
\end{equation}
On a soustrait le potentiel $H_{int}\overrightarrow{C}$ à $\overrightarrow{A}$ et donc 
\begin{equation}
g_{p}^{v_{1}}(z)+g_{p}^{v_{2}}(z+v_{1})-g_{p}^{v_{1}}(z+v_{2})-g_{p}^{v_{2}}(z)=0 .
\end{equation}
{\bf $2^{eme}$ étape :}\\
On souhaite maintenant rendre le potentiel vecteur $\overrightarrow{A}^{p}$ périodique par un changement de jauge. Supposons que l'on sache résoudre les équations suivantes où $\eta$ est l'inconnue
\begin{equation}\label{equations-a-resoudre}
\begin{array}{l}
\left\lbrace\begin{array}{l}
g_{p}^{v_{1}}(z)=\eta(z+v_{1})-\eta(z)\\
g_{p}^{v_{2}}(z)=\eta(z+v_{2})-\eta(z)
\end{array}\right.\\
\mbox{avec~}g_{p}^{t}\in H^{2}_{loc}(\R^2,\R)\mbox{~et~} \eta\in H^{2}_{loc}(\R^2,\R).
\end{array}
\end{equation}
Si $\eta$ est solution de~(\ref{equations-a-resoudre}) alors le potentiel $\overrightarrow{A}^{p}-\overrightarrow{\nabla}\,\eta(z)$ est périodique et il ne restera plus à faire que des opérations sur le domaine $\Omega$.\\
Les équations~(\ref{equations-a-resoudre}) n'ont pas une solution unique, on peut en effet rajouter à $\eta$ une fonction ${\cal L}$-périodique pour obtenir une autre solution de~(\ref{equations-a-resoudre}). La première idée qui vient à l'esprit pour résoudre ces équations est la suivante:\\
{\it Posons $\eta=0$ sur un domaine fondamental de ${\cal L}$ alors les équations~(\ref{equations-a-resoudre}) sont uniquement solubles puisque $g_{p}^{v_{1}}(z)+g_{p}^{v_{2}}(z+v_{1})-g_{p}^{v_{1}}(z+v_{2})-g_{p}^{v_{2}}(z)=0$.\\
Le problème est que cette fonction $\eta$ n'appartient pas a priori à $H^2_{loc}(\R^2,\R)$, ce qui est pourtant indispensable à cause des problèmes qui apparaissent au bord de $\Omega$.\\
}
Il faut donc trouver un procédé permettant de résoudre dans $H^{2}_{loc}(\R^2,\R)$.\\
Examinons d'abord comment on peut résoudre en dimension~$1$ le problème.
\begin{proposition}.\label{exemple-resolu}
Soit $m\in \N$, $1\leq p \leq \infty$ et $f\in W^{p}_{m,loc}(\R)$ alors  $\exists g\in W^{p}_{m,loc}(\R)$ tel que $f(x)=g(x+1)-g(x)$
\end{proposition}
{\it Preuve.} On construit deux fonctions $C^{\infty}$ vérifiant \mbox{$1=h_{1}+h_{2}$},  $h_{1}$ et $h_{2}$ \mbox{$1$-périodiques}. On suppose de plus que la fonction $h_{1}$ a un support compact inclus dans $]0,1[$ et ses translatés et que la fonction $h_{2}$ a un support compact inclus dans $]1/2, 3/2[$ et ses translatés. Le dessin ci-dessous \index{$h_i$}montre plus clairement le comportement de ces fonctions\\
\resizebox{12cm}{!}{\includegraphics{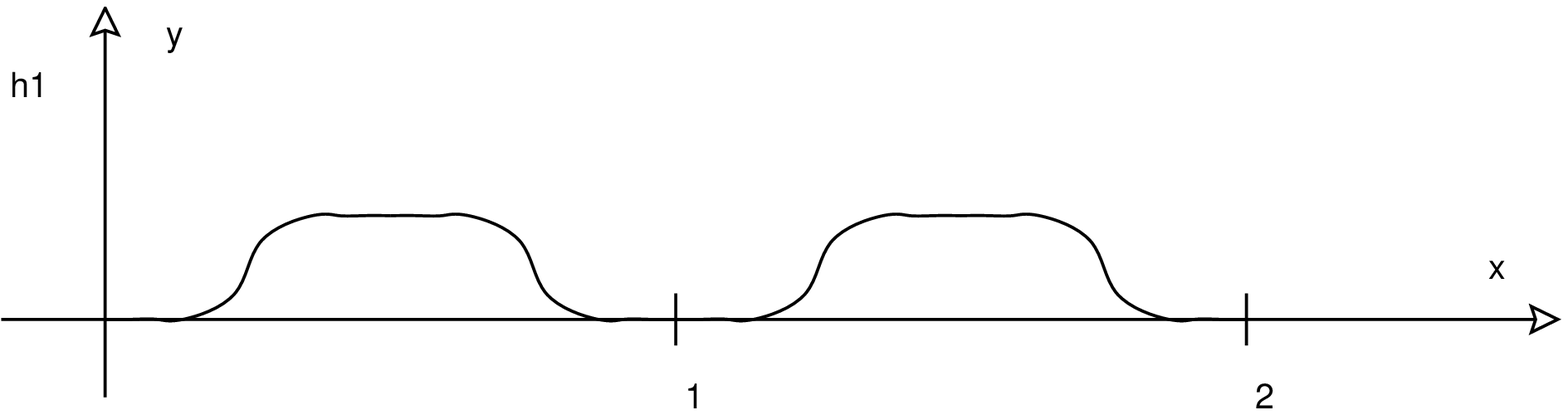}}\\
\resizebox{12cm}{!}{\includegraphics{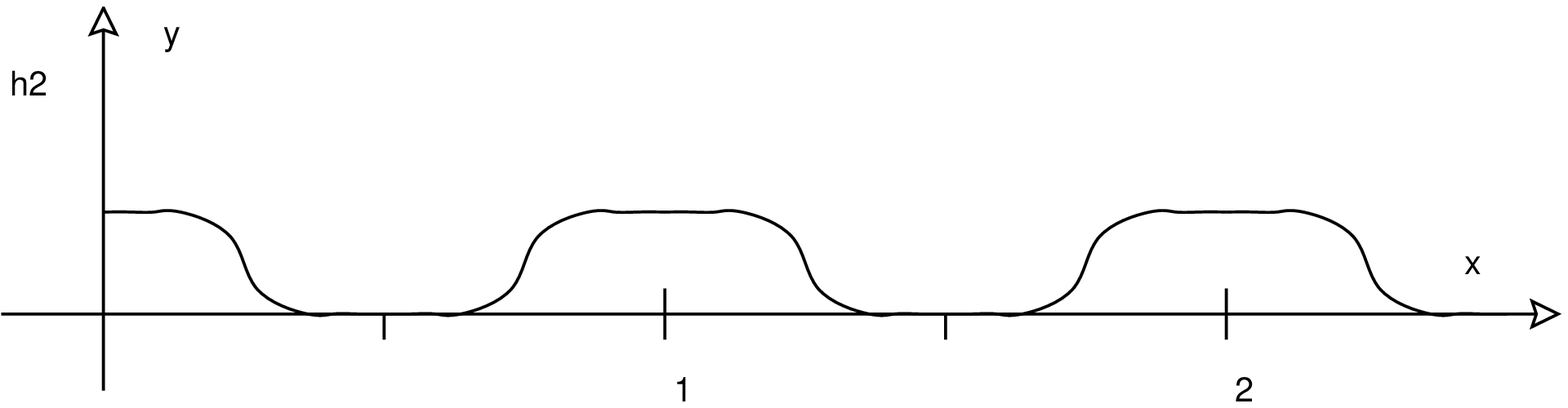}}.\\
On considère les fonctions $f_{1}=fh_{1}$, $f_{2}=fh_{2}$ et on résout le système d'équations
\begin{equation}\label{systeme-pour-g}
\left\lbrace\begin{array}{l}
f_{1}(x)=g_{1}(x+1)-g_{1}(x),\, g_{1}=0\mbox{~sur~} [0,1],\\
f_{2}(x)=g_{2}(x+1)-g_{2}(x),\, g_{2}=0\mbox{~sur~} [1/2,3/2].
\end{array}\right.
\end{equation}
Il est possible de résoudre le système~(\ref{systeme-pour-g}) en obtenant la m\^eme régularité pour $g_i$ que $f_i$.\\
La solution du problème est alors $g_{1}+g_{2}$. $\hfill{\bf CQFD}$\\
\\
Revenons à notre cas en dimension $2$.\\
On construit une \index{partition $C^{\infty}$ de l'unité}partition $C^{\infty}$ de l'unité $\phi_{ij}$ ($i,j=1,2$) comme suit: on écrit $z$ sous la forme $t_1v_1+t_2v_2$ avec $t_i\in\R$\index{$t_i$} et on \index{$\phi_{ij}$}définit
\begin{equation}
\phi_{ij}(z)=h_{i}(t_1)h_{j}(t_2),
\end{equation}
avec $h_1$ et $h_2$ définis dans les dessins précédents.\\
Montrons que le problème~(\ref{equations-a-resoudre}) est soluble pour les fonctions suivantes
\begin{equation}\label{fonction-cut-off}
\left\lbrace\begin{array}{l}
g_{ij}^{v_{1}}(z)=\phi_{ij}(z)g_{p}^{v_{1}}(z),\\
g_{ij}^{v_{2}}(z)=\phi_{ij}(z)g_{p}^{v_{2}}(z).
\end{array}\right.
\end{equation}
Les fonctions $\phi_{ij}$ étant périodiques selon le réseau ${\cal L}$, les fonctions $g_{ij}$ vérifient l'équation algébrique suivante
\begin{equation}\label{equation-sans-bord}
g_{ij}^{v_{1}}(z)+g_{ij}^{v_{2}}(z+v_{1})-g_{ij}^{v_{1}}(z+v_{2})-g_{ij}^{v_{2}}(z)=0 .
\end{equation}
On peut résoudre les équations~(\ref{equation-sans-bord}) dans la régularité $H^2_{loc}$ puisque les fonctions $g_{ij}$ sont nulles au voisinage du bord de $\Omega$ et sont de régularité $H^2_{loc}$. On appelle \index{$\eta_{ij}$}$\eta_{ij}$ une solution de classe $H^2$ des équations~(\ref{equation-sans-bord}). Alors la fonction $\eta=\eta_{11}+\eta_{12}+\eta_{21}+\eta_{22}$ est solution du problème~(\ref{equations-a-resoudre}). On a prouvé le résultat souhaité.\\
{\bf $3^{eme}$ étape :}\\
On a donc réussi à rendre le potentiel vecteur périodique selon ${\cal L}$. Montrons que l'on peut rendre sa divergence nulle par un changement de jauge.\\
Le potentiel $\overrightarrow{A}^p-\overrightarrow{\nabla}\eta$ est périodique par rapport à ${\cal L}$, par conséquent sa divergence est périodique par rapport à ${\cal L}$ et son coefficient de Fourier d'indice~$(0,0)$ est nul. La fonction $\divergence\, [\overrightarrow{A}^p-\overrightarrow{\nabla}\eta]$ est donc orthogonale au noyau du laplacien sur le tore. On peut donc par série de Fourier trouver un unique $\rho$ dans $H^{2}(\Tore{\cal L},\R)$ d'intégrale nulle tel que $\Delta\,\rho=\divergence\,[\overrightarrow{A}^p-\overrightarrow{\nabla}\eta]$. Nous reviendrons sur l'étude du laplacien dans la section~\ref{sec:operateur-L} .\\
Après le changement de jauge
\begin{equation}
\left\lbrace\begin{array}{ccl}
\phi&\rightarrow&\phi e^{-ik(\rho+\eta)}, \\
\overrightarrow{A}&\rightarrow&\overrightarrow{A}-{\overrightarrow{\nabla}}\,(\rho+\eta)
\end{array}\right.
\end{equation}
on obtient un nouveau potentiel vecteur de divergence nulle et $\overrightarrow{A}$ est de la forme $H_{int}\overrightarrow{C}+\overrightarrow{P}$ avec $\overrightarrow{C}=\frac{1}{2}\left(\begin{array}{c}
-y\\
x
\end{array}\right)$ et $\overrightarrow{P}$ périodique.\\
%Par ces différents changements de jauge on s'est ramené à la situation où $\overrightarrow{A}=H_{int}\overrightarrow{C}+\overrightarrow{P}$.
\begin{remarque}.\label{difference-deux-methode}
Dans cette réduction du potentiel vecteur, on a fait deux opérations: une réduction topologique en montrant que l'on peut se ramener à un potentiel périodique plus un champ magnétique constant quantifié et une réduction analytique qui consiste à prendre la jauge de Coulomb ($\divergence\, \overrightarrow{A}=0$).\\
La  méthode de l'article \cite{Barany-Golubitsky} consistait à faire d'abord une réduction analytique, puis une réduction topologique. Leur méthode (ou plutôt l'appendice de \cite{Barany-Golubitsky}) met en évidence des problèmes à coins qui n'apparaissent pas ici.
\end{remarque}
{\bf $4^{eme}$ étape :}\\
L'invariance de jauge par translation s'écrit
\begin{equation}\label{changement}
\left\lbrace\begin{array}{rcl}
\phi(z+v_{i})&=&e^{ikg^{v_{i}}(z)}\,\phi(z)\\
(H_{int}\overrightarrow{C}+\overrightarrow{P})(z+v_{i})&=&(H_{int}\overrightarrow{C}+\overrightarrow{P})(z)+{\overrightarrow{\nabla}}\,g^{v_{i}}(z) \mbox{~pour~}i=1,2 .
\end{array}\right.
\end{equation}
On obtient donc en simplifiant la deuxième équation de~(\ref{changement})
\begin{equation}\label{variation-de-g}
\frac{H_{int}}{2}{\left(\begin{array}{c}
-\eta_{i}\\
\xi_{i}
\end{array}\right)}={\overrightarrow{\nabla}}\, g^{v_{i}}(z)\mbox{~avec~}i\in\lbrace 1, 2 \rbrace .
\end{equation}
En intégrant l'équation~(\ref{variation-de-g}), on obtient
\begin{equation}\label{expression-de-g}
g^{v_{i}}(z)=\beta_i+\frac{H_{int}}{2}(\xi_{i}y-\eta_{i}x)\mbox{~avec~} i\in\lbrace 1, 2 \rbrace .
\end{equation}
Maintenant on souhaite éliminer les termes $\beta_{i}$ tout en gardant la moyenne du potentiel vecteur $\overrightarrow{P}$ égale à $0$ sur $\Omega$. Pour cela on applique une double transformation: d'abord une translation du couple $(\phi,\overrightarrow{P})$ puis un changement de jauge linéaire (Si on fait d'abord le changement linéaire puis la translation on obtient un système non linéaire).\\
Soit $u=u_x+iu_y$ le nombre complexe selon lequel on translate le couple $(\phi, \overrightarrow{P})$ et soit $h(x,y)=h_x x+h_y y$ la fonction de changement de jauge.\\
L'application de ces deux transformations donne alors le nouveau couple $(\phi',\overrightarrow{P}')$
\begin{equation}\label{dernier-changement-de-jauge}
\left\lbrace\begin{array}{rcl}
\overrightarrow{P'}(z)&=&\overrightarrow{P}(z+u)+\left(\begin{array}{c}
-\frac{H_{int}}{2}u_y+h_x\\
\frac{H_{int}}{2}u_x+h_y
\end{array}\right)\\
\phi'(z)&=&\phi(z+u)e^{ik[h_x x+h_y y]} .
\end{array}\right.
\end{equation}
La fonction $\phi'$ se transforme sur le réseau ${\cal L}$ par:
\begin{equation}
\begin{array}{rcl}
\phi'(z+v_{i})&=&e^{ik[h_x(x+\xi_{i})+h_y(y+\eta_{i})]}\phi(z+v_{i}+u)\\
&=&e^{ik[h_x(x+\xi_{i})+h_y(y+\eta_{i})+\beta_{i}+\frac{H_{int}}{2}(\xi_{i}(y+u_y)-\eta_{i}(x+u_x))]}\phi(z+u)\\
&=&e^{ik[h_x \xi_{i}+h_y \eta_{i}+\beta_{i}+\frac{H_{int}}{2}(\xi_{i}(y+u_y)-\eta_{i}(x+u_x))]}\phi'(z)\\
&=&e^{ik[\frac{H_{int}}{2}(\xi_{i}y-\eta_{i}x)]}e^{ik[h_x \xi_{i}+h_y \eta_{i}+\beta_{i}+\frac{H_{int}}{2}(\xi_{i}u_y-\eta_{i}u_x)]}\phi'(z) .
\end{array}
\end{equation}
La moyenne de la partie périodique du potentiel vecteur sur $\Omega$ est égale à
\begin{equation}\label{moyenne-potentiel-vecteur}
\frac{1}{|\Omega|}\int_{\Omega}\overrightarrow{P}+\left(\begin{array}{c}
-\frac{H_{int}}{2}u_y+h_x\\
\frac{H_{int}}{2}u_x+h_y
\end{array}\right) .
\end{equation}
La fonction $\phi'$ doit se comporter comme énoncé dans le théorème~\ref{reduction-de-jauge}, cela implique modulo $\frac{2\pi}{k}$
\begin{equation}\label{le-systeme-final-the-final-battle}
\left\lbrace\begin{array}{l}
h_x \xi_{1}+h_y \eta_{1}=-\beta_{1}-\frac{H_{int}}{2}(\xi_{1}u_y-\eta_{1}u_x)\\
h_x \xi_{2}+h_y \eta_{2}=-\beta_{2}-\frac{H_{int}}{2}(\xi_{2}u_y-\eta_{2}u_x) .
\end{array}\right. 
\end{equation}
Si on pose $(u_x,u_y)=(0,0)$, le système (\ref{le-systeme-final-the-final-battle}) devient une équation d'inconnues $(h_x,h_y)$.\\
Ce système possède une solution unique puisque $v_{1}$ et $v_{2}$ forment une base de $\R^2$. La solution $\phi'$ pour ces choix vérifie la relation voulue:
\begin{equation}\label{equation-de-changement-de-phi}
\phi'(z+v_{i})=e^{ik\frac{H_{int}}{2}(\xi_{i}y-\eta_{i}x)}\phi'(z)~.
\end{equation}
Les autres relations exigées sont conservées; on a prouvé le résultat souhaité.\\
Si $H_{int}\not=0$ on exige que la partie périodique du potentiel vecteur ait une moyenne nulle; cela nous donne 
\begin{equation}
\left\lbrace\begin{array}{l}
\frac{H_{int}}{2}u_y=h_x+C_x\\
\frac{H_{int}}{2}u_x=-h_y+C_y
\end{array}\right. 
\end{equation}
o\`u $C_x$ et $C_y$ sont des réels positifs. Le système à résoudre est alors
\begin{equation}\label{le-systeme-final-the-final-battle-second-stage}
\left\lbrace\begin{array}{l}
2h_x \xi_{1}+2h_y \eta_{1}=-\beta_{1}-\xi_{1}C_x+\eta_{1}C_y\\
2h_x \xi_{2}+2h_y \eta_{2}=-\beta_{2}-\xi_{2}C_x+\eta_{2}C_y .
\end{array}\right. 
\end{equation}
Ce système possède une solution unique puisque $2v_{1}$ et $2v_{2}$ forment une base de $\R^2$. $\hfill{\bf CQFD}$\\
\\
Maintenant il nous faut ramener le problème fonctionnel au tore $\Tore{\cal L}$; pour le potentiel vecteur, cela ne pose pas de problème puisque l'espace des potentiels possibles est un espace affine et que le potentiel $\overrightarrow{P}$ est périodique. Mais pour la fonction d'onde la réduction est plus délicate.

\begin{theorem}.\label{il-sagit-dun-fibre}
Les fonctions $\phi\in C^{\infty}(\R^2,\C)$ vérifiant 
\begin{equation}
\phi(z+t)=e^{i\frac{\pi d}{|\Omega|}(t_{x}y-t_{y}x)}\phi(z),\,\,\forall t\in {\cal L},\,\, \forall z\in \R^2
\end{equation}
avec $d=\frac{kH_{int}|\Omega|}{2\pi}\in \Z$ forment les sections d'un fibré vectoriel sur le tore $\Tore{\cal L}$. On appelle ce fibré \index{$E_1$}\index{$E_d$}$E_{d}$. Si $d\not= 0$ alors il est non trivial.
\end{theorem}
{\it Preuve.} Voir \cite{Griffiths-Harris}, p.~307. $\hfill{\bf CQFD}$
\begin{corollaire}.\label{annulation-section}
Si $\phi$ est une section continue de $E_d$ avec $d\not=0$ alors $\exists z\in\Tore{\cal L}$ tel que $\phi(z)=0$.
\end{corollaire}
{\it Preuve.} D'après le théorème \ref{il-sagit-dun-fibre}, le fibré $E_d$ est non trivial.\\
Si la fonction $\phi$ ne s'annule pas alors cette section définie une trivialisation du fibré. C'est impossible, donc il existe $z$ tel que $\phi(z)=0$.  $\hfill{\bf CQFD}$\\
\\
Le fibré étant $C^{\infty}$, on peut définir des espaces de sections de classe $H^{1}$ à partir des sections $C^{\infty}$ (voir par exemple \cite{spin-geom}, p.~170).\\
Le théorème~\ref{reduction-de-jauge} joint au théorème~\ref{il-sagit-dun-fibre} nous permet d'introduire un autre espace fonctionnel plus adapté à notre problème de minimisation et qui fait intervenir le tore uniquement.\\
Rappelons que $\overrightarrow{C}=\frac{1}{2}\left(\begin{array}{c}
-y\\
x
\end{array}\right)$ et que $H_{int}$ vérifie la quantification \index{quantification}$2\pi d=kH_{int}|\Omega|$. Pour $d\in\Z$, on pose \index{${\cal B}_d$}
\begin{equation}\label{avant-dernier-espace}
{\cal B}_{d}=\left\lbrace 
\begin{array}{c}
(\phi,\overrightarrow{A})\in H^{1}(E_{d})\times H^{1}(\Tore{\cal L},\R^2) \mbox{~tel ~que}\\
\phi(z+v_{i})=e^{i\frac{\pi d}{|\Omega|}(\xi_{i}y-\eta_{i}x)}\phi(z), \\
\overrightarrow{A}=H_{int}\overrightarrow{C}+\overrightarrow{P} \mbox{~avec~} \overrightarrow{P} \mbox{~périodique~}\\
\divergence\,\overrightarrow{P}=0\mbox{~et~}\int_{\Omega}\overrightarrow{P}=0\mbox{~si~}d\not= 0 .
\end{array}\right\rbrace
\end{equation}
On utilisera la notation\index{${\cal B}$}
\begin{equation}
{\cal B}=\cup_{d\in\Z}{\cal B}_{d}
\end{equation}
La fonctionnelle utilisée est toujours notée $E_n^2$ car ${\cal B}\subset {\cal C}$ o\`u ${\cal C}$ est défini à l'équation~(\ref{definition-de-espace-C}).\\
Par le théorème~\ref{il-sagit-dun-fibre}, cet espace s'identifie aux sections $H^1$ d'un fibré vectoriel de dimension réelle $4$ sur le tore.\\
L'espace ${\cal B}_{d}$ est un espace affine, ce qui simplifie l'analyse du problème. Il est intéressant de remarquer que l'espace ${\cal D}$ défini au début du chapitre n'est pas un espace vectoriel ni même un espace affine, ceci provenant de la structure assez complexe du groupe de jauge.

\saction{Modélisation finale du problème traité}\label{sec:modelisation-finale}
\noindent Dans les sections précédentes, on a transformé un problème de minimisation sur $\R^3$ associé à des conditions de périodicité en un problème défini sur un fibré vectoriel de dimension $4$ au dessus du tore $\Tore{\cal L}$.\\
Dans cette section, on montre quelques résultats élémentaires sur cette fonctionnelle et on réduit le problème de minimisation en deux sous problèmes plus simples.\\
On note $E^{int,d}$\index{$E^{int,d}$} la fonctionnelle définie sur ${\cal B}_d$ qui a pour expression
\begin{equation}\label{expression-dernier}
\begin{array}{rcl}
E^{int,d}(\phi,\overrightarrow{P})&=&\frac{1}{|\Omega|}\int_{\Omega}
\frac{1}{2}\Vert ik^{-1}\overrightarrow{\nabla}\phi+(H_{int}\overrightarrow{C}+\overrightarrow{P})\phi\Vert ^2+\frac{1}{4}(1-|\phi|^2)^2\\
&+&\frac{1}{|\Omega|}\int_{\Omega}\frac{1}{2}(\rot\,\overrightarrow{P})^2
\end{array}
\end{equation}
avec
\begin{equation}
\overrightarrow{C}=\frac{1}{2}\left(\begin{array}{c}
-y\\
x
\end{array}\right) .
\end{equation}
On calcule l'énergie magnétique $\int_{\Omega}(\rot\,\overrightarrow{A}-H_{ext})^2dxdy$ avec l'expression
\begin{equation}\label{expression-simplifie-du-champ-magnetique}
{\overrightarrow{A}}=\overrightarrow{P}+H_{int}\overrightarrow{C},\,
\rot\, {\overrightarrow{A}}-H_{ext}=\frac{\partial P_{y}}{\partial x}-\frac{\partial P_{x}}{\partial y}+H_{int}-H_{ext} .
\end{equation}
Puisque les fonctions $P_{x}$ et $P_{y}$ sont périodiques et de classe $H^1$, elles admettent un développement en série de Fourier. Les fonctions $\frac{\partial P_{y}}{\partial x}$ et $\frac{\partial P_{x}}{\partial y}$ ont un développement de Fourier sans terme constant. Cela signifie que la fonction $(\frac{\partial P_{y}}{\partial x}-\frac{\partial P_{x}}{\partial y})$ est orthogonale aux constantes et donc que
\begin{equation}\label{energie-magnetique-simplifie}
\int_{\Omega}(\rot\,\overrightarrow{A}-H_{ext})^2=\int_{\Omega}(\frac{\partial P_{y}}{\partial x}-\frac{\partial P_{x}}{\partial y})^2+(H_{int}-H_{ext})^2|\Omega| .
\end{equation}
La fonctionnelle $E_{n}^{2}$ se décompose comme ceci:
\begin{equation}\label{wrong-reference}
E_n^2(\phi, \overrightarrow{A})=E^{int,d}(\phi, \overrightarrow{P})+\frac{1}{2}(H_{int}-H_{ext})^2 .
\end{equation}
Nous indiquons ci-dessous la résolution du problème de minimisation dans le cas $d=0$.
\begin{proposition}.\label{energie-croissante}
Si on impose $d=0$, autrement dit que $H_{int}=0$, alors\\
$\inf_{(\phi,\overrightarrow{A})\in {\cal B}_{0}} E^{2}_{n}(\phi, \overrightarrow{A})=E^{2}_n(1,0)=\frac{1}{2}H_{ext}^2$.
\end{proposition}
{\it Preuve.} Ici $\overrightarrow{C}=0$, il suffit d'appliquer la formule~(\ref{wrong-reference}) pour obtenir une inégalité sur la fonctionnelle
\begin{equation}
E^{2}_{n}(\phi, \overrightarrow{A})\geq \frac{1}{2}(H_{int}-H_{ext})^2=\frac{1}{2}H_{ext}^2=E^2_{n}(1,0)
\end{equation}
~$\hfill{\bf CQFD}$
\begin{proposition}.\label{une-majoration-simple}
On a l'inégalité suivante sur l'énergie
\begin{equation}\label{bornage-du-minimum}
\inf_{(\phi,\overrightarrow{A})\in{\cal B}} E^2_{n}(\phi, \overrightarrow{A})\leq \frac{1}{4}+\frac{1}{2}\min_{d\in\Z}(\frac{2\pi d}{k|\Omega|}-H_{ext})^2\leq \frac{1}{4}+\frac{\pi^2}{2k^2|\Omega|^2} .
\end{equation}
De plus il existe $H_{0}$ tel que si $H_{ext}>H_{0}$, les solutions minimisantes vérifient $H_{int}\not=0$.
\end{proposition}
{\it Preuve.} On prend $\phi=0$ et $\overrightarrow{A}=H_{int}\overrightarrow{C}$ et le premier résultat en découle.\\
Si $H_{ext}>\sqrt{\frac{1}{2}+\frac{\pi^2}{k^2|\Omega|^2}}$, alors on a
\begin{equation}
\inf_{(\phi,\overrightarrow{A})\in {\cal B}_{0}} E^{2}_{n}(\phi, \overrightarrow{A})=E^{2}_n(1,0)=\frac{1}{2}H_{ext}^2>\frac{1}{4}+\frac{\pi^2}{2k^2|\Omega|^2}\geq \inf_{(\phi,\overrightarrow{A})\in{\cal B}} E^2_{n}(\phi, \overrightarrow{A})
\end{equation}
Par conséquent, le minimum n'est pas atteint par un état vérifiant $H_{int}=0$. On a donc le résultat avec 
\begin{equation}
H_0=\sqrt{\frac{1}{2}+\frac{\pi^2}{k^2|\Omega|^2}} .
\end{equation}
 $\hfill{\bf CQFD}$

\begin{proposition}.\label{etat-supra}
Si $H_{ext}=0$ alors le minimum de $E_n^2$ est atteint pour le couple $(\phi, \overrightarrow{P})=(1,0)$ avec $d=0$. On a  $E^2_{n}(1,0)=0$.
\end{proposition}
{\it Preuve.} On a toujours $E^2_{n}\geq 0=E_{n}^2(1,0)$ par une majoration triviale. $\hfill{\bf CQFD}$\\
\\
De ces propositions on peut retenir que le supraconducteur ne peut pas empêcher la pénétration du champ magnétique à l'intérieur du matériau pour toutes les valeurs du champ extérieur.\\
Nous démontrerons au théorème~\ref{vanishing-theorem} qu'il existe un champ $H_{c2}(k)$ tel que si $H_{ext}>H_{c2}(k)$ alors le seul couple $(\phi,\overrightarrow{A})$ réalisant le minimum est le couple $(0,0)$.
%fonctions d'onde $\phi$ du couple réalisant le minimum s'annule; on a alors un retour à l'état normal.

\subsection{Hypothèse: Quantification choisie}\label{hypothese_de_quantification}
\noindent La valeur de $d$ dans $\Z$ qui minimise le problème variationnel est inconnue à priori.\\
La quantification transforme le problème de la minimisation de $E_n^2$ en un autre problème: La minimisation des fonctionnelles $E^{int,d}$ sur les espaces ${\cal B}_d$.\\
%La quantification scinde le problème de la minimisation de la fonctionnelle $E_n^2$ en uneinfinité dénombrable de problèmes classiques de minimisation de la fonctionnalle $E^{int}$ sur un espace de Banach indexée par $d$.\\
On a trouvé le minimum de cette fonctionnelle pour $d=0$ à la proposition~\ref{energie-croissante}, on peut donc se limiter au cas $d\not=0$.\\
Ensuite il faudrait minimiser sur $d$ mais nous n'effectuons pas cette étude.\index{hypothèse H1}
\begin{hypoth}.\label{hypothese-H1}
Dans la suite des calculs, on va supposer que l'on est dans la configuration $d=1$.
\end{hypoth}
L'hypothèse \ref{hypothese-H1} simplifie l'analyse de la bifurcation. C'est une hypothèse qui est faite dans pratiquement tous les articles sur la question, voir par exemple les articles \cite{odehI}, \cite{odehII} et \cite{Barany-Golubitsky}. Cependant l'article \cite{pirate-I} aborde légèrement la minimisation en quantification supérieure: $d>1$.\\
C'est une hypothèse très naturelle, que l'expérience physique semble confirmer puisque l'on voit apparaître des cellules hexagonales avec un seul vortex par cellule. Cependant la raison essentielle de cette hypothèse est que les calculs sont plus simples dans ce cadre.

\subsection{Hypothèse $2$: choix de la géométrie du réseau}\label{choix-geometrie}
\begin{hypoth}.\label{hypothèse-H2}
On fixe l'angle entre les deux vecteurs $v_1$ et $v_2$ ainsi que le quotient de leurs longueurs.
\end{hypoth}
\begin{proposition}.
Les hypothèses \index{hypothèse H2}\ref{hypothese-H1} et \ref{hypothèse-H2} fixent la forme du réseau à isométrie près.
\end{proposition}
{\it Preuve.} Par l'hypothèse \ref{hypothese-H1}, on a $d=1$ et on a trouvé en~(\ref{relation-s-Hint}) la relation $d=\frac{kH_{int}|\Omega|}{2\pi}$. On obtient 
\begin{equation}
|\Omega|=\frac{2\pi}{k H_{int}} .
\end{equation}
L'hypothèse \ref{hypothèse-H2} fixe l'angle entre les vecteurs et le quotient de leurs longueurs. La forme du réseau est donc fixée à isométrie près. $\hfill{\bf CQFD}$\\
\\
Les deux vecteurs qui définissent le réseau ${\cal L}$ sont
\begin{equation}
\left\lbrace\begin{array}{l}
v_{1}=(\xi_{1},\eta_{1})\\
v_{2}=(\xi_{2},\eta_{2})
\end{array}\right.
\end{equation}
tandis que $\Omega$ est défini à l'équation~(\ref{definition-domaine}). On rappelle que $|\Omega|$ désigne \index{l'aire du domaine fondamental}l'aire du domaine fondamental.\\
Modulo une isométrie du plan on peut supposer que le vecteur $v_1$ est parallèle à l'axe $Ox$. Il existe alors deux vecteurs $v'_1$ et $v'_2$ tel que
\begin{equation}
\left\lbrace\begin{array}{rcccl}
v'_{1}&=&(r,0)&=&(\xi'_1,\eta'_1)\\
v'_{2}&=&(w,u)&=&(\xi'_2,\eta'_2)
\end{array}\right.
\end{equation}
avec $r>0$,  $u>0$, $ru=1$ et 
\begin{equation}\label{formule-de-changement-de-variable}
\left\lbrace\begin{array}{rcl}
v_1&=&\sqrt{\frac{2\pi}{k H_{int}}}v'_1\\
v_2&=&\sqrt{\frac{2\pi}{k H_{int}}}v'_2 .
\end{array}\right.
\end{equation}
On note \index{${\cal L'}$}${\cal L'}$ le réseau de $\R^2$ engendré par les vecteurs $v'_i$.\\
On appelle \index{$\Omega'$}$\Omega'$ le domaine fondamental de ce réseau, il vérifie $|\Omega'|=1$.

\subsection{Changement de problème}\label{subsect:renorm}
\noindent On pose\index{$\lambda$}\index{$\overrightarrow{A}_{0}$}
\begin{equation}\label{quelques-definitions}
\left\lbrace\begin{array}{l}
\lambda=\frac{2\pi k}{H_{int}}\\
\overrightarrow{A}_{0}=\pi\left(\begin{array}{c}
-y\\
x
\end{array}\right) .
\end{array}\right.
\end{equation}

\begin{definition}.
On définit un nouvel espace fonctionnel ${\cal A'}$\index{${\cal A'}$} par:
\begin{equation}\label{dernier-espace-prime}
{\cal A}'=\left\lbrace 
\begin{array}{c}
(\phi,\overrightarrow{a})\in H^{1}(E_{1})\times H^{1}(\Tore{\cal L'},\R^2)\\
\mbox{~avec~} \phi(z+v'_{i})=e^{i\pi(\xi'_{i}y-\eta'_{i}x)}\phi(z), \\
\overrightarrow{a} \mbox{~${\cal L'}$-périodique,~} \divergence\,\overrightarrow{a}=0\mbox{~et~}\int_{\Omega'}\overrightarrow{a}=0 .
\end{array}\right\rbrace
\end{equation}
\end{definition}
Le fait que l'on se limite à des états de quantification $1$ est codé dans cette définition par le fait que $ru=1$\index{$ru=1$}.

\begin{definition}.
On définit une fonctionnelle $F_{\lambda,k}$\index{$F_{\lambda,k}$} sur ${\cal A}'$:
\begin{equation}\label{fonctionelle-Odeh-simplifie}
\begin{array}{l}
\forall(\phi,\overrightarrow{a})\in{\cal A}'\\
F_{\lambda,k}(\phi, \overrightarrow{a})=\int_{\Omega'}\frac{1}{2}\Vert i\overrightarrow{\nabla}\phi+(\overrightarrow{A}_{0}+\overrightarrow{a})\phi\Vert^2+\frac{1}{4}(\lambda-|\phi|^2)^2+\frac{k^2}{2}|\rot\, \overrightarrow{a}|^2 .
\end{array}
\end{equation}
\end{definition}

\begin{definition}.\label{definition-pour-toute-la-suite}
On définit la fonction $D_{\lambda,k}$\index{$D_{\lambda,k}$} par
\begin{equation}\label{definition-Dklambdaphia}
\begin{array}{rcl}
D_{\lambda,k}(\phi,\overrightarrow{a})&=&\frac{1}{\lambda^2}F_{\lambda,k}(\sqrt{\lambda}\phi,\overrightarrow{a})\\
&=&\int_{\Omega'}\frac{1}{2\lambda}\Vert i\overrightarrow{\nabla}\phi+(\overrightarrow{A}_{0}+\overrightarrow{a})\phi\Vert^2+\frac{1}{4}(1-|\phi|^2)^2+\frac{k^2}{2\lambda^2}|\rot\, \overrightarrow{a}|^2 .
\end{array}
\end{equation}
On s'intéresse au minimum de ces deux fonctionelles \index{$m_F(\lambda,k)$}\index{$m_D(\lambda,k)$}, on pose donc
\begin{equation}\label{definition-mF-mD}
\begin{array}{l}
m_F(\lambda,k)=\inf_{(\phi,\overrightarrow{a})\in{\cal A}'}F_{\lambda,k}(\phi,\overrightarrow{a}),\\
m_D(\lambda,k)=\inf_{(\phi,\overrightarrow{a})\in{\cal A}'}D_{\lambda,k}(\phi,\overrightarrow{a}).
\end{array}
\end{equation}
\end{definition}
Pour alléger les calculs effectués dans cette section, on pose
\begin{equation}\label{definition-alpha-beta}
\left\lbrace\begin{array}{l}
\alpha=\sqrt{\frac{2\pi k}{H_{int}}}\\
\beta=\sqrt{\frac{2\pi}{k H_{int}}} .
\end{array}\right.
\end{equation}

\begin{proposition}.\label{application-D}
On pose \index{$W_i$}
\begin{equation}\label{definition-W1-W2}
\left\lbrace\begin{array}{rcl}
W_1\phi&=&\alpha\phi(\beta z)\\
W_2\overrightarrow{A}&=&\alpha [\overrightarrow{A}-H_{int}\overrightarrow{C}](\beta z) .
\end{array}\right.
\end{equation}
Alors l'application $W$ de ${\cal B}_1$ vers ${\cal A}'$
\begin{equation}
\left\lbrace\begin{array}{rcl}
W:{\cal B}_1&\mapsto&{\cal A}'\\
(\phi,\overrightarrow{A})&\mapsto&(W_1\phi,W_2\overrightarrow{A})
\end{array}\right.
\end{equation}
est bijective.
\end{proposition}
{\it Preuve.} Vérifions que le champ $[\overrightarrow{A}-H_{int}\overrightarrow{C}](\beta z)$ est périodique.\\
On sait que $(\phi,\overrightarrow{A})\in{\cal B}_1$; donc on peut écrire $\overrightarrow{A}=H_{int}\overrightarrow{C}+\overrightarrow{P}$, o\`u $\overrightarrow{P}$ est un champ périodique pour $v_1$ et $v_2$. On a alors
\begin{equation}
\begin{array}{rcl}
W_2\overrightarrow{A}(z+v'_i)&=&\alpha \overrightarrow{P}(\beta[z+v'_i])\\
&=&\alpha \overrightarrow{P}(\beta z+\beta v'_i)\\
&=&\alpha \overrightarrow{P}(\beta z+v_i)\\
&=&\alpha \overrightarrow{P}(\beta z)\\
&=&W_2\overrightarrow{A}(z) .
\end{array}
\end{equation}
Calculons maintenant pour $\phi$:
\begin{equation}
\begin{array}{rcl}
W_1\phi(z+v'_i)&=&\alpha\phi(\beta[z+v'_i])\\
&=&\alpha\phi(\beta z+v_i)\\
&=&\alpha e^{i\frac{\pi}{|\Omega|}\beta(\xi_i y-\eta_i x)}\phi(\beta z)\\
&=&\alpha e^{i\frac{\pi}{|\Omega|}\frac{2\pi}{k H_{int}}(\xi'_i y-\eta'_i x)}\phi(\beta z)\\
&=&\alpha e^{i\frac{\pi}{|\Omega|}|\Omega|(\xi'_i y-\eta'_i x)}\phi(\beta z)\\
&=&\alpha e^{i\pi(\xi'_i y-\eta'_i x)}\phi(\beta z)\\
&=&e^{i\pi(\xi'_i y-\eta'_i x)}W_1\phi(z) .
\end{array}
\end{equation}
Par conséquent, l'application $W$ est bien à valeur dans ${\cal A}'$. Le fait qu'elle soit bijective est évident. $\hfill{\bf CQFD}$

\begin{theorem}.\label{erreurs-pendant-2-ans}
Soit $(\phi,\overrightarrow{A})$ un couple appartenant à ${\cal B}_1$ on a alors
\begin{equation}\label{a-montrer-horreur-terrible}
\lambda^2E^{int,1}(\phi,\overrightarrow{P})=F_{\lambda,k}(W_1\phi,W_2\overrightarrow{A})
\end{equation}
avec $\overrightarrow{A}=H_{int}\overrightarrow{C}+\overrightarrow{P}$.
\end{theorem}
{\it Preuve.} On appelle $T$\index{$T$} le second membre de l'équation~(\ref{a-montrer-horreur-terrible})
\begin{equation}
\begin{array}{rcl}
T&=&\int_{\Omega'}\frac{1}{2}\Vert i\overrightarrow{\nabla}W_1\phi+(\overrightarrow{A}_0+W_2\overrightarrow{A})W_1\phi\Vert^2+\frac{(\lambda-|W_1\phi|^2)^2}{4}+\frac{k^2}{2}(\rot\,W_2\overrightarrow{A})^2\\[2mm]
&=&\int_{\Omega'}\frac{1}{2}\Vert i\frac{2\pi}{H_{int}}\overrightarrow{\nabla}\phi(\beta z)+(\overrightarrow{A}_0+\alpha\overrightarrow{P}(\beta z))\alpha\phi(\beta z)\Vert^2\\
&+&\int_{\Omega}\frac{(\lambda-\lambda|\phi(\beta z)|^2)^2}{4}+\frac{k^2}{2}(\frac{2\pi}{H_{int}}[\rot\,\overrightarrow{P}](\beta z))^2\\[2mm]
&=&\int_{\Omega'}\frac{1}{2}\Vert i\frac{2\pi}{H_{int}}\overrightarrow{\nabla}\phi+(\sqrt{\frac{k H_{int}}{2\pi}}\overrightarrow{A}_0+\alpha\overrightarrow{P})\alpha\phi\Vert^2(\beta z)\\
&+&\frac{(\lambda-\lambda|\phi(\beta z)|^2)^2}{4}+\frac{k^2}{2}(\frac{2\pi}{H_{int}}[\rot\,\overrightarrow{P}](\beta z))^2\\[2mm]
&=&(\frac{2\pi k}{H_{int}})^2\{\int_{\Omega'}\frac{1}{2}\Vert\frac{i}{k}\overrightarrow{\nabla}\phi+(\frac{H_{int}}{2\pi}\overrightarrow{A}_0+\overrightarrow{P})\phi\Vert^2(\beta z)\\
&+&\frac{(1-|\phi(\beta z)|^2)^2}{4}+\frac{1}{2}([\rot\,\overrightarrow{P}](\beta z))^2dxdy\} .
\end{array}
\end{equation}
On effectue maintenant le changement de variable $z'=\beta z$ dans l'intégrale et l'on se ramène alors à une intégrale sur $\Omega$ (cf~(\ref{formule-de-changement-de-variable})). On obtient
\begin{equation}
\begin{array}{rcl}
T&=&\lambda^2\{\int_{\Omega}\frac{1}{2}\Vert\frac{i}{k}\overrightarrow{\nabla}\phi+(H_{int}\overrightarrow{C}+\overrightarrow{P})\phi\Vert^2\\
&+&\frac{(1-|\phi|^2)^2}{4}+\frac{1}{2}(\rot\,\overrightarrow{P})^2dxdy\}(\frac{kH_{int}}{2\pi})\\
&=&\lambda^2|\Omega|E^{int,1}(\phi,\overrightarrow{P})(\frac{kH_{int}}{2\pi})\mbox{~par~definition~de~$E^{int,1}$}\\
&=&\lambda^2E^{int,1}(\phi,\overrightarrow{P}) .
\end{array}
\end{equation}
On a utilisé dans la dernière équation la relation $k|\Omega|H_{int}=2\pi$.\\
On obtient donc l'égalité voulue. $\hfill{\bf CQFD}$

\begin{definition}.\label{definition-EVkHext}
On définit l'énergie volumique $E^{V}_{k,H_{ext}}$\index{$E^{V}_{k,H_{ext}}$} par 
\begin{equation}\label{definition-EVkHext-compl}
E^{V}_{k,H_{ext}}(H_{int},\phi, \overrightarrow{a})=\frac{1}{\lambda^2} F_{\lambda,k}(\phi, \overrightarrow{a})+\frac{1}{2}(H_{int}-H_{ext})^2
\end{equation}
avec $\lambda=\frac{2\pi k}{H_{int}}$. On définit aussi le minimum partiel $G^{V}_{k,H_{ext}}$\index{$G^{V}_{k,H_{ext}}$} par
\begin{equation}\label{definition-EVkHext-restreint}
G^{V}_{k,H_{ext}}(H_{int})=\frac{1}{\lambda^2} m_F(\lambda,k)+\frac{1}{2}(H_{int}-H_{ext})^2 .
\end{equation}
o\`u $m_F$ est définie à l'équation~(\ref{definition-mF-mD}). Le minimum qui nous intéresse est ${\cal E}^{V}_{k,H_{ext}}$\index{${\cal E}^{V}_{k,H_{ext}}$}, il est définit par:
\begin{equation}\label{infimum-Hint}
{\cal E}^{V}_{k,H_{ext}}=\inf_{H_{int}>0}G^{V}_{k,H_{ext}}(H_{int}) .
\end{equation}
\end{definition}
On verra par la suite que l'infimum (\ref{infimum-Hint}) peut ne pas \^etre atteint sur $]0,+\infty[$.\\
Dans la suite de la thèse on enlève les ' pour simplifier l'écriture à $v'_i$, $\xi'_i$, $\eta'_i$, $\Omega'$, ${\cal L}'$, $\phi'$,  $\overrightarrow{a}'$ et ${\cal A}'$.
\begin{definition}.(Résumé des définitions précédentes)\\
Dans la suite de la thèse nous utiliserons les notations suivantes:\\
Le réseau ${\cal L}$ de périodicité est engendré par 
\begin{equation}
\left\lbrace\begin{array}{rcl}
v_{1}&=&(r,0)\\
v_{2}&=&(w,u)
\end{array}\right.
\end{equation}
et le domaine fondamental associé $\Omega$ a une aire égale à $1$.\index{${\cal A}$}
\begin{equation}\label{dernier-espace}
{\cal A}=\left\lbrace 
\begin{array}{c}
(\phi,\overrightarrow{a})\in H^{1}(E_{1})\times H^{1}(\Tore{\cal L},\R^2)\\
\mbox{avec~}\phi(z+v_{i})=e^{i\pi(\xi_{i}y-\eta_{i}x)}\phi(z), \\
\overrightarrow{a} \mbox{~${\cal L}$-périodique,~} \divergence\,\overrightarrow{a}=0\mbox{~et~}\int_{\Omega}\overrightarrow{a}=0 .
\end{array}\right\rbrace
\end{equation}
La fonctionnelle $F_{\lambda,k}$ ayant pour expression
\begin{equation}\label{expression-Flambda-k}
F_{\lambda,k}(\phi, \overrightarrow{a})=\int_{\Omega}\frac{1}{2}\Vert i\overrightarrow{\nabla}\phi+(\overrightarrow{A}_{0}+\overrightarrow{a})\phi\Vert^2+\frac{1}{4}(\lambda-|\phi|^2)^2+\frac{k^2}{2}|\rot\, \overrightarrow{a}|^2 .
\end{equation}
\end{definition}
Le problème \ref{probleme-de-la-these} prend une forme différente dans le nouveau système de coordonnées
\begin{probleme}.\label{quelles-peine-pour-ca}
Le problème qui nous occupera dans la suite de la thèse est le suivant. Trouver
\begin{equation}
{\cal E}^V_{k,H_{ext}}=\inf_{\begin{array}{l}
H_{int}\in\R_+^*\\
(\phi,\overrightarrow{a})\in{\cal A}
\end{array}}E^{V}_{k,H_{ext}}(H_{int},\phi,\overrightarrow{a})
\end{equation}
et décrire les couples $(\phi,\overrightarrow{a})$ qui réalisent ce minimum.
\end{probleme}
Le problème~\ref{quelles-peine-pour-ca} est équivalent au problème~\ref{probleme-de-la-these} par le théorème~\ref{erreurs-pendant-2-ans} et la proposition~\ref{application-D}.\\
Les sections $\phi$ vérifient $\phi(z+v_{i})=e^{i\pi(\xi_{i}y-\eta_{i}x)}\phi(z)$. Ci-dessous, on représente l'allure du domaine fondamental dans les nouvelles coordonnées.
\resizebox{8cm}{!}{\includegraphics{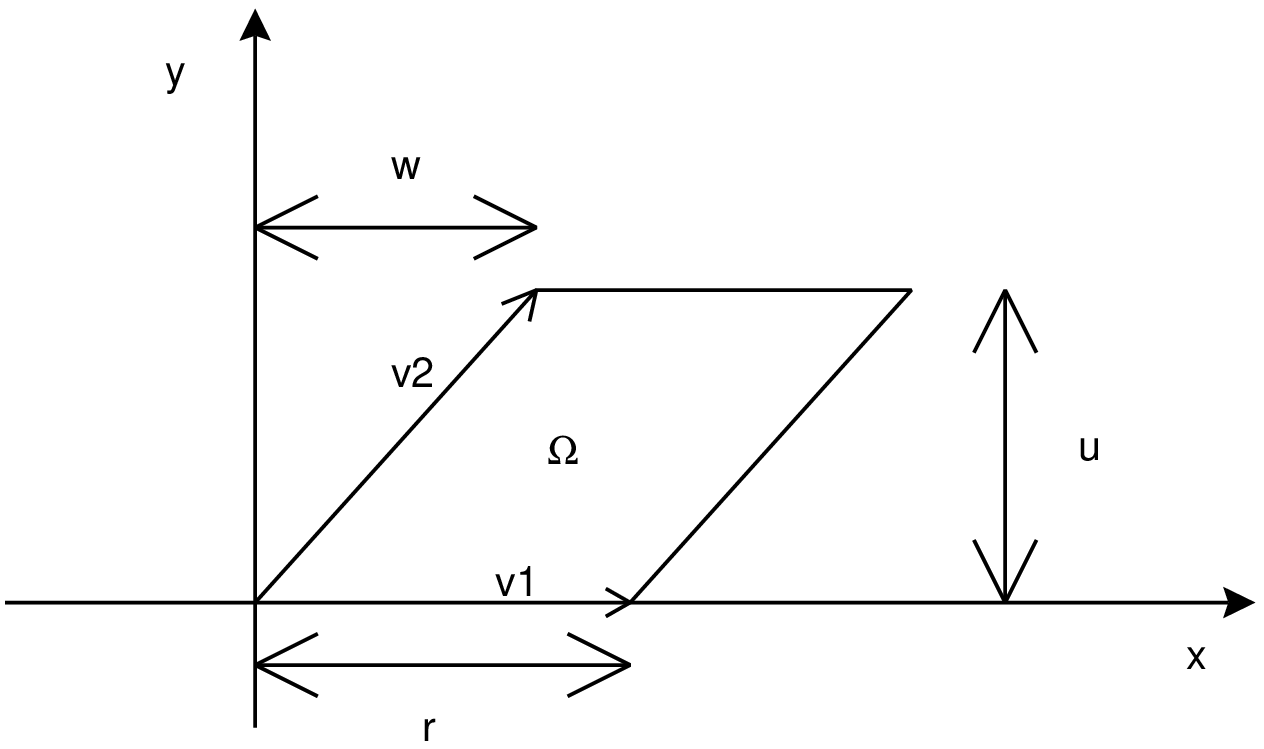}}\\
%Cela permet d'écrire le potentiel vecteur sous la forme $\overrightarrow{A}=\overrightarrow{A}_{0}+\overrightarrow{a}$ avec $\overrightarrow{a}$ périodique. On a alors $\rot\,\overrightarrow{A}-H_{int}=\rot\,\overrightarrow{a}$.\\

\begin{proposition}.\label{premieres-propriete-mF-mD}
On a les propriétés suivantes:
\begin{equation}
\begin{array}{l}
0\leq m_F(\lambda,k)\leq \frac{\lambda^2}{4},\\
0\leq m_D(\lambda,k)\leq\frac{1}{4},\\
m_D(\lambda,k)=\frac{1}{\lambda^2}m_F(\lambda,k),
\end{array}
\end{equation}
o\`u $m_F$ et $m_D$ sont définis à l'équation~(\ref{definition-mF-mD}).
\end{proposition}
{\it Preuve.} On sait que $F_{\lambda,k}$ est positif, par conséquent $m_F(\lambda,k)$ est positif.\\
On a $F_{\lambda,k}(0,0)=\frac{\lambda^2}{4}$ par conséquent, puisque $m_F$ est un minimum, on a l'inégalité $m_F(\lambda,k)\leq \frac{\lambda^2}{4}$.\\
L'égalité $D_{\lambda,k}(\phi,\overrightarrow{a})=\frac{1}{\lambda^2}F_{\lambda,k}(\sqrt{\lambda}\phi,\overrightarrow{a})$ implique l'égalité $m_D(\lambda,k)=\frac{1}{\lambda^2}m_F(\lambda,k)$.\\
On obtient alors facilement l'encadrement: $0\leq m_D(\lambda,k)\leq\frac{1}{4}$.  $\hfill{\bf CQFD}$

\begin{definition}.\label{definition-probleme-reduit}
On appellera ``problème réduit''\index{``problème réduit''} le problème de la minimisation de $F_{\lambda,k}$.
\end{definition}
\begin{definition}.\label{caracterisation-etat-possibles}
\begin{itemize}
\item On dit que le supraconducteur est à l'état normal\index{état normal} si $(H_{int},\phi,\overrightarrow{a})=(H_{ext},0,0)$ et si le minimum de la fonctionnelle $E^V_{k,H_{ext}}$ (cf~(\ref{definition-EVkHext})) est réalisé en ce point. L'énergie ${\cal E}^V_{k,H_{ext}}$ est alors égale à $\frac{1}{4}$. Cet état est désigné par la lettre $N$.
\item On dit que le supraconducteur est dans un état mixte\index{état mixte} si $\phi\not=0$ et $H_{int}>0$ et si le minimum de la fonctionnelle $E^V_{k,H_{ext}}$ (cf~(\ref{definition-EVkHext})) est réalisé en ce point. Cet état est désigné par la lettre $M$.
\item On dit que le supraconducteur est à l'état pur\index{état supraconducteur pur} si $H_{int}=0$ et $\phi$ est constante et si le minimum de la fonctionnelle $E^{V}_{k,H_{ext}}$ est réalisé en ce point. L'énergie ${\cal E}^V_{k,H_{ext}}$ est alors égale à $\frac{H_{ext}^2}{2}$. Cet état est désigné par la lettre $P$.
\end{itemize}
Le diagramme des phases est alors l'ensemble des points $(k,H_{ext})$ o\`u l'on indique sur chaque point quel est sa nature.
\end{definition}
L'énergie $\frac{H_{ext}^2}{2}$ correspond à la quantification $0$ (cf~\ref{energie-croissante}) et on verra plus loin que $\lim_{H_{int}\rightarrow 0}G^V_{k,H_{ext}}(H_{int})=\frac{H_{ext}^2}{2}$.

\saction{Structure de l'espace des réseaux}\label{sec:struct-reseau-space}
\noindent On sait par les calculs de la sous-section~\ref{subsect:renorm} que l'on peut se ramener à la situation où 
\begin{equation}
\left\lbrace\begin{array}{rcc}
v_{1}&=&(r,0)\\
v_{2}&=&(w,u)
\end{array}\right.
\end{equation}
avec
\begin{equation}
ru=1 .
\end{equation}
Un système de coordonnées possible est
\begin{equation}
\left\lbrace\begin{array}{rcl}
u&\in &\R_{+}^{*},\\
w&\in &\R\, .
\end{array}\right.
\end{equation}
On note \index{${\cal M}$}${\cal M}$ l'espace des réseaux possibles, c'est à dire des réseaux engendrés par des couples de vecteurs $(v_1,v_2)$ tel que $\mbox{dét}(v_1,v_2)=1$ (chaque réseau est isomorphe à $\Z^2$).\\
Dans la suite de la thèse, on prendra ces coordonnées et on identifiera ${\cal M}$ à
\begin{equation}\label{espace-module}
{\cal M}=\R_+^*\times \R\, .
\end{equation}
\`A tout élément $m=(u,w)\in{\cal M}$ est associé le réseau de $\R^2$ engendré par \index{$v_{i,m}$}$v_{1,m}=(r,0)$ et $v_{2,m}=(w,u)$, on l'appelle \index{${\cal L}_m$}${\cal L}_m$.\\
\begin{definition}.
Soit $m=(u,w)\in {\cal M}$, on définit les espaces \index{${\cal E}_{0,m}$}\index{${\cal E}_{1,m}$}\index{${\cal V}_{m}$}\index{${\cal A}_{m}$} suivants
\begin{equation}
\begin{array}{rcl}
{\cal E}_{0,m}&=&\lbrace f \in {\cal D}'(\R^2,\R)\mbox{~tel~que~}\forall(x,y)\in\R^2,\\
&&f(x+r, y)=f(x, y),\,\,f(x+w, y+u)=f(x,y) \rbrace\\
{\cal E}_{1,m}&=&\lbrace \phi \in {\cal D}'(\R^2,\R) \mbox{~tel~que~}\forall(x,y)\in\R^2,\\
&&\phi(x+r, y)=e^{i\pi ry}\phi(x, y),\,\,\phi(x+w, y+u)=e^{i\pi(wy-xu)}\phi(x,y) \rbrace\\
{\cal V}_{m}&=&\lbrace \overrightarrow{a} \in {\cal D}'(\R^2,\R^2) \mbox{~potentiel~vecteur~tel~que~}\forall(x,y)\in\R^2,\,\,\\
&&\overrightarrow{a}(x+r, y)=\overrightarrow{a}(x, y),\,\overrightarrow{a}(x+w, y+u)=\overrightarrow{a}(x,y) \rbrace\\
{\cal A}_{m}&=&\lbrace (\phi,\overrightarrow{a})\in{\cal A}\mbox{~tel~que~}\phi\in{\cal E}_{1,m}\mbox{~et~}\overrightarrow{a}\in{\cal V}_{m}\rbrace
\end{array}
\end{equation}
o\`u ${\cal D}'(\R^2,\R^n)$ est l'espace des distributions sur $\R^2$ à valeurs dans $\R^n$.
\end{definition}
Dans la définition précédente ${\cal E}_{0,m}$ et ${\cal V}_{m}$ sont des espaces de fonctions périodiques selon le réseau ${\cal L}$. On a aussi une écriture intrinsèque pour ${\cal E}_{1,m}$:
\begin{equation}
\phi\in{\cal E}_{1,m}\Leftrightarrow \forall v\in{\cal L}_m,\,\phi(x+v)=e^{i\pi \determinant(v,x)}\phi(x)\mbox{~avec~}x=(x_1,x_2) .
\end{equation}

\begin{definition}.
On note \index{$\tau$}$\tau$ l'application
\begin{equation}
\left\lbrace\begin{array}{rcl}
\tau:{\cal M}&\mapsto&{\cal M}\\
(u,w)&\mapsto&(u,w+\frac{1}{u}).
\end{array}\right.
\end{equation}
Les réseaux définis par les coordonnées $(u, w)$ et $(u, w+r)$ sont identiques.
\end{definition}
\begin{definition}.
On note \index{$\sigma_1$}$\sigma_1$ l'application 
\begin{equation}\label{definition-sym1}
\begin{array}{rcl}
\sigma_1: {\cal M}&\mapsto&{\cal M}\\
(u,w)&\mapsto&(u,-w) .
\end{array}
\end{equation}
\end{definition}
\begin{definition}.
On note \index{$u'$} \index{$w'$} \index{$r'$}
\begin{equation}
\left\lbrace\begin{array}{l}
u'=\frac{1}{\sqrt{w^2+u^2}},\\
w'=\frac{wr}{\sqrt{w^2+u^2}},\\
(r'=\frac{1}{u'})
\end{array}\right.
\end{equation}
et on définit \index{$\sigma_2$}$\sigma_2$ comme l'application
\begin{equation}\label{definition-sym2}
\begin{array}{rcl}
\sigma_2: {\cal M}&\mapsto&{\cal M}\\
(u,w)&\mapsto&(u',w') .
\end{array}
\end{equation}
\end{definition}
Les applications $\sigma_1$ et $\sigma_2$ sont involutives. Les trois applications $\tau$, $\sigma_1$ et $\sigma_2$ sont bijectives.
\begin{definition}.
On appelle \index{$G$}$G$ le groupe de bijections de ${\cal M}$ engendré par les applications $\tau$, $\sigma_1$ et $\sigma_2$.\\
Ce groupe est discret infini et est un sous groupe du groupe de Lie $PGL(2,\R)$.
\end{definition}
On définit maintenant deux applications linéaires orthogonales sur $\R^2$ associé aux $\sigma_i$:
\begin{equation}
\left\lbrace\begin{array}{rcl}
\Sigma_1:\R^2&\mapsto&\R^2\\
(x,y)&\mapsto&(x,-y)
\end{array}\right.\\
\end{equation}
l'applications \index{$\Sigma_1$}$\Sigma_1$ est la symétrie orthogonale par rapport à l'axe $Oy$. On définit aussi l'application \index{$\Sigma_2$}
\begin{equation}
\left\lbrace\begin{array}{rcl}
\Sigma_2:\R^2&\mapsto&\R^2\\
(x,y)&\mapsto&(\frac{1}{\sqrt{u^2+w^2}}(wx+uy),\frac{1}{\sqrt{u^2+w^2}}(ux-wy))
\end{array}\right.
\end{equation}
qui est la symétrie orthogonale par rapport à la médiatrice des vecteurs $v_{1,m}$ et $v_{2,m}$.

\begin{proposition}.
On a les relations
\begin{equation}
\begin{array}{l}
\left\lbrace\begin{array}{rcl}
\Sigma_1(v_{1,m})&=&v_{1,\sigma_1(m)}\\
\Sigma_1(v_{2,m})&=&v_{2,\sigma_1(m)}
\end{array}\right.\\
\left\lbrace\begin{array}{rcl}
\Sigma_2(v_{1,m})&=&v_{2,\sigma_2(m)}\not=v_{1,\sigma_2(m)}\\
\Sigma_2(v_{2,m})&=&v_{1,\sigma_2(m)}\not=v_{2,\sigma_2(m)}
\end{array}\right.
\end{array}
\end{equation}
On a toujours 
\begin{equation}\label{lien-sigma_i}
\Sigma_i({\cal L}_m)={\cal L}_{\sigma_i(m)} .
\end{equation}
\end{proposition}
{\it Preuve.} \'Evident. $\hfill{\bf CQFD}$\\
\\
On définit également une autre application \index{$S_m$}
\begin{equation}\label{definition-Sm}
\left\lbrace\begin{array}{rcl}
S_m:\R^2&\mapsto&\R^2\\
(x,y)&\mapsto&(rx+wy,uy)
\end{array}\right.
\end{equation}
et on a l'égalité $S_m({\cal L}_{(1,0)})={\cal L}_m$. Ce n'est pas une application orthogonale.

\subsection{Transformations induites sur les fonctions}
\noindent Si $m\in{\cal M}$ alors les réseaux correspondant à $m$ et $\tau(m)$ sont identiques. Par conséquent, les espaces de fonctions associés sont aussi identiques. Nous allons donc nous limiter à l'étude des transformations correspondant à $\sigma_1$ et $\sigma_2$.
%On note $D_{1}=\left(\begin{array}{cc}
%1 & 0\\
%0 & -1
%\end{array}\right)$. C'est une matrice de changement de variable sur le plan $\R^2$.
\begin{proposition}.
L'application\index{$\Sigma_i^*$}
\begin{equation}
\left\lbrace\begin{array}{rcl}
\Sigma^*_i:{\cal E}_{0,m}&\mapsto&{\cal E}_{0,\sigma_i(m)}\\
f&\mapsto&\overline{f(\Sigma_i(x,y))}
\end{array}\right.
\end{equation}
est un isomorphisme continu .
\end{proposition}
{\it Preuve.} Le fait que l'application soit bien à valeur dans ${\cal E}_{0,\sigma_i(m)}$ est évident vu l'équation (\ref{lien-sigma_i}).\\
La bijectivité provient de la relation $(\Sigma_i^*)^2=Id$. $\hfill{\bf CQFD}$

\begin{proposition}.
L'application
\begin{equation}
\left\lbrace\begin{array}{rcl}
\Sigma^*_i:{\cal E}_{1,m}&\mapsto&{\cal E}_{1,\sigma_i(m)}\\
\phi&\mapsto&\overline{\phi(\Sigma_i(x,y))}
\end{array}\right.
\end{equation}
est un isomorphisme continu .
\end{proposition}
{\it Preuve.} Soit $\phi\in {\cal E}_{1,m}$. Les relations de périodicité pour $\phi$ de l'équation (\ref{dernier-espace}) peuvent s'écrirent
\begin{equation}
\phi(x+v_i)=e^{i\pi \determinant(v_i,x)}\phi(x),\,\,i=1,2
\end{equation}
avec $x=(x_1,x_2)$. Soit $i,j\in\{1,2\}^2$; on a 
\begin{equation}
\begin{array}{rcl}
(\Sigma_j^*\phi)(x+\Sigma_j v_i)
&=&\overline{\phi(\Sigma_j[x+\Sigma_j v_i])}\\
&=&\overline{\phi(\Sigma_j x+\Sigma_j^2 v_i)}\\
&=&\overline{\phi(\Sigma_j x+v_i)}\\
&=&\overline{e^{i\pi \determinant(v_i,\Sigma_j x)}\phi(\Sigma_j x)}\\
&=&\overline{e^{i\pi \determinant(\Sigma_j^2 v_i,\Sigma_j x)}\phi(\Sigma_j x)}\\
&=&\overline{e^{i\pi \determinant(\Sigma_j)\determinant(\Sigma_j v_i, x)}\phi(\Sigma_j x)}\\
&=&\overline{e^{i\pi (-1)\determinant(\Sigma_j v_i, x)}}(\Sigma_j^*\phi)(x) \\
&=&e^{i\pi \determinant(\Sigma_j v_i, x)}(\Sigma_j^*\phi)(x) .
\end{array}
\end{equation}
Puisque $\Sigma_j({\cal L}_{m})={\cal L}_{\sigma_j(m)}$, la section $\Sigma_j^*\phi$ appartient à ${\cal E}_{1,\sigma_j(m)}$. La bijectivité est évidente puisque $(\Sigma_i^*)^2=Id$.  $\hfill{\bf CQFD}$

\begin{proposition}.
L'application
\begin{equation}
\left\lbrace\begin{array}{rcl}
\Sigma^*_i:{\cal V}_{m}&\mapsto&{\cal V}_{\sigma_i(m)}\\
\overrightarrow{a}&\mapsto&\Sigma_i\overrightarrow{a}(\Sigma_i(x,y))
\end{array}\right.
\end{equation}
est un isomorphisme continu .
\end{proposition}
{\it Preuve.} Cette proposition est évidente.  $\hfill{\bf CQFD}$

\begin{proposition}.\label{fonction-Sm-etoile}
Les applications\index{$S_m^*$}
\begin{equation}
\left\lbrace\begin{array}{rcl}
S_m^*:{\cal E}_{i,(0,1)}&\mapsto&{\cal E}_{i,m}\\
f&\mapsto&\overline{f(S_m(x,y))}
\end{array}\right.\mbox{~avec~}i=0,1
\end{equation}
et
\begin{equation}
\left\lbrace\begin{array}{rcl}
S_m^*:{\cal V}_{(0,1)}&\mapsto&{\cal V}_{m}\\
\overrightarrow{a}&\mapsto&S_m\overrightarrow{a}(S_m(x,y))
\end{array}\right.
\end{equation}
sont des isomorphismes continus.
\end{proposition}
{\it Preuve.} La démonstration suit celle des propositions précédentes. $\hfill{\bf CQFD}$

\begin{remarque}.
Ces applications sont bien connus en géométrie différentielle (cf~\cite{Nakahara}, p.~147).
\end{remarque}

\begin{remarque}.
Si $f\in {\cal E}_{0,m}$ et $g=\Sigma_i^* f$ alors $\overrightarrow{\nabla}g=\Sigma_i^*\overrightarrow{\nabla}f$.
\end{remarque}
\begin{proposition}.
On a aussi les relations
\begin{equation}
\left\lbrace\begin{array}{l}
\Sigma_i \overrightarrow{A}_0(\Sigma_i(x,y))=\overrightarrow{A}_0(x,y)\\
\forall \overrightarrow{a}\in {\cal V}_m\\
\rot\,\Sigma_i^*\overrightarrow{a}=\Sigma_i^*\rot\,\overrightarrow{a}
\end{array}\right.
\end{equation}
\end{proposition}
{\it Preuve.} On calcule simplement
\begin{equation}
\left\lbrace\begin{array}{rcl}
\Sigma_1^*
(\rot\,\overrightarrow{a})&=&\frac{\partial a_{x}}{\partial y}(x, -y)-\frac{\partial a_{y}}{\partial x}(x, -y)\\
&=&\frac{\partial (\Sigma_1^*\overrightarrow{a})_{x}}{\partial y}-\frac{\partial (\Sigma_1^*\overrightarrow{a})_{y}}{\partial x}\\
&=&\rot\,\Sigma_1^*\overrightarrow{a} ,
\end{array}\right.
\end{equation}
\begin{equation}
\left\lbrace\begin{array}{rcl}
\Sigma_2^*\rot\,\overrightarrow{a}&=&\frac{\partial a_{x}}{\partial y}(x', y')-\frac{\partial a_{y}}{\partial x}(x', y')\\
&=&u'[u\frac{\partial[a_{x}(x', y')]}{\partial x}-w\frac{\partial[a_{x}(x', y')]}{\partial y}]-u'[w\frac{\partial[a_{y}(x', y')]}{\partial x}+u\frac{\partial[a_{y}(x', y')]}{\partial y}]\\
&=&\frac{\partial (\Sigma_2^*\overrightarrow{a})_{x}}{\partial y}-\frac{\partial (\Sigma_2^*\overrightarrow{a})_{y}}{\partial x}\\
&=&\rot\,\Sigma_2^*\overrightarrow{a} .
\end{array}\right.
\end{equation}
On remarquera que toutes les transformations définies dans cette section, $\sigma_{i}$, $\Sigma_i$ et $\Sigma_i^*$ sont involutives.
\begin{remarque}.
En fait ces opérations sont bien connues dans le cadre des courbes elliptiques (voir par exemple \cite{Siegel}, Tome II, p.~114-119). Les différences portent sur deux points: La normalisation est faite pour que l'aire du domaine fondamental soit égale à $1$ tandis que les calculs sur les courbes elliptiques supposent eux que l'un des vecteurs de réseau est de norme $1$. La deuxième différence est que notre groupe $G$ est plus grand que le groupe $PSL(2,\Z)$ considéré par Siegel.
\end{remarque}

\subsection{Invariance de la fonctionnelle}
\begin{theorem}.\label{invariance-number-ONE}
La fonctionnelle $F_{\lambda,k}$ introduite en~(\ref{expression-Flambda-k}) est invariante par les transformations $\Sigma_i^*$ autrement dit
\begin{equation}
F_{\lambda,k}(\Sigma_i^*\phi, \Sigma_i^*\overrightarrow{a})=F_{\lambda,k}(\phi, \overrightarrow{a}) .
\end{equation}
\end{theorem}
{\it Preuve.} Par les calculs déjà fait
\begin{equation}
\Sigma_i^*[i\overrightarrow{\nabla}\phi+(\overrightarrow{A}_{0}+\overrightarrow{a})\phi]=[i\overrightarrow{\nabla}\Sigma_i^*\phi+(\overrightarrow{A}_{0}+\Sigma_i^*\overrightarrow{a})\Sigma_i^*\phi] .
\end{equation}
Par conséquent, vu la définition de $\Sigma_i^*$ et le fait que $\Sigma_i$ est un endomorphisme orthogonal on a
\begin{equation}
\Vert i\overrightarrow{\nabla}\phi+(\overrightarrow{A}_{0}+\overrightarrow{a})\phi\Vert^2(\Sigma_i(x,y))
=\Vert i\overrightarrow{\nabla}\Sigma_i^*\phi+(\overrightarrow{A}_{0}+\Sigma_i^*\overrightarrow{a})\Sigma_i^*\phi\Vert^2(x,y) .
\end{equation}
On obtient de la m\^eme façon 
\begin{equation}
\left\lbrace\begin{array}{l}
(\lambda-|\phi|^2)^2(\Sigma_i(x,y))=(\lambda-|\Sigma_i^*\phi|^2)^2(x,y),\\
|\rot\,\overrightarrow{a}|^2(\Sigma_i(x,y))=|\rot\,\Sigma_i^*\overrightarrow{a}|^2(x,y).
\end{array}\right.
\end{equation}
Par conséquent, on a
\begin{equation}
\begin{array}{rcl}
F_{\lambda,k}(\Sigma_i^*\phi,\Sigma_i^*\overrightarrow{a})
&=&\int_{\Omega}\frac{1}{2}\Vert i\overrightarrow{\nabla}\phi+(\overrightarrow{A}_{0}+\overrightarrow{a})\phi\Vert^2(\Sigma_i(x,y))dx\,dy\\
&+&\int_{\Omega}\frac{1}{4}(\lambda-|\phi|^2)^2(\Sigma_i(x,y))+\frac{k^2}{2}|\rot\,\overrightarrow{a}|^2(\Sigma_i(x,y))dx\,dy\\[2mm]
&=&\int_{\Sigma_i\Omega}\{\frac{1}{2}\Vert i\overrightarrow{\nabla}\phi+(\overrightarrow{A}_{0}+\overrightarrow{a})\phi\Vert^2+\frac{1}{4}(\lambda-|\phi|^2)^2\\
&+&\frac{k^2}{2}|\rot\,\overrightarrow{a}|^2\}|\determinant\,\Sigma_i|dx\,dy\\[2mm]
&=&F_{\lambda,k}(\phi,\overrightarrow{a}) .
\end{array}
\end{equation}
On a donc la conclusion. $\hfill{\bf CQFD}$

\subsection{Conséquences}
\begin{definition}.
On dit que le réseau \index{$m_c$}de coordonnées $m_c=(1,0)$ est le réseau carré.\\
On dit que le réseau \index{$m_h$}de coordonnées $m_h=(\sqrt[4]{\frac{3}{4}},\frac{1}{2}\sqrt[4]{\frac{4}{3}})$ est le réseau hexagonal.
\end{definition}
Voir le Tome III de \cite{Siegel}, p.~159-172 o\`u l'on parle des réseaux carré et hexagonal pour les formes automorphes.

\begin{theorem}.\label{but-de-ce-fatras}
Soit $g$ une fonction définie sur ${\cal M}$ de classe $C^1$ invariante par le groupe $G$. Les réseaux carré et hexagonal sont alors des points critiques pour $g$.
\end{theorem}
{\it Preuve.} On rappelle que $\sigma_1$ et $\sigma_2$ sont définis aux équations~(\ref{definition-sym1}) et~(\ref{definition-sym2}). Les coordonnées du réseau carré vérifient 
\begin{equation}
\begin{array}{l}
\sigma_{1}(m_c)=m_c,\\
\sigma_{2}(m_c)=m_c.
\end{array}
\end{equation}
Le réseau hexagonal vérifie 
\begin{equation}
\begin{array}{l}
\sigma_{1}(\tau(m_h))=m_h,\\
\sigma_{2}(m_h)=m_h.
\end{array}
\end{equation}
La fonction $g$ étant invariante par $\tau$, $\sigma_1$ et $\sigma_2$ elle vérifie
\begin{equation}\label{relation-invariance}
\left\lbrace\begin{array}{l}
g(u, -w)=g(u,w)\\
g(u,w+\frac{1}{u})=g(u,w)\\
g(\frac{1}{\sqrt{w^2+u^2}},\frac{w}{u\sqrt{w^2+u^2}})=g(u, w) .
\end{array}\right.
\end{equation}
On différencie les relations~(\ref{relation-invariance}) et on les évaluent sur le réseau carré
\begin{equation}
\left\lbrace\begin{array}{l}
-\frac{\partial g}{\partial w}(1,0)=\frac{\partial g}{\partial w}(1,0) ,\\
-\frac{\partial g}{\partial u}(1,0)=\frac{\partial g}{\partial u}(1,0) .
\end{array}\right.
\end{equation}
Cela entra\^ine l'annulation des dérivées partielles en $(1,0)$. Le réseau carré est donc un point critique pour la fonction $g$.\\
On différencie les relations~(\ref{relation-invariance}) et on les évaluent sur le réseau hexagonal qui correspond à $m_h=(u_{h}, w_{h})=(\sqrt[4]{\frac{3}{4}}, \frac{1}{2}\sqrt[4]{\frac{4}{3}})$
\begin{equation}\label{equation-hexagonale}
\left\lbrace\begin{array}{l}
-\frac{\partial g}{\partial w}(u_{h}, w_{h})=\frac{\partial g}{\partial w}(u_{h}, w_{h}) ,\\
\frac{\partial g}{\partial w}(u_{h}, w_{h})=-w_{h}u_{h}^3\frac{\partial g}{\partial u}(u_{h}, w_{h})+u_{h}^4\frac{\partial g}{\partial w}(u_{h}, w_{h}) .
\end{array}\right.
\end{equation}
Les équations~(\ref{equation-hexagonale}) entraînent la nullité des dérivées partielles pour les coordonnées $(u_{h},w_{h})$. Ce point est donc critique pour $g$.$\hfill{\bf CQFD}$
\begin{theorem}.
Soit $m\in{\cal M}$. Il existe un élément $g\in G$ tel que $g(m)=(u,w)$ vérifie les conditions
\begin{equation}
\begin{array}{l}
0\leq w \leq \frac{r}{2},\\
r\leq \sqrt{w^2+u^2} .
\end{array}
\end{equation}
Ces relations impliquent l'inégalité $r\leq \sqrt[4]{\frac{4}{3}}$.
\end{theorem}
On peut lire le Tome III de \cite{Siegel} pour des calculs similaires sur le groupe modulaire.\\
{\it Preuve.} Soit donc $(u, w)$ un point du réseau. Modulo la transformations $\tau$ on peut supposer que $|w|\leq \frac{r}{2}$.\\
En utilisant $\sigma_{2}$ on peut se placer systématiquement dans le cas où $\Vert v_{2}\Vert \leq \Vert v_{1}\Vert $. On a alors
\begin{equation}
\left\lbrace\begin{array}{rcl}
r^2&\geq&w^2+\frac{1}{r^2}\\
&\geq&\frac{r^2}{4}+\frac{1}{r^2} .
\end{array}\right.
\end{equation}
Cela entraîne donc $\frac{3}{4}r^2\geq \frac{1}{r^2}$.\\
Ceci s'écrit plus simplement sous la forme $r\leq \sqrt[4]{\frac{4}{3}}$.\\
Ensuite une application éventuelle de $\sigma_{1}$ permet d'obtenir $w\geq 0$.$\hfill{\bf CQFD}$
\begin{remarque}.
Puisque l'expérience physique fait appara\^itre un système de cellules hexagonales, on devrait pouvoir montrer que les énergies minimales sont atteintes par des réseaux hexagonaux.\\
Nous n'avons pas pu montrer ceci m\^eme si nous avons des résultats partiels comme la proposition \ref{invariance-FS-Fas-I-K}. Il existe d'autres études locale des énergies des solutions bifurqués, par exemple \cite{almogI}.\\
\end{remarque}
Dans la suite de la thèse, on utilisera assez rarement la géométrie du réseau, par conséquent on n'indiquera pas $(u, w)$ quand ce n'est pas utile. On voit également que l'espace des réseaux quotienté par $\tau$, $\sigma_i$ est un espace localement compact non compact, la suite $(u_n,w_n)=(\frac{1}{n},0)$ ne convergeant vers aucun réseau (géométriquement on peut voir ce réseau comme un rectangle qui s'allonge de plus en plus et se rétrécie de plus en plus en gardant son aire constante).

\chapter{Étude de la fonctionnelle $F_{\lambda,k}$}\label{chapitre-generalite-sur-Flambda-k}
\noindent Dans ce chapitre nous étudions la fonctionnelle $F_{\lambda,k}$ définie à l'équation (\ref{expression-Flambda-k}); nous montrons l'existence de solutions minimisantes ainsi que leur régularité.\\
Nous étudions comment évoluent les minima $m_F$ et $m_D$ selon $\lambda$ et $k$.\\
A la fin du chapitre nous décrivons complètement la situation si $\lambda\leq 2\pi$ et $k\geq \frac{1}{\sqrt{2}}$.

\saction{Quelques opérateurs différentiels linéaires}\label{sec:operateur-lineaire}
\noindent On rappelle que l'on note ${\cal L}$ le réseau engendré par les deux vecteurs 
\begin{equation}
\left\lbrace\begin{array}{rcc}
v_{1}&=&(r,0)\\
v_{2}&=&(w,u) .
\end{array}\right.
\end{equation}
Comme à la section \ref{sec:modelisation-finale}, la quantification est choisie égale à $1$; avec nos notations (voir la sous section \ref{choix-geometrie}), cela s'écrit $ru=1$.\\
On définit au dessus du tore $\Tore{\cal L}$ l'espace fonctionnel \index{$L^{2}_{\divergence, 0}(\Tore{\cal L}; \R^2)$}suivant:
\begin{equation}\label{espace-a}
\begin{array}{rcl}
L^{2}_{\divergence, 0}(\Tore{\cal L}; \R^2)&=&\lbrace \overrightarrow{a}:\Tore{\cal L}\mapsto \R^2,\,\overrightarrow{a}\in L^2(\Tore{\cal L}),\\
&&\divergence\,\overrightarrow{a}=0\mbox{~et~}\int_{\Omega}\overrightarrow{a}=0 \rbrace
\end{array}
\end{equation}
o\`u $\divergence\,\overrightarrow{a}=0$ est une équation au sens des distributions.\\
On note \index{$C^{\infty}(E_{1})$}$C^{\infty}(E_{1})$ l'espace des sections $C^{\infty}$ du fibré $E_{1}$ défini au théorème~\ref{il-sagit-dun-fibre}. On note aussi \index{$C^{\infty}_{\divergence, 0}(\Tore{\cal L}; \R^2)$}$C^{\infty}_{\divergence, 0}(\Tore{\cal L}; \R^2)$ le sous espace de $L^{2}_{\divergence, 0}(\Tore{\cal L}; \R^2)$ formé des potentiels vecteurs $C^{\infty}$ de divergence nulle et de moyenne nulle sur $\Omega$. On définit de manière similaire les espaces $H^{r}$.\\
Dans cette section, on étudie l'opérateur $H$\index{$H$} défini sur les sections $C^{\infty}$ de $E_{1}$ par:
\begin{equation}\label{schrodinger-magnetique}
\begin{array}{rcl}
H:C^{\infty}(E_{1})&\longmapsto&C^{\infty}(E_{1})\\
\phi&\longmapsto&[i\overrightarrow{\nabla}+\overrightarrow{A}_{0}]^*[i\overrightarrow{\nabla}+\overrightarrow{A}_{0}]\phi
\end{array}
\end{equation}
et l'opérateur $L$\index{$L$} défini sur l'espace $C^{\infty}_{\divergence, 0}(\Tore{\cal L}; \R^2)$ par:
\begin{equation}\label{double-rotationnel}
\begin{array}{rcl}
L:C^{\infty}_{\divergence, 0}(\Tore{\cal L}; \R^2)&\longmapsto&C^{\infty}_{\divergence, 0}(\Tore{\cal L}; \R^2)\\
\overrightarrow{a}&\longmapsto&\rot^{*}\rot\, \overrightarrow{a} .
\end{array}
\end{equation}
Le deuxième opérateur est effectivement bien défini sur $C^{\infty}_{\divergence, 0}(\Tore{\cal L}; \R^2)$; Le potentiel $L\overrightarrow{a}$ est de moyenne nulle car $L$ est un opérateur différentiel d'ordre $2$ à coefficients constants. $L\overrightarrow{a}$ est de divergence nulle car l'opérateur de divergence commute avec $\rot^{*}\rot$.\\
On déterminera ici le spectre d'extensions autoadjointes convenables (notées aussi $H$ et $L$) de ces opérateurs.

\subsection{L'opérateur de Schrödinger magnétique, $H$}
\noindent Dans cette section, on rappelle le calcul du spectre de l'opérateur $H$ en utilisant la méthode de l'article \cite{Barany-Golubitsky}.\\
On rappelle que 
\begin{equation}
\left\lbrace\begin{array}{l}
\partial_{\overline{z}}=\frac{1}{2}(\frac{\partial}{\partial x}+i\frac{\partial}{\partial y}),\\
\partial_{z}=\frac{1}{2}(\frac{\partial}{\partial x}-i\frac{\partial}{\partial y}) .
\end{array}\right.
\end{equation}
Notons \index{$L_+$}\index{$L_-$}
\begin{equation}
\left\lbrace\begin{array}{l}
L_{+}=\partial_{\overline{z}}+\frac{\pi}{2}z=\frac{-i}{2}[i\frac{\partial}{\partial x}-\pi y]+\frac{1}{2}[i\frac{\partial}{\partial y}+\pi x]\\
\mbox{~et~}L_{-}=\partial_{z}-\frac{\pi}{2}\overline{z}=\frac{-i}{2}[i\frac{\partial}{\partial x}-\pi y]-\frac{1}{2}[i\frac{\partial}{\partial y}+\pi x] .
\end{array}\right.
\end{equation}
Ce sont des opérateurs elliptiques d'ordre $1$. On vérifie facilement que $L_{\pm}$ envoient $H^r(E_{1})$ dans $H^{r-1}(E_{1})$, pour tout $r$.\\
Quelques calculs permettent de prouver les formules suivantes
\begin{equation}\label{Oscillateur-harmonique}
\left\lbrace
\begin{array}{l}
[L_{-},L_{+}]=\pi,\\
H=-4L_{-}L_{+}+2\pi,\\
L_{-}=-(L_{+})^* .
\end{array}
\right.
\end{equation}

\begin{theorem}.\label{spectre-de-H-C-infinite}
Le spectre de l'extension autoadjointe de $H$ est de la forme $\lambda_{n}=2\pi+4\pi n$ avec $n\in \N$. Les valeurs propres $\lambda_n$ sont simples de vecteur propre associé \index{$\phi_n$}$\phi_{n}=\frac{1}{\sqrt{{\pi}^n n!}}(L_{-})^n\phi_{0}$ o\`u $\phi_0$ vérifie \index{$\phi_0$}$L_+\phi_0=0$.
\end{theorem}
{\it Preuve.} L'opérateur $H$ est symétrique sur l'espace des sections $C^{\infty}$ de $E_{1}$.  C'est un opérateur différentiel elliptique d'ordre $2$ sur le tore $\Tore{\cal L}$ qui est une variété compacte.  Donc  l'opérateur $H$ est essentiellement autoadjoint à partir de $C^{\infty}(E_1)$ et le domaine de son extension autoadjointe est \index{${\cal D}(H)$}${\cal D}(H)=H^{2}(E_{1})$ (voir par exemple \cite{spin-geom}, théorème III.5.8). De plus, l'injection de $H^{2}(E_1)$ dans $L^2(E_1)$ étant compacte, son spectre est discret.\\
Puisque $H$ est un opérateur différentiel elliptique, ses fonctions propres sont $C^{\infty}$. Par ailleurs l'opérateur~$H$  est positif puisque 
\begin{equation}
\langle H\phi, \phi\rangle=\int_{\Omega}\Vert i\overrightarrow{\nabla}\phi+ \overrightarrow{A}_0\phi\Vert^2\geq 0
\end{equation}
si $\phi$ est $C^{\infty}$; cette propriété s'étend à ${\cal D}(H)$ par continuité et densité.\\
Soit $\phi\in C^{\infty}(E_{1})$ une fonction de norme $1$. On peut donc effectuer le petit calcul suivant
\begin{equation}\label{little-calculus}
\begin{array}{rcl}
\langle H\phi,\phi\rangle
&=&-4\langle L_{-}L_{+}\phi,\phi\rangle+2\pi\Vert\phi\Vert_{L^2}\\
&=&4\Vert L_{+}\phi\Vert ^2_{L^{2}(E_{1})}+2\pi\Vert\phi\Vert_{L^2}\\
&\geq& 2\pi\Vert\phi\Vert_{L^2}=2\pi .
\end{array}
\end{equation}
\begin{lemme}.\label{lemme-de-non-degenerescence}
Le noyau de $L_{+}$ est de dimension $1$.
\end{lemme}
\begin{remarque}.
Plus généralement, en quantification $d$ avec $d>0$, la dimension de $Ker\,L_+$ est $d$.
\end{remarque}
\begin{remarque}.
Certains auteurs (\cite{lasher}, repris par \cite{odehII}) laissent parfois entendre que la multiplicité de la plus petite valeur propre dépend de la nature de ${\cal L}$. C'est sans doute du au fait qu'ils ne précisaient pas sur quel fibré ils travaillent.
% Nous montrons ici qu'il n'en est rien.
\end{remarque}
Soit $\phi_0$ dans le noyau de $L_+$. Le vecteur $\phi_{0}$ vérifie $L_{+}\phi_{0}=0$.\\
Posons $\phi_{0}=e^{-(x^2+y^2)\frac{\pi}{2}}g$. La condition $L_{+}\phi_{0}=0$ équivaut aux relations de Cauchy-Riemann pour $g$. La fonction $g$ est donc holomorphe.\\
Sur le plan complexe, les relations de périodicité s'écrivent, en notant:\\
$t_1=r=\frac{1}{u}$ et $t_2=w+iu$,
\begin{equation}\label{periodicite-phi}
\phi_{0}(z+t_i)=\exp[\frac{\pi}{2}(\overline{t_i}z-t_i\overline{z})]\phi_{0}(z)
\end{equation}
et
\begin{equation}\label{periodicite-g}
g(z+t_i)=\exp(\frac{\pi}{2}\overline{t_i}(t_i+2z))g(z) .
\end{equation}
On pose \index{$t_s$}
\begin{equation}
\left\lbrace\begin{array}{l}
g(z)=\exp(\frac{\pi}{2}z^2)u(\frac{z}{r}),\\
t_s=\frac{w+iu}{r}=wu+iu^2
\end{array}\right.
\end{equation}
et on obtient
\begin{equation}\label{periodicite-u}
\left\lbrace\begin{array}{l}
u(z+1)=u(z)\\
u(z+t_s)=\exp(-2\pi i[z+\frac{t_s}{2}])u(z) .
\end{array}\right.
\end{equation}
Il existe une fonction holomorphe unique $h$\index{$h$} définie sur $\C^*$ définie par $u(z)=h(e^{2\pi iz})$. La fonction $h$ étant holomorphe sur $\C^*$ on peut faire un développement en série de Laurent
\begin{equation}
\left\lbrace\begin{array}{l}
h(v)=\sum_{n\in \Z}a_{n}v^n\\
u(z)=\sum_{n\in \Z}a_{n}e^{2\pi inz} .
\end{array}\right.
\end{equation}
La deuxième condition de l'équation~(\ref{periodicite-u}) sur $u$ est équivalente à l'équation
\begin{equation}
a_ne^{2\pi in t_s}=a_{n+1}e^{-\pi i t_s}
\end{equation}
qui se résout et donne la forme suivante pour $u$
\begin{equation}\label{serie-de-u}
u(z)=C\sum_{n\in \Z}e^{\pi it_s n^2+2\pi i n z} .
\end{equation}
Cette série converge dans $H^{2}(E_{1})$.\\
On peut déterminer la constante $C$ à une phase près en imposant la condition $\Vert \phi_0\Vert_{L^2}=1$. $\hfill{\bf CQFD}$\\
\\
Les solutions de l'équation $H\phi=2\pi\phi$ correspondent par les calculs (\ref{little-calculus}) et (\ref{Oscillateur-harmonique}) aux solutions de $L_+\phi=0$. Ces solutions forment un espace vectoriel de dimension $1$; donc on a bien prouvé que la dégénérescence du niveau fondamental est égale à $1$.
%Le calcul~(\ref{little-calculus}) nous dit que le niveau fondamental $\lambda_{0}$ est supérieur ou égal à $2\pi$. Nous allons voir que \mbox{$dim\lbrace Ker\,L_{+}\rbrace=1$} ce qui prouvera l'égalité $\lambda_{0}=2\pi$. On prendra alors pour $\phi_{0}$ un générateur de norme $1$ de cet espace vectoriel.

Si $\phi$ est un vecteur propre non trivial pour $H$ de valeur propre $\lambda$ alors $L_{-}\phi$ est vecteur propre non trivial de $H$ de valeur propre $\lambda+4\pi$ (ceci peut-\^etre démontré en utilisant les propriétés~(\ref{Oscillateur-harmonique}) qui sont celles d'un oscillateur harmonique).\\
Par contre $L_{+}\phi$ n'est un vecteur propre non trivial de $H$ de valeur propre $\lambda-4\pi$ que si $\lambda\not=2\pi$.\\
Le théorème spectral affirme que $L^2(E_{1})$ est somme hilbertienne des espaces propres de $H$. Le calcul précédent montre que $L_{-}^n\phi_{0}$ est vecteur propre de l'opérateur $H$ de valeur propre $2\pi+4\pi n$, donc que l'ensemble $2\pi+4\pi\N$ est inclus dans le spectre de $H$.\\
En appliquant l'opérateur $L_{+}$ on peut prouver l'inclusion inverse. Le théorème est prouvé.$\hfill{\bf CQFD}$

\begin{theorem}.\label{etat-non-trivial}
Si $\lambda>2\pi$, les couples minimisants la fonctionnelle $F_{\lambda,k}$ sont différents du couple $(0,0)$.
\end{theorem}
{\it Preuve.} Soit la courbe formée des couples $(t\phi_{0}, 0)_{t\in \R}$; alors 
\begin{equation}
\frac{d}{dt}F_{\lambda,k}(t\phi_{0}, 0)(t=0)=0 .
\end{equation}
Il faut donc calculer \`a l'ordre $2$ pour obtenir le résultat

\begin{eqnarray}\label{derivee-seconde}
\frac{d^2}{dt^2}F_{\lambda,k}(t\phi_{0}, 0)(t=0) &=& \int_{\Omega}\Vert i\overrightarrow{\nabla}\phi_{0}+\overrightarrow{A}_{0}\phi_{0}\Vert^2-\lambda|\phi_{0}|^2\nonumber\\
&=& \langle H\phi_{0}, \phi_{0}\rangle-\lambda\langle\phi_{0}, \phi_{0}\rangle\nonumber\\
&=& 2\pi-\lambda<0 .
\end{eqnarray}
Le  Hessien de la fonctionnelle en $(0,0)$ possède donc une valeur propre strictement négative, il est donc impossible que ce soit un minimum. $\hfill{\bf CQFD}$

\begin{proposition}.\label{unique-zero}
La section $\phi_0$ possède un unique zéro dans la variété $\Tore{\cal L}$.
\end{proposition}
{\it Preuve.} La fonction $\phi_0$ a pour expression $\phi_0(x,y)=e^{-(x^2+y^2)\frac{\pi}{2}}g(x+iy)$ où $g$ est holomorphe.\\
L'ensemble des zéros de $\phi_0$ est égal à l'ensemble des zéros de $g$. Or cet ensemble est discret car $g$ est holomorphe et non identiquement nulle. Donc il existe $z_s\in\C$ tel que le parallélogramme \index{${\cal P}$}${\cal P}$ de sommets
\begin{equation}
z_s,\,\,z_s+r,\,\,z_s+r+(w+iu)\mbox{~et~}z_s+(w+iu)
\end{equation}
ne rencontre aucun zéro de la fonction $g$.\\
Le nombre de zéros de la fonction $g$ situé à l'intérieur du parallélogramme ${\cal P}$ est fini et est donné par le théorème de Rouché-Fontené sous la forme suivante:
\begin{equation}\label{Nombre-de-zeros}
\frac{1}{2\pi i}\int_{\cal P}\frac{g'(z)}{g(z)}dz .
\end{equation}
En utilisant l'équation (\ref{periodicite-g}), on obtient si $t\in{\cal L}$
\begin{equation}
\frac{g'(z+t)}{g(z+t)}=\frac{g'(z)}{g(z)}+\pi\bar{t}
\end{equation}
On obtient donc 
\begin{equation}
\begin{array}{rcl}
\int_{\cal P}\frac{g'(z)}{g(z)}dz&=&\int_0^1\frac{g'(z_s+rh)}{g(z_s+rh)}rdh-\int_0^1\frac{g'(z_s+(w+iu)+rh)}{g(z_s+(w+iu)+rh)}rdh\\
&+&\int_0^1\frac{g'(z_s+r+(w+iu)h)}{g(z_s+r+(w+iu)h)}(w+iu)dh-\int_0^1\frac{g'(z_s+(w+iu)h)}{g(z_s+(w+iu)h)}(w+iu)dh\\[2mm]
&=&\int_0^1[\frac{g'(z_s+rh)}{g(z_s+rh)}-\frac{g'(z_s+(w+iu)+rh)}{g(z_s+(w+iu)+rh)}]rdh\\
&+&\int_0^1[\frac{g'(z_s+r+(w+iu)h)}{g(z_s+r+(w+iu)h)}-\frac{g'(z_s+(w+iu)h)}{g(z_s+(w+iu)h)}](w+iu)dh\\[2mm]
&=&\int_0^1[-\pi\overline{w+iu}]rdh+\int_0^1[\pi\overline{r}](w+iu)dh\\
&=&-\pi(w-iu)r+\pi r(w+iu)\\
&=&-\pi wr+i\pi+\pi rw+i\pi\\
&=&2i\pi\, .
\end{array}
\end{equation}
Par conséquent, l'intégrale~(\ref{Nombre-de-zeros}) est égale à $1$. On a bien un unique zéro pour la fonction $g$ dans le parallélogramme ${\cal P}$. Donc en passant au quotient, la section $\phi_0$ a un unique zéro sur $\Tore{\cal L}$. $\hfill{\bf CQFD}$
\begin{remarque}
Dans le preprint \cite{Takac} il est montré que les solutions bifurquées définies dans le chapitre IV ont un seul zéro. Il est aussi montré que ce zéro est dans le cas d'un réseau triangulaire ou carré situé à un endroit que l'on peut expliciter de $\Tore{\cal L}$.\\
Dans l'article \cite{matamo-qi} il est montré que les zéros de $\phi$ sont isolés dans le cas d'un ouvert simplement connexe de $\R^2$.
\end{remarque}

\subsection{L'opérateur $L$}\label{sec:operateur-L}
\noindent L'opérateur $L$ défini à l'équation~(\ref{double-rotationnel}) a pour expression 
\begin{equation}
L\left(\begin{array}{c}
a_{x}\\
a_{y}
\end{array}\right)=\left(\begin{array}{c}
-\frac{\partial}{\partial y}(\frac{\partial a_x}{\partial y}-\frac{\partial a_y}{\partial x})\\
\frac{\partial}{\partial x}(\frac{\partial a_x}{\partial y}-\frac{\partial a_y}{\partial x})
\end{array}\right) .
\end{equation}
On vérifie facilement que l'opérateur $L$ envoie l'espace $C^{\infty}_{\divergence, 0}(\Tore{\cal L}; \R^2)$  dans lui même.\\
L'espace $L^2_{\divergence,0}(\Tore{\cal L}; \R^2)$ est l'espace des potentiels vecteurs ${\cal L}$-périodiques sur $\R^2$ localement intégrable d'intégrale nulle sur $\Omega$ et de divergence distributionelle nulle. On définit de la m\^eme façon \index{$H^2_{\divergence,0}(\Tore{\cal L}; \R^2)$}$H^2_{\divergence,0}(\Tore{\cal L}; \R^2)$.
\begin{proposition}.\label{spectre-L-theo}
L'extension autoadjointe de l'opérateur $L$ a pour domaine $H^2_{\divergence, 0}(\Tore{\cal L}; \R^2)$. Son spectre est discret et on a l'égalité
\begin{equation}
sp(L)=\{\lambda\in\R\mbox{~tel~que~}\exists(n_1,n_2)\not=(0,0),\,\,\lambda=(2\pi)^2[(n_2u)^2+(n_{1}r+n_{2}w)^2]\} .
\end{equation}
L'opérateur $L$ est inversible.
\end{proposition}
{\it Preuve.} L'opérateur $L$ opère sur l'espace des champs de vecteurs de divergence nulle, on peut donc l'écrire sous la forme
\begin{equation}\label{Laplacien-en-veux-tu-en-voila}
L\left(\begin{array}{c}
a_{x}\\
a_{y}
\end{array}\right)=\left(\begin{array}{c}
-\Delta\, a_{x}\\
-\Delta\, a_{y}
\end{array}\right) .
\end{equation}
L'opérateur différentiel $L$ est symétrique sur l'espace $C^{\infty}_{\divergence, 0}(\Tore{\cal L}; \R^2)$. Le tore $\Tore{\cal L}$ est compact sans bord et l'opérateur $L$ est elliptique donc il est essentiellement autoadjoint (voir par exemple \cite{spin-geom}, théorème III.5.8)  et son extension autoadjointe a pour domaine \index{${\cal D}(L)$}${\cal D}(L)=H^{2}_{\divergence, 0}(\Tore{\cal L}; \R^2)$.\\
L'opérateur $-\Delta$ est autoadjoint sur $L^2(\Tore{\cal L},\C)$ de domaine $H^2(\Tore{\cal L},\C)$. Il est diagonalisé par les fonctions propres $e_{n_1,n_2}(x,y)=e^{2\pi i[x n_{2}u+y(n_{1}r+n_{2} w)]}$ de valeur propres $(2\pi)^2[(n_2u)^2+(n_{1}r+n_{2}w)^2]$.\\
Par la formule (\ref{Laplacien-en-veux-tu-en-voila}) L'opérateur $L$ est la restriction de l'opérateur $(-\Delta,-\Delta)$ à l'espace $H^{2}_{\divergence, 0}(\Tore{\cal L}; \R^2)$:
%\begin{equation}
%\left\lbrace\begin{array}{rcl}
%(-\Delta,-\Delta)H^2(\Tore{\cal L},\C)\times H^2(\Tore{\cal L},\C)&\mapsto&L^2(\Tore{\cal L},\C)\times L^2(\Tore{\cal L},\C)\\
%(f,g)&\mapsto&(-\Delta f,-\Delta g)
%\end{array}\right.
%\end{equation}
\begin{equation}
\begin{array}{rcl}
(f,g)&\mapsto&(-\Delta f,-\Delta g) .
\end{array}
\end{equation}
Si $(n_1,n_2)=(0,0)$ on prend pour base orthonormée de $\R^2$ $\overrightarrow{A}_{n_1,n_2}=(1,0)$ et $\overrightarrow{B}_{n_1,n_2}=(1,0)$. Si $(n_1,n_2)\not=(0,0)$ on prend pour base orthonormée de $\R^2$
\begin{equation}
\begin{array}{c}
\overrightarrow{A}_{n_1,n_2}=\frac{1}{\sqrt{(n_2w+n_1r)^2+(n_2 u)^2}}\left(\begin{array}{c}
n_2w+n_1r\\
-n_2 u
\end{array}\right)\mbox{~et~}\\
\overrightarrow{B}_{n_1,n_2}=\frac{1}{\sqrt{(n_2w+n_1r)^2+(n_2 u)^2}}\left(\begin{array}{c}
n_2 u\\
n_2w+n_1r
\end{array}\right) .
\end{array}
\end{equation}
Les champs de vecteurs $\overrightarrow{A}_{n_1,n_2}e_{n_1,n_2}(x,y)$ et $\overrightarrow{B}_{n_1,n_2}e_{n_1,n_2}(x,y)$ forment donc une base hilbertienne de $L^2(\Tore{\cal L},\C)\times L^2(\Tore{\cal L},\C)$.\\
Tout champ de vecteur $\overrightarrow{V}$ de $L^2(\Tore{\cal L},\C)\times L^2(\Tore{\cal L},\C)$ peut s'écrire sous la forme 
\begin{equation}
\overrightarrow{V}=\sum_{n_1,n_2}\alpha_{n_1,n_2}\overrightarrow{A}_{n_1,n_2}e_{n_1,n_2}+\beta_{n_1,n_2}\overrightarrow{B}_{n_1,n_2}e_{n_1,n_2} .
\end{equation}
Maintenant, si le champ de vecteur $\overrightarrow{V}$ appartient à $L^{2}_{\divergence, 0}(\Tore{\cal L}; \R^2)$, on a en plus les conditions
\begin{equation}
\left\lbrace\begin{array}{l}
\int_{\Omega}\overrightarrow{V}=0, \\
\divergence\,\overrightarrow{V}=0, \\
\overrightarrow{V}\mbox{~champ~de~vecteur~réel.}
\end{array}\right.
\end{equation}
La première condition nous donne $\alpha_{0,0}=\beta_{0,0}=0$.\\
Si le champ $\overrightarrow{V}$ appartient à $H^1$, alors on peut calculer sa divergence:
\begin{equation}
\begin{array}{rcl}
\divergence\,\overrightarrow{V}&=&\sum_{n_1,n_2}\frac{\alpha_{n_1,n_2}[2\pi in_2u(n_2w+n_1r)+2\pi i(n_1r+n_2w)(-n_2u)]}{\sqrt{(n_2w+n_1r)^2+(n_2 u)^2}}e_{n_1,n_2}\\
&+&\sum_{n_1,n_2}\frac{\beta_{n_1,n_2}[2\pi in_2u(n_2u)+2\pi i(n_1r+n_2w)(n_2w+n_1r)]}{\sqrt{(n_2w+n_1r)^2+(n_2 u)^2}}e_{n_1,n_2}\\
&=&\sum_{n_1,n_2}2\pi i\beta_{n_1,n_2}\sqrt{(n_2w+n_1r)^2+(n_2 u)^2}e_{n_1,n_2}
\end{array}
\end{equation}
Par conséquent, si $\divergence\,\overrightarrow{V}=0$, alors $\beta_{n_1,n_2}=0$ si $(n_1,n_2)\not=(0,0)$.\\
Le champ de vecteur $\overrightarrow{V}$ étant réel, cela nous donne $\overline{\alpha_{n_1,n_2}}=\alpha_{-n_1,-n_2}$.\\
Cela montre que les fonctions
\begin{equation}\label{fonctions-propres}
\left\lbrace\begin{array}{c}
\overrightarrow{b}_{n_{1}, n_{2}}(x,y)=\sqrt{2}\overrightarrow{A}_{n_1,n_2}\cos(2\pi[x n_{2}u+y(n_{1}r+n_{2} w)])\\
\overrightarrow{c}_{n_{1}, n_{2}}(x,y)=\sqrt{2}\overrightarrow{A}_{n_1,n_2}\sin(2\pi[x n_{2}u+y(n_{1}r+n_{2} w)])
\end{array}\right.
\end{equation}
forment une base hilbertienne de $L^{2}_{\divergence, 0}(\Tore{\cal L}; \R^2)$ avec $(n_{1},n_{2})\in \Z\times\Z-(0,0)$.\\
Par ailleurs les fonctions $\overrightarrow{b}_{n_{1}, n_{2}}$ et $\overrightarrow{c}_{n_{1}, n_{2}}$ sont fonctions propres de ${\cal L}$ de valeur propre $(2\pi)^2[(n_2u)^2+(n_{1}r+n_{2}w)^2]$.\\
Cette valeur propre est donc au moins double. La multiplicité de cette valeur propre peut \^etre plus grande que $2$ car $(n_2u)^2+(n_{1}r+n_{2}w)^2$ peut à priori \^etre atteint par différents couples $(n_1,n_2)$.\\
Le réel $0$ n'est pas dans le spectre de $L$; par conséquent le spectre étant discret l'opérateur $L$ est inversible.\\
Les valeurs propres de $L$ peuvent se réécrire $(2\pi)^2\Vert n_{1}v_{1}+n_{2}v_{2}\Vert ^2$.
% On s'est restreint sans perdre en généralité (voir la section~\ref{sec:struct-reseau-space}) au cas o\`u 
Si $w\geq 0$ et $\Vert v_{2}\Vert \geq \Vert v_{1}\Vert $ alors le minimum de $(2\pi)^2[(n_2u)^2+(n_{1}r+n_{2}w)^2]$ avec $(n_1,n_2)\not=(0,0)$ est égal à $(2\pi r)^2$.\\
Pour les réseaux carré et hexagonal, cela nous donne:
\begin{equation}
\left\lbrace\begin{array}{ccl}
\mbox{~modèle~carré:~}&r=1\mbox{~et~}w=0&\lambda_{min}^{car}=(2\pi)^2;\\
\mbox{~modèle~hexagonal:~}&r=(\frac{4}{3})^{1/4}\mbox{~et~}w=\frac{r}{2}&\lambda_{min}^{h}=(2\pi)^2\sqrt{\frac{4}{3}} .
\end{array}\right.
\end{equation}
$\hfill{\bf CQFD}$

\begin{lemme}.\label{lemme-laplacien}
Il existe une constante $c_0>0$ telle que, pour toute fonction $f_0\in L^2(\Tore{\cal L})$ avec $\int_{\Omega}f_0=0$, il existe une unique solution $u_0\in H^{2}(\Tore{\cal L})$ des équations
\begin{equation}
-\Delta u_0=f_0\mbox{~et~}\int_{\Omega}u_0=0 .
\end{equation}
Cette solution vérifie:
\begin{equation}
\Vert u_0\Vert_{L^{\infty}}\leq c_0\Vert f_0\Vert_{L^2} .
\end{equation}
\end{lemme}
{\it Preuve.} L'application
\begin{equation}
\begin{array}{rcl}
-\Delta: C^{\infty}_0(\Tore{\cal L},\R)&\mapsto&C^{\infty}_0(\Tore{\cal L},\R)\\
f&\mapsto&-\Delta\,f
\end{array}
\end{equation}
o\`u $C^{\infty}_0(\Tore{\cal L},\R)$ est l'espace des fonctions $C^{\infty}$, ${\cal L}$-périodique d'intégrale nulle sur $\Omega$ est symétrique. Le tore $\Tore{\cal L}$ est compact sans bord et l'opérateur $-\Delta$ est elliptique donc il est essentiellement autoadjoint (voir par exemple \cite{spin-geom}, théorème III.5.8) et son extension autoadjointe encore noté $-\Delta$ a pour domaine ${\cal D}(-\Delta)=H^2_0(\Tore{\cal L},\R)$.\\
L'injection $H^2_0(\Tore{\cal L})\hookrightarrow L^2_0(\Tore{\cal L})$ nous indique que la fonction $(-\Delta+i)^{-1}$ est compacte donc de spectre discret. L'opérateur $-\Delta$ possède donc un spectre discret.\\
L'égalité 
\begin{equation}
\forall f\in H^2_0(\Tore{\cal L},\R),\,\langle -\Delta\,f,f\rangle=\int_{\Omega}\Vert \overrightarrow{\nabla}f\Vert^2
\end{equation}
nous indique que le noyau de $-\Delta$ est constitué de fonctions constantes. Or notre opérateur est défini sur des fonctions de moyenne nulle, le noyau est donc nul. L'opérateur $-\Delta$ est inversible.\\
L'estimée $L^{\infty}$ provient de l'injection de sobolev $H^2(\Tore{\cal L})\hookrightarrow C^0(\Tore{\cal L})$ (voir théorème~\ref{sobolev-imbedding}). $\hfill{\bf CQFD}$

\begin{proposition}.\label{annulation-une-integrale}
Soit $\overrightarrow{a}\in H^1(\Tore{\cal L},\R^2)$ un potentiel vecteur de divergence nulle et d'intégrale nulle sur $\Omega$. Si $\rot\,\overrightarrow{a}=0$ alors $\overrightarrow{a}=0$
\end{proposition}
{\it Preuve.} Le domaine de l'opérateur $L$ est $H^2_{\divergence,0}(\Tore{\cal L},\R^2)$. La forme quadratique associée est
\begin{equation}
\left\lbrace\begin{array}{rcl}
q_L:H^{1}_{\divergence,0}(\Tore{\cal L},\R^2)&\mapsto&\R\\
\overrightarrow{a}&\mapsto&\langle L\overrightarrow{a},\overrightarrow{a}\rangle=\int_{\Omega}|\rot\,\overrightarrow{a}|^2
\end{array}\right.
\end{equation}
et elle est définie positive. Puisque le spectre de $L$ est inclus dans $[c,\infty[$ avec $c>0$, la forme quadratique $q_L$ est définie positive.\\
Si le champ de vecteur $\overrightarrow{a}$ est comme dans l'énoncé, on a 
\begin{equation}
\begin{array}{rcl}
q_L(\overrightarrow{a})
&=&\langle L\overrightarrow{a},\overrightarrow{a}\rangle\\
&=&\langle 0,\overrightarrow{a}\rangle\\
&=&0 .
\end{array}
\end{equation}
On a donc $\overrightarrow{a}=0$. $\hfill{\bf CQFD}$

\saction{Existence de solutions minimisantes pour la fonctionnelle $F_{\lambda,k}$.}\label{sec:functional-analysis}
\noindent Dans cette section, on montre que la fonctionnelle $F_{\lambda,k}$ atteint son minimum. Ces résultats sont évoqués dans l'article \cite{pirate-I} qui contient une démonstration incomplète. Ces points sont détaillés ici pour la commodité du lecteur.\\
La fonctionnelle $F_{\lambda,k}$ a été définie à l'équation~(\ref{expression-Flambda-k}) et pour $(\phi, \overrightarrow{a}) \in {\cal A}$ par
\begin{equation}\label{seconde-definition-Flambdak}
F_{\lambda,k}(\phi, \overrightarrow{a})=\frac{1}{2}\int_{\Omega}\Vert i\overrightarrow{\nabla}\phi+(\overrightarrow{A}_{0}+\overrightarrow{a})\phi\Vert^2+\frac{1}{4}\int_{\Omega}(\lambda-|\phi|^2)^2+\frac{k^2}{2}\int_{\Omega}|\rot\, \overrightarrow{a}|^2 .
\end{equation}
On montre maintenant que la fonctionnelle $F_{\lambda,k}$ est coercive sur ${\cal A}$ et que le minimum est atteint sur ${\cal A}$.

\begin{lemme}.\label{tend-vers-l-infini}
Pour tout $E$ dans $\R$, il existe $C=C(E)$ tel que, $\forall(\phi,\overrightarrow{a})\in {\cal A}$ vérifiant $F_{\lambda,k}(\phi,\overrightarrow{a})<E$, on ait
\begin{equation}
\Vert \overrightarrow{a}\Vert _{H^{1}}+\Vert \phi\Vert _{H^{1}(E_{1})}<C(E) .
\end{equation}
\end{lemme}
{\it Preuve.} L'estimée sur le potentiel $\overrightarrow{a}$ s'obtient facilement. On observe en effet que:
\begin{equation}\label{estimee-pour-A}
\int_{\Omega}\vert \rot\, \overrightarrow{a}\vert ^2\leq \frac{2}{k^2}F_{\lambda,k}(\phi, \overrightarrow{a}) .
\end{equation}
Maintenant, la norme $L^2$ de $\rot\,\overrightarrow{a}$ s'écrit aussi
\begin{eqnarray}\label{majoration-de-a}
\int_{\Omega}\vert \rot\,\overrightarrow{a}\vert ^2 &=& \langle \rot^*\rot\,\overrightarrow{a},\overrightarrow{a}\rangle\nonumber\\
&=& \langle -\Delta\, \overrightarrow{a},\overrightarrow{a}\rangle .
\end{eqnarray}
On a utilisé le fait que $\divergence\,\overrightarrow{a}=0$ à la fin du calcul~(\ref{majoration-de-a}). 
On sait aussi que $\int_{\Omega}\overrightarrow{a}=0$. L'étape $3$ du théorème~\ref{reduction-de-jauge} permet de prouver qu'il existe des constantes $C$, $C'$ telle que
\begin{equation}\label{ellipticite-du-laplacien}
C\Vert \overrightarrow{a}\Vert ^2_{H^{1}}\leq \langle -\Delta\, \overrightarrow{a}, \overrightarrow{a}\rangle \leq C'\Vert \overrightarrow{a}\Vert ^2_{H^{1}} .
\end{equation}
Donc il existe une constante $C$ telle que 
\begin{equation}\label{inegalite-pour-A}
\Vert \overrightarrow{a}\Vert _{H^1}\leq C\sqrt{F_{\lambda,k}(\phi,\overrightarrow{a})} .
\end{equation}
La première partie du lemme est prouvée.\\
On utilise ensuite le théorème~\ref{sobolev-imbedding} pour obtenir une estimée $L^4$ sur $\overrightarrow{a}$ et donc sur $\overrightarrow{A}=\overrightarrow{A}_{0}+\overrightarrow{a}$.\\
On obtient une estimée $L^4$ sur $\phi$ facilement, en écrivant
\begin{equation}\label{estimée-L4}
\int_{\Omega}|\phi|^4dxdy\leq\int_{\Omega}[2\lambda^2+2(\lambda-|\phi|^2)^2]dxdy\leq 2\lambda^2+8F_{\lambda,k}(\phi, \overrightarrow{a}) .
\end{equation}
Maintenant il faut une estimée pour $\Vert \phi\Vert _{H^{1}}$. Pour cela on développe l'énergie magnétique
\begin{equation}
\int_{\Omega}\Vert i\overrightarrow{\nabla}\phi\Vert ^2dxdy\leq 2\int_{\Omega}\Vert i\overrightarrow{\nabla}\phi+\overrightarrow{A}\phi\Vert ^2dxdy+2\int_{\Omega}\overrightarrow{A}^2|\phi|^2dxdy .
\end{equation}
On majore le dernier terme en utilisant l'inégalité de Hölder
\begin{equation}\label{majoration-terme-energie-H1}
\int_{\Omega}\overrightarrow{A}^2|\phi|^2dxdy\leq \Vert \overrightarrow{A}\Vert ^{2}_{L^4} \Vert \phi\Vert ^2_{L^4} .
\end{equation}
Donc la norme $L^2$ de  $\overrightarrow{\nabla}\phi$ est contrôlée grâce à~(\ref{inegalite-pour-A}) et~(\ref{estimée-L4}).\\
La norme $L^4$ de $\phi$ est contr\^olée; il en est de même de sa norme $L^2$. La norme $H^{1}$ de $\phi$ est donc contrôlée. $\hfill{\bf CQFD}$

\begin{lemme}.\label{resultat-standard-style-brezis}
Si $f_{n}\rightharpoonup f$ dans $L^2$ et si $g_{n}\rightarrow g$ dans $L^2$ alors $f_{n}g_{n}\rightarrow fg$ dans $L^1$.
\end{lemme}
{\it Preuve.} On écrit simplement la différence
\begin{equation}
\begin{array}{rcl}
\int_{\Omega}|f_{n}g_{n}-fg|dxdy&\leq&\int_{\Omega}|f_{n}|\times |g_{n}-g|dxdy+\int_{\Omega}(f_{n}-f)g\,dxdy\\
&\leq&\Vert f_{n}\Vert _{L^2}\Vert g_{n}-g\Vert _{L^2}+\int_{\Omega}(f_{n}-f)g\,dxdy\\
&\leq&\sup_{n}\Vert f_{n}\Vert _{L^2}\Vert g_{n}-g\Vert _{L^2}+\int_{\Omega}(f_{n}-f)g\,dxdy .
\end{array}
\end{equation}
On a $\sup_{n}\Vert f_{n}\Vert _{L^2}<\infty$ car une suite faiblement convergente est bornée (cf~\cite{brezis}, III.5(iii)).\\
Combiné avec le fait que $g_{n}$ tend vers $g$ dans $L^2$, cela implique que le terme $\int_{\Omega}f_{n}(g_{n}-g)dxdy$ tend vers $0$.\\
Le terme $\int_{\Omega}(f_{n}-f)g\,dxdy$ tend vers $0$, en vertu de la convergence faible de $f_{n}$ vers $f$. $\hfill{\bf CQFD}$

\begin{lemme}.\label{semi-continuite-inferieure-pour-la-convergence-faible}
La fonctionnelle $F_{\lambda,k}(\phi, \overrightarrow{a})$ est semicontinue inférieurement pour la topologie de la convergence faible sur \mbox{$V$}.
\end{lemme}
{\it Preuve.} Soit $(\phi_{n}, \overrightarrow{a}_{n})$ dans ${\cal A}$ convergeant faiblement vers $(\phi, \overrightarrow{a})$ dans ${\cal A}$. Il faut estimer la limite supérieure de la quantité suivante
\begin{equation}
\begin{array}{rcl}
F_{\lambda,k}(\phi_n, \overrightarrow{a}_n)
&=&\frac{1}{2}\int_{\Omega}\Vert i\overrightarrow{\nabla}\phi_n\Vert^2+\int_{\Omega}\Rez\,i\overrightarrow{\nabla}\phi_n\overline{\phi_n}(\overrightarrow{A}_0+\overrightarrow{a}_n)\phi_n\\
&+&\frac{1}{2}\int_{\Omega}(\overrightarrow{A}_0+\overrightarrow{a}_{n})^{2}|\phi_{n}|^2+\frac{1}{4}\int_{\Omega}(\lambda-|\phi_{n}|^2)^2+\frac{k^2}{2}\int_{\Omega}|\rot\,\overrightarrow{a}_n|^2 .
\end{array}
\end{equation}
Pour la topologie faible de $H^{1}$, on a
\begin{equation}\label{convergence-faible}
\left\lbrace\begin{array}{ccc}
\phi_{n}&\rightharpoonup&\phi,\\
\overrightarrow{a}_{n}&\rightharpoonup&\overrightarrow{a}.
\end{array}\right. 
\end{equation}
Grâce au théorème~\ref{sobolev-imbedding}, on a, pour tout $p\in[1,\infty[$, la convergence forte suivante:
\begin{equation}\label{norme-Lp}
\left\lbrace\begin{array}{ccl}
\phi_{n}\rightarrow \phi&\mbox{dans}&L^{p},\\
\overrightarrow{a}_{n}\rightarrow\overrightarrow{a} &\mbox{dans}&L^{p} .
\end{array}\right.
\end{equation}
Cela entraîne donc trivialement
\begin{equation}\label{second-terme}
\int_{\Omega}(\lambda-|\phi_{n}|^2)^2dxdy\rightarrow \int_{\Omega}(\lambda-|\phi|^2)^2dxdy .
\end{equation}
La convergence forte~(\ref{norme-Lp}) combinée au lemme~\ref{resultat-standard-style-brezis} entraîne aussi la convergence du terme suivant
\begin{equation}\label{convergence-du-terme-A^2phi^2}
\int_{\Omega}(\overrightarrow{A}_0+\overrightarrow{a}_{n})^{2}|\phi_{n}|^2\rightarrow\int (\overrightarrow{A}_0+\overrightarrow{a})^2|\phi|^2 .
\end{equation}
La propriété~(\ref{convergence-faible}) entraîne, pour la topologie faible de $L^2$
\begin{equation}\label{convergence-faible-pour-rot}
\left\lbrace\begin{array}{c}
\overrightarrow{\nabla}\phi_{n}\rightharpoonup \overrightarrow{\nabla}\phi ,\\
\rot\,\overrightarrow{a}_{n}\rightharpoonup \rot\,\overrightarrow{a} .
\end{array}\right.
\end{equation}
La norme $\Vert .\Vert _{L^2}:\,\, L^2\mapsto \R$ est semicontinue inférieurement pour la topologie de la convergence faible sur $L^{2}$ (cf~\cite{brezis}). Donc on obtient la semicontinuité de certains termes de la fonctionnelle
\begin{equation}\label{semi-continuite-inferieure}
\left\lbrace\begin{array}{l}
\Vert i\overrightarrow{\nabla}\phi\Vert _{L^2}\leq \underline{\lim}_{n}\Vert i\overrightarrow{\nabla}\phi_{n}\Vert _{L^2},\\
\Vert \rot\,\overrightarrow{a}\Vert _{L^2}\leq \underline{\lim}_{n}\Vert \rot\,\overrightarrow{a}_{n}\Vert _{L^2} .
\end{array}\right.
\end{equation}
On doit prouver une continuité forte sur le terme $\int_{\Omega}2\Rez[\overline{\phi}(\overrightarrow{A}_{0}+\overrightarrow{a}).i\overrightarrow{\nabla}\phi]$ pour ensuite obtenir une semicontinuité de la fonctionnelle.\\
Les sections $\overrightarrow{a}_{n}$ (respectivement $\phi_{n}$) convergent dans $L^4$ vers $\overrightarrow{a}$ (respectivement $\phi$). Donc le produit $\phi_{n}\overrightarrow{a}_{n}$ converge vers $\phi\overrightarrow{a}$ dans $L^2$ en vertu de l'inégalité de Hölder.\\
L'application du lemme~\ref{resultat-standard-style-brezis} avec  $f_{n}=\phi_{n}(\overrightarrow{A}_{0}+\overrightarrow{a}_{n})$ et $g_{n}=\overrightarrow{\nabla}\phi_{n}$ entra\^ine la convergence de la suite $\int_{\Omega}2\Rez[\overline{\phi_{n}}(\overrightarrow{A}_{0}+\overrightarrow{a_{n}}).i\overrightarrow{\nabla}\phi_{n}]$. $\hfill{\bf CQFD}$

\begin{theorem}.\label{existence-d-un-minimum}
La fonctionnelle $F_{\lambda,k}(\phi, \overrightarrow{a})$ atteint son minimum sur ${\cal A}$.
\end{theorem}
{\it Preuve.} Soit $F_{min}=\inf_{(\phi, \overrightarrow{a})} F_{\lambda,k}(\phi, \overrightarrow{a})$ l'infimum de la fonctionnelle sur l'espace ${\cal A}$. Cet infimum existe car la fonctionnelle est positive. Par définition, il existe une suite de $(\phi_{\nu}, \overrightarrow{a}_{\nu})\in{\cal A}$ telle que 
\begin{equation}
F_{min}=\lim_{\nu\rightarrow \infty}F_{\lambda,k}(\phi_{\nu},\overrightarrow{a}_{\nu}) .
\end{equation}
Le lemme~\ref{tend-vers-l-infini} nous dit que cette suite est bornée dans ${\cal A}$. Il existe donc $(\phi,\overrightarrow{a})\in {\cal A}$ qui soit valeur d'adhérence de la suite $(\phi_{\nu}, \overrightarrow{a}_{\nu})$ pour la topologie faible de ${\cal A}$. Puisque la fonctionnelle $F_{\lambda,k}$ est semicontinue inférieurement pour la topologie de la convergence faible, $F_{\lambda,k}(\phi, \overrightarrow{a})\leq \underline{\lim}_{\nu}F_{\lambda,k}(\phi_{\nu},\overrightarrow{a}_{\nu})$ et donc $F_{min}=F_{\lambda,k}(\phi,\overrightarrow{a})$; l'infimum est donc un minimum. $\hfill{\bf CQFD}$

\saction{Régularité des solutions}\label{sec:regularite}
\noindent On montre ici que les couples minimisants la fonctionnelle $F_{\lambda,k}$ vérifient un système d'équations aux dérivées partielles et que les solutions de cette équation sont $C^{\infty}$.\\
On utilise pour cela l'ellipticité du système d'équations avec le théorème suivant. La notion d'ellipticité utilisée est celle de l'inversibilité du symbole principal avec une estimée sur l'inverse (cf~\cite{spin-geom}, p.~188).
\begin{theorem}.\label{crave-for-it-too}
Soit $X$ une variété compacte sans bord; soient $E$ et $F$ des fibrés vectoriels $C^{\infty}$ sur $X$ et $K$ un opérateur différentiel linéaire elliptique de ${\cal D}'(E)$ vers ${\cal D}'(F)$ d'ordre $n\geq 0$. Si $f\in {\cal D}'(E)$ vérifie
\begin{equation}
Kf \in W_{m}^p(F)
\end{equation}
alors $f$ appartient à $W_{m+n}^p(E)$.
\end{theorem}
{\it Preuve.} Par le calcul symbolique, il suffit de montrer que, si $U$ est un opérateur pseudodifférentiel elliptique d'ordre $-n$ de $F$ vers $E$ et si $f_2\in W_{m+n}^p(E)$, alors $U(f_2)\in W_{m}^p(F)$.\\
Le cas particulier o\`u $E$ et $F$ sont égaux au fibré trivial $\R\times X$ est traité dans le théorème 2.5 (p.~268) de \cite{taylor}.\\
En combinant les résultats de ce livre (en particulier la proposition 4.5 p.~278 de \cite{taylor} qui concerne la régularité $L^p$ locale) avec les théorèmes du livre \cite{spin-geom} (p.~177), on obtient notre résultat.  $\hfill{\bf CQFD}$

\begin{theorem}.
La fonctionnelle $F_{\lambda,k}$ est $C^1$ et on a, en tout point $(\phi,\overrightarrow{a})$ de ${\cal A}$, et pour tout $(\phi',\overrightarrow{a}')\in T_{(\phi,\overrightarrow{a})}{\cal A}$ (identifié à ${\cal A}$):
\begin{equation}\label{expression-differentielle}
\begin{array}{rcl}
D_{(\phi,\overrightarrow{a})}F_{\lambda,k}(\phi', \overrightarrow{a}')&=&\int_{\Omega}\Rez[\overline{\phi'}\lbrace[i\overrightarrow{\nabla}+\overrightarrow{A}_{0}+\overrightarrow{a}]^2\phi-(\lambda-|\phi|^2)\phi\rbrace]\\
&+&\int_{\Omega}\overrightarrow{a}'\lbrace \Rez[\overline{\phi}(i\overrightarrow{\nabla}\phi+(\overrightarrow{A}_0+\overrightarrow{a})\phi)]+k^2L\,\overrightarrow{a}\rbrace
\end{array}
\end{equation}
o\`u $T_{(\phi,\overrightarrow{a})}{\cal A}$ est l'espace tangent, $D_{(\phi,\overrightarrow{a})}F_{\lambda,k}$ est la différentielle de $F_{\lambda,k}$ et $L$ est défini en (\ref{double-rotationnel}).
\end{theorem}
{\it Preuve.} Nous allons montrer que la fonctionnelle est dérivable au sens de Fréchet et on notera la différentielle \index{$DF$}$DF$.\\
Soit $(\phi, \overrightarrow{a})\in {\cal A}$ et $(\phi', \overrightarrow{a}')$ un élément de l'espace tangent $T_{(\phi, \overrightarrow{a})}{\cal A}$ qui s'identifie à ${\cal A}$. On omet dans cette démonstration la référence à $(\lambda,k)$ dans $F_{\lambda,k}$.\\
On donne tout de suite l'expression de la différentielle, et on vérifiera un peu plus loin que c'est bien la différentielle au sens de Fréchet
\begin{equation}
\left\lbrace\begin{array}{rcl}
DF_{(\phi,\overrightarrow{a})}(\phi', \overrightarrow{a}')&=&\int_{\Omega}\Rez\lbrace\overline{[i\overrightarrow{\nabla}\phi'+\overrightarrow{A}\phi']}.[i\overrightarrow{\nabla}\phi+\overrightarrow{A}\phi]\rbrace\\
&+&\int_{\Omega}k^2\rot\,\overrightarrow{a}'\rot\,\overrightarrow{a}-\Rez[\phi'(\lambda-|\phi|^2)\phi]\\
&+&\int_{\Omega}\overrightarrow{a}'\Rez[\overline{\phi}(i\overrightarrow{\nabla}\phi+\overrightarrow{A}\phi)]
\end{array}\right.
\end{equation}
avec $\overrightarrow{A}=\overrightarrow{A}_0+\overrightarrow{a}$.\\
Une intégration par partie donne une autre forme à ces équations:
\begin{equation}\label{expression-differentielle-par-partie}
\left\lbrace\begin{array}{rcl}
DF_{\phi, \overrightarrow{a}}(\phi', \overrightarrow{a}')
&=&\int_{\Omega}\Rez[\overline{\phi'}\lbrace[i\overrightarrow{\nabla}+\overrightarrow{A}_{0}+\overrightarrow{a}]^2\phi-(\lambda-|\phi|^2)\phi\rbrace]\\
&+&\int_{\Omega}\overrightarrow{a}'\lbrace \Rez[\overline{\phi}(i\overrightarrow{\nabla}\phi+\overrightarrow{A}\phi)]+k^2L\,\overrightarrow{a}\rbrace .
\end{array}\right.
\end{equation}
Pour tout $(\phi,\overrightarrow{a})\in{\cal A}$ l'application $DF_{\phi, \overrightarrow{a}}$ est continue pour la norme de ${\cal A}$, il suffit d'appliquer les estimées $L^p$ (\ref{sobolev-imbedding}) et l'inégalité de Hölder.\\
On introduit la variation seconde $W$ de $F$:
\begin{equation}\label{differentielle-frechet-vanishing}
F(\phi+\phi',\overrightarrow{a}+\overrightarrow{a}')=F(\phi,\overrightarrow{a})+DF_{(\phi,\overrightarrow{a})}(\phi',\overrightarrow{a}')+W .
\end{equation}
On peut alors calculer explicitement $W$
\begin{equation}\label{vanishing-explicite}
\begin{array}{rl}
W&=\int_{\Omega}\Rez[\overrightarrow{a}'\overline{\phi'}\lbrace i\overrightarrow{\nabla}(\phi+\phi')+(\overrightarrow{A}_{0}+\overrightarrow{a})(\phi+\phi')+\overrightarrow{a}'\phi\rbrace]\\
&+\int_{\Omega}\frac{1}{2}\Vert i\overrightarrow{\nabla}\phi'+(\overrightarrow{A}_{0}+\overrightarrow{a})\phi'+\overrightarrow{a}'\phi\Vert^2+\frac{1}{2}|\overrightarrow{a}'\phi'|^2\\
&+\int_{\Omega}-\frac{|\phi'|^2}{2}(\lambda-|\phi|^2)+(\Rez[\phi\overline{\phi'}]+\frac{|\phi'|^2}{2})^2+\frac{k^2}{2}|\rot\,\overrightarrow{a}'|^2 .
\end{array}
\end{equation}
En utilisant les inégalités de Sobolev du théorème \ref{sobolev-imbedding} et l'inégalité de Hölder, on obtient l'estimée suivante
\begin{equation}\label{majoration-vanishing}
\Vert W\Vert_{H^1}\leq (\Vert \phi'\Vert _{H^{1}}+\Vert \overrightarrow{a}'\Vert _{H^{1}})^2 C(\Vert \phi\Vert _{H^{1}},\Vert \overrightarrow{a}\Vert _{H^{1}},\Vert \phi'\Vert _{H^{1}},\Vert \overrightarrow{a}'\Vert _{H^{1}}) .
\end{equation}
Dans l'équation~(\ref{majoration-vanishing}), $C(\alpha,\beta,\gamma,\delta)$ est une fonction continue de $\alpha,\, \beta,\, \gamma$ et $\delta$. Par conséquent, la fonction $F_{\lambda,k}$ est de classe $C^1$. $\hfill{\bf CQFD}$

\begin{remarque}.\label{derivation-fonctionnelle}
On peut en fait montrer que $F_{\lambda,k}$ est indéfiniment Fréchet dérivable. En particulier la différentielle seconde \index{$(D^2 F_{\lambda,k})_{(\phi,\overrightarrow{a})}$}$(D^2 F_{\lambda,k})_{(\phi,\overrightarrow{a})}$ est définie par
\begin{equation}
\begin{array}{rcl}
{\cal A}\times{\cal A}&\mapsto&\R\\
(\phi_1,\overrightarrow{a}_1),(\phi_2,\overrightarrow{a}_2)&\mapsto&
\int_{\Omega}\Rez\{\overline{\phi_1} [i\overrightarrow{\nabla}+\overrightarrow{A}]^2\phi_2\}\\
&&+2\Rez[\int_{\Omega}[\overrightarrow{a}_2\overline{\phi_1}+\overrightarrow{a}_1\overline{\phi_2}].(i\overrightarrow{\nabla}\phi+\overrightarrow{A}\phi)]\\
&&+\int_{\Omega}-(\lambda-|\phi|^2)\Rez\,\overline{\phi_1}\phi_2+2\Rez(\overline{\phi}\phi_2)\Rez(\overline{\phi}\phi_1)\\
&&+\int_{\Omega} \overrightarrow{a}_1.\overrightarrow{a}_2|\phi|^2+k^2 \rot\, \overrightarrow{a}_1 \rot\, \overrightarrow{a}_2 .
\end{array}
\end{equation}

\end{remarque}

\begin{definition} .
On définit l'espace \index{${\cal A}_2$}
\begin{equation}
{\cal A}_2=\left\lbrace 
\begin{array}{c}
(\phi,\overrightarrow{a})\in H^{1}(E_{1})\times H^{1}(\Tore{\cal L},\R^2)\\
\mbox{~avec~}\phi(z+v_{i})=e^{i\pi(\xi_{i}y-\eta_{i}x)}\phi(z)\\
\mbox{et~}\overrightarrow{a} \mbox{~${\cal L}$-périodique} .
\end{array}\right\rbrace
\end{equation}
On a enlevé à la définition de ${\cal A}$ (\ref{dernier-espace}) les contraintes $\int_{\Omega}\overrightarrow{a}=0$ et $\divergence\,\overrightarrow{a}=0$.
\end{definition}

\begin{proposition}.\label{proposition-representation-Aprime-A2}
L'espace $H^2_0(\Tore{\cal L})$ est l'espace des fonctions périodique de moyenne nulle sur $\Omega$ et de classe $H^2$. On définit l'espace ${\cal A'}={\cal A}\times H^{2}_0(\Tore{\cal L},\R)\times \R^2$. L'application
\begin{equation}
\left\lbrace\begin{array}{rcl}
S:{\cal A'}&\mapsto&{\cal A}_2\\
(\phi,\overrightarrow{a},f,t)&\mapsto&\left(\begin{array}{c}
\phi(z+t)e^{if(z+t)+i\pi[t_x y-t_y x]}\\
\overrightarrow{a}(z+t)+\overrightarrow{\nabla}f(z+t)+2\pi\left(\begin{array}{c}
-t_y\\
t_x
\end{array}\right)
\end{array}\right)
\end{array}\right.
\end{equation}
est bijective. Par ailleurs $F_{\lambda,k}(S(\phi,\overrightarrow{a},f,t))=F_{\lambda,k}(\phi,\overrightarrow{a})$.
\end{proposition}

\begin{remarque}.
L'application $S$ n'est pas différentiable au sens de Fréchet car elle contient des translations qui ne sont pas différentiables.
\end{remarque}
{\it Preuve.} Montrons que l'application $S$ est bien à valeur dans ${\cal A}_2$. Soit $(\phi',\overrightarrow{a}')=S(\phi,\overrightarrow{a},f,t)$; on a 
\begin{equation}
\begin{array}{rcl}
\phi'(z+t_i)&=&\phi(z+t+t_i)e^{if(z+t_i+t)+i\pi[t_x (y+t_{i,y})-t_y (x+t_{i,x})]}\\
&=&\phi(z+t)e^{i\pi[t_{i,x}(y+t_y)-t_{i,y}(x+t_x)]}e^{if(z+t)+i\pi[t_x y-t_y x]+i\pi[t_x t_{i,y}-t_y t_{i,x}]}\\
&=&\phi(z+t)e^{i\pi[t_{i,x}y-t_{i,y}x]}e^{i\pi[t_{i,x}t_y-t_{i,y}t_x]}\\
&&e^{if(z+t)+i\pi[t_x y-t_y x]+i\pi[t_x t_{i,y}-t_y t_{i,x}]}\\[1mm]
&=&\phi(z+t)e^{i\pi[t_{i,x}y-t_{i,y}x]}e^{if(z+t)+i\pi[t_x y-t_y x]}\\
&=&\phi'(z)e^{i\pi[t_{i,x}y-t_{i,y}x]}
\end{array}
\end{equation}
donc $\phi'\in H^1(E_1)$. Pour $\overrightarrow{a}'$ la vérification est triviale.\\
On a facilement en intégrant $\overrightarrow{a}'$ sur $\Omega$ la relation
\begin{equation}\label{definition-de-t}
\int_{\Omega}\overrightarrow{a}'=2\pi\left(\begin{array}{c}
-t_y\\
t_x
\end{array}\right) .
\end{equation}
La divergence de $\overrightarrow{a}'$ vérifie
\begin{equation}
\divergence\,\overrightarrow{a}'=\Delta\,f(z+t) .
\end{equation}
La fonction $f$ est de moyenne nulle sur $\Omega$ et est périodique. On peut donc par le lemme \ref{lemme-laplacien} inverser l'équation précédente avec $t$ donné par (\ref{definition-de-t}):
\begin{equation}
\Delta^{-1}[\divergence\,\overrightarrow{a}'](z-t)=f(z) .
\end{equation}
o\`u $\Delta^{-1}$ est l'application
\begin{equation}
\left\lbrace\begin{array}{rcl}
\Delta^{-1}:L^{2}(\Tore{\cal L},\R)&\mapsto&H^{2}_0(\Tore{\cal L},\R)\\
f&\mapsto&S^{-1}(f) .
\end{array}\right.
\end{equation}
On a donc une fonction réciproque de $S$.\\
Calculons maintenant l'énergie de $(\phi',\overrightarrow{a}')$; on a la formule de changement de jauge
\begin{equation}
\begin{array}{rcl}
\phi(z+t)e^{if(z+t)}&=&\phi'e^{-iu(z)}\\
\overrightarrow{a}(z+t)+\overrightarrow{\nabla}f(z+t)+\pi\left(\begin{array}{c}
-t_y\\
t_x
\end{array}\right)&=&\overrightarrow{a}'-\overrightarrow{\nabla}u
\end{array}
\end{equation}
avec $u(z)=\pi[t_x y-xt_y]$. Cela nous donne donc
\begin{equation}
F_{\lambda,k}(\phi',\overrightarrow{a}')=F_{\lambda,k}(\phi(z+t)e^{if(z+t)},\overrightarrow{a}(z+t)+\overrightarrow{\nabla}f(z+t)+\pi\left(\begin{array}{c}
-t_y\\
t_x
\end{array}\right)) .
\end{equation}
L'intégrande de $F_{\lambda,k}$ est périodique selon le réseau ${\cal L}$. Puisque 
\begin{equation}
(\overrightarrow{A}_0+\overrightarrow{a}(z+t)+\overrightarrow{\nabla}f(z+t)+\pi\left(\begin{array}{c}
-t_y\\
t_x
\end{array}\right))=(\overrightarrow{A}_0+\overrightarrow{a}+\overrightarrow{\nabla}f)(z+t)
\end{equation}
on obtient l'égalité
\begin{equation}
F_{\lambda,k}(\phi(z+t)e^{if(z+t)},\overrightarrow{a}(z+t)+\overrightarrow{\nabla}f(z+t)+\pi\left(\begin{array}{c}
-t_y\\
t_x
\end{array}\right))=F_{\lambda,k}(\phi e^{if},\overrightarrow{a}+\overrightarrow{\nabla}f) .
\end{equation}
Par changement de jauge on obtient 
\begin{equation}
F_{\lambda,k}(\phi e^{if},\overrightarrow{a}+\overrightarrow{\nabla}f)=F_{\lambda,k}(\phi,\overrightarrow{a})
\end{equation}
Ces trois égalités nous donnent le résultat.  $\hfill{\bf CQFD}$

%\begin{equation}
%\begin{array}{rcl}
%F_{\lambda,k}(\phi',\overrightarrow{a}')
%&=&F_{\lambda,k}(\phi(z+t)e^{if(z+t)+i\pi[t_x y-xt_y]},\overrightarrow{a}(z+t)+\overrightarrow{\nabla}f(z+t)+2\pi\left(\begin{array}{c}
%-t_y\\
%t_x
%\end{array}\right))\\
%&=&F_{\lambda,k}(\phi(z+t)e^{if(z+t)},\overrightarrow{a}(z+t)+\overrightarrow{\nabla}f(z+t)+\pi\left(\begin{array}{c}
%-t_y\\
%t_x
%\end{array}\right))\\
%&=&\int_{\Omega}\frac{1}{2}\Vert i\overrightarrow{\nabla}(\phi(z+t)e^{if(z+t)})+(\overrightarrow{A_0}+\pi\left(\begin{array}{c}
%-t_y\\
%t_x
%\end{array}\right)+\overrightarrow{a}(z+t)+\overrightarrow{\nabla}f(z+t))(\phi(z+t)e^{if(z+t)})|^2+\frac{1}{4}(\lambda-|\phi(z+t)e^{if(z+t)}|^2)^2+\frac{k^2}{2}|\rot\,(\overrightarrow{a}(z+t)+\overrightarrow{\nabla}f(z+t))|^2\\
%&=&\int_{\Omega}\frac{1}{2}|i\overrightarrow{\nabla}(\phi e^{if})+(\overrightarrow{A_0}+\overrightarrow{a}+\overrightarrow{\nabla}f)(\phi e^{if})|^2+\frac{1}{4}(\lambda-|\phi e^{if}|^2)^2+\frac{k^2}{2}|\rot\,(\overrightarrow{a}+\overrightarrow{\nabla}f)|^2(z+t)\\
%&=&F_{\lambda,k}(\phi e^{if},\overrightarrow{a}+\overrightarrow{\nabla}f)\\
%&=&F_{\lambda,k}(\phi,\overrightarrow{a})
%\end{array}
%\end{equation}
%On a donc l'égalité voulue.  $\hfill{\bf CQFD}$

\begin{theorem}.
Si $(\phi,\overrightarrow{a})\in{\cal A}$ est un point critique de la fonctionnelle $F_{\lambda,k}$, alors le couple vérifie les équations
\begin{equation}\label{Odeh-equations}
\left\lbrace\begin{array}{rcl}
[i\overrightarrow{\nabla}+\overrightarrow{A}_0+\overrightarrow{a}]^2\phi&=&(\lambda-|\phi|^2)\phi,\\
-\rot^*\,\rot\, \overrightarrow{a}&=&\frac{1}{k^2}\Rez[\overline{\phi}(i\overrightarrow{\nabla}\phi+(\overrightarrow{A}_0+\overrightarrow{a})\phi)] .
\end{array}\right.
\end{equation}
\end{theorem}

\begin{remarque}.
En particulier si $(\phi,\overrightarrow{a})\in{\cal A}$ minimise la fonctionnelle $F_{\lambda,k}$ alors il vérifie les équations (\ref{Odeh-equations}).
\end{remarque}
{\it Preuve.} Si $(\phi,\overrightarrow{a})\in{\cal A}$ est un point critique pour $F_{\lambda,k}$ défini sur l'espace ${\cal A}$, alors on a pour tout $(\phi',\overrightarrow{a}')\in{\cal A}$ 
\begin{equation}
\begin{array}{rcl}
0&=&\int_{\Omega}\Rez[\overline{\phi'}\lbrace[i\overrightarrow{\nabla}+(\overrightarrow{A}_{0}+\overrightarrow{a})]^2\phi\rbrace-(\lambda-|\phi|^2)\phi\rbrace]\\
&+&\int_{\Omega}\overrightarrow{a}'\lbrace \Rez[\overline{\phi}(i\overrightarrow{\nabla}\phi+(\overrightarrow{A}_0+\overrightarrow{a})\phi)]+k^2L\,\overrightarrow{a}\rbrace .
\end{array}
\end{equation}
On peut donc affirmer que 
\begin{equation}
\begin{array}{l}
[i\overrightarrow{\nabla}+(\overrightarrow{A}_{0}+\overrightarrow{a})]^2\phi-(\lambda-|\phi|^2)\phi\in(H^1(E_1))^{\perp}\\
\Rez[\overline{\phi}(i\overrightarrow{\nabla}\phi+(\overrightarrow{A}_0+\overrightarrow{a})\phi)]+k^2L\,\overrightarrow{a}\in(H^1_{\divergence,0}(\Tore{\cal L},\R^2))^{\perp} .
\end{array}
\end{equation}
Puisque $(H^1(E_1))^{\perp}=\lbrace 0\rbrace$ on obtient la premère équation
\begin{equation}
[i\overrightarrow{\nabla}+(\overrightarrow{A}_{0}+\overrightarrow{a})]^2\phi-(\lambda-|\phi|^2)\phi=0 .
\end{equation}
Cependant $(H^1_{\divergence,0}(\Tore{\cal L},\R^2))^{\perp}$ n'est pas réduit au potentiel nul. Par exemple tous les champs constants sont dans cet espace, il nous faut donc réfléchir un peu plus; on pose
\begin{equation}
\overrightarrow{V}=\Rez[\overline{\phi}(i\overrightarrow{\nabla}\phi+(\overrightarrow{A}_0+\overrightarrow{a})\phi)]+k^2L\,\overrightarrow{a} .
\end{equation}
On peut écrire 
\begin{equation}
\overrightarrow{V}=\overrightarrow{X}+\overrightarrow{d}+\overrightarrow{\nabla}\,f
\end{equation}
avec $\overrightarrow{d}\in H^1_{\divergence,0}(\Tore{\cal L},\R^2)$,  $f\in H^2(\Tore{\cal L},\R)$, $\int_{\Omega}f=0$ et $\overrightarrow{X}$ est un vecteur constant.
\begin{remarque}.
Il s'agit en fait d'une décomposition de Hodge (cf~\cite{Griffiths-Harris}, p.~84).
\end{remarque}
Cette écriture se trouve facilement en dérivant
\begin{equation}
\divergence\,\overrightarrow{V}=\Delta\,f
\end{equation}
que l'on inverse facilement en utilisant le lemme \ref{lemme-laplacien}, on décompose ensuite
\begin{equation}
\overrightarrow{V}-\overrightarrow{\nabla}\,f
\end{equation}
en un champ constant plus un champ d'intégrale nulle.\\
Le couple $(0,\overrightarrow{d})$ appartient à l'espace ${\cal A}$; donc, puisque le couple $(\phi,\overrightarrow{a})$ est critique, on obtient en utilisant l'expression différentielle (\ref{expression-differentielle})
\begin{equation}
\int_{\Omega}\overrightarrow{d}.\overrightarrow{V}=0 .
\end{equation}
L'expression différentielle (\ref{expression-differentielle}) est aussi valable sur ${\cal A}_2$, on obtient alors si $u\in\R$:
\begin{equation}
F_{\lambda,k}(\phi e^{iuf},\overrightarrow{a}+u\overrightarrow{\nabla}\,f)=F_{\lambda,k}(\phi,\overrightarrow{a})+u\int_{\Omega}\overrightarrow{V}.\overrightarrow{\nabla}f+o(u) .
\end{equation}
Or on sait depuis la proposition \ref{proposition-representation-Aprime-A2} que
\begin{equation}
F_{\lambda,k}(\phi e^{iuf},\overrightarrow{a}+u\overrightarrow{\nabla}f)=F_{\lambda,k}(\phi,\overrightarrow{a}) .
\end{equation}
On obtient donc
\begin{equation}
\int_{\Omega}\overrightarrow{V}.\overrightarrow{\nabla}f=0 .
\end{equation}
Cela nous donne donc
\begin{equation}
\int_{\Omega}(\overrightarrow{V}-\overrightarrow{X}).\overrightarrow{V}=0 .
\end{equation}
Le champ $\overrightarrow{V}-\overrightarrow{X}$ est d'intégrale nulle, cela nous donne
\begin{equation}
\int_{\Omega}||\overrightarrow{V}-\overrightarrow{X}||^2=0 .
\end{equation}
On a donc $\overrightarrow{V}=\overrightarrow{X}$, le champ est constant.\\
On pose
\begin{equation}
\begin{array}{rcl}
(\phi_{t},\overrightarrow{a}_{t})
&=&S(\phi,\overrightarrow{a},0,t)\\
&=&(\phi(z+t)e^{i\pi[t_x y-xt_y]}, \overrightarrow{a}(z+t)+2\pi\left(\begin{array}{c}
-t_y\\
t_x
\end{array}\right)) .
\end{array}
\end{equation}
La propriété d'invariance de la proposition \ref{proposition-representation-Aprime-A2} nous donne
\begin{equation}\label{penible-prem}
F_{\lambda,k}(\phi_{t},\overrightarrow{a}_{t})=F_{\lambda,k}(\phi,\overrightarrow{a}) .
\end{equation}
L'intégrande est périodique selon ${\cal L}$ et
\begin{equation}
(\overrightarrow{A}_0+\overrightarrow{a}_t)(z)=(\overrightarrow{A}_0+\overrightarrow{a}+\pi\left(\begin{array}{c}
-t_y\\
t_x
\end{array}\right))(z+t)
\end{equation}
on obtient donc 
\begin{equation}\label{penible-deuxio}
\begin{array}{rcl}
F_{\lambda,k}(\phi_{t},\overrightarrow{a}_{t})
&=&F_{\lambda,k}(\phi e^{i\pi[t_x y-xt_y]},\overrightarrow{a}+\pi\left(\begin{array}{c}
-t_y\\
t_x
\end{array}\right)) .
\end{array}
\end{equation}
L'expression différentielle (\ref{expression-differentielle}) nous donne
\begin{equation}\label{penible-tertio}
F_{\lambda,k}(\phi e^{i\pi[t_x y-xt_y]},\overrightarrow{a}+\pi\left(\begin{array}{c}
-t_y\\
t_x
\end{array}\right))=F_{\lambda,k}(\phi,\overrightarrow{a})+\int_{\Omega}\pi\left(\begin{array}{c}
-t_y\\
t_x
\end{array}\right).\overrightarrow{V}+o(|t|)
\end{equation}
En combinant les équations (\ref{penible-prem}), (\ref{penible-deuxio}) et (\ref{penible-tertio}), on obtient 
\begin{equation}
\int_{\Omega}\pi\left(\begin{array}{c}
-t_y\\
t_x
\end{array}\right).\overrightarrow{X}=0
\end{equation}
pour tout complexe $t\in\C$. Cela ne peut \^etre possible que si $\overrightarrow{X}=0$ et donc $\overrightarrow{V}=0$. $\hfill{\bf CQFD}$\\
\\
Les deux équations~(\ref{Odeh-equations}) sont appelées \index{équations de Ginzburg-Landau}équations de Ginzburg-Landau.\\
Le résultat suivant est très classique.

%\begin{proposition}.\label{image-F1-F2-etat-critique}
%{\it (On utilise les notations de la section \ref{sec:struct-reseau-space})}Soit $m\in{\cal M}$. Soit $(\phi,\overrightarrow{a})\in{\cal A}_{m}$ un point critique de la fonctionnelle $F_{\lambda,k}$.\\
%Alors $\Sigma_i^*\phi,\Sigma_i^*\overrightarrow{a})\in {\cal A}_{\sigma_i(m)}$ est un point critique de la fonctionnelle $F_{\lambda,k}$
%\end{proposition}
%{\it Preuve.} Les applications $\Sigma_i^*$ sont des difféomorphismes, la proposition est alors évidente. $\hfill{\bf CQFD}$

\begin{theorem}.\label{regularite}
Si un couple $(\phi, \overrightarrow{a})$ est solution des équations~(\ref{Odeh-equations}) alors ce couple est $C^{\infty}$. En particulier les couples minimisants la fonctionnelle $F_{\lambda,k}$ sont $C^{\infty}$.
\end{theorem}
{\it Preuve.}\\
{\bf $1^{ere}$ étape :}\\
On va prouver que $(\phi, \overrightarrow{a})$ est $C^{\infty}$ en montrant que $\forall n\in \N$, $(\phi, \overrightarrow{a})\in H^{n}(E_{1})\times H^{n}_{\divergence,0}(\Tore{\cal L}; \R^2)$. On conclut ensuite grâce aux théorèmes d'injection de Sobolev.\\
Le système elliptique dont on souhaite montrer la régularité est le suivant
\begin{equation}\label{premier-systeme-elliptique}
\left\lbrace\begin{array}{rcl}
H\phi&=&\phi(\lambda-|\phi|^2)-2\overrightarrow{a}.(i\overrightarrow{\nabla}\phi+\overrightarrow{A}_{0}\phi)-\overrightarrow{a}^2\phi\\
\Delta\, \overrightarrow{a}&=&\frac{1}{k^2}\Rez[\overline{\phi}(i\overrightarrow{\nabla}\phi+\overrightarrow{A}_{0}\phi)]+\frac{1}{k^2}\overrightarrow{a}|\phi|^2 .
\end{array}\right.
\end{equation}
On a utilisé le fait que $\divergence\,\overrightarrow{a}=0$ pour remplacer $\rot^{*}\,\rot$ par $-\Delta$ et pour développer $[i\overrightarrow{\nabla}+\overrightarrow{A}_{0}+\overrightarrow{a}]^2$.\\
Il n'est pas clair à priori que le membre de droite de l'équation~(\ref{premier-systeme-elliptique}) appartienne à $L^{2}$, ce qui est pourtant le moyen le plus simple pour montrer que $(\phi,\overrightarrow{a})\in H^2$.\\
{\bf $2^{eme}$ étape :}\\
Si $(\phi,\overrightarrow{a})\in H^{1}$, alors $(\phi,\overrightarrow{a})\in W^{\frac{3}{2}}_{2}$.\\
En effet une application répétée de l'inégalité de Hölder permet de prouver que le membre de droite de l'équation~(\ref{premier-systeme-elliptique}) appartient à $L^{\frac{3}{2}}$.\\
L'équation~(\ref{premier-systeme-elliptique}) nous dit alors précisément que $H\phi$ et $L\overrightarrow{a}$ appartiennent à $L^{\frac{3}{2}}$. Ensuite on applique le théorème~\ref{crave-for-it-too} pour obtenir que \mbox{$(\phi, \overrightarrow{a})\in W^{\frac{3}{2}}_{2}\times W^{\frac{3}{2}}_{2}$}.\\
{\bf $3^{eme}$ étape :}\\
On échange la régularité contre un contrôle en norme $L^p$.\\
Le théorème~\ref{sobolev-imbedding} nous donne l'inclusion
\begin{equation}\label{sobolev-imbedding-dim2}
W^{\frac{3}{2}}_{1}\subset L^{6} .
\end{equation}
L'inclusion~(\ref{sobolev-imbedding-dim2}) prouve que $(\phi,\overrightarrow{a})\in W_{1}^{6}$. Cette inclusion de Sobolev est décisive car elle va nous permettre de montrer que le côté droit de l'équation~(\ref{premier-systeme-elliptique}) appartient à $L^2$, ce qui nous manquait au début.\\
{\bf $4^{eme}$ étape :}\\
On prouve que $(\phi, \overrightarrow{a})\in H^{2}$.\\
Par la régularité elliptique, il suffit de prouver que le membre de droite de~(\ref{premier-systeme-elliptique}) appartient à $L^{2}$. On utilise pour cela les inégalités de Hölder et les résultats précédents
\begin{equation}\label{majoration-Linfini-en-vue}
\begin{array}{rcl}
\int_{\Omega}|\overrightarrow{a}.\overrightarrow{\nabla}\phi|^2dxdy&\leq&\Vert \overrightarrow{\nabla}\phi\Vert _{L^6}^2\Vert \overrightarrow{a}\Vert _{L^3}^2\\
\int_{\Omega}\Vert\overline{\phi}\overrightarrow{\nabla}\phi\Vert^2dxdy&\leq&\Vert \overrightarrow{\nabla}\phi\Vert _{L^6}^2\Vert \phi\Vert _{L^3}^2 .
\end{array}
\end{equation}
Les autres termes ne posant pas de problème nouveau, nous ne les écrivons pas. On a donc prouvé que $(\phi,\overrightarrow{a})\in H^{2}$.  Une application du théorème~\ref{sobolev-imbedding} nous donne que $(\phi,\overrightarrow{a})\in L^{\infty}$.\\
{\bf $5^{eme}$ étape :}\\
Supposons que $(\phi,\overrightarrow{a})$ est de classe $H^{n+1}$ avec $n\geq 1$.\\
Soit $D_{n}$ un opérateur différentiel d'ordre $n$, on l'applique aux équations~(\ref{premier-systeme-elliptique})
\begin{equation}\label{differenciation-de-premier-systeme-elliptique}
\left\lbrace\begin{array}{rcl}
D_{n}[H\phi]&=&-2\overrightarrow{a}.iD_{n}(\overrightarrow{\nabla}\phi)+P_{n}(\phi,\overrightarrow{a})\\
D_{n}[-\Delta\, \overrightarrow{a}]&=&\frac{1}{k^2}\Rez[\overline{\phi}iD_{n}(\overrightarrow{\nabla}\phi)]+Q_{n}(\phi, \overrightarrow{a}) .
\end{array}\right.
\end{equation}
Les termes $P_{n}(\phi, \overrightarrow{a})$ et $Q_{n}(\phi, \overrightarrow{a})$ désignent des polynômes en les $n$ premières dérivées de $\phi$ et $\overrightarrow{a}$. Donc tous les termes des deux polynômes appartiennent à $H^{1}$ donc à $L^{p}$ pour $p<\infty$. Donc $P_{n}$ et $Q_{n}$ sont de carré intégrable grâce à l'inégalité de Hölder.\\
Le couple $(\phi,\overrightarrow{a})$ appartient à $L^{\infty}$. La section $D_{n}(\overrightarrow{\nabla}\phi)$ appartient à $L^{2}$ par l'hypothèse de récurrence, le membre de gauche de l'équation~(\ref{differenciation-de-premier-systeme-elliptique}) appartient alors à $L^{2}$.\\
D'après la définition \ref{espace-W} des espaces de Sobolev, $H\phi$ et $-\Delta\, \overrightarrow{a}$ appartiennent à $H^{n}$.\\
L'ellipticité des opérateurs $H$ et $-\Delta$ nous donne $(\phi, \overrightarrow{a})\in H^{n+2}\times H^{n+2}$. Par récurrence $(\phi,\overrightarrow{a})$ est de classe $H^n$ pour tout $n$ donc $C^{\infty}$. $\hfill{\bf CQFD}$

\saction{Propriétés du minimum de la fonctionnelle $F_{\lambda,k}$}\label{sec:universelle}
\noindent On montre dans cette section divers résultats sur les minimums $m_F$ et $m_D$ définis à l'équation (\ref{definition-mF-mD}) des fonctionnelles $F_{\lambda,k}$ et $D_{\lambda,k}$ définies aux équations (\ref{seconde-definition-Flambdak}) et (\ref{definition-Dklambdaphia}).\\
On utilisera les propriétés \ref{premieres-propriete-mF-mD} de $m_F$ et $m_D$. Commençons par un résultat asymptotique.
\begin{lemme}.\label{prolongement-par-continuite}
Pour tout $k>0$, on a les limites suivantes
\begin{equation}
\left\lbrace\begin{array}{l}
\lim_{\lambda\rightarrow\infty}m_D(\lambda,k)=0\\
\lim_{H_{int}\rightarrow 0}G^V_{k,H_{ext}}(H_{int})=\frac{H_{ext}^2}{2}
\end{array}\right.
\end{equation}
\end{lemme}
{\it Preuve.} En revenant aux définitions (\ref{definition-Dklambdaphia}) et (\ref{definition-mF-mD}), on a l'égalité
\begin{equation}\label{autre-formulation}
\begin{array}{rcl}
m_D(\lambda,k)
&=&\min_{(\phi, \overrightarrow{a})\in {\cal A}}\int_{\Omega}\frac{1}{2\lambda}\Vert i\overrightarrow{\nabla}\phi+(\overrightarrow{A}_0+\overrightarrow{a})\phi\Vert^2\\
&&+\int_{\Omega}\frac{1}{4}(1-|\phi|^2)^2+\frac{k^2}{2\lambda^2}|\rot\,\overrightarrow{a}|^2 .
\end{array}
\end{equation}
L'ensemble des zéros de $\phi_0$ est $z_0+{\cal L}$ et ils sont tous de multiplicité $1$ (cf~proposition \ref{unique-zero}). On peut donc choisir un domaine fondamental $\Omega'$ tel que $\phi_0$ possède un unique zéro dans l'intérieur de $\Omega'$ et ne possède pas d'autres zéro dans $\overline{\Omega'}$.\\
Soit \index{$f_{\delta}$}$f_{\delta}$ la fonction définie sur $\R^2$ privée de $B(z_0,\delta)$ et de ses translatés et de valeur $\frac{1}{|\phi_0(z)|}$. La fonction $f_{\delta}$ est ${\cal L}$-périodique et de norme $L^{\infty}$ bornée par $\frac{D}{\delta}$.\\
On ne peut pas à priori prolonger $f_{\delta}$ en une fonction ${\cal L}$-périodique $C^{\infty}$ sur $\R^2$ encore notée $f_{\delta}$ de norme $\frac{D}{\delta}$. Cependant on peut le faire si on autorise une norme légèrement plus grande, $\frac{D+1}{\delta}$ par exemple.\\
On a alors l'estimée
\begin{equation}
\int_{\Omega'}(1-|f_{\delta}\phi_0|^2)^2=O(\delta^2) .
\end{equation}
Pour tout $\epsilon>0$, il existe donc une section $\phi_{\epsilon}$ de classe $C^{\infty}$ vérifiant
\begin{equation}
\int_{\Omega'}(1-|\phi_{\epsilon}|^2)^2<\epsilon .
\end{equation}
Pour toute section $\phi\in H^1(E_1)$, on a:
\begin{equation}
\lim_{\lambda\rightarrow\infty}\int_{\Omega'}\frac{1}{2\lambda}\Vert i\overrightarrow{\nabla}\phi+\overrightarrow{A}_0\phi\Vert^2=0 .
\end{equation}
Il existe donc $\lambda_{M}(\epsilon)>0$ tel que, $\forall \lambda>\lambda_{M}(\epsilon)$, 
\begin{equation}
\int_{\Omega'}\frac{1}{2\lambda}\Vert i\overrightarrow{\nabla}\phi_{\epsilon}+\overrightarrow{A}_0\phi_{\epsilon}\Vert^2<\epsilon .
\end{equation}
On obtient donc, $\forall \lambda>\lambda_M(\epsilon)$:
\begin{equation}
0\leq \frac{1}{\lambda^2}m_F(\lambda,k)\leq \frac{1}{\lambda^2}F_{\lambda,k}(\phi_{\epsilon},0)\leq \frac{3}{2}\epsilon .
\end{equation}
On a donc bien la limite voulue. La deuxième limite provient de l'égalité $G^V_{k,H_{ext}}(H_{int})=m_D(\frac{2\pi k}{H_{int}},k)+\frac{1}{2}(H_{int}-H_{ext})^2$. $\hfill{\bf CQFD}$\\
\\
Le théorème ci-dessous utilise des idées de \cite{riviere}, p.~253; la seule différence est que l'on opère sur une variété compacte.
\begin{theorem}.\label{principe-maximum}
Si un couple $(\phi, \overrightarrow{a})$ est solution des équations~(\ref{Odeh-equations}), alors la fonction $\phi$ vérifie l'inégalité:
\begin{equation}\label{inegalite-universelle}
|\phi|\leq\sqrt{\lambda}=\sqrt{\frac{2\pi k}{H_{int}}} .
\end{equation}
\end{theorem}
\begin{remarque}.
Pour le problème initial introduit en~(\ref{probleme-de-la-these}), l'inégalité s'écrit $|\phi|\leq 1$ et résulte des transformations (\ref{definition-alpha-beta}) et (\ref{definition-W1-W2}). La densité d'électrons supraconducteurs $n=|\phi|^2$ est donc bornée par $1$.
\end{remarque}
{\it Preuve.} La fonction $|\phi|^2$ définie sur $\R^2$ est ${\cal L}$-périodique et définit par passage au quotient une fonction sur $\Tore{\cal L}$.
On va montrer qu'aux points où $|\phi|^2$ est maximum l'inégalité~(\ref{inegalite-universelle}) est vérifiée.
L'équation~(\ref{Odeh-equations}) peut s'écrire
\begin{equation}\label{developpement-de-ginzburg-landau-1}
-\Delta\phi+2i\overrightarrow{A}.\overrightarrow{\nabla}\phi+\overrightarrow{A}^2\phi=\phi(\lambda-|\phi|^2) .
\end{equation}
On calcule maintenant le laplacien de $|\phi|^2$ comme dans l'article \cite{riviere} .
\begin{equation}\label{calcul-laplacien}
\begin{array}{rcl}
\Delta\,|\phi|^2 &=& \overline{\phi}\Delta\phi+\phi\Delta\overline{\phi}+2\Vert \overrightarrow{\nabla}\,\phi\Vert^2\\
&=& -2|\phi|^2(\lambda-|\phi|^2)+2\overrightarrow{A}^2|\phi|^2\\
&+&4\Rez[\overline{\phi}\overrightarrow{A}.i\overrightarrow{\nabla}\phi]+2\Vert i\overrightarrow{\nabla}\phi\Vert^2\\[1mm]
&=& -2|\phi|^2(\lambda-|\phi|^2)+2\Vert i\overrightarrow{\nabla}\phi+\overrightarrow{A}\phi\Vert^2 .
\end{array}
\end{equation}
Soit $z$ un point de $\Tore{\cal L}$ où la fonction $|\phi|^2$ atteint son maximum. On obtient alors $[\Delta(|\phi|^2)](z)\leq 0$. Donc, par le calcul~(\ref{calcul-laplacien}), on a $|\phi|(z)\leq \sqrt{\lambda}$ et, puisque $z$ est un maximum, l'inégalité est vraie partout.$\hfill{\bf CQFD}$
\begin{theorem}.\label{Borne-sur-m}
On a, pour tout $\lambda,\lambda'\in(\R_+^*)^2$, l'inégalité
\begin{equation}\label{estimee-de-continuite}
|(m_F(\lambda,k)-\frac{\lambda^2}{4})-(m_F(\lambda',k)-\frac{\lambda'^2}{4})|\leq \frac{1}{2}|\lambda-\lambda'|\max(|\lambda|,|\lambda'|) .
\end{equation}
En particulier la fonction $m_F$ est continue par rapport à $\lambda$ (localement lipschitzienne).
\end{theorem}
{\it Preuve.} On a l'égalité suivante
\begin{equation}
(F_{\lambda,k}(\phi, \overrightarrow{a})-\frac{\lambda^2}{4})-(F_{\lambda',k}(\phi, \overrightarrow{a})-\frac{\lambda'^2}{4})=-\frac{1}{2}(\lambda-\lambda')\int_{\Omega}|\phi|^2 .
\end{equation}
Soit $(\phi, \overrightarrow{a})$ un couple qui minimise la fonctionnelle $F_{\lambda,k}$, un tel couple existe grâce au théorème~\ref{existence-d-un-minimum}. On a alors la relation suivante qui découle du théorème~\ref{principe-maximum}
\begin{equation}
|(m_F(\lambda,k)-\frac{\lambda^2}{4})-(F_{\lambda',k}(\phi, \overrightarrow{a})-\frac{\lambda'^2}{4})|\leq\frac{1}{2}|\lambda-\lambda'|\lambda .
\end{equation}
Par conséquent
\begin{equation}
(F_{\lambda',k}(\phi, \overrightarrow{a})-\frac{\lambda'^2}{4})-(m_F(\lambda,k)-\frac{\lambda^2}{4}) \leq \frac{1}{2}|\lambda-\lambda'|\max(|\lambda|,|\lambda'|) .
\end{equation}
De par la définition de $m_F(\lambda',k)$ comme minimum de $F_{\lambda',k}$, on déduit
\begin{equation}
(m_F(\lambda',k)-\frac{\lambda'^2}{4})-(m_F(\lambda,k)-\frac{\lambda^2}{4}) \leq \frac{1}{2}|\lambda-\lambda'|\max(|\lambda|,|\lambda'|) .
\end{equation}
On intervertit $\lambda$ et $\lambda'$ et on obtient l'inégalité~(\ref{estimee-de-continuite}). La continuité de $m_F$ découle trivialement de cette inégalité.$\hfill{\bf CQFD}$

\begin{remarque}.
Nous prolongeons la fonction $G^{V}_{k,H_{ext}}(H_{int})$ en $0$ par continuité en posant:
\begin{equation}
G^{V}_{k,H_{ext}}(0)=\frac{H_{ext}^2}{2} .
\end{equation}
\end{remarque}

\begin{theorem}.\label{infimum_minimum}
Il existe un champ $H_{int}^a\geq 0$ tel que: 
\begin{equation}
{\cal E}^V_{k,H_{ext}}=G^V_{k,H_{ext}}(H_{int}^a) .
\end{equation}
\end{theorem}
{\it Preuve.} Si l'on revient à la définition (\ref{infimum-Hint}) et utilisant la positivité de $F_{\lambda,k}$, on a l'inégalité évidente
\begin{equation}
G^{V}_{k,H_{ext}}(H_{int})\geq\frac{1}{2}(H_{int}-H_{ext})^2 .
\end{equation}
Par conséquent: 
\begin{equation}
\lim_{H_{int}\rightarrow+\infty}G^{V}_{k,H_{ext}}(H_{int})=+\infty .
\end{equation}
La fonction 
\begin{equation}
\left\lbrace\begin{array}{rcl}
[0,+\infty[&\mapsto&\R\\
H_{int}&\mapsto&G^V_{k,H_{ext}}(H_{int})
\end{array}\right.
\end{equation}
est continue et tend vers $+\infty$ en $+\infty$.\\
Par conséquent, elle atteint son minimum ${\cal E}^{V}_{k,H_{ext}}$ en au moins un point de $[0,+\infty[$.\\
Si ${\cal E}^{V}_{k,H_{ext}}=\frac{H_{ext}^2}{2}$, alors le minimum est atteint en $H_{int}=0$ et cette énergie peut \^etre atteinte par l'état pur.\\
Si ${\cal E}^{V}_{k,H_{ext}}=\frac{1}{4}$, alors le minimum de $G^{V}_{k,H_{ext}}(H_{int},\phi,\overrightarrow{a})$ est atteint par $H_{int}=H_{ext}$ et $(\phi,\overrightarrow{a})=(0,0)$; l'énergie du supraconducteur peut \^etre atteinte par l'état normal.\\
Dans les autres cas, l'énergie minimale ${\cal E}^V_{k,H_{ext}}$ est atteinte par un état mixte de la forme $(H_{int},\phi,\overrightarrow{a})$ avec $(\phi,\overrightarrow{a})\not=(0,0)$. $\hfill{\bf CQFD}$

\begin{remarque}.
Comme souvent dans les transitions de phase, il existe certains cas o\`u la m\^eme énergie minimale peut \^etre atteinte par deux états différents. Ce cas se produit effectivement dans notre étude, comme on peut le voir au lemme \ref{lemme-preparatoire} ou au théorème \ref{usage-theoreme-monotonie}.
\end{remarque}

%\begin{lemme}.\label{annulation-une-integrale}
%Soit $\overrightarrow{a}$ est un potentiel vecteur de classe $H^1$ d'intégrale nulle et de divergence nulle; on a 
%\begin{equation}
%\int_{\Omega}|\rot\,\overrightarrow{a}|^2=0\Leftrightarrow \rot\,\overrightarrow{a}=0\Leftrightarrow \overrightarrow{a}=0 .
%\end{equation}
%\end{lemme}
%{\it Preuve.} Supposons
%\begin{equation}
%\int_{\Omega}|\rot\,\overrightarrow{a}|^2=0;
%\end{equation}
%on a alors $\rot\,\overrightarrow{a}=0$. Puisque $L\overrightarrow{a}=\rot^* \rot\,\overrightarrow{a}$, on a $L\overrightarrow{a}=0$.\\
%Or on a vu (cf~\ref{spectre-L-theo}) que $L$ est inversible donc $\overrightarrow{a}=0$.\\
%La réciproque est immédiate. $\hfill{\bf CQFD}$

\begin{lemme}.\label{annulation-deux-integrale}
Si $(\phi, \overrightarrow{a})\in {\cal A}$ vérifie :
\begin{equation}
\int_{\Omega}|\rot\,\overrightarrow{a}|^2=0\mbox{~et~}\int_{\Omega}\Vert i\overrightarrow{\nabla}\phi+(\overrightarrow{A}_0+\overrightarrow{a})\phi\Vert^2=0
\end{equation}
alors $\phi=0$ et $\overrightarrow{a}=0$ .
\end{lemme}
{\it Preuve.} Par la proposition \ref{annulation-une-integrale}, on a $\overrightarrow{a}=0$; il reste donc l'équation
\begin{equation}
\langle H\phi,\phi\rangle=\int_{\Omega}\Vert i\overrightarrow{\nabla}\phi+\overrightarrow{A}_0\phi\Vert^2=0 .
\end{equation}
Puisque $\phi$ est de carré intégrable, on peut effectuer un développement de $\phi$ sur les fonctions propres normalisées définies au théor\`eme~\ref{spectre-de-H-C-infinite}:
\begin{equation}
\phi=\sum_{i=0}^{\infty}\alpha_i\phi_i, 
\end{equation}
on a alors
\begin{equation}
\begin{array}{rcl}
\langle H\phi,\phi\rangle
&=&\langle \sum_{i=0}^{\infty}\alpha_iH\phi_i,\sum_{i=0}^{\infty}\alpha_i\phi_i\rangle\\
&=&\langle \sum_{i=0}^{\infty}\alpha_i(2\pi+4\pi i)\phi_i,\sum_{i=0}^{\infty}\alpha_i\phi_i\rangle\\
&=&\sum_{i=0}^{\infty}|\alpha_i|^2(2\pi+4\pi i)\langle \phi_i,\phi_i\rangle\\
&=&\sum_{i=0}^{\infty}|\alpha_i|^2(2\pi+4\pi i) .
\end{array}
\end{equation}
Cette quantité ne peut \^etre nulle que si pour tout $i\geq 0$, $\alpha_i=0$, c'est à dire $\phi=0$. $\hfill{\bf CQFD}$

\begin{lemme}.\label{champ-a-determine}
Si le couple $(\phi, \overrightarrow{a})\in {\cal A}$ vérifie les équations de Ginzburg-Landau~(\ref{Odeh-equations}) et $\phi$ n'est pas identiquement nul alors $\overrightarrow{a}$ n'est pas identiquement nul.
\end{lemme}
{\it Preuve.} Par le théorème \ref{regularite}, le couple $(\phi,\overrightarrow{a})$ est $C^{\infty}$. Supposons par l'absurde que $\overrightarrow{a}= 0$; on a alors 
\begin{equation}\label{annulation-a-utiliser}
\Rez\,\overline{\phi}[i\overrightarrow{\nabla}\phi+\overrightarrow{A}_0\phi]=0 .
\end{equation}
La fonction $\phi$ n'est pas identiquement nulle; par conséquent l'ensemble
\begin{equation}
C=\lbrace z\in\Omega\mbox{~tel~que~}\phi(z)\not=0 \rbrace
\end{equation}
est un ouvert non vide.\\
Soit donc $B(z_a,\delta)$ une boule incluse dans $C$; si $\delta$ est assez petit, alors on peut écrire sur $B(z_a,\delta)$
\begin{equation}\label{idee-helffer-encore}
\phi(z)=r(z)e^{i\theta(z)}
\end{equation}
avec $r$ et $\theta$ fonctions réelles de classe $C^{\infty}$ et $r>0$.\\
Réécrivons l'équation (\ref{annulation-a-utiliser}) avec l'écriture (\ref{idee-helffer-encore}); on a 
\begin{equation}
\begin{array}{rcl}
0&=&\Rez\,\overline{\phi}[i\overrightarrow{\nabla}\phi+\overrightarrow{A}_0\phi]\\
&=&\Rez\,r e^{-i\theta(z)}[i\{\overrightarrow{\nabla}re^{i\theta}+ir\overrightarrow{\nabla}\theta e^{i\theta}\}+\overrightarrow{A}_0r e^{i\theta}]\\
&=&\Rez\,r[i\{\overrightarrow{\nabla}r+ir\overrightarrow{\nabla}\theta \}+\overrightarrow{A}_0r ]\\
&=&\Rez\,r[i\overrightarrow{\nabla}r-r\overrightarrow{\nabla}\theta+\overrightarrow{A}_0r ]\\
&=&r[-r\overrightarrow{\nabla}\theta+\overrightarrow{A}_0r ]\\
&=&r^2[-\overrightarrow{\nabla}\theta+\overrightarrow{A}_0 ]
\end{array}
\end{equation}
On obtient donc l'égalité $\overrightarrow{A}_0=\overrightarrow{\nabla}\theta$ car $r>0$.\\
En appliquant le rotationnel, on obtient 
$\rot\,\overrightarrow{\nabla}\theta=2\pi$, car d'après (\ref{quelques-definitions}), $rot\,\overrightarrow{A}_0=2\pi$. C'est absurde car $\rot\,\overrightarrow{\nabla}\theta=0$ et donc $\phi=0$. $\hfill{\bf CQFD}$

\begin{proposition}.\label{majoration-concert-Divine-comedy}
Soit $k'>k$; si $(\phi_k,\overrightarrow{a}_k)$ minimise $F_{\lambda,k}$ et $(\phi_{k'},\overrightarrow{a}_{k'})$ minimise $F_{\lambda,k'}$, alors on a l'inégalité:
\begin{equation}
\int_{\Omega}|\rot\,\overrightarrow{a}_{k'}|^2\leq \int_{\Omega}|\rot\,\overrightarrow{a}_{k}|^2.
\end{equation}
Si de plus $m_F(\lambda,k)<\frac{\lambda^2}{4}$, alors l'inégalité est stricte.
\end{proposition}
{\it Preuve.} Montrons la première assertion. Supposons par l'absurde que $\int_{\Omega}|\rot\,\overrightarrow{a}_{k'}|^2 > \int_{\Omega}|\rot\,\overrightarrow{a}_{k}|^2$; on a
\begin{equation}
\begin{array}{rcl}
F_{\lambda,k'}(\phi_{k'},\overrightarrow{a}_{k'})
&=&F_{\lambda,k}(\phi_{k'},\overrightarrow{a}_{k'})+\frac{k'^2-k^2}{2}\int_{\Omega}|\rot\,\overrightarrow{a}_{k'}|^2\\
&>&F_{\lambda,k}(\phi_{k'},\overrightarrow{a}_{k'})+\frac{k'^2-k^2}{2}\int_{\Omega}|\rot\,\overrightarrow{a}_{k}|^2\\
&>&F_{\lambda,k}(\phi_{k},\overrightarrow{a}_{k})+\frac{k'^2-k^2}{2}\int_{\Omega}|\rot\,\overrightarrow{a}_{k}|^2\\
&>&F_{\lambda,k'}(\phi_{k},\overrightarrow{a}_{k}) .
\end{array}
\end{equation}
Par conséquent, le couple $(\phi_{k'},\overrightarrow{a}_{k'})$ ne minimise pas la fonctionnelle, ce qui est absurde.\\
Montrons la deuxième assertion, puisque $m_F(\lambda,k)<\frac{\lambda^2}{4}$ on a $(\phi_k,\overrightarrow{a}_k)\not=(0,0)$; donc par la proposition \ref{champ-a-determine} le champ $\overrightarrow{a}_k$ est différent de $0$.\\
Supposons par l'absurde que $\int_{\Omega}|\rot\,\overrightarrow{a}_{k'}|^2\geq \int_{\Omega}|\rot\,\overrightarrow{a}_{k}|^2$; on a 
\begin{equation}
\begin{array}{rcl}
m_F(\lambda,k')
&=&F_{\lambda,k'}(\phi_{k'},\overrightarrow{a}_{k'})\\
&=&F_{\lambda,k}(\phi_{k'},\overrightarrow{a}_{k'})+\frac{k'^2-k^2}{2}\int_{\Omega}|\rot\,\overrightarrow{a}_{k'}|^2\\
&\geq&F_{\lambda,k}(\phi_{k'},\overrightarrow{a}_{k'})+\frac{k'^2-k^2}{2}\int_{\Omega}|\rot\,\overrightarrow{a}_{k}|^2\\
&\geq&F_{\lambda,k}(\phi_{k},\overrightarrow{a}_{k})+\frac{k'^2-k^2}{2}\int_{\Omega}|\rot\,\overrightarrow{a}_{k}|^2\\
&\geq&F_{\lambda,k'}(\phi_{k},\overrightarrow{a}_{k}) .
\end{array}
\end{equation}
Par conséquent, le couple $(\phi_{k},\overrightarrow{a}_{k})$ minimise aussi la fonctionnelle $F_{\lambda,k'}$; on a alors par les équations de Ginzburg-Landau
\begin{equation}
\left\lbrace\begin{array}{rcl}
L\overrightarrow{a}_{k}&=&-\frac{1}{k^2}\Rez\,\overline{\phi_{k}}(i\overrightarrow{\nabla}\phi_k+(\overrightarrow{A}_0+\overrightarrow{a}_k)\phi_k)\\
L\overrightarrow{a}_{k}&=&-\frac{1}{k'^2}\Rez\,\overline{\phi_{k}}(i\overrightarrow{\nabla}\phi_k+(\overrightarrow{A}_0+\overrightarrow{a}_k)\phi_k) .
\end{array}\right.
\end{equation}
Puisque $k'>k$, on obtient $L\overrightarrow{a}_{k}=0$; donc $\overrightarrow{a}_{k}=0$ car $L$ est injectif (cf~\ref{spectre-L-theo}). C'est absurde. $\hfill{\bf CQFD}$

\begin{theorem}.(Comportement par rapport à $k$ de $m_F$)
La fonction $k\mapsto m_F(\lambda,k)$ est monotone croissante. De plus
\begin{itemize}
\item Si $m_F(\lambda,k)=\frac{\lambda^2}{4}$ alors
\begin{equation}
\forall k'>k,\,m_F(\lambda,k')=\frac{\lambda^2}{4}.
\end{equation}
\item Si $m_F(\lambda,k) < \frac{\lambda^2}{4}$ alors
\begin{equation}
\forall k'>k,\,m_F(\lambda,k')>m_F(\lambda,k) .
\end{equation}
\end{itemize}
\end{theorem}
{\it Preuve.} La fonction 
\begin{equation}
F_{\lambda,k}(\phi,\overrightarrow{a})=\int_{\Omega}\frac{1}{2}\Vert i\overrightarrow{\nabla}\phi+(\overrightarrow{A}_0+\overrightarrow{a})\phi\Vert^2+\frac{(\lambda-|\phi|^2)^2}{4}+\frac{k^2}{2}|\rot\,\overrightarrow{a}|^2
\end{equation}
est croissante par rapport à $k$ par conséquent la fonction $m_F$ est également croissante par rapport à $k$.\\
Montrons le premier résultat; si $k'>k$, on a par les hypothèses et le raisonnement précédent
\begin{equation}
m_F(\lambda,k)=\frac{\lambda^2}{4}\mbox{~et~}m_F(\lambda,k')\geq m_F(\lambda,k) .
\end{equation}
Or $m_F(\lambda,k')\leq\frac{\lambda^2}{4}$ par la proposition~\ref{premieres-propriete-mF-mD}; par conséquent, on a 
\begin{equation}
\forall k' >k,\,\,m_F(\lambda,k')=\frac{\lambda^2}{4} .
\end{equation}
Montrons le second résultat: Si $m_F(\lambda,k')=\frac{\lambda^2}{4}$ alors on a le résultat. Si $m_F(\lambda,k')\not=\frac{\lambda^2}{4}$ on a par la proposition~\ref{premieres-propriete-mF-mD} 
\begin{equation}
m_F(\lambda,k')<\frac{\lambda^2}{4} .
\end{equation}
Il existe un couple $(\phi_e, \overrightarrow{a_e})\in{\cal A}$ qui est forcément différent du couple nul tel que $m_F(\lambda,k')=F_{\lambda,k'}(\phi_e,\overrightarrow{a_e})$. Ce couple vérifie les équations de Ginzburg-Landau.\\
Le lemme~\ref{champ-a-determine} implique que le champ $\overrightarrow{a_e}$ est différent de $0$.\\
On a alors 
\begin{equation}
\begin{array}{rcl}
m_F(\lambda,k')&=&F_{\lambda,k'}(\phi_e,\overrightarrow{a_e})\\
&>&F_{\lambda,k}(\phi_e,\overrightarrow{a_e})\geq m_F(\lambda,k) .
\end{array}
\end{equation}
On a bien le résultat voulu dans les deux cas. $\hfill{\bf CQFD}$

%\begin{theorem}.\label{decroissance}
%La fonction $m_D(\lambda,k)$ est une fonction décroissante par rapport à $\lambda$ et croissante par rapport à $k$.
%\end{theorem}
%{\it Preuve.} En utilisant l'équation~(\ref{autre-formulation}) on obtient
%\begin{equation}
%\begin{array}{rcl}
%m_D(\lambda,k)=\frac{1}{\lambda^2}m_F(\lambda,k)&=&\inf_{(\phi, \overrightarrow{a})\in{\cal A}}D_{\lambda, k}(\phi, \overrightarrow{a}) .
%\end{array}
%\end{equation}
%L'expression à l'intérieur du minimum (cf~(\ref{definition-Dklambdaphia})) est décroissante par rapport à $\lambda$ et croissante par rapport à $k$; il en est donc de même du minimum.$\hfill{\bf CQFD}$

\begin{theorem}.\label{stricte-decroissance-cas-lambda}
(Comportement par rapport à $\lambda$)\\
La fonction $\lambda\mapsto m_D(\lambda,k)$ est monotone décroissante. De plus
\begin{itemize}
\item Si $m_D(\lambda,k)=\frac{1}{4}$ alors
\begin{equation}
\forall \lambda'<\lambda,\,m_D(\lambda',k)=\frac{1}{4}.
\end{equation}
\item Si $m_D(\lambda,k)<\frac{1}{4}$ alors
\begin{equation}
\forall \lambda'<\lambda,\,m_D(\lambda',k)>m_D(\lambda,k) .
\end{equation}
\end{itemize}
\end{theorem}
{\it Preuve.} On a 
\begin{equation}
m_D(\lambda,k)=\inf_{(\phi,\overrightarrow{a})\in{\cal A}}\int_{\Omega}\frac{1}{2\lambda}\Vert i\overrightarrow{\nabla}\phi+(\overrightarrow{A}_{0}+\overrightarrow{a})\phi\Vert^2+\frac{1}{4}(1-|\phi|^2)^2+\frac{k^2}{2\lambda^2}|\rot\, \overrightarrow{a}|^2 .
\end{equation}
L'expression à l'intérieur du minimum (cf~(\ref{definition-Dklambdaphia})) est décroissante par rapport à $\lambda$; il en est donc de même du minimum.\\
Montrons le premier résultat. On sait que la fonction $m_D(\lambda,k)$ est décroissante par rapport à $\lambda$ donc 
\begin{equation}\label{inegalite-dans-un-sens}
\forall \lambda'<\lambda,\,m_D(\lambda',k)\geq\frac{1}{4}.
\end{equation}
Mais on sait (cf~\ref{premieres-propriete-mF-mD}) que
\begin{equation}\label{inegalite-autre-sens}
\forall \lambda'>0,\,\,m_D(\lambda',k)\leq\frac{1}{4} .
\end{equation}
Les équations~(\ref{inegalite-dans-un-sens}) et~(\ref{inegalite-autre-sens}) impliquent donc une égalité.\\
Supposons maintenant que $m_D(\lambda,k)<\frac{1}{4}$.\\
%L'inégalité~(\ref{inegalite-autre-sens}) est en fait vraie dans tous les cas. Par conséquent, si $m_D(\lambda,k)\not=\frac{1}{4}$ alors
%\begin{equation}
%m_D(\lambda,k)<\frac{1}{4}
%\end{equation}
On sait par le théorème~\ref{existence-d-un-minimum} qu'il existe un couple $(\phi_e, \overrightarrow{a_e})\in{\cal A}$ tel que 
\begin{equation}
m_D(\lambda,k)=D_{\lambda, k}(\phi_e, \overrightarrow{a_e}) .
\end{equation}
Le minimum de la fonctionnelle $F_{\lambda,k}$ est alors atteint sur un couple différent du couple $(0,0)$ puisque si c'était le cas, on aurait $m_D(\lambda,k)=\frac{1}{4}$.\\
On a alors par la contraposée du lemme~\ref{annulation-deux-integrale}

\begin{equation}\label{un-ou-l-autre}
\begin{array}{c}
\int_{\Omega}\Vert i\overrightarrow{\nabla}\phi_e+(\overrightarrow{A}_0+\overrightarrow{a_e})\phi_e\Vert^2\not=0\\
\mbox{~ou~}\int_{\Omega}|\rot\,\overrightarrow{a_e}|^2\not=0 .
\end{array}
\end{equation}
Une inspection de la formule~(\ref{definition-Dklambdaphia}) nous montre que $D_{\lambda, k}(\phi, \overrightarrow{a})$ est alors strictement décroissant par rapport à $\lambda$. Soit $\lambda'<\lambda$, on a alors
\begin{equation}
\begin{array}{rcl}
m_D(\lambda',k)&=&\inf_{(\phi, \overrightarrow{a})\in{\cal A}}D_{\lambda', k}(\phi, \overrightarrow{a})\\
&=&D_{\lambda', k}(\phi_e, \overrightarrow{a_e})\\
&>&D_{\lambda, k}(\phi_e, \overrightarrow{a_e})\geq m_D(\lambda,k) .
\end{array}
\end{equation}
On a donc bien montré qu'il y a une stricte décroissance.$\hfill{\bf CQFD}$

\begin{theorem}.\label{ecrantement}
Si le champ $H^a_{int}$ réalise le minimum de la fonction $H_{int}\mapsto G^{V}_{k,H_{ext}}(H_{int})$, alors $H^a_{int}\leq H_{ext}$
\end{theorem}
{\it Preuve.} Par le théorème~\ref{stricte-decroissance-cas-lambda}, la fonction 
\begin{equation}
H_{int}\mapsto(\frac{H_{int}}{2\pi k})^2m_F(\frac{2\pi k}{H_{int}},k)
\end{equation}
est une fonction croissante par rapport à $H_{int}$.\\
Par conséquent, la fonction
\begin{equation}
\left\lbrace\begin{array}{rcl}
[0,+\infty[&\mapsto&\R\\
H_{int}&\mapsto&G^{V}_{k,H_{ext}}(H_{int})=(\frac{H_{int}}{2\pi k})^2m_F(\frac{2\pi k}{H_{int}},k)+\frac{1}{2}(H_{int}-H_{ext})^2
\end{array}\right.
\end{equation}
est strictement croissante sur l'intervalle $]H_{ext}, +\infty[$ comme somme d'une fonction croissante et d'une fonction strictement croissante.
Donc le minimum de $G^{V}_{k,H_{ext}}(H_{int})$ n'est pas atteint sur cet intervalle; le minimum sur $[0,+\infty[$ vérifie par conséquent $H^a_{int}\leq H_{ext}$.$\hfill{\bf CQFD}$

\saction{Formule de Bochner-Kodaira-Nakano}\label{sec:Bochner-Kodaira-Nakano}
\noindent La fonctionnelle réduite lorsque $k=\frac{1}{\sqrt{2}}$ prend la forme particulière suivante
\begin{equation}
F_{\lambda,\frac{1}{\sqrt{2}}}(\phi, \overrightarrow{a})=\int_{\Omega}\frac{1}{2}\Vert i\overrightarrow{\nabla}\phi+(\overrightarrow{A}_{0}+\overrightarrow{a})\phi\Vert^2+\frac{1}{4}(\lambda-|\phi|^2)^2+\frac{1}{4}|\rot\, \overrightarrow{a}|^2 .
\end{equation}
Dans cette section, on montre une formule de Bochner-Kodaira-Nakano (cf~\cite{Demally} pour des formules générales sur des fibrés de variétés complexes) qui va nous donner de très utiles renseignements. Cette formule est très classique et est aussi appelée formule de \index{Bogmol'nyi}Bogmol'nyi, \index{Weitzenbock}Weitzenbock, \index{Lichnerowicz}Lichnerowicz (cf~\cite{Acker} et \cite{Lichne}) selon les différentes écoles scientifiques.\\
On introduit pour cela une nouvelle \index{$A_+$}fonctionnelle
\begin{equation}\label{fonctionnelle-critique-2}
A_{+}(\phi, \overrightarrow{a})=\int_{\Omega}\frac{1}{2}\vert D_{+}\phi\vert^2+\frac{1}{4}|\rot\,\overrightarrow{A_{0}}+\rot\,\overrightarrow{a}-(\lambda-|\phi|^2)|^2,
\end{equation}
où \index{$D_+$}
\begin{equation}
\begin{array}{rcl}
D_{+}\phi&=&2L_{+}\phi+(a_{y}-ia_{x})\phi\\
&=&\frac{\partial\phi}{\partial x}+i\frac{\partial\phi}{\partial y}+A_{y}\phi-iA_{x}\phi\\
&=&-i[i\frac{\partial\phi}{\partial x}+A_x\phi]+[i\frac{\partial\phi}{\partial y}+A_y\phi] .
\end{array}
\end{equation}
Le résultat suivant est fondamental.
\begin{theorem}.\label{Bochner-Kodaira-Nakano}
Si $(\phi, \overrightarrow{a})\in {\cal A}$, alors
\begin{equation}
F_{\lambda,\frac{1}{\sqrt{2}}}(\phi, \overrightarrow{a})=\lambda\pi-\pi^2+A_{+}(\phi, \overrightarrow{a}) .
\end{equation}
\end{theorem}
{\it Preuve.} On fait le calcul sur les fonctions régulières et on conclut par densité.
\begin{equation}
\begin{array}{lcl}
2(A_{+}-F)&=&\int_{\Omega}[(\frac{\partial\phi}{\partial x}-iA_{x}\phi)(\overline{i\frac{\partial\phi}{\partial y}+A_{y}\phi})+(\overline{\frac{\partial\phi}{\partial x}-iA_{x}\phi})(i\frac{\partial\phi}{\partial y}+A_{y}\phi)\\
&-&(\rot\,\overrightarrow{A})(\lambda-|\phi|^2)]dxdy+\frac{1}{2}\int_{\Omega}\lbrace|\rot\,\overrightarrow{A}|^2-|\rot\,\overrightarrow{a}|^2\rbrace dxdy\\[2mm]
&=&\int_{\Omega}[i(\partial_{x}{\overline{\phi}}\partial_{y}\phi-\partial_{x}\phi\partial_{y}{\overline{\phi}})+(A_{y}\phi\partial_{x}{\overline{\phi}}-A_{x}{\overline{\phi}}\partial_{y}\phi)\\
&+&(A_{y}\overline{\phi}\partial_{x}{\phi}-A_{x}{\phi}\partial_{y}\overline{\phi})+\partial_{x}A_{y}|\phi|^2-\partial_{y}A_{x}|\phi|^2]dxdy\\
&+&2\pi^2-2\pi\lambda\\[2mm]
&=&\int_{\Omega}i(\partial_{x}[{\overline{\phi}}\partial_{y}\phi]-\partial_{y}[{\overline{\phi}}\partial_{x}\phi])+\partial_{x}[A_{y}|\phi|^2]-\partial_{y}[A_{x}|\phi|^2]\\
&+&2\pi^2-2\pi\lambda\\[2mm]
&=&\int_{\Omega}\partial_{x}[{\overline{\phi}}(i\partial_{y}\phi+A_{y}\phi)]-\partial_{y}[{\overline{\phi}}(i\partial_{x}\phi+A_{x}\phi)]\\
&+&2\pi^2-2\pi\lambda\\[2mm]
&=&\int_{\Omega}\divergence\,\overrightarrow{W}dxdy+2\pi^2-2\pi\lambda .
\end{array}
\end{equation}
Le champ de vecteur 
\begin{equation}
\overrightarrow{W}=\left(\begin{array}{c}
\overline{\phi}(i\frac{\partial\phi}{\partial y}+A_y\phi)\\
-\overline{\phi}(i\frac{\partial\phi}{\partial x}+A_x\phi)
\end{array}\right)
\end{equation}
est ${\cal L}$-périodique sur $\R^2$ donc est bien défini sur $\Tore{\cal L}$. Par conséquent, la dernière intégrale est nulle puisque cet espace est compact sans bord. $\hfill{\bf CQFD}$

\begin{remarque}.
Compte tenu de la positivité de $A_+$, on a pour tout $(\phi, \overrightarrow{a})\in {\cal A}$ 
\begin{equation}
F_{\lambda,\frac{1}{\sqrt{2}}}(\phi, \overrightarrow{a})\geq \lambda\pi-\pi^2 .
\end{equation}
\end{remarque}
%\begin{remarque}.
%On a une formule similaire pour la fonctionnelle 
%\begin{equation}
%A_{-}(\phi, \overrightarrow{a})=\int_{\Omega}\frac{1}{2}\Vert -2L_{-}\phi+(a_{y}+ia_{x})\phi\Vert^2+\frac{1}{4}|\rot\,\overrightarrow{A_0}+\overrightarrow{a}+(\lambda-|\phi|^2)|^2
%\end{equation}
%La formule étant $F_{\lambda,\frac{1}{\sqrt{2}}}=A_{-}-\lambda\pi-\pi^2$. Cette formule ne donne rien d'intéressant malheureusement.
%\end{remarque}
Le théorème~\ref{Bochner-Kodaira-Nakano} nous permet de donner des résultats très précis sur le problème réduit.
\begin{theorem}.\label{merveille-praguoise}
Si $k\geq\frac{1}{\sqrt{2}}$ et si $\lambda\leq 2\pi$ alors 
\begin{equation}
\inf_{(\phi, \overrightarrow{a})\in{\cal A}}F_{\lambda,k}(\phi, \overrightarrow{a})=\frac{\lambda^2}{4} .
\end{equation}
De plus ce minimum n'est atteint que pour le couple $(0,0)$.
\end{theorem}
\begin{remarque}.
Le théorème précédent ne résout pas le problème \ref{quelles-peine-pour-ca}. Ce problème sera résolu pour une large classe de cas dans les théorèmes \ref{usage-theoreme-monotonie} et \ref{apparition-intervalle2} et la démonstration n'utilisera pas le théorème précédent.
\end{remarque}
{\it Preuve.} On utilise la fonctionnelle~(\ref{fonctionnelle-critique-2}).\\
On développe $F_{\lambda,k}$:
\begin{equation}\label{emerveillement-a-prague}
\begin{array}{rcl}
F_{\lambda,k}(\phi, \overrightarrow{a})
&=&\frac{1}{2}(k^2-\frac{1}{2})\int_{\Omega}|\rot\,\overrightarrow{a}|^2+F_{\lambda,\frac{1}{\sqrt{2}}}(\phi, \overrightarrow{a})\\
&=&\frac{1}{2}(k^2-\frac{1}{2})\int_{\Omega}|\rot\,\overrightarrow{a}|^2+\lambda\pi-\pi^2+A_{+}(\phi, \overrightarrow{a})\\
&=&\lambda\pi-\pi^2+\frac{1}{4}\int_{\Omega}(2\pi-\lambda)^2+2(2\pi-\lambda)(\rot\,\overrightarrow{a}+|\phi|^2)\\
&+&\int_{\Omega}|\rot\,\overrightarrow{a}+|\phi|^2|^2+\frac{1}{2}(k^2-\frac{1}{2})\int_{\Omega}|\rot\,\overrightarrow{a}|^2+\frac{1}{2}\int_{\Omega}\Vert D_+\phi\Vert^2\\[2mm]
&=&\frac{\lambda^2}{4}+2\int_{\Omega}(2\pi-\lambda)|\phi|^2\\
&+&\int_{\Omega}|\rot\,\overrightarrow{a}+|\phi|^2|^2+\frac{1}{2}(k^2-\frac{1}{2})\int_{\Omega}|\rot\,\overrightarrow{a}|^2+\frac{1}{2}\int_{\Omega}\Vert D_+\phi\Vert^2
\end{array}
\end{equation}
Ce calcul nous donne si $k\geq\frac{1}{\sqrt{2}}$ et $\lambda\leq 2\pi$ l'inégalité
\begin{equation}
F_{\lambda,k}(\phi, \overrightarrow{a})\geq \frac{\lambda^2}{4} .
\end{equation}
Supposons que $F_{\lambda,k}(\phi, \overrightarrow{a})=\frac{\lambda^2}{4}$; alors la formule~(\ref{emerveillement-a-prague}) nous donne les égalités suivantes:
\begin{equation}\label{On-reprend-tout}
\begin{array}{rcl}
0&=&(2\pi-\lambda)\int_{\Omega}|\phi|^2 ,\\
0&=&(k^2-\frac{1}{2})\int_{\Omega}|\rot\,\overrightarrow{a}|^2 ,\\
0&=&\int_{\Omega}|\rot\,\overrightarrow{a}+|\phi|^2|^2 ,\\
0&=&\int_{\Omega}\Vert D_+\phi\Vert^2 .
\end{array}
\end{equation}
La troisième égalité implique 
\begin{equation}
\rot\,\overrightarrow{a}+|\phi|^2=0 .
\end{equation}
En intégrant sur $\Omega$, on obtient
\begin{equation}
\int_{\Omega}|\phi|^2=-\int_{\Omega}\rot\,\overrightarrow{a}=0 
\end{equation}
et par conséquent $\phi=0$.\\
Cette relation implique alors que $\rot\,\overrightarrow{a}=0$; la proposition \ref{annulation-une-integrale} nous donne la dernière annulation $\overrightarrow{a}=0$. $\hfill{\bf CQFD}$

\chapter{La bifurcation d'Abrikosov}\label{chapitre-bifurcation-abrikosov}
\noindent Ce chapitre est consacré exclusivement à l'étude de la bifurcation d'Abrikosov. Dans la première section on définit la bifurcation et, dans les deux sections qui suivent, on étudie la stabilité de ce minimum.\\
Dans  la dernière section on étudie le comportement quand $k$ tend vers $\infty$ des solutions des équations de Ginzburg-Landau et on montre que les solutions qui apparaissent sont les solutions bifurquées et la solution triviale.
\saction{Réduction de Lyapunov-Schmidt et bifurcation}\label{sec:lyapunov-schmidt}
\noindent La bifurcation d'Abrikosov consiste à trouver les solutions des équations de Ginzburg-Landau dans un voisinage de l'état normal $(H_{ext},\phi,\overrightarrow{a})$.\\
Pour cela on va d'abord étudier dans ce chapitre les solutions des équations de Ginzburg-Landau pour la fonctionnelle $F_{\lambda,k}$; cette étude ne dépend pas de $H_{ext}$ mais de $H_{int}$ et $k$. On étudiera alors dans le chapitre suivant le minimum de
\begin{equation}
G^{V}_{k,H_{ext}}(H_{int})=\frac{1}{\lambda^2}m_{F}(\lambda,k)+\frac{1}{2}(H_{int}-H_{ext})^2
\end{equation}
avec $\lambda=\frac{2\pi k}{H_{int}}$. On montrera que l'hypothèse d'Abrikosov concernant les solutions minimisantes est pertinente dans un certain domaine du plan $(k,H_{ext})$.\\
Par commodité on utilisera la variable $\lambda=\frac{2\pi k}{H_{int}}$ qui est plus pratique pour écrire nos calculs.\\
Commençons par déterminer la nature du couple $(0,0)$ du point de vue de la théorie de Morse.
\begin{proposition}.
Le couple $(0,0)$ est un point critique pour la fonctionnelle $F_{\lambda,k}$. Ce point est non dégénéré si et seulement si $\lambda\not=2\pi+4\pi n$ ($\forall n\in\N$).\\
La forme quadratique $D^2F_{(0,0)}(\delta\phi,\overrightarrow{\delta a})$ est définie positive si et seulement si $\lambda<2\pi$.
\end{proposition}
{\it Preuve.} La différentielle première au point $(\phi,\overrightarrow{a})$ est la forme linéaire
\begin{equation}
\begin{array}{rcl}
D_{(\phi,\overrightarrow{a})}F_{\lambda,k}(\phi', \overrightarrow{a}')&=&\int_{\Omega}\Rez[\overline{\phi'}\lbrace[i\overrightarrow{\nabla}+(\overrightarrow{A}_{0}+\overrightarrow{a})]^2\phi\rbrace-(\lambda-|\phi|^2)\phi\rbrace]\\
&+&\int_{\Omega}\overrightarrow{a}'\lbrace \Rez[\overline{\phi}(i\overrightarrow{\nabla}\phi+(\overrightarrow{A}_0+\overrightarrow{a})\phi)]+k^2L\,\overrightarrow{a}\rbrace
\end{array}
\end{equation}
o\`u $L$ est l'opérateur étudié dans la section \ref{sec:operateur-L}. On a facilement
\begin{equation}
D_{(0,0)}F_{\lambda,k}=0 .
\end{equation}
La différentielle seconde au point $(\phi,\overrightarrow{a})$ est la forme bilinéaire de ${\cal A}\times{\cal A}$ dans $\R$ définie par:
\begin{equation}
\begin{array}{rcl}
%D^2_{(\phi,\overrightarrow{a})}F_{\lambda,k}:{\cal A}\times{\cal A}&\mapsto&\R\\
(\phi_1,\overrightarrow{a}_1),(\phi_2,\overrightarrow{a}_2)&\mapsto&
\int_{\Omega}\Rez\{\overline{\phi_1} [i\overrightarrow{\nabla}+\overrightarrow{A}]^2\phi_2\}\\
&&+2\Rez[\int_{\Omega}[\overrightarrow{a}_2\overline{\phi_1}+\overrightarrow{a}_1\overline{\phi_2}].(i\overrightarrow{\nabla}\phi+\overrightarrow{A}\phi)]\\
&&+\int_{\Omega}-(\lambda-|\phi|^2)\Rez\,\overline{\phi_1}\phi_2+2\Rez(\overline{\phi}\phi_2)\Rez(\overline{\phi}\phi_1)\\
&&+\int_{\Omega} \overrightarrow{a}_1.\overrightarrow{a}_2|\phi|^2+k^2 \rot\, \overrightarrow{a}_1 .\rot\, \overrightarrow{a}_2 .
\end{array}
\end{equation}
On a facilement
\begin{equation}
\begin{array}{rcl}
D^2_{(0,0)}F_{\lambda,k}:{\cal A}\times{\cal A}&\mapsto&\R\\
(\phi_1,\overrightarrow{a}_1),(\phi_2,\overrightarrow{a}_2)&\mapsto&
\int_{\Omega}\Rez\{\overline{\phi_1} ([i\overrightarrow{\nabla}+\overrightarrow{A}_0]^2-\lambda)\phi_2\}\\
&&+\int_{\Omega} k^2 \rot\, \overrightarrow{a}_1 \rot\, \overrightarrow{a}_2 .
\end{array}
\end{equation}
La forme quadratique associée $D^2_{(0,0)}F_{\lambda,k}(\phi', \overrightarrow{a}')$ a pour expression
\begin{equation}
D^2_{(0,0)}F_{\lambda,k}(\phi', \overrightarrow{a}')=\int_{\Omega}\overline{\phi'} ([i\overrightarrow{\nabla}+\overrightarrow{A}_0]^2-\lambda)\phi'+k^2 |\rot\, \overrightarrow{a}'|^2 .
\end{equation}
La famille $(\phi_n)_{n\in\N}$ (cf~\ref{spectre-de-H-C-infinite}) est une base hilbertienne de $L^2(E_1)$ formée de vecteur propres pour $H$. Si 
\begin{equation}
\phi'=\sum_{n=0}^{\infty}\mu'_n\phi_n ,
\end{equation}
alors cette forme quadratique peut s'exprimer sous la forme:
\begin{equation}
D^2_{(0,0)}F_{\lambda,k}(\phi', \overrightarrow{a}')=
\sum_{n=0}^{\infty}(\mu'_n)^2(2\pi+4\pi n-\lambda)+k^2 \int_{\Omega}|\rot\, \overrightarrow{a}'|^2 .
\end{equation}
On sait que $\rot\,\overrightarrow{a}=0$ équivaut à $\overrightarrow{a}=0$ (cf~\ref{annulation-une-integrale}). La différentielle seconde est donc non dégénérée si et seulement si $2\pi+4\pi n-\lambda\not=0$ ($\forall n\in\N$).\\
La différentielle seconde est positive si et seulement si $2\pi+4\pi n-\lambda>0$ ($\forall n\in\N$), c'est à dire $\lambda<2\pi$. $\hfill{\bf CQFD}$\\
\\
On va maintenant étudier ce qui se passe au voisinage de $\lambda=2\pi$. Tous les calculs que nous effectuons sont pour $(\phi,\overrightarrow{a})$ dans ${\cal A}$. Pour $k>0$ fixé, on va montrer qu'il existe une courbe de couples $\lambda\mapsto (\phi(\lambda),\overrightarrow{a}(\lambda))$ solution des équations (\ref{Odeh-equations}) qui ne soit pas $C^{\infty}$ en $2\pi$. Ces couples seront dits bifurqués et on dira qu'il y a une bifurcation en $2\pi$.\\
%\begin{definition}.
%Soit $H(\alpha, \lambda)$ est une fonction analytique réelle définie sur un voisinage de $(0,0)$ dans $\R^2$ et s'annulant en $(0,0)$. On dit que $\eta(\lambda)$ est une solution bifurquée de l'équation $H=0$ si $\eta$ est continu vérifie l'équation $H(\eta(\lambda), \lambda)=0$ et $\eta(0)=0$ mais que la fonction $\eta$ n'est pas analytique.
%\end{definition}
%Dans cette section, on montre qu'il y a une bifurcation de la solution
%\begin{equation}
%(\phi, \overrightarrow{a})=(0,\overrightarrow{0})
%\end{equation}
%vers des couples où $\phi$ est non nul pour une valeur remarquable du champ magnétique. 
Cette étude a été entreprise pour la première fois par Odeh dans \cite{odehII}. Malheureusement ces premiers travaux souffrent d'un défaut de rigueur, aucun espace fonctionnel n'apparaissant très nettement dans la démonstration.\\
Dans la suite de cette section, on écrit\index{$\epsilon$}
\begin{equation}\label{approx-premier-niveau}
\left\lbrace\begin{array}{l}
\phi=\alpha\phi_0+\theta\mbox{~où~}\theta\in\phi_0^{\perp}\\
\lambda=\lambda_0+\epsilon=2\pi+\epsilon .
\end{array}\right.
\end{equation}
La valeur propre $\lambda_0$ vaut $2\pi$ et a été introduite au théorème~\ref{spectre-de-H-C-infinite}. Puisque l'étude se fait au voisinage de $\lambda=\lambda_0$, on a isolé la composante sur $\phi_0$ de $\phi$. Dans notre cadre les physiciens appellent cette décomposition << l'approximation du premier niveau de Landau >>. Dans d'autres contextes, c'est la méthode de Lyapunov-Schmidt (ou encore la méthode de projection de Feschbach, ou la méthode  dite du problème de Grushin).
\begin{remarque}.\label{remarque-tres-intelligente}
Le paramètre $\alpha$ est a priori complexe, mais on voit facilement que l'on peut se ramener à l'étude du cas o\`u $\alpha$ est réel, car si le couple $(\phi,\overrightarrow{a})$ est solution des équations~(\ref{Odeh-equations}) et $\omega\in\C$ avec $|\omega|=1$, alors le couple $(\omega\phi,\overrightarrow{a})$ est solution de la même équation. On parle d'action de $S^1$ laissant invariante la fonctionnelle. Si on prend $\alpha$ réel on obtiendra des solutions par paires car si $(\phi,\overrightarrow{a})$ est solution alors $(-\phi,\overrightarrow{a})$ est aussi solution. Dans toute la suite de cette section, on supposera que $\alpha$ est réel.
\end{remarque}
%Les couples $(\phi, \overrightarrow{a})=(0,0)$ sont solutions du système~(\ref{Odeh-equations}) mais il y a également d'autres solutions qui elles sont bifurquées, c'est à dire non analytique.\\
On va utiliser la méthode de Lyapunov-Schmidt qui consiste à transformer un problème de bifurcation de dimension infinie en un problème plus simple de dimension finie en projetant sur des espaces adéquats. On cherchera alors à mettre en évidence l'existence de solutions $(\phi,\overrightarrow{a})$ telles que $\phi\not=0$.\\
%On rappelle que l'on note ${\cal M}$ l'espace des réseaux possibles qui est défini à l'équation~(\ref{espace-module}). On montrera que la dépendance des solutions bifurquées par rapport à  $m$ est controlée sur les parties compactes de ${\cal M}$. 
On cherche aussi à garder un contr\^ole sur $k$ quand $k\rightarrow\infty$, on pose donc \index{$\beta$}
\begin{equation}
\beta=\frac{1}{k}
\end{equation}
et on réécrit les équations~(\ref{Odeh-equations}) avec $\beta$; on obtient:
\begin{equation}\label{Odeh-equations-bis}
\left\lbrace\begin{array}{rcl}
[i\overrightarrow{\nabla}+\overrightarrow{A}_0+\overrightarrow{a}]^2\phi&=&(\lambda-|\phi|^2)\phi,\\
-\rot^*\,\rot\, \overrightarrow{a}&=&\beta^2\Rez[\overline{\phi}(i\overrightarrow{\nabla}\phi+(\overrightarrow{A}_0+\overrightarrow{a})\phi)] .
\end{array}\right.
\end{equation}
Ci-dessous on effectue la réduction à la dimension finie.
\begin{theorem}.\label{reduc-lyapunov}
Soit $\beta_{0}>0$ il existe un voisinage ${\cal V}$ de $(0,0,0)$ dans ${\cal A}\times \R$ de la forme
\begin{equation}
\left\lbrace\begin{array}{l}
\Vert \overrightarrow{a}\Vert _{H^{1}}<\delta,\\
\Vert \phi\Vert_{H^{1}(E_{1})}<\delta\mbox{~et~}\\
|\epsilon|<\delta
\end{array}\right.
\end{equation}
avec $\delta>0$ tel que pour tout $\beta\in[-\beta_0,\beta_{0}]$, tout triplet $(\phi,\overrightarrow{a},\epsilon)$ dans ${\cal A}\times\R$ solution de l'équation~(\ref{Odeh-equations-bis}) s'exprime sous la forme \index{$\Phi$}\index{$\overrightarrow{\Psi}$}
\begin{equation}
\left\lbrace\begin{array}{rcl}
\phi&=&\alpha\phi_{0}+\Phi(\alpha,\epsilon,\beta),\\
\overrightarrow{a}&=&\overrightarrow{\Psi}(\alpha,\epsilon,\beta)
\end{array}\right.
\end{equation}
o\`u $\Phi$ et $\overrightarrow{\Psi}$ sont des fonctions analytiques de $\alpha$, $\epsilon$ et $\beta$ où $\alpha$ est solution d'une équation \index{$g$}
\begin{equation}\label{definition-fonction-g}
g(\alpha, \epsilon,\beta)=0,
\end{equation}
o\`u $g$ fonction analytique par rapport à $\alpha$, $\epsilon$ et $\beta$ est donnée en (\ref{equation-reduite}).\\
Ces fonctions vérifient $\Phi(0,0,\beta)=0$, $\overrightarrow{\Psi}(0,0,\beta)=0$ et $\langle \phi_0,\Phi(\alpha,\epsilon,\beta)\rangle=0$.
\end{theorem}
{\it Preuve.} On pose\index{$F_{\beta}$}\index{$G_{\beta}$}
\begin{equation}\label{reduction-en-cours}
\left\lbrace
\begin{array}{rcl}
F_{\beta}(\phi,\overrightarrow{a})&=&-\phi|\phi|^2-2 \overrightarrow{a}.(i\overrightarrow{\nabla}\phi+\overrightarrow{A}_0\phi)-\overrightarrow{a}^2\phi,\\
G_{\beta}(\phi,\overrightarrow{a})&=&-\beta^2\Rez(\overline{\phi}(i\overrightarrow{\nabla}\phi+(\overrightarrow{A}_0+\overrightarrow{a})\phi).
\end{array}\right.
\end{equation}
Les équations de Ginzburg-Landau~(\ref{Odeh-equations-bis}) s'écrivent alors
\begin{equation}
\left\lbrace
\begin{array}{l}
[H-\lambda]\phi=F_{\beta}(\phi,\overrightarrow{a}),\\
L\overrightarrow{a}=G_{\beta}(\phi,\overrightarrow{a}).
\end{array}\right.
\end{equation}
On réécrit ces équations dans la décomposition~(\ref{approx-premier-niveau})
\begin{equation}\label{expansion-des-inconnues}
\left\lbrace
\begin{array}{rcl}
-\epsilon\alpha\phi_{0}+[H-\lambda_0]\theta-\epsilon\theta&=&F_{\beta}(\alpha\phi_{0}+\theta,\overrightarrow{a}),\\
L\overrightarrow{a}&=&G_{\beta}(\alpha\phi_0+\theta,\overrightarrow{a}).
\end{array}\right.
\end{equation}
\begin{lemme}.\label{divers-sur-R0}
Il existe une application linéaire noté $R_0$\index{$R_0$} telle que
\begin{equation}\label{definition-R0}
\begin{array}{rcll}
R_0&=&(H-\lambda_0)^{-1}&\mbox{~sur~}\{\phi_0\}^{\perp}\\
&=&0&\mbox{~sur~}Vect\,\{\phi_0\} .
\end{array}
\end{equation}
L'application $R_0$ est définie sur $L^2(E_1)$ et à valeur dans $H^2(E_1)$. Cette application est autoadjointe.\\
Par ailleurs $R_0[H-\lambda_0]=[H-\lambda_0]R_0=P_0$ o\`u $P_0$ est la projection orthogonale sur $\{\phi_0\}^{\perp}$.
\end{lemme}
{\it Preuve.} Si $\phi=\alpha\phi_0+\theta$ avec $\theta\in\{\phi_0\}^{\perp}$ alors $\theta$ a un antécédent par $[H-\lambda_0]$ on a donc
\begin{equation}\label{equation-definissante-pour-R0}
R_0(\phi)=(H-\lambda_0)^{-1}\theta .
\end{equation}
L'opérateur $R_0$ est à valeur dans $H^2(E_1)$ car $R_0$ est un opérateur elliptique d'ordre $-2$. 
On projette sur $\phi_{0}^{\perp}$ la première équation et on utilise l'opérateur $R_0$ (cf~(\ref{definition-R0})).\\
La nature autoadjointe de $R_0$ est évidente dans la base hilbertienne $(\phi_n)_{n\in\N}$ définie au théorème \ref{spectre-de-H-C-infinite}.\\
La dernière assertion est triviale par l'équation (\ref{equation-definissante-pour-R0}). $\hfill{\bf CQFD}$\\
\\
On sait (cf~\ref{spectre-L-theo}) que l'opérateur $L$ est inversible; on note \index{$M$}
\begin{equation}\label{definition-de-M}
M=L^{-1}
\end{equation}
cet inverse qui est un opérateur d'ordre $-2$. Puisque $L$ est positif, l'opérateur $M$ est aussi positif. L'opérateur $M$ a pour espace de départ $L^2_{\divergence,0}(\Tore{\cal L},\R^2)$ et pour image $H^2_{\divergence,0}(\Tore{\cal L},\R^2)$.\\
On inverse les opérateurs $L$ et $[H-\lambda_0]$ dans l'équation~(\ref{expansion-des-inconnues})

\begin{equation}\label{systeme-elliptique}
\left\lbrace
\begin{array}{rcl}
\theta&=&R_{0}[\epsilon\theta+F_{\beta}(\alpha\phi_{0}+\theta,\overrightarrow{a})],\\
\overrightarrow{a}&=&MG_{\beta}(\alpha\phi_{0}+\theta,\overrightarrow{a}).
\end{array}\right.
\end{equation}
On note \index{$w$}
\begin{equation}
w=\left(\begin{array}{c}
\theta\\
\overrightarrow{a}
\end{array}\right)
\end{equation}
et \index{$N(w,\alpha, \epsilon,\beta)$}$N(w,\alpha, \epsilon,\beta)$ la fonction qui apparaît dans le second membre de l'équation~(\ref{systeme-elliptique})
\begin{equation}
N(w,\alpha,\epsilon,\beta)=\left(\begin{array}{l}
R_{0}[\epsilon\theta+F_{\beta}(\alpha\phi_{0}+\theta,\overrightarrow{a})]\\
MG_{\beta}(\alpha\phi_{0}+\theta,\overrightarrow{a})
\end{array}\right) .
\end{equation}
Un triplet $(\alpha, \epsilon,\beta)$ étant donné et le paramètre $\beta$ étant fixé dans un intervalle $[-\beta_0,\beta_0]$, on cherche  donc pour trouver des solutions de l'équation de Ginzburg-Landau, à trouver une solution $w$ dans ${\cal A}$ telle que 
\begin{equation}\label{equation-typique-TFI}
w=N(w,\alpha,\epsilon,\beta) .
\end{equation}

\begin{lemme}.\label{differentiabilité-de-Psi-phi}
Si $\beta_{0}>0$ alors il existe un voisinage de $(w, \alpha,\epsilon)=(0,0,0)$ dans $(H^{1}(E_1)\cap \{\phi_0\}^{\perp})\times H^1(\Tore{\cal L},\R^2)\times\R^2$ tel que, pour tout $\beta\in[-\beta_0,\beta_0]$, le système~(\ref{systeme-elliptique}) admette une solution unique $w_s(\alpha,\epsilon,\beta)$ dans la classe $H^{1}$. Cette solution est réelle analytique en $\alpha$, $\epsilon$ et $\beta$. La solution obtenue est à valeur dans $(C^{\infty}(E_{1})\cap \{\phi_0\}^{\perp})\times C^{\infty}_{\divergence, 0}(\Tore{\cal L}; \R^2)$.
\end{lemme}
{\it Preuve.}\\
{\bf $1^{ere}$ étape :}\\
On supposera toujours que $|\beta|\leq\beta_0$.\\
Nous allons montrer que la fonction $N$ est $C^{\infty}$ pour la topologie normique de l'espace $(H^{1}(E_1)\cap \{\phi_0\}^{\perp})\times H^1(\Tore{\cal L},\R^2)\times\R^2$.\\
%en fonction de $w, \alpha,\epsilon, \beta$ pour la norme
%\begin{equation}
%\Vert w\Vert_{H^{1}(E_{1})\oplus H^{1}}+|\alpha|+|\epsilon| .
%\end{equation}
Le fait que $N$ dépende de façon $C^{\infty}$ de $\alpha$, $\epsilon$ et $\beta$ est évident.\\
Les termes $-\phi|\phi|^2$, $\overrightarrow{a}^2\phi$, $\overline{\phi}(\overrightarrow{A}_0+\overrightarrow{a})\phi$ et $-2\overrightarrow{a}.\overrightarrow{A}_0\phi$ appartiennent à $L^2$; donc leurs images par les opérateurs $R_{0}$ et $M$ appartiennent à $H^2$ donc a fortiori à $H^1$.\\
Il reste à prouver que les termes $R_{0}[\overrightarrow{a}.\overrightarrow{\nabla}\phi]$ et $M[\overline{\phi}\overrightarrow{\nabla}\phi]$ appartiennent à $H^{1}$.\\
Le terme $\overrightarrow{\nabla}\phi$ appartient à $L^2$ tandis que le potentiel $\overrightarrow{a}$ appartient à $H^1$. Donc une combinaison des estimées du théorème~\ref{sobolev-imbedding} et de l'inégalité de Hölder montre que $\overrightarrow{a}.\overrightarrow{\nabla}\phi$ appartient à $L^\frac{3}{2}$. Les estimées elliptiques \ref{crave-for-it-too} montrent que $R_{0}[\overrightarrow{a}.\overrightarrow{\nabla}\phi]$ appartient à  $W_{2}^{\frac{3}{2}}$. Une deuxième application du théorème~\ref{sobolev-imbedding} montre que $R_{0}[\overrightarrow{a}.\overrightarrow{\nabla}\phi]$ appartient à $W_{1}^{6}$ donc à $H^{1}$.\\
L'application $N$ est donc bien à valeur dans 
\begin{equation}
(H^{1}(E_{1})\cap \{\phi_0\}^{\perp})\times H^{1}_{\divergence, 0}(\Tore{\cal L}; \R^2) .
\end{equation}
Le fait que cette application soit en fait $C^{\infty}$ sur l'espace 
\begin{equation}
(H^{1}(E_1)\cap \{\phi_0\}^{\perp})\times H^{1}_{\divergence, 0}(\Tore{\cal L}; \R^2)\times \R\times \R\times \R
\end{equation}
se montre de la même manière.\\
{\bf $2^{eme}$ étape :}\\
En utilisant le théorème des fonctions implicites dans les espaces de Banach, on montre qu'il existe $\delta>0$ tel que, pour tout $\beta\in[-\beta_0,\beta_0]$, on ait dans la boule :
\begin{equation}
\Vert \theta\Vert _{H^1}+\Vert \overrightarrow{a}\Vert _{H^1}+|\alpha|+|\epsilon| < \delta
\end{equation}
une solution unique des équations~(\ref{systeme-elliptique}) qui dépende de façon $C^{\infty}$ de $\alpha$, $\epsilon$ et $\beta$.\\
On note cette solution \index{$w_s$}
\begin{equation}\label{definition-de-phi-psi}
w_s(\alpha,\epsilon,\beta)=(\Phi(\alpha,\epsilon,\beta), \overrightarrow{\Psi}(\alpha,\epsilon,\beta)) .
\end{equation}
Le théorème~\ref{regularite} permet de prouver que pour tout $\alpha$, $\epsilon$ et $\beta$ le couple $w_{s}(\alpha,\epsilon,\beta)\in{\cal A}$ est $C^{\infty}$ sur $\Tore{\cal L}$.\\
{\bf $3^{eme}$ étape :}\\
On suppose $\alpha$ réel et on développe le système~(\ref{systeme-elliptique}). Ceci donne l'expression suivante
\begin{equation}\label{systeme-elliptique-complexifie}
\left\lbrace
\begin{array}{l}
\begin{array}{l}
\theta=R_0[\epsilon\theta-(\alpha\phi_{0}+\theta)(\alpha^2|\phi_{0}|^2+2\alpha \Rez[\overline{\phi_{0}}\theta]+|\theta|^2)\\
\,\,\,\,\,\,-2\overrightarrow{a}.(i\overrightarrow{\nabla}+\overrightarrow{A}_{0})(\alpha\phi_{0}+\theta)-\overrightarrow{a}^2(\alpha\phi_{0}+\theta)],
\end{array}\\
\overrightarrow{a}=-\beta^2M[(\alpha\overline{\phi_{0}}+\overline{\theta})(i\overrightarrow{\nabla}+\overrightarrow{A}_0+\overrightarrow{a})(\alpha\phi_{0}+\theta)].
\end{array}\right.
\end{equation}
On peut alors considérer que les variables $\alpha$, $\epsilon$ et $\beta$ sont complexes et effectuer des calculs de fonctions analytiques par rapport à $\alpha$,  $\epsilon$ et $\beta$.\\
On calcule maintenant $(\frac{\partial\Phi}{\partial\overline{\alpha}},\frac{\partial\overrightarrow{\Psi}}{\partial\overline{\alpha}})$ à partir du système complexifié~(\ref{systeme-elliptique-complexifie})
\begin{equation}\label{anal-I}
\left\lbrace\begin{array}{ccl}
\frac{\partial\Phi}{\partial\overline{\alpha}}&=&P_{11}(\Phi, \overrightarrow{\Psi})\frac{\partial\Phi}{\partial\overline{\alpha}}+P_{12}(\Phi, \overrightarrow{\Psi})\frac{\partial\overrightarrow{\Psi}}{\partial\overline{\alpha}} ,\\
\frac{\partial\overrightarrow{\Psi}}{\partial\overline{\alpha}}&=&P_{21}(\Phi, \overrightarrow{\Psi})\frac{\partial\Phi}{\partial\overline{\alpha}}+P_{22}(\Phi, \overrightarrow{\Psi})\frac{\partial\overrightarrow{\Psi}}{\partial\overline{\alpha}} .
\end{array}\right.
\end{equation}
Les opérateurs $P_{ij}$ sont continus pour la norme $H^1$, dépendent continument de $(\alpha,\epsilon,\beta)$ et s'annulent pour $(\alpha, \epsilon)=(0,0)$. Par conséquent, quitte à diminuer $\delta$, on peut supposer que l'opérateur $(\delta_{ij}-P_{ij})$ est inversible.\\
L'inversion du système~(\ref{anal-I}) donne l'égalité $(\frac{\partial\Phi}{\partial\overline{\alpha}},\frac{\partial\overrightarrow{\Psi}}{\partial\overline{\alpha}})=(0,\overrightarrow{0})$. On prouve le même résultat pour $(\frac{\partial\Phi}{\partial\overline{\epsilon}},\frac{\partial\overrightarrow{\Psi}}{\partial\overline{\epsilon}})$ et $(\frac{\partial\Phi}{\partial\overline{\beta}},\frac{\partial\overrightarrow{\Psi}}{\partial\overline{\beta}})$.\\
Les résultats élémentaires sur les fonctions analytiques permettent alors de montrer que $(\Phi, \overrightarrow{\Psi})$ sont analytiques en $(\alpha, \epsilon,\beta)$. $\hfill{\bf CQFD}$\\
\\
On obtient l'équation réduite à partir de l'équation~(\ref{expansion-des-inconnues}) en projetant sur $\phi_{0}$. Cette équation s'écrit $g(\alpha,\epsilon,\beta)=0$ où $g$ \index{$g$}est la fonction suivante
\begin{equation}\label{equation-reduite}
g(\alpha,\epsilon,\beta)=\epsilon\alpha+\langle\phi_0,F_{\beta}(\alpha\phi_0+\Phi(\alpha,\epsilon,\beta),\overrightarrow{\Psi}(\alpha,\epsilon,\beta))\rangle .
\end{equation}
Puisque les fonctions $\Phi$ et $\Psi$ sont analytiques par rapport à $\alpha$, $\epsilon$ et $\beta$ la fonction $g$ est analytique par rapport à ces variables.
% Par ailleurs la fonction $g$ est $C^{\infty}$ par rapport à $\beta,\alpha$ et $\epsilon$.

\begin{lemme}.\label{realite-merci-bernard}
La fonction $g$ est à valeur réelle.
\end{lemme}
{\it Preuve.} Pour tout couple $(\phi,\overrightarrow{a})\in{\cal A}$ on a 
\begin{equation}
\langle \phi, F_{\beta}(\phi,\overrightarrow{a})\rangle\in\R
\end{equation}
En effet la fonction $F_{\beta}$ définie à l'équation (\ref{reduction-en-cours}) peut s'écrire 
\begin{equation}
F_{\beta}(\phi,\overrightarrow{a})=-\phi|\phi|^2+[i\overrightarrow{\nabla}+\overrightarrow{A}_0]^2\phi-[i\overrightarrow{\nabla}+\overrightarrow{A}_0+\overrightarrow{a}]^2\phi .
\end{equation}
Ceci nous donne
\begin{equation}
\begin{array}{rcl}
\langle \phi, F_{\beta}(\phi,\overrightarrow{a})\rangle
&=&-\Vert\phi^2\Vert^2_{L^2}+\Vert i\overrightarrow{\nabla}\phi+\overrightarrow{A}_0\phi\Vert^2_{L^2}-\Vert i\overrightarrow{\nabla}\phi+(\overrightarrow{A}_0+\overrightarrow{a})\phi\Vert^2_{L^2} .
\end{array}
\end{equation}
Montrons maintenant que
\begin{equation}
\langle \theta, F_{\beta}(\alpha\phi_0+\theta,\overrightarrow{a})\rangle\in\R
\end{equation}
si $\theta=\Phi(\alpha,\epsilon,\beta)$ et $\overrightarrow{a}=\overrightarrow{\Psi}(\alpha,\epsilon,\beta)$.\\
La section $\theta$ vérifie l'équation 
\begin{equation}
\theta=R_{0}[\epsilon\theta+F_{\beta}(\alpha\phi_{0}+\theta,\overrightarrow{a})] .
\end{equation}
On obtient donc en appliquant $[H-\lambda_0]$
\begin{equation}
[H-\lambda_0]\theta=P_{0}[\epsilon\theta+F_{\beta}(\alpha\phi_{0}+\theta,\overrightarrow{a})]
\end{equation}
o\`u $P_0$ est la projection orthogonale sur $\{\phi_0\}^{\perp}$ (cf~\ref{divers-sur-R0}); cela nous donne puisque $\theta\in\{\phi_0\}^{\perp}$
\begin{equation}
\begin{array}{rcl}
\langle \theta, [H-\lambda_0]\theta\rangle&=&\langle \theta, P_{0}[\epsilon\theta+F_{\beta}(\alpha\phi_{0}+\theta,\overrightarrow{a})]\rangle\\
&=&\langle\theta, \epsilon\theta+F_{\beta}(\alpha\phi_{0}+\theta,\overrightarrow{a})\rangle\\
&=&\epsilon||\theta||^2_{L^2}+\langle \theta, F_{\beta}(\alpha\phi_{0}+\theta,\overrightarrow{a})\rangle .
\end{array}
\end{equation}
On a donc 
\begin{equation}
\langle \theta, F_{\beta}(\alpha\phi_{0}+\theta,\overrightarrow{a})\rangle=\langle \theta, [H-\lambda_0]\theta\rangle-\epsilon||\theta||^2_{L^2}\in\R
\end{equation}
car $H-\lambda_0$ est un opérateur autoadjoint.\\
Puisque $\phi=\alpha\phi_0+\theta$, on obtient
\begin{equation}
\langle F_{\beta}(\alpha\phi_0+\Phi(\alpha,\epsilon,\beta),\overrightarrow{\Psi}(\alpha,\epsilon,\beta)),\alpha\phi_0\rangle \in\R .
\end{equation}
La fonction $g$ est donc réelle si $\alpha$ est différent de $0$. Cette fonction est continue donc elle est toujours réelle. $\hfill{\bf CQFD}$

\begin{lemme}.\label{parite-imparite}
La fonction $\Phi$ vérifie $\Phi(-\alpha,\epsilon,\beta)=-\Phi(\alpha,\epsilon,\beta)$.\\
La fonction $\overrightarrow{\Psi}$ vérifie $\overrightarrow{\Psi}(-\alpha,\epsilon,\beta)=\overrightarrow{\Psi}(\alpha,\epsilon,\beta)$.\\
La fonction $g$ vérifie $g(-\alpha,\epsilon,\beta)=-g(\alpha,\epsilon,\beta)$.
\end{lemme}
{\it Preuve.} Les fonctions $\Phi$ et $\overrightarrow{\Psi}$ vérifient les équations:
\begin{equation}\label{systeme-elliptique-sol}
\left\lbrace
\begin{array}{rcl}
\Phi(\alpha,\epsilon,\beta)&=&R_{0}[\epsilon\Phi(\alpha,\epsilon,\beta)+F_{\beta}(\alpha\phi_{0}+\Phi(\alpha,\epsilon,\beta),\overrightarrow{\Psi}(\alpha,\epsilon,\beta))],\\
\overrightarrow{\Psi}(\alpha,\epsilon,\beta)&=&MG_{\beta}(\alpha\phi_{0}+\Phi(\alpha,\epsilon,\beta),\overrightarrow{\Psi}(\alpha,\epsilon,\beta)).
\end{array}\right.
\end{equation}
On obtient donc les deux systèmes suivants:
\begin{equation}
\left\lbrace
\begin{array}{rcl}
\Phi(-\alpha,\epsilon,\beta)&=&R_{0}[\epsilon\Phi(-\alpha,\epsilon,\beta)\\
&&+F_{\beta}(-\alpha\phi_{0}+\Phi(-\alpha,\epsilon,\beta),\overrightarrow{\Psi}(-\alpha,\epsilon,\beta))],\\
\overrightarrow{\Psi}(-\alpha,\epsilon,\beta)&=&MG_{\beta}(-\alpha\phi_{0}+\Phi(-\alpha,\epsilon,\beta),\overrightarrow{\Psi}(-\alpha,\epsilon,\beta))
\end{array}\right.
\end{equation}
et
\begin{equation}
\left\lbrace
\begin{array}{rcl}
-\Phi(\alpha,\epsilon,\beta)&=&R_{0}[-\epsilon\Phi(\alpha,\epsilon,\beta)+F_{\beta}(-\alpha\phi_{0}-\Phi(\alpha,\epsilon,\beta),\overrightarrow{\Psi}(\alpha,\epsilon,\beta))],\\
\overrightarrow{\Psi}(\alpha,\epsilon,\beta)&=&MG_{\beta}(-\alpha\phi_{0}-\Phi(\alpha,\epsilon,\beta),\overrightarrow{\Psi}(\alpha,\epsilon,\beta)).
\end{array}\right.
\end{equation}
On a utilisé entre autre la propriété:
\begin{equation}
\begin{array}{l}
F_{\beta}(-\phi,\overrightarrow{a})=-F_{\beta}(\phi,\overrightarrow{a}),\\
G_{\beta}(-\phi,\overrightarrow{a})=G_{\beta}(\phi,\overrightarrow{a}).
\end{array}
\end{equation}
Vu qu'il y a unicité des solutions localement, on obtient les relations:
\begin{equation}
\left\lbrace
\begin{array}{rcl}
-\Phi(\alpha,\epsilon,\beta)&=&\Phi(-\alpha,\epsilon,\beta),\\
\overrightarrow{\Psi}(\alpha,\epsilon,\beta)&=&\overrightarrow{\Psi}(-\alpha,\epsilon,\beta) .
\end{array}\right.
\end{equation}
Le résultat sur $g$ provient alors directement de l'expression~(\ref{equation-reduite}) de $g$.  $\hfill{\bf CQFD}$\\
\\
Les fonctions $(\Phi, \overrightarrow{\Psi})$ vérifient les équations 
\begin{equation}\label{equation-assez-hard}
\left\lbrace\begin{array}{l}
\Phi=R_{0}[\epsilon\Phi+F_{\beta}(\alpha\phi_{0}+\Phi,\overrightarrow{\Psi})] ,\\
\overrightarrow{\Psi}=MG_{\beta}(\alpha\phi_{0}+\Phi,\overrightarrow{\Psi}) .
\end{array}\right.
\end{equation}
Si $(\alpha, \epsilon)$ est solution de l'équation~(\ref{definition-fonction-g}), alors la paire
\begin{equation}
\left(\begin{array}{l}
\alpha\phi_{0}+\Phi(\alpha, \epsilon,\beta)\\
\overrightarrow{\Psi}(\alpha, \epsilon,\beta)
\end{array}\right)
\end{equation}
est l'unique solution de l'équation~(\ref{Odeh-equations-bis}) dans le voisinage 
\begin{equation}
\Vert \theta\Vert _{H^1}+\Vert \overrightarrow{a}\Vert _{H^1}+|\alpha|+|\epsilon| < \delta .
\end{equation}
La fonction $g$ est analytique en $\alpha$, $\epsilon$ et $\beta$.\\
Dans la suite de la thèse, nous aurons besoin des constantes \index{$I$}\index{$K$}\index{$I_m$}\index{$K_m$}suivantes:
\begin{equation}\label{defcon-I}
I_{m}=\int_{\Omega}|\phi_{0,m}|^4dxdy
\end{equation}
et
\begin{equation}\label{defcon-K}
\begin{array}{rcl}
K_{m}&=&\langle \Rez[\overline{\phi_{0,m}}(i\overrightarrow{\nabla}\phi_{0,m}+\overrightarrow{A}_0\phi_{0,m})],M\lbrace \Rez[\overline{\phi_{0,m}}(i\overrightarrow{\nabla}\phi_{0,m}+\overrightarrow{A}_0\phi_{0,m})]\rbrace\rangle\\
&=&\Rez\langle M\lbrace \Rez[\overline{\phi_{0,m}}(i\overrightarrow{\nabla}\phi_{0,m}+\overrightarrow{A}_0\phi_{0,m})]\rbrace.(i\overrightarrow{\nabla}\phi_{0,m}+\overrightarrow{A}_0\phi_{0,m}),\phi_{0,m}\rangle .
\end{array}
\end{equation}
Dans l'équation précédente, $M$ est l'inverse de $L$. La section $\phi_0$ est définie dans le théorème \ref{spectre-de-H-C-infinite}. Comme auparavant, si la dépendance de $I$ ou $K$ par rapport à $m\in{\cal M}$ (cf la section \ref{sec:struct-reseau-space}) est sans importance on omettra d'écrire $m$.
\begin{lemme}.\label{positivite-K-et-I}
Les constantes $K$ et $I$ sont strictement positives.
\end{lemme}
{\it Preuve.} La section $\phi_0$ est non identiquement nulle, par conséquent $\int_{\Omega}|\phi_0|^4$ est strictement positive.\\
L'opérateur $M$ est défini positif, par conséquent $K$ est positive. Supposons par l'absurde $K=0$ on a alors puisque $M$ est strictement positif
\begin{equation}
\Rez [\overline{\phi_{0,m}}(i\overrightarrow{\nabla}\phi_{0,m}+\overrightarrow{A}_0\phi_{0,m})]=0 .
\end{equation}
Il suffit alors de reprendre la démonstration du lemme \ref{champ-a-determine} à partir de l'équation (\ref{annulation-a-utiliser}). $\hfill{\bf CQFD}$

\begin{hypoth}.\label{hypothese-suite}
Dans toute la suite de la thèse on suppose que $I-2\beta^2K\not=0$ ce qui s'écrit aussi $I-\frac{2}{k^2}K\not=0$
\end{hypoth}

\begin{theorem}.\label{existence-de-la-bifurcation}
Les solutions différentes de $(0,0)$ du système~(\ref{Odeh-equations-bis}) s'expriment sous la forme $(\omega\phi_+,\overrightarrow{a_+})$ o\`u $\omega\in\C$, $|\omega|=1$ et $(\phi_+,\overrightarrow{a_+})$ \index{$\phi_+$}\index{$\overrightarrow{a_+}$}est défini ci dessous:
\begin{equation}
(\phi_+,\overrightarrow{a_+})=(\sqrt{W(\epsilon,\beta)}\phi_0+\Phi(\sqrt{W(\epsilon,\beta)},\epsilon,\beta),\overrightarrow{\Psi}(\sqrt{W(\epsilon,\beta)},\epsilon,\beta))
\end{equation}
avec \index{$W(\epsilon,\beta)$}
\begin{equation}
W(\epsilon,\beta)=\left\lbrace\begin{array}{rcl}
0&\mbox{~si~}&\frac{\epsilon}{I-2\beta^2K}<0\\
D(\epsilon,\beta)&\mbox{~si~}&\frac{\epsilon}{I-2\beta^2K}>0
\end{array}\right.
\end{equation}
et $D(x,\beta)$ est une fonction analytique.
\end{theorem}
\begin{remarque}.
Si $I-\frac{2}{k^2}K>0$ alors la bifurcation appara\^it si $H_{int}<k$.
\end{remarque}
{\it Preuve.} On continue d'utiliser $\beta=\frac{1}{k}$. Pour déterminer la nature de cette bifurcation, on va calculer certaines dérivées partielles de $g$. L'identité $g(-\alpha,\epsilon,\beta)=-g(\alpha,\epsilon,\beta)$ obtenue au lemme \ref{parite-imparite} permet d'écrire $g$ sous la forme
\begin{equation}\label{changement-utile}
g(\alpha, \epsilon,\beta)=\alpha h(\alpha^2, \epsilon,\beta)
\end{equation}
avec $h$ fonction dépendant analytiquement de $\beta$, $\alpha$ et $\epsilon$.\\
Pour montrer qu'il y a des solutions bifurquées, nous avons besoin de calculer les dérivées partielles premières de $h$ par rapport aux deux premières variables encore notées $\alpha$ et $\epsilon$. Si on montre que ces deux dérivées partielles sont non nulles le théorème des fonctions implicites pour les fonctions analytique (cf~\cite{spe-M'-dolbeault}, p.~186) nous donnera une expression de la forme $\alpha^2=D(\epsilon,\beta)$ o\`u $D(x,\beta)$ est une fonction analytique.\\
On a les relations
\begin{equation}
\left\lbrace\begin{array}{l}
\frac{\partial h}{\partial \epsilon}(0,0,\beta)=\frac{\partial^2 g}{\partial\alpha\partial\epsilon}(0,0,\beta),\\
\frac{\partial h}{\partial \alpha}(0,0,\beta)=\frac{1}{6}\frac{\partial^3 g}{\partial\alpha^3}(0,0,\beta) .\\
\end{array}\right.
\end{equation}
Si on différencie les équations~(\ref{equation-assez-hard}) par rapport à $\epsilon$, $\alpha$ et que l'on évalue en $(\alpha, \epsilon,\beta)=(0,0,\beta)$, on obtient:
\begin{equation}\label{ordre-un-des-derivees}
\left\lbrace\begin{array}{rcc|rcc}
\frac{\partial \Phi}{\partial\alpha}(0,0,\beta)&=&0&\frac{\partial \Phi}{\partial\epsilon}(0,0,\beta)&=&0\\
\frac{\partial \overrightarrow{\Psi}}{\partial\alpha}(0,0,\beta)&=&0&\frac{\partial \overrightarrow{\Psi}}{\partial\epsilon}(0,0,\beta)&=&0
\end{array}\right\rbrace .
\end{equation}
Ces relations impliquent, par l'expression de $g$ et les formules (\ref{ordre-un-des-derivees}), (\ref{changement-utile}) et (\ref{equation-reduite}):
\begin{equation}
\frac{\partial h}{\partial \epsilon}(0,0,\beta)=1 .
\end{equation}
Il faut maintenant dériver deux fois par rapport à $\alpha$ les équations~(\ref{equation-assez-hard}):
\begin{equation}\label{ordre-deux-des-derivees}
\left\lbrace\begin{array}{rcl}
\frac{\partial^2 \Phi}{\partial\alpha^2}(0,0,\beta)&=&0\\
\frac{\partial^2 \overrightarrow{\Psi}}{\partial\alpha^2}(0,0,\beta)&=&-2M[\beta^2\Rez(\overline{\phi_{0}} (i\overrightarrow{\nabla}+\overrightarrow{A}_0)\phi_{0})]
\end{array}\right\rbrace .
\end{equation}
Cela suffit pour calculer la dérivée troisième de $g$ par rapport à $\alpha$. On commence d'abord par calculer la dérivée troisième de $F_{\beta}(\Phi(\alpha, \epsilon), \overrightarrow{\Psi}(\alpha,\epsilon,\beta))$:
\begin{equation}
[\frac{\partial^3 F_{\beta}}{\partial\alpha^3}(\Phi(\alpha, \epsilon), \overrightarrow{\Psi}(\alpha,\epsilon,\beta))](0,0,\beta)=-6\phi_0|\phi_0|^2-6\frac{\partial^2 \overrightarrow{\Psi}}{\partial\alpha^2}(0,0,\beta).(i\overrightarrow{\nabla}+\overrightarrow{A}_0)\phi_{0} .
\end{equation}
En dérivant la fonction $g$ par rapport à $\alpha$ et en évaluant en $(0,0,\beta)$, on trouve:
\begin{equation}\label{definition-implicite-R}
\begin{array}{rcl}
\frac{\partial^3g}{\partial\alpha^3}(0,0,\beta)&=&-6I+12\beta^2R
\end{array}
\end{equation}
avec \index{$R$}$R=\langle M[\Rez(\overline{\phi_0} (i\overrightarrow{\nabla}+\overrightarrow{A}_0)\phi_0)].(i\overrightarrow{\nabla}+\overrightarrow{A}_0)\phi_0, \phi_0 \rangle$.\\
$g$ et $I$ étant réels, la quantité $R$ est également réelle. Or on a les relations suivantes:
\begin{equation}
\left\lbrace\begin{array}{rcl}
R&=&\langle M[\Rez(\overline{\phi_0} (i\overrightarrow{\nabla}+\overrightarrow{A}_0)\phi_0)].(i\overrightarrow{\nabla}+\overrightarrow{A}_0)\phi_0, \phi_0 \rangle\\
&=&\int_{\Omega}M[\Rez(\overline{\phi_0} (i\overrightarrow{\nabla}+\overrightarrow{A}_0)\phi_0)].(i\overrightarrow{\nabla}+\overrightarrow{A}_0)\phi_0\overline{\phi_0}\\
&=&\int_{\Omega}M[\Rez(\overline{\phi_0} (i\overrightarrow{\nabla}+\overrightarrow{A}_0)\phi_0)].\\
&&[\Rez(\overline{\phi_0}(i\overrightarrow{\nabla}+\overrightarrow{A}_0)\phi_0)+i\Imz(\overline{\phi_0}(i\overrightarrow{\nabla}+\overrightarrow{A}_0)\phi_0)]\\[2mm]
&=&K+\int_{\Omega}M[\Rez(\overline{\phi_0} (i\overrightarrow{\nabla}+\overrightarrow{A}_0)\phi_0)].[i\Imz(\overline{\phi_0}(i\overrightarrow{\nabla}+\overrightarrow{A}_0)\phi_0)]\\
&=&K+i\langle \Imz(\overline{\phi_0}(i\overrightarrow{\nabla}+\overrightarrow{A}_0)\phi_0),M[\Rez(\overline{\phi_0} (i\overrightarrow{\nabla}+\overrightarrow{A}_0)\phi_0)]\rangle .
\end{array}\right.
\end{equation}
La quantité $R$ ne peut \^etre réelle que si
\begin{equation}
\langle \Imz(\overline{\phi_0}(i\overrightarrow{\nabla}+\overrightarrow{A}_0)\phi_0),M[\Rez(\overline{\phi_0} (i\overrightarrow{\nabla}+\overrightarrow{A}_0)\phi_0)]\rangle=0 .
\end{equation}
On a alors $R=K$ et donc:
\begin{equation}
\begin{array}{rcl}
\frac{\partial^3g}{\partial\alpha^3}&=&-6I+12\beta^2K .
\end{array}
\end{equation}
Ceci nous donne pour $h$:
\begin{equation}
\frac{\partial h}{\partial \alpha}(0,0,\beta)=-I+2\beta^2K.
\end{equation}
On cherche maintenant les solutions de l'équation $g=0$ dans un voisinage de $(0,0,\beta)$.\\
L'équation~(\ref{changement-utile}) nous donne la racine $\alpha=0$ de l'équation réduite.\\
On considère donc maintenant l'équation $h(x, \epsilon,\beta)=0$. Le calcul des dérivées de $g$ nous donne:
\begin{equation}
\left\lbrace\begin{array}{rcl}
\frac{\partial h}{\partial \alpha}(0,0,\beta)&=&\frac{1}{6}\frac{\partial^3g}{\partial\alpha^3}(0,0,\beta)=-I+2\beta^2K,\\
\frac{\partial h}{\partial \epsilon}(0,0,\beta)&=&1.
\end{array}\right.
\end{equation}
Si $I-2\beta^2K\not=0$, alors on peut utiliser le théorème des fonctions implicites et écrire l'équation sous la forme:
\begin{equation}
\begin{array}{c}
\alpha^2=D(\epsilon,\beta)\\
\mbox{~avec~} (\frac{\partial D}{\partial \epsilon})(0,\beta)=\frac{1}{I-2\beta^2K} ,
\end{array}
\end{equation}
o\`u $D$ est une fonction analytique de $\epsilon$ et $\beta$.\\
Les solutions de l'équation $g(\alpha,\epsilon,\beta)=0$ ont pour expression
\begin{equation}
\begin{array}{l}
\alpha_0=0\, ,\\
\alpha_+=\left\lbrace\begin{array}{rcl}
0&\mbox{~si~}&\frac{\epsilon}{I-\frac{2}{k^2}K}<0\\
\sqrt{D(\epsilon,\beta)}&\mbox{~si~}&\frac{\epsilon}{I-\frac{2}{k^2}K}>0
\end{array}\right.\\
\mbox{~et~}\alpha_-=-\alpha_+ .
\end{array}
\end{equation}
Les différentes solutions ont le comportement asymptotique suivant quand $\epsilon\rightarrow 0$
\begin{equation}
\begin{array}{l}
\alpha_0=0\, ,\\
\alpha_{+}\simeq\left\lbrace\begin{array}{rcl}
0&\mbox{~si~}&\frac{\epsilon}{I-\frac{2}{k^2}K}<0\\
\sqrt{\frac{\epsilon}{I-2\beta^2K}}&\mbox{~si~}&\frac{\epsilon}{I-\frac{2}{k^2}K}>0
\end{array}\right.\\
\mbox{~et~}\alpha_-=-\alpha_+ .
\end{array}
\end{equation}
On obtient ainsi toutes les solutions non nulles sous la forme $(\omega\phi_+,\overrightarrow{a_+})$. $\hfill{\bf CQFD}$\\
\\
Dans les deux sections qui suivent l'expression ci-dessous sera appelée \index{couple bifurqué}couple bifurqué
\begin{equation}\label{definition-etat-bifurqué}
(\phi_{+}(\epsilon,k), \overrightarrow{a_+}(\epsilon,k))=
\left(\begin{array}{c}
\sqrt{W(\epsilon,\beta)}\phi_{0}+\Phi(\sqrt{W(\epsilon,\beta)},\epsilon,\beta)\\
\overrightarrow{\Psi}(\sqrt{W(\epsilon,\beta)},\epsilon,\beta)
\end{array}\right)
\end{equation}
avec $\beta=\frac{1}{k}$. Ce couple vérifie: $\langle\phi_+(\epsilon,k),\phi_{0}\rangle\geq 0$. On note aussi dans la suite \index{$\alpha$}
\begin{equation}\label{definition-alpha}
\alpha=\sqrt{W(\epsilon,\beta)},\mbox{~}F_S(\epsilon,k)=F_{\lambda,k}(\phi_{+}(\epsilon,k), \overrightarrow{a}_{+}(\epsilon,k)) .
\end{equation}
La dépendance par rapport à  $\beta$ de $\alpha$ est analytique.\\
La question de l'analyticité, que nous avons résolue ici, avait été posée par Lasher dans l'article \cite{lasher}.

\begin{proposition}.
La fonction $\epsilon\mapsto F_S(\epsilon,k)$ restreinte à $\R_+$ est analytique par rapport à $\epsilon$
\end{proposition}
{\it Preuve.} Dans le domaine considéré, la fonction $W(\epsilon,\beta)$ est analytique par rapport à $\epsilon$. La fonction $\overrightarrow{\Psi}(\alpha,\epsilon,\beta)$ étant analytique par rapport à $\alpha$ (cf~\ref{reduc-lyapunov}) et paire par rapport à $\alpha$ (cf~\ref{parite-imparite}), elle peut s'écrire sous la forme $\overrightarrow{\Psi_i}(\alpha^2, \epsilon,\beta)$.\\
De m\^eme la fonction $\Phi(\alpha,\epsilon,\beta)$ peut s'écrire $\alpha \Phi_i(\alpha^2,\epsilon,\beta)$; on a 
\begin{equation}
\begin{array}{rcl}
F_S(\epsilon,k)
&=&F_{\lambda,k}(\phi_{+}(\epsilon,k), \overrightarrow{a}_{+}(\epsilon,k))\\
&=&F_{\lambda,k}(\alpha[\phi_{0}+\Phi_i(\alpha^2,\epsilon,\beta)],\overrightarrow{\Psi_i}(\alpha^2,\epsilon,\beta))
\end{array}
\end{equation}
avec $\alpha=\sqrt{W(\epsilon,\beta)}$. En utilisant l'expression de $F_{\lambda,k}$ (cf~(\ref{expression-Flambda-k})), on voit que la variable $\alpha$ appara\^it au carré dans l'expression de $F_S(\epsilon,k)$.\\
La fonction $W(\epsilon,\beta)$ est analytique par rapport à $\epsilon$ et donc $F_S(\epsilon,k)$ est aussi analytique.  $\hfill{\bf CQFD}$

\begin{proposition}.
%On peut représenter tous les couples $(\phi,\overrightarrow{a})\in{\cal A}_m$ sous la forme $(S_m^*\phi_{c,m},S_m^*\overrightarrow{a_{c,m}})$ (voir (\ref{definition-Sm}) et (\ref{fonction-Sm-etoile})) avec $(\phi_{c,m},\overrightarrow{a_{c,m}})\in{\cal A}_c$ ou ${\cal A}_c$ est l'espace fonctionnel associé au réseau carré.\\
Pour tout $m_0\in{\cal M}$, il existe un voisinage compact ${\cal V}_{m_0}$ de $m_0$ et une constante $\delta>0$ tel que
\begin{equation}
\left\lbrace\begin{array}{l}
\forall m\in{\cal V}_{m_0},\\
\forall (\phi,\overrightarrow{a})\in{\cal A}_m\mbox{~avec~}\Vert\phi\Vert_{H^1}+\Vert\overrightarrow{a}\Vert_{H^1}\leq \delta\mbox{~et}\\
|\epsilon|\leq \delta
\end{array}\right.
\end{equation}
les équations de Ginzburg-Landau possèdent comme solution le couple $(0,0,\beta)$ et les solutions bifurquées $(\omega\phi_{+,m},\overrightarrow{a}_{+,m})$.\\
Dans ce voisinage on exprime les solutions bifurquées $(\omega\phi_{+,m},\overrightarrow{a}_{+,m})$ sous la forme $(S_m^*\omega\phi_{+,c,m},S_m^*\overrightarrow{a}_{+,c,m})$ (voir (\ref{definition-Sm}) et (\ref{fonction-Sm-etoile})) avec $(\omega\phi_{+,c,m},\overrightarrow{a}_{+,c,m})\in{\cal A}_c$ ou ${\cal A}_c$ est l'espace fonctionnel associé au réseau carré.\\
Le couple $(\phi_{+,c,m},\overrightarrow{a}_{+,c,m})$, les fonctions $W_{m,k}(\epsilon)$, $g_{m,k}$, les constantes  $I_{m}$, $K_{m}$ et l'énergie bifurquée $F_{S,m}$ ont une dépendance $C^{\infty}$ par rapport à $m$.
\end{proposition}
{\it Preuve.} Avec la représentation $(\phi,\overrightarrow{a})=(S_m^*\phi_m,S_m^*\overrightarrow{a_m})$, les équations de Ginzburg-Landau se réécrivent sous la forme:
\begin{equation}\label{equation-sur-Ac}
\left\lbrace\begin{array}{rcl}
-\Delta_m\,\overrightarrow{a_{m}}&=&-\frac{1}{k^2}\Rez[\,\overline{\phi_{m}}(i\overrightarrow{\nabla}\phi_{m}+(\overrightarrow{A}_{0,m}+\overrightarrow{a})\phi_{m})]\\
\phi_m(\lambda-|\phi_m|^2)&=&-\Delta_m\,\phi_m+(S_m\overrightarrow{A_{0,m}}+S_m\overrightarrow{a_{m}}).iS_m\overrightarrow{\nabla}\phi_m\\
&+&\Vert S_m\overrightarrow{A_{0,m}}+S_m\overrightarrow{a_{m}}\Vert^2\phi_m
\end{array}\right. ,
\end{equation}
avec 
\begin{equation}
\left\lbrace\begin{array}{rcl}
\overrightarrow{A_{0,m}}(x,y)&=&S_m^{-1}\overrightarrow{A_{0}}(S_m^{-1}(x,y))\\
\Delta_m&=&(\frac{1}{u^2}+w^2)\frac{\partial^2}{\partial x^2}+2wu\frac{\partial^2}{\partial x\partial y}+u^2\frac{\partial^2}{\partial y^2} .
\end{array}\right.
\end{equation}
La dépendance de ces équations par rapport à $u$ et $w$ est $C^{\infty}$. Tous les résultats de cette section peuvent s'exprimer dans l'espace ${\cal A}_{c}$ gr\^ace à $S_m$. Par le théorème des fonctions implicites, la dépendance des solutions $(\Phi_m,\overrightarrow{\Psi_m})$ par rapport à $m$ est $C^{\infty}$.\\
 La dépendance de $g_{m,k}$ est aussi $C^{\infty}$. Par conséquent $h_{m}$ et $W_{m}$ est aussi $C^{\infty}$ par rapport à $m$. Les constantes $I_{m}$ et $K_{m}$ sont $C^{\infty}$ par rapport à $m$. L'énergie bifurquée $F_S$ est aussi $C^{\infty}$ par rapport à $m$. $\hfill{\bf CQFD}$

\saction{Évaluation des énergies sur les différentes courbes de couples pour $H_{int}<k$}\label{sec:calcul-energie}
\noindent On souhaite estimer l'énergie le long de la courbe bifurquée par rapport à $H_{int}$ pour voir si elle est plus basse que celle du couple $(0,0)$. La variable utilisée ici est comme auparavant\index{$\epsilon$}
\begin{equation}\label{definition-epsilon}
\epsilon=\lambda-\lambda_{0} .
\end{equation}
On utilise la fonctionnelle simplifiée~(\ref{expression-Flambda-k}) qui s'écrit
\begin{equation}
F_{\lambda,k}(\phi,\overrightarrow{a})=\int_{\Omega}\frac{1}{2}\Vert i\overrightarrow{\nabla}\phi+(\overrightarrow{A}_{0}+\overrightarrow{a})\phi\Vert^2+\frac{1}{4}(\lambda-|\phi|^2)^2+\frac{k^2}{2}|\rot\, \overrightarrow{a}|^2 .
\end{equation}
Pour le couple $(\phi,\overrightarrow{a})=(0,\overrightarrow{0})$, l'énergie vaut $F_{N}=\frac{\lambda^2}{4}$.\\
On note \index{$F_N$}\index{$F_S$}
\begin{equation}\label{definition-FN-FS}
\left\lbrace\begin{array}{l}
F_{S}(\epsilon,k)=F_{\lambda,k}(\phi_{+}(\epsilon,k), \overrightarrow{a_+}(\epsilon,k)), \\
F_{N}(\epsilon,k)=F_{\lambda,k}(0, 0)=\frac{\lambda^2}{4} .
\end{array}\right.
\end{equation}
Le couple bifurqué $(\phi_{+},\overrightarrow{a_+})$ est défini au théorème~\ref{existence-de-la-bifurcation}.\\
Dans cette section, on va calculer la différence d'énergie $F_{S}-F_{N}$ le long de la courbe de bifurcation.\\
Avant d'étudier notre situation réelle, considérons une situation modèle qui met bien en évidence la nature de la question. La fonctionnelle réduite modèle est la suivante
\begin{equation}\label{fonctionnelle-over-simplifie}
I(\alpha,\epsilon)=\frac{1}{4}I\alpha^4-\frac{1}{2}\alpha^2 \epsilon+H(\epsilon) \mbox{~avec~}I>0 .
\end{equation}
Le paramètre $\epsilon$ est lié au champ magnétique et est imposé au système (voir l'équation~(\ref{definition-epsilon})). La minimisation se fait par rapport à $\alpha$. La fonction $H(\epsilon)$ est obtenue par intégration et n'a aucune importance dans le calcul des couples minimisants la fonctionnelle.\\
Ses solutions bifurquées sont
\begin{equation}
\begin{array}{rcl}
\alpha_{\pm}:\R&\mapsto&\R\\
\epsilon&\mapsto&\left\lbrace\begin{array}{l}
0\mbox{~si~}\epsilon\leq 0\\
\pm\sqrt{\frac{\epsilon}{I}}\mbox{~si~}\epsilon>0 .
\end{array}\right.
\end{array}
\end{equation}
Leur énergie minimale est
\begin{equation}
\begin{array}{rcl}
I_m:\R&\mapsto&\R\\
\epsilon&\mapsto&I(\alpha_{\pm}(\epsilon),\epsilon)=\left\lbrace\begin{array}{l}
H(\epsilon)\mbox{~si~}\epsilon\leq 0\\
H(\epsilon)-\frac{\epsilon^2}{4I}\mbox{~si~}\epsilon>0
\end{array}\right.
\end{array}
\end{equation}
et elles sont trivialement stables.\\
Nous allons voir dans cette section et dans la suivante dans quelle mesure notre fonctionnelle a les mêmes propriétés.

\begin{theorem}.\label{evolution-de-F}
On a le résultat asymptotique
\begin{equation}\label{resultat-asymptotique-energie}
\lim_{\epsilon\rightarrow 0,\,\frac{\epsilon}{I-\frac{2}{k^2}K}>0}\frac{(F_{S}-F_{N})(\epsilon,k)}{\epsilon^2}=-\frac{1}{4(I-\frac{2}{k^2}K)}
\end{equation}
avec $K$ défini à l'équation~(\ref{defcon-K}) et $I$ défini à l'équation~(\ref{defcon-I}).
\end{theorem}
{\it Preuve.} Pour la fonctionnelle modèle et $\epsilon>0$ on a $I_{min}=I(\sqrt\frac{\epsilon}{I},\epsilon)=H(\epsilon)-\frac{\epsilon^2}{4I}$. Ceci indique que l'on doit faire les calculs à l'ordre deux en $\epsilon$ pour la fonctionnelle exacte. Soit $(\phi,\overrightarrow{a})\in{\cal A}$ une solution des équations de Ginzburg-Landau (\ref{Odeh-equations}); on a:
\begin{equation}\label{Gap-de-transition-de-phase}
\begin{array}{rcl}
F_{\lambda,k}(\phi,\overrightarrow{a})-F_{\lambda,k}(0,0)
&=&\frac{1}{2}\int_{\Omega}\Vert i\overrightarrow{\nabla}\phi+(\overrightarrow{A}_{0}+\overrightarrow{a})\phi\Vert^2\\
&+&\frac{1}{4}\int_{\Omega}-2\lambda|\phi|^2+|\phi|^4+\frac{k^2}{2}\int_{\Omega}|\rot\, \overrightarrow{a}|^2 .
\end{array}
\end{equation}
Utilisons maintenant que $(\phi, \overrightarrow{a})$ est solution de l'équation~(\ref{Odeh-equations}); on obtient l'égalité
\begin{equation}
\begin{array}{rcl}
\int_{\Omega}\Vert i\overrightarrow{\nabla}\phi+(\overrightarrow{A}_0+\overrightarrow{a})\phi\Vert^2&=&\langle [i\overrightarrow{\nabla}+(\overrightarrow{A}_0+\overrightarrow{a})]\phi,[i\overrightarrow{\nabla}+(\overrightarrow{A}_0+\overrightarrow{a})]\phi\rangle\\
&=&\langle [i\overrightarrow{\nabla}+(\overrightarrow{A}_0+\overrightarrow{a})]^2\phi,\phi\rangle\\
&=&\langle \phi(\lambda-|\phi|^2),\phi\rangle\\
&=&\int_{\Omega}(\lambda|\phi|^2-|\phi|^4)dxdy .
\end{array}
\end{equation}
On en déduit que:
\begin{equation}\label{Gap-de-transition-de-phase-expr-2}
F_{\lambda,k}(\phi,\overrightarrow{a})-F_{\lambda,k}(0,0)
=-\frac{1}{4}\int_{\Omega}|\phi|^4+\frac{k^2}{2}\int_{\Omega}|\rot\, \overrightarrow{a}|^2 ,
\end{equation}
pour toute solution $(\phi, \overrightarrow{a})$ de l'équation de Ginzburg-Landau~(\ref{Odeh-equations}).\\
Considérons maintenant la solution $(\phi_{+}, \overrightarrow{a_+})$ et calculons le développement de $(F_{S}-F_{N})$ en fonction de $\epsilon$. Il faut donc des dérivées de
\begin{equation}
\Phi\mbox{,~}\overrightarrow{\Psi}\mbox{~et~} \alpha
\end{equation}
fonctions qui sont définies aux équations~(\ref{definition-de-phi-psi}) et~(\ref{definition-alpha}).\\
On écrit le développement limité de $g$ tel qu'on le connait depuis le théorème~\ref{existence-de-la-bifurcation}
\begin{equation}
g(\alpha,\epsilon,\frac{1}{k})=\alpha\lbrace\epsilon-[I-\frac{2}{k^2}K]\alpha^2+J\alpha^4+o(\epsilon\alpha^2)+o(\alpha^4)\rbrace,
\end{equation}
o\`u $J$ est une constante réelle.\\
Ce calcul permet d'obtenir le développement de $\alpha$ en fonction de $\epsilon$ à l'ordre $\frac{3}{2}$. On écrit ci-dessous les différents développements limités utilisés. On les démontre en utilisant la formule de Taylor-Young pour les fonctions à valeur vectorielle et les calculs du théorème~\ref{existence-de-la-bifurcation}, plus précisément les formules~(\ref{ordre-un-des-derivees}) et~(\ref{ordre-deux-des-derivees}) .
\begin{equation}\label{developpements-limites}
\left\lbrace
\begin{array}{rcl}
\D\alpha&=&\sqrt{\frac{\epsilon}{I-\frac{2}{k^2}K}}(1+\frac{J\epsilon}{2I^2}+O(\epsilon^2))\\
\Phi(\epsilon)&=&O(|\epsilon|^{3/2})\\
\overrightarrow{\Psi}(\epsilon)&=&\frac{-\epsilon}{Ik^2-2K}M[\Rez(\overline{\phi_0}(i\overrightarrow{\nabla}+\overrightarrow{A}_0)\phi_0)]+O(\epsilon^2) .
\end{array}\right.
\end{equation}
On insère tous ces développements dans la différence d'énergie. Le développement limité de l'énergie s'écrit:
\begin{equation}\label{Gap-de-transition-de-phase-final}
\begin{array}{rcl}
(F_{S}-F_{N})(\epsilon,k)&=&-\frac{1}{4}\alpha^4\int_{\Omega}|\phi_0|^4+\frac{k^2}{2}\alpha^4\frac{1}{k^4}K+o(\epsilon^2)\\
&=&\frac{\epsilon^2}{(I-\frac{2}{k^2}K)^2}[-\frac{1}{4}I+\frac{K}{2k^2}]+o(\epsilon^2)\\
&=&-\frac{1}{4(I-\frac{2}{k^2}K)}\epsilon^2+o(\epsilon^2) .
\end{array}
\end{equation}
Le résultat du calcul est donc bien~(\ref{resultat-asymptotique-energie}):
\begin{equation}
\lim_{\epsilon\rightarrow 0,\frac{\epsilon}{I-\frac{2}{k^2}K}>0}\frac{(F_{S}-F_{N})(\epsilon,k)}{\epsilon^2}=-\frac{1}{4(I-\frac{2}{k^2}K)} .
\end{equation}
On a donc bien $F_{S}-F_{N}<0$, si $\epsilon$ est strictement positif et est assez petit. Ceci montre que le couple bifurqué trouvé est d'énergie plus basse que le couple $(0,0)$. $\hfill{\bf CQFD}$\\
\\
On note, \index{$F_{as}$}pour $I\not=\frac{2K}{k^2}$
\begin{equation}
F_{as}=-\frac{1}{4(I-\frac{2}{k^2}K)} .
\end{equation}

\begin{corollaire}.
On suppose $k > \sqrt{\frac{2K}{I}}$. Il existe $\epsilon_{0}>0$ tel que si $0<\epsilon<\epsilon_{0}$ alors le couple bifurqué est d'énergie plus basse que celle du couple $(0,0)$. Pour le problème réduit, cela signifie qu'il existe $H_{0}<k$ tel que, si $H_0<H_{int}<k$, alors l'énergie du couple bifurqué est plus basse que celle du couple $(0,0)$.
\end{corollaire}

\begin{proposition}.\label{invariance-FS-Fas-I-K}
Les fonctions sur ${\cal M}$: $m\mapsto F_{S,m}(\epsilon)$, $F_{as,m}$, $I_{m}$ et $K_{m}$ sont invariantes par les transformations $\sigma_i$ sur ${\cal M}$. De plus, elles sont critiques pour le réseau carré et le réseau hexagonal.
\end{proposition}
{\it Preuve.} Montrons que la fonction $F_S$ est invariante par $\sigma_i$. Les fonctions $\Sigma_i^*$ sont bijectives, involutives et continues pour la norme $H^1$. Par conséquent, un voisinage de $(0,0)$ est envoyé sur un voisinage de $(0,0)$.\\
Ce sont des difféomorphismes donc si $(\phi,\overrightarrow{a})$ est un point critique de $F_{\lambda,k}$ sur ${\cal A}_m$ alors $(\Sigma_i^*\phi,\Sigma_i^*\overrightarrow{a})$ est un point critique de $F_{\lambda,k}$ sur ${\cal A}_{\sigma_i(m)}$.\\
Par conséquent, si $(\phi_{+,m},\overrightarrow{a}_{+,m})$ est le couple bifurqué pour $F_{\lambda,k}$ sur ${\cal A}_m$ alors $(\Sigma_i^*\phi_{+,m},\Sigma_i^*\overrightarrow{a}_{+,m})$ est un point critique de $F_{\lambda,k}$ sur ${\cal A}_{\sigma_i(m)}$.\\
Vu le théorème \ref{existence-de-la-bifurcation}, il existe $\omega$ de module $1$ tel que 
\begin{equation}
(\Sigma_i^*\phi_{+,m},\Sigma_i^*\overrightarrow{a}_{+,m})=(\omega\phi_{+,\sigma_i(m)},\overrightarrow{a}_{+,\sigma_i(m)}) .
\end{equation}
On obtient alors la relation 
\begin{equation}
\begin{array}{rcl}
F_{S,m}&=&F_{\lambda,k}(\phi_{+,m},\overrightarrow{a}_{+,m})\\
&=&F_{\lambda,k}(\Sigma_i^*\phi_{+,m},\Sigma_i^*\overrightarrow{a}_{+,m})\\
&=&F_{\lambda,k}(\omega\phi_{+,\sigma_i(m)},\overrightarrow{a}_{+,\sigma_i(m)})\\
&=&F_{\lambda,k}(\phi_{+,\sigma_i(m)},\overrightarrow{a}_{+,\sigma_i(m)})\\
&=&F_{S,\sigma_i(m)} .
\end{array}
\end{equation}
La fonction $F_S$ est donc invariante par $\sigma_i$.\\
La fonction $F_{as}(m,k)$ est limite simple de fonctions invariantes et est donc invariante.\\
La fonction $F_{as}(m,k)$ est invariante pour tout $k>\sqrt{\frac{2K}{I}}$. Vu l'expression de $F_{as}$, cela ne peut se faire que si $K_{m}$ et $I_{m}$ sont invariants.\\
Il suffit ensuite d'appliquer le théorème \ref{but-de-ce-fatras}. $\hfill{\bf CQFD}$

\saction{Stabilité du couple bifurqué}\label{sec:stabilite-etat-bifurque}
\noindent Dans cette section, on s'intéresse au problème de la stabilité locale des solutions bifurquées.
\begin{theorem}.\label{stabilite}
Si $k>\sqrt{\frac{2K}{I}}$ alors il existe $\epsilon_{0}>0$ tel que, si $0<\epsilon<\epsilon_{0}$, alors le couple bifurqué défini au théorème~\ref{reduc-lyapunov} est un minimum local de la fonctionnelle $F_{\lambda,k}$.\\
Si $k<\sqrt{\frac{2K}{I}}$, alors il existe $\epsilon_{0}>0$ tel que, si $0<-\epsilon<\epsilon_{0}$, le couple bifurqué défini au théorème~\ref{reduc-lyapunov} n'est pas un minimum local de la fonctionnelle $F_{\lambda,k}$.
\end{theorem}
\begin{remarque}
On trouve des études physiques de stabilité dans \cite{almogI} et \cite{pirate-I}. Ces études utilisent des développements en série et ne se limitent pas à une quantification de $1$ comme nous.
\end{remarque}
{\it Preuve.} Pour traiter tous les cas d'un seul coup, on utilise la fraction
\begin{equation}
\frac{\epsilon}{I-\frac{2}{k^2}K}
\end{equation}
qui est toujours positive.\\
Pour prouver des propriétés de minimum local, il suffit de prouver que la différentielle seconde est définie positive sur le couple bifurqué $(\phi_+,\overrightarrow{a_+})\in{\cal A}$ défini à l'équation~(\ref{definition-etat-bifurqué}).\\
La différentielle seconde de la fonctionnelle calculée au point $(\phi, \overrightarrow{a})$ a pour expression
\begin{equation}\label{variation-seconde}
\begin{array}{rcl}
D^2_{\phi, \overrightarrow{a}}F_{\lambda,k}(\delta\phi,\overrightarrow{\delta a})&=&\int_{\Omega}\Vert i\overrightarrow{\nabla}\delta\phi+(\overrightarrow{A}_0+\overrightarrow{a})\delta\phi\Vert^2-\int_{\Omega}(\lambda-|\phi|^2)^2|\delta\phi|^2\\
&+&4\int_{\Omega}\overrightarrow{\delta a}.\Rez[\overline{\delta\phi}(i\overrightarrow{\nabla}\phi+(\overrightarrow{A}_{0}+\overrightarrow{a})\phi)]\\
&+&k^2\int_{\Omega}|\rot\, \overrightarrow{\delta a}|^2+\int_{\Omega}(\overrightarrow{\delta a})^2|\phi|^2+2\int_{\Omega}(\Rez\,\overline{\phi}\,\delta\phi)^2 .
\end{array}
\end{equation}
On a utilisé ici la relation:
\begin{equation}
\int_{\Omega}\Rez[\overline{\phi}(i\overrightarrow{\nabla}\phi'+(\overrightarrow{A}_0+\overrightarrow{a})\phi')]=\int_{\Omega}\Rez[\overline{\phi'}(i\overrightarrow{\nabla}\phi+(\overrightarrow{A}_0+\overrightarrow{a})\phi)] ,
\end{equation}
qui est vraie si $\phi$, $\phi'$ et $\overrightarrow{a}$ sont de classe $H^1$ et que l'on obtient par intégration par parties.\\
Dans la suite on abrégera $D^2_{\phi_+, \overrightarrow{a_+}}F_{\lambda,k}(\delta\phi,\overrightarrow{\delta a})$ en $D^2F$.\\
Si $(\phi,\overrightarrow{a})\in{\cal A}$ est un point critique de la fonctionnelle $F_{\lambda,k}$, alors on a le développement limité
\begin{equation}
F_{\lambda,k}(\phi+\delta\phi,\overrightarrow{a}+\overrightarrow{\delta a})=F_{\lambda,k}(\phi,\overrightarrow{a})+\frac{1}{2}D^2_{\phi, \overrightarrow{a}}F_{\lambda,k}(\delta\phi,\overrightarrow{\delta a})+O(||(\delta\phi,\overrightarrow{\delta a})||_{H^1})^3 .
\end{equation}
Comme dans la section~\ref{sec:lyapunov-schmidt}, on écrira:
\begin{equation}
\left\lbrace\begin{array}{rcl}
\phi_+&=&\alpha\phi_0+\Phi,\\
\overrightarrow{a_+}&=&\overrightarrow{\Psi} .
\end{array}\right.
\end{equation}
On rappelle que l'on a montré que $\alpha=O(\sqrt{\epsilon})$, $\Vert\Phi\Vert_{H^1}=O(\epsilon^{\frac{3}{2}})$ et $\Vert\overrightarrow{\Psi}\Vert_{H^1}=O(\epsilon)$. Voir l'équation~(\ref{developpements-limites}) pour plus de détails.\\
On écrit 
\begin{equation}
\delta\phi=\sqrt{\frac{I-\frac{2}{k^2}K}{\epsilon}}\mu\phi_0+w
\end{equation}
avec $w\in \phi_0^{\perp}$ et $\mu\in\R$. La variable $\mu$ est réelle puisque l'on se limite à des sections telles que $\langle\phi,\phi_{0}\rangle$ est réel, voir la remarque~\ref{remarque-tres-intelligente}.\\
On évalue d'abord la partie de la forme~(\ref{variation-seconde}) qui ne dépend que de $\delta\phi$:
\begin{equation}
Q(\delta\phi)=\int_{\Omega}\Vert i\overrightarrow{\nabla}\delta\phi+(\overrightarrow{A}_{0}+\overrightarrow{a_+})\delta\phi\Vert^2-(\lambda-|\phi_+|^2)|\delta\phi|^2+2(\Rez\,\overline{\phi_+}\,\delta\phi)^2 .
\end{equation}
On développe $Q(\delta\phi)$
\begin{equation}\label{Qdeltaphi-expression-developpee}
\begin{array}{rcl}
Q(\delta\phi)&=&\frac{2\pi-\lambda}{\epsilon}[I-\frac{2}{k^2}K]\mu^2+\langle([i\overrightarrow{\nabla}+\overrightarrow{A}_0]^2-\lambda)w,w\rangle\\
&+&2\Rez\langle\overrightarrow{\Psi}.(i\overrightarrow{\nabla}+\overrightarrow{A}_0)\delta\phi,\delta\phi\rangle+\langle\overrightarrow{\Psi}^2\delta\phi,\delta\phi\rangle\\
&+&\int_{\Omega}|\phi_+|^2|\delta\phi|^2+2\int_{\Omega}(\Rez\,\overline{\phi_+}\delta\phi)^2 .
\end{array}
\end{equation}
On estime l'avant dernier terme de l'équation~(\ref{Qdeltaphi-expression-developpee})
\begin{equation}
\begin{array}{rcl}
\int_{\Omega}|\phi_+|^2|\delta\phi|^2&=&\int_{\Omega}|\alpha\phi_0+\Phi|^2|\mu\phi_0+w|^2\\
&=&I\mu^2+\epsilon^{\frac{1}{2}}\lbrace\Vert w\Vert _{L^2}|\mu|O(1)+\Vert w\Vert ^2_{L^2}O(\epsilon^{\frac{1}{2}})+\mu^2O(\epsilon^{\frac{1}{2}})\rbrace .
\end{array}
\end{equation}
On trouve la même estimation pour le dernier terme de l'équation~(\ref{Qdeltaphi-expression-developpee}):
\begin{equation}
\int_{\Omega}(\Rez\,\overline{\phi_+}\delta\phi)^2=I\mu^2+\epsilon^{\frac{1}{2}}\lbrace\Vert w\Vert _{L^2}|\mu|O(1)+\Vert w\Vert ^2_{L^2}O(\epsilon^{\frac{1}{2}})+\mu^2O(\epsilon^{\frac{1}{2}})\rbrace .
\end{equation}
Dans l'estimation qui suit, on isole les termes d'ordre inférieur à $1$ en $\epsilon$.
\begin{equation}
\begin{array}{rcl}
\D Q(\delta\phi)&=&\langle([i\overrightarrow{\nabla}+\overrightarrow{A}_0]^2-\lambda)w,w\rangle+3I\mu^2\\[0.5mm]
&-&\D\frac{2\epsilon}{Ik^2-2K}\Rez\langle M[\Rez(\overline{\phi_0}(i\overrightarrow{\nabla}+\overrightarrow{A}_0)\phi_0)].(i\overrightarrow{\nabla}+\overrightarrow{A}_0)\delta\phi,\delta\phi\rangle\\[1.5mm]
&+&\D\epsilon^{\frac{1}{2}}\lbrace\Vert w\Vert _{L^2}|\mu|O(1)+\Vert w\Vert ^2_{L^2}O(\epsilon^{\frac{1}{2}})+\mu^2O(\epsilon^{\frac{1}{2}})\rbrace\\[2mm]
&=&\D\langle([i\overrightarrow{\nabla}+\overrightarrow{A}_0]^2-\lambda)w,w\rangle+\mu^2[3I-\frac{2}{k^2}K]\\[1mm]
&+&\D\epsilon^{\frac{1}{2}}\lbrace\Vert w\Vert _{L^2}|\mu|O(1)+\Vert w\Vert ^2_{L^2}O(\epsilon^{\frac{1}{2}})+\mu^2O(\epsilon^{\frac{1}{2}})\rbrace .
\end{array}
\end{equation}
Reprenons l'estimation de la variation seconde de l'énergie en introduisant les développements en $\epsilon$ de $\phi$ et $\overrightarrow{a}$. On a; pour $(D^2F)(\delta\phi,\overrightarrow{\delta a})$:
\begin{equation}
\begin{array}{rcl}
D^2F&=&\D\int_{\Omega}\overrightarrow{\delta a}.4\Rez[\overline{\delta\phi}(i\overrightarrow{\nabla}+(\overrightarrow{A}_0+\overrightarrow{a_+}))\phi_+]+k^2\int_{\Omega}|\rot\, \overrightarrow{\delta a}|^2\\[1mm]
&+&\D\langle([i\overrightarrow{\nabla}+\overrightarrow{A}_0]^2-\lambda)w,w\rangle+\mu^2[3I-\frac{2}{k^2}K]+\int_{\Omega}(\overrightarrow{\delta a})^2|\phi_+|^2\\[1mm]
&+&\D\epsilon^{\frac{1}{2}}\lbrace\Vert w\Vert _{L^2}|\mu|O(1)+\Vert w\Vert ^2_{L^2}O(\epsilon^{\frac{1}{2}})+\mu^2O(\epsilon^{\frac{1}{2}})\rbrace\\[3mm]
&=&\D 4\mu\int_{\Omega}\overrightarrow{\delta a}. \Rez[\overline{\phi_0}(i\overrightarrow{\nabla}+\overrightarrow{A}_0)\phi_0]+k^2\int_{\Omega}|\rot\, \overrightarrow{\delta a}|^2\\[1mm]
&+&\D\langle([i\overrightarrow{\nabla}+\overrightarrow{A}_0]^2-\lambda)w,w\rangle+\mu^2[3I-\frac{2}{k^2}K]+\Vert \overrightarrow{\delta a}\Vert_{L^2}^2O(\epsilon)\\[1mm]
&+&\D\Vert \overrightarrow{\delta a}\Vert_{L^2}O(\vert\epsilon\vert)(\vert\mu\vert+\epsilon^{\frac{1}{2}}\Vert w\Vert_{L^2})+\Vert \overrightarrow{\delta a}\Vert_{L^2}O(\vert\epsilon\vert^{\frac{1}{2}})\Vert w\Vert_{L^2}\\[1mm]
&+&\D\epsilon^{\frac{1}{2}}\lbrace\Vert w\Vert _{L^2}|\mu|O(1)+\Vert w\Vert ^2_{L^2}O(\epsilon^{\frac{1}{2}})+\mu^2O(\epsilon^{\frac{1}{2}})\rbrace .
\end{array}
\end{equation}
Le terme de degré $1$ en $\overrightarrow{\delta a}$ est difficile à minorer. C'est là que des conditions sur $k$ vont apparaître. L'opérateur $M$ (cf~(\ref{definition-de-M})) étant positif, il existe $W$\index{$W$} tel que $M=W^2$, on pose
\begin{equation}
\overrightarrow{\delta a}=W(\overrightarrow{\delta c})
\end{equation}
et on obtient:
\begin{equation}\label{forme-quad-not-completed}
\begin{array}{rcl}
D^2F&=&\D\mu^2[3I-\frac{2}{k^2}K]+k^2\int_{\Omega}\Vert\overrightarrow{\delta c}\Vert^2+\langle([i\overrightarrow{\nabla}+\overrightarrow{A}_0]^2-\lambda)w,w\rangle\\
&+&\D 4\mu\int_{\Omega}W(\overrightarrow{\delta c}).\Rez[\overline{\phi_0}(i\overrightarrow{\nabla}+\overrightarrow{A}_0)\phi_0]\\
&+&\D\Vert \overrightarrow{\delta c}\Vert_{H^{-1}}O(\vert\epsilon\vert)(\vert\mu\vert+\epsilon^{\frac{1}{2}}\Vert w\Vert_{L^2})+\Vert \overrightarrow{\delta c}\Vert_{H^{-1}}O(\vert\epsilon\vert^{\frac{1}{2}})\Vert w\Vert_{L^2}\\
&+&\D\Vert \overrightarrow{\delta c}\Vert_{H^{-1}}^2O(\epsilon)+\epsilon^{\frac{1}{2}}\lbrace\Vert w\Vert _{L^2}|\mu|O(1)+\Vert w\Vert ^2_{L^2}O(\epsilon^{\frac{1}{2}})+\mu^2O(\epsilon^{\frac{1}{2}})\rbrace .
\end{array}
\end{equation}
On estime maintenant chaque terme 
\begin{equation}\label{forme-quad-completed}
\begin{array}{rcl}
D^2F&=&\D\mu^2[3I-\frac{2}{k^2}K]+k^2\Vert \overrightarrow{\delta c}\Vert ^2_{L^2}+\langle([i\overrightarrow{\nabla}+\overrightarrow{A}_0]^2-\lambda)w,w\rangle\\
&+&\D 4\mu\int_{\Omega}W(\overrightarrow{\delta c}).\Rez[\overline{\phi_0}(i\overrightarrow{\nabla}+\overrightarrow{A}_0)\phi_0]\\
&+&\D\epsilon^{\frac{1}{2}}\lbrace \Vert \overrightarrow{\delta c}\Vert_{L^{2}}^2O(\epsilon^{\frac{1}{2}})+\Vert \overrightarrow{\delta c}\Vert_{L^{2}}O(1)(\vert\mu\vert+\Vert w\Vert_{L^2})\\
&+&\D\mu^2O(\epsilon^{\frac{3}{2}})+\Vert w\Vert _{L^2}|\mu|O(1)+\Vert w\Vert _{L^2}^2O(\epsilon^{\frac{1}{2}})\rbrace .
\end{array}
\end{equation}
On a utilisé le fait que l'opérateur $W$ est continu sur $L^2$.\\
On estime le terme linéaire en $W(\overrightarrow{\delta c})$.
\begin{equation}\label{maj-forme-lin}
\begin{array}{rcl}
|\langle W(\overrightarrow{\delta c}),\Rez[\overline{\phi_0}(i\overrightarrow{\nabla}+\overrightarrow{A}_0)\phi_0]\rangle|
&=&|\langle\overrightarrow{\delta c},W(\Rez[\overline{\phi_0}(i\overrightarrow{\nabla}+\overrightarrow{A}_0)\phi_0]\rangle|\\[1mm]
&\leq&\Vert \overrightarrow{\delta c}\Vert _{L^2}.\Vert W(\Rez[\overline{\phi_0}(i\overrightarrow{\nabla}+\overrightarrow{A}_0)\phi_0]))\Vert _{L^2}\\[1mm]
&\leq&\sqrt{K}\Vert \overrightarrow{\delta c}\Vert _{L^2} .
\end{array}
\end{equation}
Pour avoir la stabilité, on veut que la forme quadratique:
\begin{equation}\label{forme-quad}
q(x, y)=[3I-\frac{2}{k^2}K]x^2-4\sqrt{K}xy+k^2y^2 ,
\end{equation}
soit définie positive. Cela équivaut à:
\begin{equation}\label{conditions-forme-quad-pos-dim2}
\left\lbrace
\begin{array}{l}
k^2>0, \\
(2\sqrt{K})^2<[3I-\frac{2K}{k^2}][k^2] .
\end{array}\right.
\end{equation}
Les conditions~(\ref{conditions-forme-quad-pos-dim2}) sont équivalentes à $\sqrt{\frac{2K}{I}}< k$.\\
Si $\sqrt{\frac{2K}{I}}< k$, alors la forme quadratique~(\ref{forme-quad}) est définie positive; donc il existe $c>0$, $d>0$ et $f>0$ tel que les deux premières lignes de~(\ref{forme-quad-completed}) soient minorées par le terme:
\begin{equation}\label{minorante}
c\mu^2+d\Vert w\Vert ^{2}_{H^1}+f\Vert \overrightarrow{\delta c}\Vert ^2_{L^2} .
\end{equation}
Le terme entre accolades des deux dernières lignes de~(\ref{forme-quad-completed}) est majoré par une constante multipliée par~(\ref{minorante}). Donc pour $\epsilon$ assez petit, la différentielle seconde $D^2F(\delta\phi,\overrightarrow{\delta a})$ est minorée par une forme quadratique définie positive et est donc positive. Par conséquent ce couple est stable.\\
Nous pouvons démontrer maintenant l'autre partie du théorème~\ref{stabilite}, qui concerne le cas $\sqrt{\frac{2K}{I}}>k$.\\
La majoration effectuée à l'équation~(\ref{maj-forme-lin}) est optimale, puisque la norme d'une application linéaire sur un espace de Hilbert de la forme $l(x)=\langle x, b\rangle$ est égale à $\Vert b\Vert $ et est atteinte sur $b$.\\
On souhaite montrer que $D^2F(\delta\phi, \overrightarrow{\delta a})$ n'est pas positive pour les solutions bifurquées, il suffit donc de trouver un couple $(\delta\phi,\overrightarrow{\delta a})$ tel que $D^2F(\delta\phi,\overrightarrow{\delta a})<0$ pour démontrer le théorème.\\
On pose donc 
\begin{equation}
\left\lbrace\begin{array}{l}
\overrightarrow{\delta a}_{0}=\Rez[\overline{\phi_0}(i\overrightarrow{\nabla}+\overrightarrow{A}_0)\phi_0], \\
\overrightarrow{\delta c}_{0}=W(\Rez[\overline{\phi_0}(i\overrightarrow{\nabla}+\overrightarrow{A}_0)\phi_0]) .
\end{array}\right.
\end{equation}
On évalue maintenant la fonctionnelle sur un couple bien choisi
\begin{equation}\label{forme-speciale}
\begin{array}{ccl}
D^2F(\mu\sqrt{\frac{\epsilon}{I-\frac{2}{k^2}K}}\phi_{0},s\overrightarrow{\delta a}_{0})
&=&\mu^2[3I-\frac{2}{k^2}K]+\mu^2O(\epsilon^2)\\
&+&4\mu s\Vert \overrightarrow{\delta c}_{0}\Vert _{L^2}+s^2k^2\Vert \overrightarrow{\delta c}_{0}\Vert ^2_{L^2}\\
&=&\mu^2[3I-\frac{2}{k^2}K+O(\epsilon)]\\
&+&4\mu s\Vert \overrightarrow{\delta c}_{0}\Vert _{L^2}+s^2k^2\Vert \overrightarrow{\delta c}_{0}\Vert ^2_{L^2}
\end{array}
\end{equation}
Les conditions de positivité pour la forme quadratique~(\ref{forme-speciale}) s'écrivent
\begin{equation}
\left\lbrace
\begin{array}{l}
[3I-\frac{2}{k^2}K+O(\epsilon)]>0\\
(2\sqrt{K})^2<[3I-\frac{2}{k^2}K+O(\epsilon)][k^2]
\end{array}\right.
\end{equation}
et elles ne sont pas vérifiées si $\epsilon$ est assez petit puisque $\sqrt{\frac{2K}{I}}>k$. $\hfill{\bf CQFD}$

\saction{Analyse asymptotique des solutions\\
 de l'équation de Ginzburg-Landau}\label{sec:analyse-asymptotique}
\noindent Dans cette section, on calcule toutes les solutions des équations de Ginzburg-Landau
\begin{equation}
\left\lbrace\begin{array}{rcl}
[i\overrightarrow{\nabla}+(\overrightarrow{A}_{0}+\overrightarrow{a})]^2\phi&=&\phi(\lambda-|\phi|^2)\\
-L\overrightarrow{a}&=&\frac{1}{k^2}\Rez[\overline{\phi}(i\overrightarrow{\nabla}\phi+(\overrightarrow{A}_{0}+\overrightarrow{a})\phi)]
\end{array}\right. 
\end{equation}
en supposant que $\lambda\in [0, 2\pi+\delta_{0}[$ avec $\delta_{0}$ assez petit et $k$ assez grand.
\begin{lemme}.\label{lemme-un-asymptotique}
Si $(\phi,\overrightarrow{a})\in{\cal A}$ vérifie les équations de Ginzburg-Landau~(\ref{Odeh-equations}) pour $\lambda,k$ donnés, alors:
\begin{equation}
\Vert\Delta\,\overrightarrow{a}\Vert_{L^2}^2\leq \frac{\lambda^3}{4k^2} .
\end{equation}
\end{lemme}
{\it Preuve.} On se rappelle que $\divergence\,\overrightarrow{a}=0$ et on calcule
\begin{equation}
\begin{array}{rcl}
\Vert\Delta\,\overrightarrow{a}\Vert_{L^2}^2
&=&\int_{\Omega}\Vert\Delta\,\overrightarrow{a}\Vert^2\\
&=&\int_{\Omega}\Vert \rot^*\,\rot\, \overrightarrow{a}\Vert^2\\
&=&\frac{1}{k^4}\int_{\Omega}\Vert \Rez[\overline{\phi}(i\overrightarrow{\nabla}+\overrightarrow{A_0}+\overrightarrow{a})\phi]\Vert^2 ,
\end{array}
\end{equation}
en utilisant la deuxième équation de GL. On continue à majorer:
\begin{equation}
\begin{array}{rcl}
\Vert\Delta\,\overrightarrow{a}\Vert_{L^2}^2
&\leq&\D\frac{1}{k^4}\int_{\Omega}\Vert\overline{\phi}(i\overrightarrow{\nabla}+\overrightarrow{A_0}+\overrightarrow{a})\phi\Vert^2\\
&\leq&\D\frac{\lambda}{k^4}\int_{\Omega}\Vert (i\overrightarrow{\nabla}+\overrightarrow{A_0}+\overrightarrow{a})\phi\Vert^2\mbox{~,~en~utilisant l'inégalité~(\ref{inegalite-universelle})}\\
&\leq&\D\frac{\lambda}{k^4}\langle (i\overrightarrow{\nabla}+\overrightarrow{A_0}+\overrightarrow{a})\phi, (i\overrightarrow{\nabla}+\overrightarrow{A_0}+\overrightarrow{a})\phi\rangle\\
&\leq&\D\frac{\lambda}{k^4}\langle[i\overrightarrow{\nabla}+\overrightarrow{A}_0+\overrightarrow{a}]^2\phi, \phi\rangle\\
&\leq&\D\frac{\lambda}{\kappa^2}\langle\phi(\lambda-|\phi|^2),\phi\rangle\mbox{~~en~utilisant~la~première~équation~de~GL}\\
&\leq&\D\frac{\lambda}{k^4}\int_{\Omega}|\phi|^2(\lambda-|\phi|^2)\\
&\leq&\D\frac{\lambda}{k^4}\frac{\lambda^2}{4}=\frac{\lambda^3}{4k^4} .
\end{array}
\end{equation}
A la fin, on a utilisé l'égalité $\sup_{x\in[0,\lambda]}x(\lambda-x)=\frac{\lambda^2}{4}$ et $0\leq |\phi|^2\leq \lambda$. $\hfill{\bf CQFD}$

\begin{lemme}.\label{lemme-deux-asymtotique}
Si $\overrightarrow{a}\in L^{\infty}$, alors le niveau fondamental de l'opérateur 
\begin{equation}
[i\overrightarrow{\nabla}+(\overrightarrow{A}_0+\overrightarrow{a})]^2
\end{equation}
est minoré par
\begin{equation}
2\pi(1-\frac{2}{\sqrt{2\pi}}\Vert \overrightarrow{a}\Vert_{L^{\infty}}) .
\end{equation}
\end{lemme}
{\it Preuve.} L'opérateur $H$ défini à l'équation~(\ref{schrodinger-magnetique}) a un niveau fondamental égal à $2\pi$. Cela nous donne
\begin{equation}
\Vert \phi\Vert_{L^2}^2\leq \frac{1}{2\pi}\Vert (i\overrightarrow{\nabla}+\overrightarrow{A}_0)\phi\Vert_{L^2}^2=\frac{1}{2\pi}\langle H\phi,\phi \rangle .
\end{equation}
On fait alors le calcul suivant
\begin{equation}
\begin{array}{rcl}
|\langle\overrightarrow{a}\phi.(i\overrightarrow{\nabla}\phi+\overrightarrow{A_0}\phi)\rangle|&\leq&\Vert\overrightarrow{a}\Vert_{L^{\infty}}\int_{\Omega} |\phi| \Vert i\overrightarrow{\nabla}\phi+\overrightarrow{A_0}\phi\Vert\\
&\leq&\Vert\overrightarrow{a}\Vert_{L^{\infty}}\Vert \phi\Vert _{L^2}\sqrt{\int_{\Omega} \Vert i\overrightarrow{\nabla}\phi+\overrightarrow{A_0}\phi\Vert^2}\\
&\leq&\Vert\overrightarrow{a}\Vert_{L^{\infty}}\Vert \phi\Vert _{L^2}\sqrt{\langle H\phi,\phi \rangle}\\
&\leq&\frac{1}{\sqrt{2\pi}}\Vert\overrightarrow{a}\Vert_{L^{\infty}}\langle H\phi,\phi \rangle.
\end{array}
\end{equation}
Ceci permet de minorer la première valeur propre de l'opérateur de Schrödinger magnétique
\begin{equation}
\begin{array}{ccl}
\langle[i\nabla+\overrightarrow{A}_{0}+\overrightarrow{a}]^2\phi,\phi\rangle
&\geq&\D\langle H\phi,\phi\rangle+2\Rez \langle\overrightarrow{a}\phi,(i\nabla+\overrightarrow{A}_{0})\phi\rangle\\
&\geq&\D\langle H\phi,\phi\rangle-\frac{2}{\sqrt{2\pi}}\Vert\overrightarrow{a}\Vert_{L^{\infty}}\langle H\phi,\phi \rangle\\
&\geq&\D (1-\frac{2}{\sqrt{2\pi}}\Vert \overrightarrow{a}\Vert_{L^{\infty}})\langle H\phi,\phi\rangle\\
&\geq&\D 2\pi(1-\frac{2}{\sqrt{2\pi}}\Vert \overrightarrow{a}\Vert_{L^{\infty}})\Vert \phi\Vert^2 _{L^2} . 
\end{array}
\end{equation}
On a donc bien la minoration voulue.  $\hfill{\bf CQFD}$

\begin{lemme}.\label{lemme-trois-asymtotique}
Soit $(\phi,\overrightarrow{a})\in{\cal A}$ une solution des équations de Ginzburg-Landau~(\ref{Odeh-equations}) pour $\lambda,k$ donnés. Supposons que l'opérateur $[i\nabla+\overrightarrow{A}_{0}+\overrightarrow{a}]^2$ ait un niveau fondamental supérieur à $\lambda$; alors $(\phi,\overrightarrow{a})=(0,0)$.
\end{lemme}
{\it Preuve.} Appellons \index{$\lambda_{\overrightarrow{a}}$}$\lambda_{\overrightarrow{a}}$ le niveau fondamental de $[i\nabla+\overrightarrow{A}_{0}+\overrightarrow{a}]^2$. La première équation de Ginzburg-Landau nous donne
\begin{equation}\label{equation-annullante}
[i\nabla+\overrightarrow{A}_{0}+\overrightarrow{a}]^2\phi=\phi(\lambda-|\phi|^2) .
\end{equation}
En multipliant l'équation~(\ref{equation-annullante}) par $\overline{\phi}$ et en intégrant, on obtient:
\begin{equation}
\lambda_{\overrightarrow{a}}||\phi||_{L^2}^2\leq \langle [i\nabla+\overrightarrow{A}_{0}+\overrightarrow{a}]^2\phi,\phi\rangle=\int_{\Omega}\lambda|\phi|^2-|\phi|^4 .
\end{equation}
Cette équation implique:
\begin{equation}
\int_{\Omega}|\phi|^2(\lambda-\lambda_{\overrightarrow{a}}-|\phi|^4)\geq 0.
\end{equation}
Cette inégalité n'est possible que si $\phi=0$.\\
L'équation originale~(\ref{Odeh-equations}) implique alors $L\overrightarrow{a}=\rot^*\,\rot\, \overrightarrow{a}=\overrightarrow{0}$. Or on sait que $L$ est inversible (cf~\ref{spectre-L-theo}); donc $\overrightarrow{a}=0$. $\hfill{\bf CQFD}$

\begin{lemme}.\label{lemme-quatres-asymptotique}
Soit $D>0$; l'équation 
\begin{equation}
2\pi(1-\frac{1}{\sqrt{2\pi}}D\frac{\lambda^{\frac{3}{2}}}{k^2})=\lambda
\end{equation}
possède une unique solution $\lambda(k,D)\in]0,2\pi[$.\\
De plus la fonction $k\mapsto\lambda(k,D)$ vérifie les estimations
\begin{equation}
\begin{array}{l}
\lambda(k,D) > 2\pi(1-\frac{2D\pi}{k^2}), \\
\lambda(k,D) = 2\pi+O(\frac{1}{k^2}), \mbox{~si~}k\rightarrow\infty\\
\lambda(k,D)\simeq (\frac{\sqrt{2\pi}}{D})^{\frac{2}{3}}k^{\frac{4}{3}}\mbox{~si~}k\rightarrow 0 .
\end{array}
\end{equation}
\end{lemme}
{\it Preuve.} La fonction $f_k(\lambda)=\lambda+\sqrt{2\pi}\frac{D\lambda^{\frac{3}{2}}}{k^2}$ est strictement croissante sur $[0,2\pi]$ et vérifie $f_k(0)=0$ ainsi que $f_k(2\pi)>2\pi$. Par conséquent, l'équation 
\begin{equation}
f_k(\lambda)=2\pi
\end{equation}
possède une unique solution \index{$\lambda(k,D)$}$\lambda(k,D)$ dans $]0,2\pi[$.\\
L'inégalité $0<\lambda(k,D)<2\pi$ implique 
\begin{equation}
\frac{2}{\sqrt{2\pi}}\frac{D\lambda(k,D)^{\frac{3}{2}}}{2k^2}<\frac{2D\pi}{k^2} .
\end{equation}
On obtient alors:
\begin{equation}
\begin{array}{rcl}
\lambda(k,D)&=&\D 2\pi(1-\frac{2}{\sqrt{2\pi}}\frac{D\lambda(k,D)^{\frac{3}{2}}}{2k^2})\\
&>&\D 2\pi(1-\frac{2D\pi}{k^2}) .
\end{array}
\end{equation}
On a l'inégalité:
\begin{equation}
\begin{array}{rcl}
2\pi&=&\D\lambda(k,D)+\sqrt{2\pi}\frac{D\lambda(k,D)^{\frac{3}{2}}}{k^2}\\
&<&\D\lambda(k,D)+2\pi\frac{D\lambda(k,D)}{k^2}\\
&<&\D\lambda(k,D)(1+2\pi\frac{D}{k^2}) .
\end{array}
\end{equation}
Ceci implique l'inégalité:
\begin{equation}\label{inegalite-precede-equivalent}
\frac{1}{\sqrt{\lambda(k,D)}}<\frac{1}{\sqrt{2\pi}}\sqrt{1+2\pi\frac{D}{k^2}} .
\end{equation}
L'équation définissant $\lambda(k,D)$ nous donne:
\begin{equation}
2\pi=\sqrt{2\pi}\frac{D\lambda(k,D)^{\frac{3}{2}}}{k^2}(1+\frac{k^2}{\sqrt{2\pi}D\sqrt{\lambda(k,D)}}) .
\end{equation}
L'inégalité~(\ref{inegalite-precede-equivalent}) implique
\begin{equation}
\lim_{k\rightarrow 0}\frac{k^2}{\sqrt{\lambda(k,D)}}=0
\end{equation}
ce qui nous donne $2\pi\simeq \sqrt{2\pi}\frac{D\lambda(k,D)^{\frac{3}{2}}}{k^2}$. En inversant cet équivalent on obtient,
\begin{equation}
\lambda(k,D)\simeq (\frac{\sqrt{2\pi}}{D})^{\frac{2}{3}}k^{\frac{4}{3}}\mbox{~si~}k\rightarrow 0 ,
\end{equation}
par un calcul simple. $\hfill{\bf CQFD}$

\begin{theorem}.\label{vanishing-theorem}
Il existe une constante $D>0$ tel que $\forall \overrightarrow{a}\in{\cal A}$,
\begin{equation}
\Vert \overrightarrow{a}\Vert_{L^{\infty}}\leq D\Vert \Delta\,\overrightarrow{a}\Vert_{L^2} .
\end{equation}
Pour cette constante $D$, pour $\lambda<\lambda(k,D)$ et si $(\phi,\overrightarrow{a})\in{\cal A}$ est solution des équations de Ginzburg-Landau~(\ref{Odeh-equations}) alors $(\phi,\overrightarrow{a})=(0,0)$.
\end{theorem}
{\it Preuve.}\\
{\bf $1^{ere}$ étape :}\\
Il existe une constante $C$ tel qu'on obtienne la majoration:
\begin{equation}
\forall \overrightarrow{a}\in H^2_0,\,\Vert \overrightarrow{a}\Vert ^2_{H^2}\leq C\int_{\Omega}\Vert\Delta\,\overrightarrow{a}\Vert^2 .
\end{equation}
L'espace $H^2$ se plonge continument dans $L^\infty$, il existe donc $D>0$ ne dépendant que du réseau $m$ tel que
\begin{equation}
\Vert\overrightarrow{a}\Vert_{L^{\infty}}\leq D\sqrt{\int_{\Omega}\Vert\Delta\,\overrightarrow{a}\Vert^2} .
\end{equation}
{\bf $2^{eme}$ étape :}\\
Si $(\phi,\overrightarrow{a})\in{\cal A}$ est solution de Ginzburg-Landau pour $\lambda,k$, le lemme~\ref{lemme-un-asymptotique} implique l'inégalité
\begin{equation}
\Vert\overrightarrow{a}\Vert_{L^{\infty}}\leq \frac{D\lambda^{\frac{3}{2}}}{2k^2} .
\end{equation}
Le lemme \ref{lemme-deux-asymtotique} implique que le niveau fondamental de l'opérateur $[i\overrightarrow{\nabla}+(\overrightarrow{A}_0+\overrightarrow{a})]^2$ est minoré par
\begin{equation}
2\pi-2\sqrt{2\pi}(\frac{D\lambda^{\frac{3}{2}}}{2k^2})=2\pi-\sqrt{2\pi}(\frac{D\lambda^{\frac{3}{2}}}{k^2}) .
\end{equation}
La condition $\lambda_{\overrightarrow{a}}>\lambda$ du lemme \ref{lemme-trois-asymtotique} est donc vérifiée si
\begin{equation}
\lambda>2\pi-\sqrt{2\pi}\frac{D\lambda^{\frac{3}{2}}}{k^2} .
\end{equation}
Cela équivaut à $\lambda<\lambda(k,D)$ par le lemme \ref{lemme-quatres-asymptotique}.\\
Les hypothèses du lemme~\ref{lemme-trois-asymtotique} étant vérifiées, on a $(\phi,\overrightarrow{a})=(0,0)$. $\hfill{\bf CQFD}$

\begin{theorem}.
Il existe $k_{0}>0$ et $h>0$ tel que, si $k>k_{0}$ et $\lambda\in[0,2\pi+h[$, alors les seules solutions de l'équation de Ginzburg-Landau sont la solution triviale et celles construites au théorème~\ref{reduc-lyapunov}.
\end{theorem}
{\it Preuve.}\\
{\bf $1^{ere}$ étape :}\\
Le théorème~\ref{reduc-lyapunov} montre qu'il existe un voisinage de $\epsilon=0$ de la forme $]-\eta,\eta[$ et un voisinage de la solution triviale $(0,0)$ de la forme $\Vert \phi\Vert _{H^1}+\Vert \overrightarrow{a}\Vert _{H^1}<\eta$ sur lequel on peut calculer les solutions des équations de Ginzburg-Landau par perturbation. Ce voisinage ne dépend que de $k_0$.\\
Par le lemme~\ref{lemme-un-asymptotique}, il existe $k'_0>0$ tel que, si $k\geq k'_0$, alors les solutions $(\phi,\overrightarrow{a})$ des équations de Ginzburg-Landau vérifient $\Vert \overrightarrow{a}\Vert _{H^1}<\frac{\eta}{2}$.\\
{\bf $2^{eme}$ étape :}\\
Montrons qu'il suffit d'obtenir une estimée $L^2$ sur $\phi$, l'estimée $H^1$ en découlant pour obtenir $\Vert \phi\Vert _{H^1}<\frac{\eta}{2}$.\\
Pour cela on utilise l'ellipticité du système d'équations de Ginzburg-Landau. Les calculs suivants sont possibles puisque l'on sait que $(\phi, \overrightarrow{a})$ est $C^{\infty}$. L'équation de Ginzburg-Landau pour $\phi$ s'écrit
\begin{equation}\label{majorante-preview}
[i\overrightarrow{\nabla}+\overrightarrow{A}_{0}]^2\phi=\phi(\lambda-|\phi|^2)-2\overrightarrow{a}.(i\overrightarrow{\nabla}+\overrightarrow{A}_{0})\phi-\overrightarrow{a}^2\phi .
\end{equation}
On multiplie cette formule par $\overline{\phi}$, on intègre sur $\Omega$ et on utilise la majoration sur $\phi$ du théorème~\ref{principe-maximum}.
\begin{equation}\label{majorante}
\begin{array}{rcl}
\Vert (i\overrightarrow{\nabla}+\overrightarrow{A}_{0})\phi\Vert ^2_{L^2}
&\leq&\lambda\Vert \phi\Vert ^2_{L^2}+2\Vert \overrightarrow{a}\phi\Vert _{L^2}\Vert (i\overrightarrow{\nabla}+\overrightarrow{A}_{0})\phi\Vert _{L^2}+\Vert \overrightarrow{a}\phi\Vert ^2_{L^2} .
\end{array}
\end{equation}
L'inéquation~(\ref{majorante}) implique
\begin{equation}\label{majorante-best}
\begin{array}{rcl}
\Vert (i\overrightarrow{\nabla}+\overrightarrow{A}_{0})\phi\Vert _{L^2}&\leq& \Vert \overrightarrow{a}\phi\Vert _{L^2}+\sqrt{2\Vert \overrightarrow{a}\phi\Vert ^2_{L^2}+\lambda \Vert \phi\Vert ^2_{L^2}}\\
&\leq&\Vert \overrightarrow{a}\Vert_{L^4}\Vert\phi\Vert_{L^4}+\sqrt{2\Vert \overrightarrow{a}\Vert^2_{L^4}\Vert\phi\Vert^2_{L^4}+\lambda \Vert \phi\Vert ^2_{L^2}}.
\end{array}
\end{equation}
La norme $L^{4}$ de $\overrightarrow{a}$ tend vers $0$ quand $k$ tend vers $\infty$; donc le contrôle en norme $L^2$ de $\phi$ se transforme trivialement en un contrôle en norme $H^1$ de $\phi$ par l'inégalité~(\ref{majorante-best}) et l'ellipticité de l'opérateur $i\overrightarrow{\nabla}+\overrightarrow{A}_{0}$.\\
Il existe donc $\eta_{2}>0$ tel que pour $k$ assez grand $\Vert \phi\Vert_{L^{2}}<\eta_{2}$ implique $\Vert \phi\Vert _{H^1}<\frac{\eta}{2}$. Montrons maintenant l'estimée $L^2$.\\
{\bf $3^{eme}$ étape :}\\
L'équation de Ginzburg-Landau donne la relation
\begin{equation}\label{inegalite-fond}
\begin{array}{rcl}
\int_{\Omega}|\phi|^2(\lambda-|\phi|^2)
&=&\langle\phi, \phi(\lambda-|\phi|^2)\rangle\\
&=&\langle \phi, [i\overrightarrow{\nabla}+(\overrightarrow{A}_{0}+\overrightarrow{a})]^2\phi\rangle\\
&\geq &(2\pi-w(k))\int_{\Omega}|\phi|^2 .
\end{array}
\end{equation}
Dans les équations~(\ref{inegalite-fond}) et~(\ref{inegalite-fond-reecrite}), $2\pi-w(k)$ correspond à la valeur du niveau fondamental de l'opérateur $[i\overrightarrow{\nabla}+(\overrightarrow{A}_{0}+\overrightarrow{a})]^2$. Le lemme~\ref{lemme-quatres-asymptotique} affirme que $w(k)$ tend vers $0$ quand $k$ tend vers l'infini.\\
L'inégalité~(\ref{inegalite-fond}) s'écrit alors
\begin{equation}\label{inegalite-fond-reecrite}
\int_{\Omega}|\phi|^2(\epsilon+w(k)-|\phi|^2)\geq 0 .
\end{equation}
On pose $r=\epsilon+w(k)$. L'inégalité~(\ref{inegalite-fond-reecrite}) implique
\begin{equation}
(\int_{\Omega}|\phi|^2)^2\leq \int_{\Omega}|\phi|^4\leq r \int_{\Omega}|\phi|^2 .
\end{equation}
Cette équation implique que $\int_{\Omega}|\phi|^2\leq |r|$.\\
On choisit $\epsilon\in]-\eta_{2}/2,\eta_{2}/2[$  et $k$ assez grand pour que $\Vert \phi\Vert _{L^{2}}$ soit inférieur à $\eta_{2}$.\\
Pour $\lambda\in[0,2\pi-\eta_{2}/2[$ le théorème~\ref{vanishing-theorem} et le lemme~\ref{lemme-quatres-asymptotique} nous donnent le résultat. $\hfill{\bf CQFD}$\\
\\
On a donc entièrement résolu notre problème initial dans le cas où $k$ est grand. On peut préciser un peu le théorème~\ref{stabilite}.
\begin{proposition}.
Il existe $k_{0}>0$ et $h>0$ tel que si $k>k_{0}$ et $\lambda\in[2\pi,2\pi+h[$ alors la courbe de solutions bifurquées définie au théorème~\ref{reduc-lyapunov} est le minimum de la fonctionnelle.
\end{proposition}
{\it Preuve.} Si le couple $(\phi, \overrightarrow{a})$ est un couple qui minimise la fonctionnelle de Ginzburg-Landau, alors il est solution des équations de Ginzburg-Landau donc il est égal soit à la solution nulle soit au couple bifurqué.\\
Le théorème~\ref{evolution-de-F} indique que l'énergie du couple bifurqué est plus basse que celle du couple $(0,0)$. Le couple $(\phi, \overrightarrow{a})$ est donc égal au couple bifurqué.$\hfill{\bf CQFD}$

\chapter{Analyse du minimum ${\cal E}^{V}_{k,H_{ext}}$}\label{Diagramm-phase}
\noindent Dans ce chapitre, on décrit la structure du diagramme de phase (cf~\ref{caracterisation-etat-possibles}). On utilise pour cela la fonctionnelle $E^{V}_{k,H_{ext}}(H_{int},\phi,\overrightarrow{a})$ (cf~(\ref{definition-EVkHext-compl})). On utilise aussi la distinction entre les différents états possibles effectués dans la définition \ref{caracterisation-etat-possibles}.\\
On montre d'abord des théorèmes de monotonie sur l'état du supraconducteur puis on revient sur le cas $k=\frac{1}{\sqrt{2}}$.\\
Les troisième et quatrième parties sont consacrées à l'état normal et à l'état pur défini en \ref{caracterisation-etat-possibles}.\\
On rappelle que l'énergie de l'état normal est $\frac{1}{4}$ et que l'énergie pour $H_{int}=0$ est $\frac{H_{ext}^2}{2}$ qui est l'énergie de l'état pur.

\saction{Théorèmes de monotonie}\label{sec:minimum-et-monotonie}
\noindent On utilise ici la fonctionnelle $E^V_{k,H_{ext}}(H_{int},\phi,\overrightarrow{a})$ définie à l'équation (\ref{definition-EVkHext-compl}).

\begin{theorem}.\label{monotonie-etat-pur}
Soit $(k,H_{ext})$ un point du diagramme des phases du supraconducteur o\`u ${\cal E}^V_{k,H_{ext}}=\frac{H_{ext}^2}{2}$. Alors, pour tout $k'\leq k$, tout $H'_{ext}\leq H_{ext}$, on a: ${\cal E}^V_{k',H'_{ext}}=\frac{{H'_{ext}}^2}{2}$.
\end{theorem}
\begin{remarque}.
En utilisant le langage des physiciens, on dira que, si $(k,H_{ext})$ est un point du diagramme des phases o\`u le supraconducteur est pur et si $k'\leq k$ et $H'_{ext}\leq H_{ext}$, alors le supraconducteur est aussi pur en ce point.
\end{remarque}
{\it Preuve.} Soit donc $(k, H_{ext})$ un point du diagramme des phases sur lequel ${\cal E}^V_{k,H_{ext}}=\frac{H_{ext}^2}{2}$. Cela signifie que l'énergie minimale est $\frac{H_{ext}^2}{2}$. En ce point l'énergie minimale de la fonctionnelle est atteinte par l'état $(0,\sqrt{\lambda},0)$ qui est un état de quantification $0$. Par définition cela équivaut à 
\begin{equation}\label{inegalite-definissante-etat-pur}
\begin{array}{l}
\forall H_{int}>0, \forall (\phi, \overrightarrow{a})\in{\cal A},\,\,
D_{\lambda, k}(\phi,\overrightarrow{a})+\frac{1}{2}(H_{int}-H_{ext})^2\geq \frac{H_{ext}^2}{2} .
\end{array}
\end{equation}
%Le minimum est effectué sur des états de quantification $1$ (voir l'hypothèse \ref{hypothese_de_quantification}) m\^eme si cette hypothèse ne sert pas ici.\\
L'inégalité (\ref{inegalite-definissante-etat-pur}) est équivalente à
\begin{equation}\label{equation-phase}
\begin{array}{c}
\D\forall H_{int}>0, \forall (\phi, \overrightarrow{a})\in{\cal A},\\
\D\lbrace \int_{\Omega}\frac{H_{int}}{4\pi k}\Vert i\overrightarrow{\nabla}\phi+(\overrightarrow{A}_0+\overrightarrow{a})\phi\Vert^2+\frac{1}{4}(1-|\phi|^2)^2\\
\D +\frac{H_{int}^2}{2(2\pi)^2}|\rot\,\overrightarrow{a}|^2\rbrace+\frac{H_{int}^2-2H_{int}H_{ext}}{2}\geq 0 .
\end{array}
\end{equation}
Si $k$ décro\^it alors $\frac{1}{k}$ croît; les intégrales étant positives, le membre de gauche de~(\ref{equation-phase}) croît. Si $H_{ext}$ décro\^it, alors la quantité $-2H_{int}H_{ext}$ croît et donc le membre de gauche croît. Par conséquent, si $k'\leq k$ et si $H'_{ext}\leq H_{ext}$, l'état qui minimise la fonctionnelle vérifie donc ${\cal E}^V_{k,H_{ext}}\geq \frac{H_{ext}^2}{2}$ et il y a égalité par le lemme \ref{prolongement-par-continuite}. $\hfill{\bf CQFD}$

\begin{theorem}.\label{normal-k-inegalite}
Soit $(k, H_{ext})$ un point du diagramme des phases du supraconducteur o\`u le minimum de la fonctionnelle est atteint par l'état normal. Alors, pour tout $k'\geq k$, le minimum de la fonctionnelle au point $(k',\frac{k'}{k}H_{ext})$ est atteint par l'état normal.
\end{theorem}
{\it Preuve.} L'énergie de l'état normal est égale à $\frac{1}{4}$. On pose $H'_{ext}=\frac{k'}{k}H_{ext}$.\\
On a donc, puisque l'état normal est un minimum:

\begin{equation}\label{equation-definissante-etat-normal}
\begin{array}{c}
\forall H_{int}>0, \forall (\phi, \overrightarrow{a})\in{\cal A},\\
\frac{H_{int}^2}{(2\pi k)^2}\lbrace \int_{\Omega}\frac{1}{2}\Vert i\overrightarrow{\nabla}\phi+(\overrightarrow{A}_0+\overrightarrow{a})\phi\Vert^2+\frac{1}{4}(\frac{2\pi k}{H_{int}}-|\phi|^2)^2\\
+\frac{k^2}{2}|\rot\,\overrightarrow{a}|^2\rbrace+\frac{1}{2}(H_{int}-H_{ext})^2\geq \frac{1}{4} .
\end{array}
\end{equation}
L'inégalité à démontrer est:
\begin{equation}\label{machin-a-demontrer}
\begin{array}{c}
\forall H'_{int}>0, \forall (\phi, \overrightarrow{a})\in{\cal A},\\
\frac{{H'_{int}}^2}{(2\pi k')^2}\lbrace \int_{\Omega}\frac{1}{2}\Vert i\overrightarrow{\nabla}\phi+(\overrightarrow{A}_0+\overrightarrow{a})\phi\Vert^2+\frac{1}{4}(\frac{2\pi k'}{H'_{int}}-|\phi|^2)^2\\
+\frac{k'^2}{2}|\rot\,\overrightarrow{a}|^2\rbrace+\frac{1}{2}(H'_{int}-H'_{ext})^2\geq \frac{1}{4} .
\end{array}
\end{equation}
On écrit $H'_{int}=\frac{k'}{k}H_{int}$ et on obtient
\begin{equation}
\frac{k}{H_{int}}=\frac{k'}{H'_{int}} .
\end{equation}
L'inégalité à démontrer est alors équivalente à 
\begin{equation}\label{normal-phase-equation}
\begin{array}{c}
\forall H_{int}>0, \forall (\phi, \overrightarrow{a})\in{\cal A},\\
\frac{H_{int}^2}{(2\pi k)^2}\lbrace \int_{\Omega}\frac{1}{2}\Vert i\overrightarrow{\nabla}\phi+(\overrightarrow{A}_0+\overrightarrow{a})\phi\Vert^2+\frac{1}{4}(\frac{2\pi k}{H_{int}}-|\phi|^2)^2\\
+\frac{k'^2}{2}|\rot\,\overrightarrow{a}|^2\rbrace+(\frac{k'}{k})^2\frac{1}{2}(H_{int}-H_{ext})^2\geq \frac{1}{4} .
\end{array}
\end{equation}
Cette inégalité est impliquée par l'inégalité (\ref{equation-definissante-etat-normal}) car $k'\geq k$. $\hfill{\bf CQFD}$

\begin{theorem}.\label{normal-H-inegalite}
Soit $(k, H_{ext})$ un point du diagramme des phases du supraconducteur o\`u le minimum de la fonctionnelle est atteint par l'état normal. Alors $\forall H'_{ext}\geq H_{ext}$ le minimum de la fonctionnelle est atteint par l'état normal.
\end{theorem}
{\it Preuve.} L'énergie de l'état normal est égale à $\frac{1}{4}$.\\
On a donc, puisque l'état normal est un minimum:
\begin{equation}\label{normal-phase-pour-B}
\forall H_{int}>0,\,\,\frac{1}{\lambda^2}m_F(\lambda,k)+\frac{1}{2}(H_{int}-H_{ext})^2\geq \frac{1}{4} .
\end{equation}
Rappelons que $m_F(\lambda,k)$ a été introduit en (\ref{definition-mF-mD}).\\
En particulier si $H_{int}=H_{ext}$, cette inégalité implique 
\begin{equation}\label{inegalite-particuliere}
(\frac{H_{ext}}{2\pi k})^2m_F(\frac{2\pi k}{H_{ext}},k)\geq \frac{1}{4} .
\end{equation}
C'est l'inégalité~(\ref{normal-phase-pour-B}) qu'il faut démontrer avec $H_{ext}$ remplacé par $H'_{ext}$. Si $H_{int}\leq H_{ext}$, alors on a trivialement
\begin{equation}
(H_{int}-H_{ext})^2\leq (H_{int}-H'_{ext})^2 .
\end{equation}
Par conséquent, l'inégalité~(\ref{normal-phase-pour-B}) est démontrée pour $H_{int}\leq H_{ext}$ avec $H_{ext}$ remplacé par $H'_{ext}$.\\
Si $H_{int}>H_{ext}$ alors le théorème~(\ref{stricte-decroissance-cas-lambda}) combinée à l'inégalité~(\ref{inegalite-particuliere}) nous donne
\begin{equation}
(\frac{H_{int}}{2\pi k})^2m_F(\frac{2\pi k}{H_{int}},k)\geq (\frac{H_{ext}}{2\pi k})^2m_F(\frac{2\pi k}{H_{ext}},k)\geq \frac{1}{4} .
\end{equation}
Par conséquent l'inégalité~(\ref{normal-phase-pour-B}) est démontrée si $H_{int}>H_{ext}$ par conséquent l'inégalité~(\ref{normal-phase-pour-B}) est donc démontré avec $H_{ext}$ remplacé par $H'_{ext}$.$\hfill{\bf CQFD}$
\begin{remarque}.
M\^eme si la démonstration utilise la modélisation de l'état de quantification $1$ (voir l'hypothèse \ref{hypothese_de_quantification}), la démonstration des théorèmes de monotonies n'utilise pas cette hypothèse. Ces théorèmes doivent donc \^etre vrai dans un cadre plus général.
\end{remarque}
Le théorème~\ref{Bochner-Kodaira-Nakano} nous permet de trouver le minimum ${\cal E}^{V}_{k,H_{ext}}$ dans certains cas.

\begin{lemme}.\label{lemme-preparatoire}
Si $H_{ext}=k=\frac{1}{\sqrt{2}}$, alors le minimum ${\cal E}^{V}_{k,H_{ext}}$ est égal à $\frac{1}{4}$. De plus, il est atteint par l'état normal et par l'état pur.
\end{lemme}
{\it Preuve.} En vertu du théorème~(\ref{ecrantement}), il faut démontrer l'inégalité:
\begin{equation}\label{inegalite-a-demontrer}
\forall H_{int}\leq \frac{1}{\sqrt{2}},\,\,(\frac{H_{int}}{2\pi k})^2m_F(\frac{2\pi k}{H_{int}},k)+\frac{1}{2}(H_{int}-\frac{1}{\sqrt{2}})^2\geq \frac{1}{4} .
\end{equation}
Mais le théorème~\ref{Bochner-Kodaira-Nakano} nous donne l'inégalité:
\begin{equation}
m_F(\lambda,k)\geq \lambda\pi-\pi^2 .
\end{equation}
Par conséquent, l'inégalité~(\ref{inegalite-a-demontrer}) est impliquée par l'inégalité (on utilise $\lambda=\frac{2\pi k}{H_{int}}$)
\begin{equation}
\forall H_{int}\leq \frac{1}{\sqrt{2}},\,\,
[\frac{H_{int}}{2k}-(\frac{H_{int}}{2k})^2]+\frac{1}{2}(H_{int}-\frac{1}{\sqrt{2}})^2\geq \frac{1}{4} .
\end{equation}
Or cette dernière inégalité est en fait une égalité; donc on a démontré ce que l'on voulait.\\
L'énergie de l'état normal est $\frac{1}{4}$. L'énergie de l'état pur est égale à $\frac{1}{2}(\frac{1}{\sqrt{2}})^2=\frac{1}{4}$. $\hfill{\bf CQFD}$

\begin{theorem}.\label{usage-theoreme-monotonie}
Si $k\leq \frac{1}{\sqrt{2}}$, alors
\begin{equation}
{\cal E}^{V}_{k,H_{ext}}=\left\lbrace\begin{array}{l}
\frac{H_{ext}^2}{2}\mbox{~si~}H_{ext}\leq \frac{1}{\sqrt{2}}\\
\frac{1}{4}\mbox{~si~}H_{ext}\geq \frac{1}{\sqrt{2}} .
\end{array}\right.
\end{equation}

\end{theorem}
\begin{remarque}.
En terme physique cela se dit: Si $k\leq \frac{1}{\sqrt{2}}$ alors le minimum ${\cal E}^V_{k,H_{ext}}$ est atteint par l'état supraconducteur pur si $H_{ext} < \frac{1}{\sqrt{2}}$ et par l'état normal si $H_{ext} > \frac{1}{\sqrt{2}}$ et par ces deux états si $H_{ext}=\frac{1}{\sqrt{2}}$.
\end{remarque}
{\it Preuve.} Le lemme \ref{lemme-preparatoire} combiné au théorème \ref{monotonie-etat-pur} donne le résultat dans le cas $H_{ext}\leq \frac{1}{\sqrt{2}}$.\\
Par conséquent, le minimum de la fonctionnelle est égal à $\frac{1}{4}$ si $k\leq \frac{1}{\sqrt{2}}$ et $H_{ext}=\frac{1}{\sqrt{2}}$. Il se trouve que $\frac{1}{4}$ est égal à l'énergie de l'état normal.\\
Donc le théorème \ref{normal-H-inegalite} nous donne la conclusion si $H_{ext}\geq \frac{1}{\sqrt{2}}$. $\hfill{\bf CQFD}$

\begin{theorem}.\label{usage-theoreme-monotonie-2}
Si $H_{ext}\geq k\geq \frac{1}{\sqrt{2}}$, alors:
\begin{equation}
{\cal E}^{V}_{k,H_{ext}}=\frac{1}{4} .
\end{equation}
Autrement dit le théorème nous dit que le minimum de la fonctionnelle est atteint par l'état normal.
\end{theorem}
{\it Preuve.} Le lemme \ref{lemme-preparatoire}, le théorème \ref{normal-k-inegalite} et le théorème \ref{normal-H-inegalite} nous donnent le résultat. $\hfill{\bf CQFD}$

\saction{Analyse du cas $k=\frac{1}{\sqrt{2}}$ (d'après \cite{merci-MXK})}\label{sec:salamon-merci}
\noindent Dans cette section, on souhaite étudier en profondeur la fonctionnelle $F_{\lambda,k}$ dans le cas $k=\frac{1}{\sqrt{2}}$. Dans ce but, on montre que l'équation $A_+(\phi, \overrightarrow{a})=0$ possède une solution pour tout $\lambda>2\pi$.\\
On a besoin d'un résultat d'analyse sur les variétés dans cette partie; on reprend ici l'appendice $4$ du preview \cite{merci-MXK} qui n'est semble t'il pas publié. Ce résultat occupe la première sous section qui est reprise de \cite{merci-MXK}.\\
On peut alors résoudre l'équation $A_+=0$ et trouver ainsi les couples minimisants la fonctionnelle $F_{\lambda,\frac{1}{\sqrt{2}}}$. On déduit de ce résultat une égalité remarquable sur les constantes $I$ et $K$ définies aux équations (\ref{defcon-I}) et (\ref{defcon-K}). Cette partie n'est pas complètement originale, il y a des travaux similaires dans \cite{caffarelli}, \cite{Garcia-prada}, \cite{Taubes}.

\subsection{L'équation de Kazdan-Warner (d'après \cite{merci-MXK})}\label{Recopie-Dietmar-Salamon}
\noindent Rappelons que ${\cal L}$ est un réseau de $\R^2$ et $\Omega$ un domaine fondamental pour ce réseau. Les fonctions ${\cal L}$-périodiques s'identifient aux fonctions sur $\Tore{\cal L}$.
\begin{theorem}.\label{Kazdan-Warner-Salamon-Auroux}
Soit $h$ une fonction positive non nulle appartenant à $H^n(\Tore{\cal L})$ avec $n\geq 2$. Si $A$ est un réel strictement positif, alors l'équation
\begin{equation}\label{equation-Kazdan-Warner}
-\Delta u+e^u h=A
\end{equation}
possède une unique solution $u$ dans $H^{n+2}(\Tore{\cal L})$
\end{theorem}
{\it Preuve.} On utilise l'ensemble ouvert\index{${\cal H}$}
\begin{equation}
{\cal H}=\{h\in H^n(\Tore{\cal L})\mbox{~avec~}h\geq 0\mbox{~et~}\int_{\Omega}h>0\} .
%{\cal H}=L^{\frac{p}{2}}(\Tore{\cal L})-\{0\} \subset L^{\frac{p}{2}}(\Tore{\cal L}) .
\end{equation}
On \index{${\cal U}$}note:
\begin{equation}
{\cal U}=\{ (h,u)\in {\cal H}\times H^{n+2}(\Tore{\cal L})|-\Delta\,u+e^u h=A\} ,
\end{equation}
l'espace des solutions de l'équation~(\ref{equation-Kazdan-Warner}). Le lemme suivant donne ses propriétés.
\begin{lemme}.\label{variete-banach}
L'espace ${\cal U}\subset H^{n}(\Tore{\cal L})\times H^{n+2}(\Tore{\cal L})$ est une variété $C^{\infty}$ de Banach et la projection\index{$\pi$}
\begin{equation}
\pi: {\cal U}\mapsto {\cal H}
\end{equation}
définie par $\pi(h,u)=h$ est un difféomorphisme local au voisinage de tous les points de ${\cal U}$.
\end{lemme}
{\it Preuve.} Considérons l'application \index{$P$}$P: H^n(\Tore{\cal L})\times H^{n+2}(\Tore{\cal L})\mapsto H^n(\Tore{\cal L})$ définie par
\begin{equation}
P(h,u)=-\Delta u+e^u h-A .
\end{equation}
Sa différentielle au point $(h,u)$ est :
\begin{equation}
dP_{(h,u)}(\hat{h},\hat{u})=-\Delta\,\hat{u}+e^uh\hat{u}+e^u\hat{h} .
\end{equation}
\'Etudions l'opérateur 
\begin{equation}
\begin{array}{rcl}
R:C^{\infty}(\Tore{\cal L},\R)&\mapsto&C^{\infty}(\Tore{\cal L},\R)\\
f&\mapsto&Rf=-\Delta f+e^uhf .
\end{array}
\end{equation}
Cet opérateur est symétrique sur l'espace $C^{\infty}(\Tore{\cal L},\R)$. C'est un opérateur différentiel elliptique d'ordre $2$ sur le tore $\Tore{\cal L}$ qui est une variété compacte. Donc l'opérateur $R$ possède une extension autoadjointe de domaine ${\cal D}(R)=H^2(\Tore{\cal L},\R)$ que l'on continue à appeller $R$ (voir par exemple \cite{spin-geom}, théorème III.5.8). De plus, l'injection de $H^2(\Tore{\cal L})$ dans $L^2(\Tore{\cal L})$ étant compacte, son spectre est discret.\\
Si $(h,u)\in H^{n+2}(\Tore{\cal L},\R)$ alors l'opérateur $R$ restreint à $H^{n}(\Tore{\cal L},\R)$ a pour image $H^{n+2}(\Tore{\cal L},\R)$.\\
Il suffit pour montrer que $R$ est inversible de montrer que $R$ est injectif car $R$ possède un spectre discret.\\
Soit donc $\hat{u}\in H^{2}(\Tore{\cal L})$ tel que 
\begin{equation}\label{equation-noyau}
-\Delta\,\hat{u}+e^uh\hat{u}=0 .
\end{equation}
En faisant le produit scalaire de l'équation~(\ref{equation-noyau}) par $\hat{u}$ on obtient l'équation
\begin{equation}
\int_{\Omega}(\Vert\overrightarrow{\nabla}\hat{u}\Vert^2+e^u h |\hat{u}|^2)=0
\end{equation}
qui implique que la fonction $\hat{u}$ s'annule sur un ensemble de mesure positive (là o\`u la fonction $h$ est différente de $0$). De plus $\overrightarrow{\nabla}\hat{u}=0$ donc la fonction $\hat{u}$ est constante. Elle est donc nulle. Par conséquent, l'opérateur $R$ est injectif donc surjectif et il est inversible pour toute paire $(h,u)\in {\cal H}\times H^{n+2}(\Tore{\cal L}\,)$.\\
Cela montre que $0$ est une valeur régulière de la fonction $P$; donc $P^{-1}(0)$ est une variété de Banach par le théorème des fonctions implicites dans les espaces de Banach. Or ${\cal U}=P^{-1}(0)$; donc ${\cal U}$ est une variété de Banach.\\
L'espace tangent à ${\cal U}$ en $(h,u)$ est \index{$T_{(h,u)}{\cal U}$}le noyau de $dP(h,u)$:
\begin{equation}
T_{(h,u)}{\cal U}=\{(\hat{h},\hat{u})|(-\Delta+e^uh)\hat{u}+e^u\hat{h}=0\} .
\end{equation}
Maintenant le linéarisé de l'opérateur de projection:
\begin{equation}
d\pi(h,u):T_{(h,u)}{\cal U}\mapsto H^n(\Tore{\cal L}),
\end{equation}
est donnée par $(\hat{h},\hat{u})\mapsto \hat{h}$. Cet opérateur est bijectif si et seulement si, pour tout ${\hat h}\in H^n(\Tore{\cal L})$, il existe un unique $\hat{u}\in H^{n+2}(\Tore{\cal L})$ tel que $(\hat{h},\hat{u})\in T_{(h,u)}{\cal U}$. Cela provient de la bijectivité de l'opérateur $-\Delta+e^uh$. Maintenant le théorème d'inversion locale montre que chaque paire $(h,u)\in{\cal U}$ possède un voisinage ${\cal V}$ tel que la restriction de $\pi$ à ${\cal V}$ est un difféomorphisme de ${\cal V}$ sur $\pi({\cal V})$. $\hfill{\bf CQFD}$\\
\\
La proposition suivante est la clé du théorème, il donne des estimées à priori sur les solutions de l'équation~(\ref{equation-Kazdan-Warner}).
\begin{proposition}.\label{lemme-majoration-a-priori}
Il existe une fonction continue $\phi_A: ]0,\infty[^2\rightarrow ]0,\infty[$ telle que 
\begin{equation}
\Vert u \Vert_{L^{\infty}}\leq \phi_A(\Vert h \Vert_{L^{2}}, \int_{\Omega} h)
\end{equation}
pour toute solution $(h,u)$ de l'équation~(\ref{equation-Kazdan-Warner}).
\end{proposition}
\begin{remarque}.
La preuve de la proposition montrera que l'on peut prendre pour $\phi_A$ la fonction \index{$\phi_A$}
\begin{equation}\label{definition-phiA}
\phi_A(t,B)=\vert \log(\frac{A}{B})\vert+2c_0A+2\frac{c_0At}{B}\exp(4\frac{c_0At}{B})
\end{equation}
o\`u $c_0$ est la constante introduite au lemme \ref{lemme-laplacien}.
\end{remarque}
{\it Preuve.}\\
{\bf $1^{ere}$ étape :} {\it Si $(h,u)\in C^{\infty}(\Tore{\cal L})^2$ et sont solution de l'équation~(\ref{equation-Kazdan-Warner}), alors:
\begin{equation}\label{step-1}
u(x)\leq 4\frac{c_0At}{B}+\log(\frac{A}{B})
\end{equation}
pour tout $x\in \R^2$, avec $t=\Vert h \Vert_{L^{2}}$, $B=\int_{\Omega} h$.}\\
La fonction $h_0=h-B$ a une moyenne nulle; par conséquent, d'après le lemme~\ref{lemme-laplacien}, il existe une unique fonction $v_0\in C^{\infty}(\Omega)$ telle que:
\begin{equation}
-\Delta\,v_0=-h_0=B-h\mbox{~et~}\int_{\Omega}v_0=0 .
\end{equation}
Le lemme~\ref{lemme-laplacien} donne la majoration:
\begin{equation}\label{uniforme-maj}
\Vert v_0\Vert_{L^{\infty}}\leq c_0\Vert h_0\Vert_{L^2}\leq c_0(\Vert h\Vert_{L^2}+B)\leq 2c_0 t .
\end{equation}
Considérons maintenant la fonction \index{$w_{\epsilon}$}$w$ ${\cal L}$-périodique sur $\R^2$, définie par:
\begin{equation}
w_{\epsilon}(x)=\log(\frac{A+\epsilon}{B})+\frac{A+\epsilon}{B}(v_0(x)+2c_0t)-u(x)\mbox{~avec~}\epsilon>0 .
\end{equation}
Si $x_{\epsilon}$ est un point de $\R^2$ o\`u elle atteint son minimum, alors:
\begin{equation}
w_{\epsilon}(x_{\epsilon})=\inf_{\Omega}w_{\epsilon}\mbox{~et~}(-\Delta\,w_{\epsilon})(x_{\epsilon})\leq 0 .
\end{equation}
Cette inégalité nous donne:
\begin{equation}
\begin{array}{rcl}
0&\geq&(-\Delta w_{\epsilon})(x_{\epsilon})\\
&=&\frac{A+\epsilon}{B}(-\Delta\,v_0(x_{\epsilon}))+\Delta\,u(x_\epsilon)\\
&=&\frac{A+\epsilon}{B}(B-h(x_{\epsilon}))+e^{u(x_{\epsilon})}h(x_{\epsilon})-A\\
&=&\epsilon+h(x_{\epsilon})(e^{u(x_{\epsilon})}-\frac{A+\epsilon}{B}) .
\end{array}
\end{equation}
Cela implique:
\begin{equation}
h(x_{\epsilon})\not= 0\mbox{~et~} u(x_{\epsilon})<\log (\frac{A+\epsilon}{B}) .
\end{equation}
Par conséquent, on a $w_{\epsilon}(x_{\epsilon})\geq 0$ et donc $w_{\epsilon}(x)\geq 0$, pour tout $x\in\R^2$. La limite $\epsilon\rightarrow 0$ donne l'inégalité
\begin{equation}
u(x)\leq \log(\frac{A}{B})+\frac{A}{B}(v_0(x)+2c_0t) .
\end{equation}
Cette inégalité, combinée à l'inégalité~(\ref{uniforme-maj}), donne l'inégalité~(\ref{step-1}).\\
{\bf $2^{eme}$ étape :} {\it Si $(h,u)\in C^{\infty}(\Tore{\cal L})^2$ et sont solution de l'équation~(\ref{equation-Kazdan-Warner}), alors la fonction $A-e^uh$ a une moyenne nulle. L'unique solution $u_0\in C^{\infty}(\Tore{\cal L})$ de
\begin{equation}
-\Delta\,u_0=A-e^uh\mbox{~et~}\int_{\Omega}u_0=0
\end{equation}
vérifie}
\begin{equation}\label{conclu-troisieme-etape}
\Vert u_0\Vert_{L^{\infty}}\leq c_0A+\frac{c_0At}{B}\exp(4\frac{c_0At}{B}) .
\end{equation}
La fonction $u$ vérifie $-\Delta\,u=A-e^uh$, elle est donc de moyenne nulle.\\
Par le lemme~\ref{lemme-laplacien}, on a l'inégalité:
\begin{equation}
\begin{array}{rcl}
\Vert u_0\Vert_{L^{\infty}}&\leq&\D c_0\Vert A-e^{u}h\Vert_{L^{2}}\\
&\leq&\D c_0(\Vert A\Vert_{L^{2}}+\Vert e^{u}h\Vert_{L^{2}})\\
&\leq&\D c_0(A+te^{\sup\,u}) .
\end{array}
\end{equation}
La conclusion provient alors de la première étape.\\
{\bf $3^{eme}$ étape :} {\it Si $(h,u,u_0)$ sont comme dans l'étape $2$, alors
\begin{equation}\label{constante-a-trouver}
u=u_0-\log(\frac{1}{A}\int_{\Omega}e^{u_0}h)
\end{equation}
et
\begin{equation}\label{inegalite-infini}
\Vert u\Vert_{L^{\infty}}\leq 2\Vert u_0\Vert_{L^{\infty}}+\vert \log(\frac{A}{B})\vert .
\end{equation}
}\\
Puisque $\Delta\,(u-u_0)=0$, il s'ensuit que: $u=u_0+c$, o\`u $c$ est une constante. La valeur de $c$ est déterminée par le fait que la fonction $A-e^{u_0+c}h$ a une moyenne nulle. Par conséquent, on a:
\begin{equation}
c=-\log(\frac{1}{A}\int_{\Omega}e^{u_0}h) ,
\end{equation}
ce qui prouve la relation~(\ref{constante-a-trouver}). Maintenant, on remarque que:
\begin{equation}
\exp(-\Vert u_0\Vert_{L^{\infty}})h\leq e^{u_0}h\leq \exp(\Vert u_0\Vert_{L^{\infty}})h .
\end{equation}
En intégrant sur $\Omega$, on obtient:
\begin{equation}
\exp(-\Vert u_0\Vert_{L^{\infty}})\frac{B}{A}\leq \frac{1}{A}\int_{\Omega}e^{u_0}h\leq \exp(\Vert u_0\Vert_{L^{\infty}})\frac{B}{A} .
\end{equation}
En prenant les logarithmes, on obtient
\begin{equation}
|c|\leq \Vert u_0\Vert_{L^{\infty}} +\vert \log(\frac{A}{B})\vert .
\end{equation}
Puisque $u=u_0+c$, cela prouve la relation~(\ref{inegalite-infini}).\\
{\bf $4^{eme}$ étape :} {\it Preuve de la proposition .}\\
On déduit des étapes $2$ et $3$ que toute solution $(h,u)$ de~(\ref{equation-Kazdan-Warner}), avec $h\not=0$, vérifie:
\begin{equation}\label{lemme-pour-lisse}
\begin{array}{rcl}
\Vert u\Vert_{L^{\infty}}&\leq&\vert \log(\frac{A}{B})\vert+2\Vert u_0\Vert_{L^{\infty}}\\
&\leq&\vert \log(\frac{A}{B})\vert+2[c_0A+\frac{c_0At}{B}\exp(4\frac{c_0At}{B})]\\
&\leq&\phi_A(t,B) .
\end{array}
\end{equation}
On a utilisé l'inégalité (\ref{conclu-troisieme-etape}). Cela prouve le lemme pour les solutions $C^{\infty}$.\\
Supposons maintenant par l'absurde qu'il existe un couple $(h_0,u_0)\in{\cal U}$ tel que:
\begin{equation}
\Vert u_0\Vert_{L^{\infty}}>\phi_A(\Vert h_0\Vert_{L^{2}},\int_{\Omega}h_0) .
\end{equation}
Considérons la projection $\pi:{\cal U}\mapsto {\cal H}$ et l'inverse local
\begin{equation}
\left\lbrace\begin{array}{rcl}
{\cal H}&\mapsto&H^{n+2}(\Tore{\cal L})\\
h&\mapsto&u_{h}
\end{array}\right .
\end{equation}
qui fait correspondre à tout élément $h$ suffisamment proche de $h_0$ l'unique solution $u=u_h$ de l'équation~(\ref{equation-Kazdan-Warner}) proche de $u_0$. Cette application est continue et donc 
\begin{equation}\label{contradiction-proche}
\Vert u_h\Vert_{L^{\infty}}>\phi_A(\Vert h\Vert_{L^{2}},\int_{\Omega}h)
\end{equation}
pour $h$ suffisamment proche de $h_0$. Prenons une fonction $h$ proche de $h_0$. Alors, par la régularité elliptique, la fonction $u$ est aussi $C^{\infty}$ donc l'inégalité~(\ref{contradiction-proche}) contredit l'inégalité~(\ref{lemme-pour-lisse}).$\hfill{\bf CQFD}$\\
\\
{\it Preuve du théorème.} Considérons la projection
\begin{equation}
\pi:{\cal U}\mapsto {\cal H}
\end{equation}
du lemme~\ref{variete-banach}. Le lemme~\ref{lemme-majoration-a-priori} montre que l'ensemble $\pi^{-1}(h)$ des solutions $u\in H^{n+2}(\Tore{\cal L})$ de l'équation~(\ref{equation-Kazdan-Warner}) est compact.\\
En effet si $u_{\nu}\in \pi^{-1}(h)$ est une suite, elle vérifie une estimée uniforme
\begin{equation}
\sup_{\nu}\Vert u_{\nu}\Vert_{L^{\infty}}\leq c_1 .
\end{equation}
Puisque $-\Delta\,u_{\nu}=A-e^{u_{\nu}}h$ les estimées elliptiques dans les espaces $L^2$ (cf~(\ref{crave-for-it-too})) donnent
\begin{equation}
\sup_{\nu}\Vert u_{\nu}\Vert_{H^{n+2}}\leq c_2 .
\end{equation}
Le théorème~\ref{sobolev-imbedding} nous dit que l'injection de Sobolev $H^{n+2}(\Tore{\cal L})\hookrightarrow C^0(\Tore{\cal L})$ est compacte. Par conséquent il existe une sous suite $u_{\psi(\nu)}$ qui converge en norme $L^{\infty}$. Puisque $u_{\psi(\nu)}$ est solution de l'équation~(\ref{equation-Kazdan-Warner}), on obtient par le théorème \ref{crave-for-it-too} la convergence  $u_{\psi(\nu)}$ en norme $H^{n+2}$. Cela prouve la compacité.\\
Le lemme~\ref{variete-banach} nous indique que les points de $\pi^{-1}(h)$ sont isolés pour la norme $H^{n+2}$; cela montre que $\pi^{-1}(h)$ est un ensemble fini, pour tout $h\in{\cal H}$.\\
Cela implique que la projection $\pi$ est un rev\^etement d'espace topologique (voir par exemple \cite{spe-M'-dolbeault}, p.~102). On a donc la relation
\begin{equation}
\mbox{card}\, \pi^{-1}(h_0)=\mbox{card}\, \pi^{-1}(h_1)
\end{equation}
pour tout $h_0,h_1\in{\cal H}$. Cela montre que le nombre de points dans $\pi^{-1}(h)$ est indépendant de $h\in{\cal H}$. Montrons que ce nombre est égal à $1$, en utilisant la fonction $h(x)=1$. L'équation
\begin{equation}\label{Kazdan-cas-particulier}
-\Delta\,u+e^u=A
\end{equation}
possède la solution évidente $u(x)=\log\,A$. Montrons que c'est la seule solution.\\
Soit donc $u$ une solution de l'équation~(\ref{Kazdan-cas-particulier}) et soit $x_0$ un point o\`u $u$ atteint son maximum. En ce point, on a la relation:
\begin{equation}
0\leq -\Delta\,u(x_0)=A-e^{u(x_0)} .
\end{equation}
On a donc $e^{u(x_0)}\leq A$ et donc:
\begin{equation}
e^{u(x)}\leq A,
\end{equation}
pour tout $x\in\Omega$. Par ailleurs l'équation~(\ref{Kazdan-cas-particulier}) implique:
\begin{equation}
\int_{\Omega}(A-e^u)=0
\end{equation}
et ceci n'est possible que si $e^{u(x)}=A$. $\hfill{\bf CQFD}$\\
\\
Dans la section suivante on va ramener l'équation $A_+=0$ à l'équation de Kazdan-Warner.

\subsection{Résolution de $A_+=0$ et conséquences}\label{consequence-Dietmar-Salamon}
\noindent Dans cette section on trouve toutes les sections qui minimisent la fonctionnelle $A_+$. Il existe un étude similaire dans \cite{almogII} pour un problème rectangulaire.\\
Rappelons le théorème~\ref{Bochner-Kodaira-Nakano}: si $(\phi, \overrightarrow{a})\in{\cal A}$ alors 
\begin{equation}\label{BKN}
F_{\lambda,\frac{1}{\sqrt{2}}}(\phi, \overrightarrow{a})=\lambda\pi-\pi^2+A_{+}(\phi, \overrightarrow{a}) .
\end{equation}
L'expression de $A_+$ est 
\begin{equation}\label{fonctionnelle-critique-2-ecriture-2}
A_{+}(\phi, \overrightarrow{a})=\int_{\Omega}\frac{1}{2}\vert D_{+}\phi\vert^2+\frac{1}{4}|2\pi+\rot\,\overrightarrow{a}-(\lambda-|\phi|^2)|^2,
\end{equation}
où
\begin{equation}
D_{+}\phi=2L_{+}\phi+(a_{y}-ia_{x})\phi=\frac{\partial\phi}{\partial x}+i\frac{\partial\phi}{\partial y}+A_{y}\phi-iA_{x}\phi.
\end{equation}
Dans le théorème suivant, $\phi_0$ est l'état fondamental de l'opérateur de Schrödinger magnétique $H$ défini à la section~\ref{sec:operateur-lineaire}. La fonction $\phi_0$ appartient aussi au noyau de l'opérateur $L_+$ (cf le lemme \ref{lemme-de-non-degenerescence}). On note \index{$z_0$}$z_0$ le zéro de $\phi_0$ dans $\Omega$ qui est unique par la proposition \ref{unique-zero}.\\
\begin{definition}.
Si $h=h_x+ih_y\in\C$, on définit\index{$\phi_h$} :
\begin{equation}
\phi_h(x,y)=e^{i\pi(h_y x-h_x y)}\phi_0(z-h) .
\end{equation}
\end{definition}
\begin{proposition}.
On a les propriétés suivantes:
\begin{equation}\label{propriete-phi_h}
\left\lbrace\begin{array}{l}
\phi_h\in C^{\infty}(E_1)\\
\phi_h(z_0+h)=0\\
\phi_h^{-1}(\{0\})=z_0+h+{\cal L}\\
%\phi_h(z)=0\Leftrightarrow\exists l\in{\cal L},\mbox{~tel~que~}z=z_0+h+l\\
\phi_{h=0}=\phi_0\\
L_+(\phi_h)=\pi h\phi_h
\end{array}\right.
\end{equation}
Par ailleurs si $v=n_1r+n_2(w+iu)\in{\cal L}$ alors il existe un complexe $D$ de module $1$ tel que 
\begin{equation}
\phi_{h+v}(z)=De^{2i\pi([n_2u]x-[n_1r+n_2w]y)}\phi_h(z), 
\end{equation}
la fonction $e^{2i\pi([n_2u]x-[n_1r+n_2w]y)}$ étant ${\cal L}$-périodique.
\end{proposition}
{\it Preuve.} Montrons tout d'abord que $\phi_h$ est bien une section de $E_1$; on a :
\begin{equation}
\begin{array}{rcl}
\phi_{h}(z+r)
&=&\D e^{i\pi(h_y (x+r)-h_x y)}\phi_0(z+r-h)\\
&=&\D e^{i\pi h_y r}e^{i\pi(h_y x-h_x y)}e^{i\pi r(y-h_y)}\phi_0(z-h)\\
&=&\D e^{i\pi (h_y x-h_x y)}e^{i\pi ry}\phi_0(z-h)\\
&=&\D e^{i\pi ry}\phi_h(z),\\[2mm]
\phi_{h}(z+w+iu)
&=&\D e^{i\pi(h_y (x+w)-h_x (y+u))}\phi_0(z+w+iu-h)\\
&=&\D e^{i\pi(h_y w-h_x u)}e^{i\pi(h_y x-h_x y)}e^{i\pi [w(y-h_y)-u(x-h_x)]}\phi_0(z-h)\\
&=&\D e^{i\pi(h_y x-h_x y)}e^{i\pi [wy-ux]}\phi_0(z-h)\\
&=&\D e^{i\pi [wy-ux]}\phi_h(z) .
\end{array}
\end{equation}
L'ensemble des zéros de $\phi_0$ est $z_0+{\cal L}$ par la proposition \ref{unique-zero}. Puisque l'exponentielle ne s'annule pas, l'ensemble des zéros de $\phi_h$ est exactement $z_0+h+{\cal L}$ (c'est pour cela que $\phi_h$ a été construit).\\
La section $\phi_0$ vérifie l'équation
\begin{equation}
L_+(\phi_0)(z)=\frac{\partial \phi_0}{\partial\overline{z}}(z)+\frac{\pi}{2}z\phi_0(z)=0 .
\end{equation}
Par conséquent, on obtient
\begin{equation}
\begin{array}{rcl}
L_+(\phi_h)(z)
&=&\D\frac{\partial e^{i\pi(h_y x-h_x y)}\phi_0(z-h)}{\partial\overline{z}}+\frac{\pi}{2}ze^{i\pi(h_y x-h_x y)}\phi_0(z-h)\\
&=&\D\frac{\pi}{2}ih_y\phi_h+\frac{i\pi}{2}(-ih_x)\phi_h+e^{i\pi(h_y x-h_xy)}\frac{\partial \phi_0}{\partial\overline{z}}(z-h)+\frac{\pi}{2}z\phi_h\\
&=&\D\frac{\pi}{2}h\phi_h+e^{i\pi(h_y x-h_xy)}\frac{-\pi}{2}(z-h)\phi_0(z-h)+\frac{\pi}{2}z\phi_h\\
&=&\D\frac{\pi}{2}h\phi_h+\frac{-\pi}{2}(z-h)\phi_h+\frac{\pi}{2}z\phi_h\\
&=&\D\frac{\pi}{2}h\phi_h+\frac{\pi}{2}h\phi_h\\
&=&\pi h\phi_h .
\end{array}
\end{equation}
%Si $h=0$ on a facilement 
%\begin{equation}
%\begin{array}{rcl}
%\phi_{h=0}(z)
%&=&e^{i\pi(0*x-0*y)}\phi_0(z-0)\\
%&=&\phi_0(z) .
%\end{array}
%\end{equation}
On calcule en utilisant les relations précédentes
\begin{equation}
\begin{array}{rcl}
\phi_{h+v}(z)&=&\D e^{i\pi([h_y+n_2u]x-[h_x+n_1r+n_2w]y)}\phi_0(z-h-v)\\
&=&\D e^{i\pi(h_y x-y h_x)}e^{i\pi([n_2u]x-[n_1r+n_2w]y)}\phi_0(z-h-n_1r-n_2(w+iu))\\
&=&\D e^{i\pi(h_y x-y h_x)}e^{i\pi([n_2u]x-[n_1r+n_2w]y)}\\
&&\D e^{-i\pi n_1 r(y-h_y)-i\pi n_2[w(x-h_x)-u(y-h_y)]}\phi_0(z-h)\\
&=&\D e^{2i\pi([n_2u]x-[n_1r+n_2w]y)}e^{i\pi n_1 rh_y-i\pi n_2[-wh_x+uh_y]}\phi_h(z)\\
&=&\D De^{2i\pi([n_2u]x-[n_1r+n_2w]y)}\phi_h(z)
\end{array}
\end{equation}
avec $D=e^{i\pi n_1 rh_y-i\pi n_2[-wh_x+uh_y]}$. $\hfill{\bf CQFD}$

\begin{theorem}.\label{piratage-salamon}
On suppose $\lambda>2\pi$.\\
Les solutions dans ${\cal A}$ de l'équation $A_+(\phi,\overrightarrow{a})=0$ sont de la forme
\begin{equation}\label{forme-speciale-sol-BKN}
\left\lbrace\begin{array}{l}
\phi=\phi_{0}e^{ic}e^{f}\\
\overrightarrow{a}=(\frac{\partial f}{\partial y},-\frac{\partial f}{\partial x})
\end{array}\right.
\end{equation}
o\`u $f$ est la solution régulière de l'équation
\begin{equation}
0=-\Delta f-(\lambda-2\pi)+|\phi_{0}|^2e^{2f}
\end{equation}
et $c\in\R$ .
\end{theorem}
{\it Preuve.}\\
{\bf $1^{ere}$ étape :} {\it On spécifie ici les notations, les équations à résoudre et la régularité.}\\
Soit $(\phi, \overrightarrow{a})$ un couple vérifiant $A_+(\phi, \overrightarrow{a})=0$.\\
Résoudre l'équation $A_+=0$ équivaut compte tenu de (\ref{fonctionnelle-critique-2-ecriture-2}) à résoudre le système d'équation
\begin{equation}\label{equation-bogolomyi}
\left\lbrace\begin{array}{l}
D_{+}\phi=0, \\
2\pi+\rot\,\overrightarrow{a}=\lambda-|\phi|^2 .
\end{array}\right.
\end{equation}
Puisque $F_{\lambda,\frac{1}{\sqrt{2}}}(\phi, \overrightarrow{a})=\lambda\pi-\pi^2+A_+(\phi, \overrightarrow{a})$, le couple $(\phi, \overrightarrow{a})$ minimise la fonctionnelle $F_{\lambda,\frac{1}{\sqrt{2}}}$. Par le théorème \ref{regularite}, le couple $(\phi,\overrightarrow{a})$ est $C^{\infty}$.\\
Par le théorème \ref{annulation-section}, la section $\phi$ admet un zéro que l'on écrit sous la forme $z_h=h+z_0$.\\
La fonction $\phi$ définie sur $\C$ s'annule donc au moins sur l'ensemble $z_h+{\cal L}$ tandis que la fonction $\phi_h$ s'annule exactement sur l'ensemble $z_h+{\cal L}$ (cf~(\ref{propriete-phi_h})).\\
{\bf $2^{eme}$ étape :} {\it On réécrit les équations avec $\phi_h$.}\\
On pose 
\begin{equation}
f=\frac{\phi}{\phi_h} .
\end{equation}
La fonction \index{$f$}$f$ est défini sur $\R^2\diagdown \{z_h+{\cal L}\}$. Puisque $\phi,\phi_h\in C^{\infty}(E_1)$, le quotient $f$ est une fonction ${\cal L}$-périodique. L'équation $D_+(\phi)=0$ se réécrit:
\begin{equation}
\begin{array}{rcl}
0&=&2L_+(f\phi_h)+(a_y-ia_x)f\phi_h\\
&=&2[\frac{\partial f}{\partial\overline{z}}\phi_h+\pi hf\phi_h]+(a_y-ia_x)f\phi_h\\
&=&[2[\frac{\partial f}{\partial\overline{z}}+\pi h f]+(a_y-ia_x)f]\phi_h, 
\end{array}
\end{equation}
cette équation étant vérifiée sur $\R^2\diagdown \{z_h+{\cal L}\}$.\\
Puisque $\phi_h$ ne s'annule pas sur $\R^2\diagdown \{z_h+{\cal L}\}$ qui est l'ensemble de définition de $f$, on peut simplifier l'équation sur cet ensemble:
\begin{equation}\label{equation-z-barre}
0=2[\frac{\partial f}{\partial\overline{z}}+\pi hf]+(a_y-ia_x)f .
\end{equation}
On pose \index{$w$}$w=\frac{(-a_y-2\pi h_x)+i(a_x-2\pi h_y)}{2}$; la fonction $w$ est $C^{\infty}$, ${\cal L}$-périodique et est définie sur $\R^2$.\\
On a alors
\begin{equation}
\begin{array}{rcl}
\frac{\partial f}{\partial\overline{z}}&=&-f\pi (h_x+ih_y)-\frac{(a_y-ia_x)}{2}f\\
&=&f[\frac{(-a_y-2\pi h_x)+i(a_x-2\pi h_y)}{2}]\\
&=&fw .
\end{array}
\end{equation}
{\bf $3^{eme}$ étape :} {\it On étend la fonction $f$ à $\R^2$.}\\
Par le lemme du $d''$ (qui concerne la résolution des équations $\frac{\partial u}{\partial\overline{z}}=f$ voir le texte \cite{spe-M'-dolbeault}, p.~46), il existe une fonction $k\in C^{\infty}(\R^2)$ telle que :
\begin{equation}
w=\frac{\partial k}{\partial\overline{z}} .
\end{equation}
On pose $g=fe^{-k}$ et on a 
\begin{equation}
\begin{array}{rcl}
\frac{\partial g}{\partial\overline{z}}&=&\frac{\partial fe^{-k}}{\partial\overline{z}}\\
&=&\frac{\partial f}{\partial\overline{z}}e^{-k}-\frac{\partial k}{\partial\overline{z}}e^{-k}f\\
&=&wfe^{-k}-we^{-k}f\\
&=&0 .
\end{array}
\end{equation}
Par conséquent, la fonction $g$ est analytique sur $\R^2$ privé de $z_h+{\cal L}$.\\
Si $m\in z_h+{\cal L}$ alors $\phi(m)=0$ et puisque $\phi$ est $C^{\infty}$, on a $\phi=O(z-m)$. La fonction $\phi_h$ s'annule en $m$, mais c'est un zéro simple (cf~\ref{unique-zero}) par conséquent $\phi_h^{-1}=O(|z-m|^{-1})$ et on a $f=O(1)$ au voisinage de $m$ donc $g=O(1)$.\\
Par un résultat classique sur les développements en série de Laurent, on conclut que $g$ se prolonge en une fonction analytique en $m$. On peut donc prolonger $g$ en une fonction analytique sur $\C$.\\
Puisque $f=ge^k$, la fonction $f$ s'étend à $\R^2$.\\
{\bf $4^{eme}$ étape :} {\it On montre que $g$ ne s'annule pas.}\\
La fonction $g$ est entière, elle possède donc des zéros discrets. Il existe $z_l$ tel que le parallélogramme, dont les sommets sont:
\begin{equation}
z_l,\,\,z_l+r,\,\,z_l+r+(w+iu)\mbox{~et~}z_l+(w+iu),
\end{equation}
ne rencontre aucun zéro de $g$.\\
Le parallélogramme est appellé \index{$\partial {\cal P}$}$\partial {\cal P}$; l'intérieur du parallélogramme est appellé \index{${\cal P}$}${\cal P}$.\\
Le nombre $n$ de zéros à l'intérieur du parallélogramme ${\cal P}$ est égal à :
\begin{equation}
\begin{array}{rcl}
n&=&\frac{1}{2\pi i}\int_{\partial {\cal P}}\frac{\frac{\partial g}{\partial z}(z)}{g(z)}dz\\
&=&\frac{1}{2\pi i}\int_{\partial {\cal P}}\frac{\frac{\partial f}{\partial z}(z)}{f(z)}dz\\
&+&\frac{1}{2\pi i}\int_{\partial {\cal P}}-\frac{\partial k}{\partial z}dz\\[2mm]
&=&\frac{1}{2\pi i}\int_{\partial {\cal P}}-\frac{\partial k}{\partial z}dz .
\end{array}
\end{equation}
L'intégrale de $\frac{\frac{\partial f}{\partial z}(z)}{f(z)}$ sur le bord $\partial {\cal P}$ est nulle car $f$ est une fonction ${\cal L}$-périodique.\\
Maintenant, si $\omega$ est une $1$-forme différentielle, alors:
\begin{equation}
\int_{\partial {\cal P}}\omega=\int_{\cal P} d\omega .
\end{equation}
Par conséquent, on obtient :
\begin{equation}
\begin{array}{rcl}
n&=&\D\frac{1}{2\pi i}\int_{\partial {\cal P}}-\frac{\partial k}{\partial z}dz\\
&=&\D\frac{1}{2\pi i}\int_{{\cal P}}d(-\frac{\partial k}{\partial z}dz)\\
&=&\D\frac{1}{2\pi i}\int_{{\cal P}}-\frac{\partial }{\partial \overline{z}}\frac{\partial k}{\partial z} d\overline{z}\wedge dz\\
&=&\D\frac{1}{2\pi i}\int_{{\cal P}}-\frac{\partial }{\partial z}\frac{\partial k}{\partial \overline{z}} d\overline{z}\wedge dz\\
&=&\D\frac{1}{2\pi i}\int_{{\cal P}}-\frac{\partial w}{\partial z} d\overline{z}\wedge dz .
\end{array}
\end{equation}
La fonction $w$ est une fonction ${\cal L}$-périodique; par conséquent la fonction $\frac{\partial w}{\partial z}$ est une fonction ${\cal L}$-périodique d'intégrale nulle sur ${\cal P}$. La précédente intégrale est donc nulle.\\
On a donc $n=0$ et la fonction $g$ ne s'annule pas sur $\R^2$.\\
{\bf $5^{eme}$ étape :} {\it On obtient l'écriture $\phi=\phi_{h_2}e^{v_2}$ avec $v_2$ fonction ${\cal L}$-périodique.}\\
Puisque $f=ge^{k}$, la fonction $f$ ne s'annule pas sur $\R^2$. Puisque $\R^2$ est simplement connexe, il existe une fonction $C^{\infty}$ à valeur complexe $u$ telle que: $f=e^{u}$.\\
La fonction $u$ a priori n'est pas périodique sur $\R^2$; cependant la fonction $f$ est périodique et $C^{\infty}$; on a donc :
\begin{equation}
\left\lbrace\begin{array}{l}
e^{u(x+r,y)-u(x,y)}=1\\
e^{u(x+w,y+u)-u(x,y)}=1 .
\end{array}\right.
\end{equation}
Par conséquent, il existe deux entiers relatifs $n_1$ et $n_2$ tels que :
\begin{equation}
\left\lbrace\begin{array}{l}
u(x+r,y)=u(x,y)+2\pi i n_1\\
u(x+w,y+u)=u(x,y)+2\pi i n_2 .
\end{array}\right.
\end{equation}
La fonction
\begin{equation}
v(z)=u(z)-2\pi i[n_1u x+(n_2r-n_1w)y]
\end{equation}
est ${\cal L}$-périodique. On a donc:
\begin{equation}
\begin{array}{rcl}
\phi(z)
&=&f(z)\phi_h(z)\\
&=&e^{v(z)+2\pi i[n_1u x+(n_2r-n_1w)y]}\phi_h(z)\\
&=&e^{v(z)}e^{2\pi i[n_1u x-(n_1w-n_2r)y]}\phi_h(z)\\
&=&D^{-1}e^{v(z)}\phi_{h-n_2r+n_1(w+iu)}(z)
\end{array}
\end{equation}
avec $|D|=1$. Il existe donc $d\in\C$ tel que $D=e^{d}$; on a donc:
\begin{equation}
\phi(z)=e^{v(z)-d}\phi_{h-n_2r+n_1(w+iu)}(z)
\end{equation}
On pose: \index{$h_2$}$h_2=h-n_2r+n_1(w+iu)$ et \index{$v_2$}$v_2=v-d$, et on réécrit $\phi$ sous la forme:
\begin{equation}
\phi(z)=e^{v_2(z)}\phi_{h_2}(z)
\end{equation}
o\`u $v_2$ est une fonction $C^{\infty}$, ${\cal L}$-périodique.\\
{\bf $6^{eme}$ étape :} {\it On utilise l'expression précédente pour $\phi$ et on obtient la réécriture suivante des équations (\ref{equation-bogolomyi}):}
\begin{equation}
\left\lbrace\begin{array}{rcl}
\frac{\partial v_2}{\partial \overline{z}}&=&\frac{(-a_y-\pi h_{2,x})+i(a_x-\pi h_{2,y})}{2}\\
0&=&2\pi-\lambda+|\phi_{h_2}|^2e^{2\Rez\,v_2}+\rot\,\overrightarrow{a} .
\end{array}\right.
\end{equation}
On a les relations :
\begin{equation}
\frac{\partial v_2}{\partial x}+i\frac{\partial v_2}{\partial y}=(-a_y-\pi h_{2,x})+i(a_x-\pi h_{2,y}) .
\end{equation}
On sépare la partie imaginaire et la partie réelle dans l'équation précédente:
\begin{equation}
\left\lbrace\begin{array}{rcl}
-\frac{\partial \Imz\,v_2}{\partial y}+\frac{\partial \Rez\,v_2}{\partial x}&=&-a_y-\pi h_{2,x}\\
\frac{\partial \Imz\,v_2}{\partial x}+\frac{\partial \Rez\,v_2}{\partial y}&=&a_x-\pi h_{2,y} ,
\end{array}\right.
\end{equation}
qui nous donne l'expression du potentiel vecteur:
\begin{equation}\label{expression-pour-a}
\left\lbrace\begin{array}{rcl}
a_y&=&-\pi h_{2,x}-\frac{\partial \Rez\,v_2}{\partial x}+\frac{\partial \Imz\,v_2}{\partial y}\\
a_x&=&\pi h_{2,y}+\frac{\partial \Rez\,v_2}{\partial y}+\frac{\partial \Imz\,v_2}{\partial x} .
\end{array}\right.
\end{equation}
{\bf $7^{eme}$ étape :} {\it On précise la forme de $v_2$.}\\
Par hypothèse $(\phi,\overrightarrow{a})\in{\cal A}$ et donc $\divergence\,\overrightarrow{a}=0$; calculons cette divergence avec l'expression précédente:
\begin{equation}
\begin{array}{rcl}
0&=&\divergence\,\overrightarrow{a}\\
&=&\frac{\partial a_x}{\partial x}+\frac{\partial a_y}{\partial y}\\
&=&\frac{\partial \frac{\partial \Rez\,v_2}{\partial y}+\frac{\partial \Imz\,v_2}{\partial x}}{\partial x}+\frac{\partial -\frac{\partial \Rez\,v_2}{\partial x}+\frac{\partial \Imz\,v_2}{\partial y}}{\partial y}\\
&=&\frac{\partial^2 \Rez\,v_2}{\partial x\partial y}-\frac{\partial^2 \Rez\,v_2}{\partial y\partial x}+\frac{\partial^2 \Imz\,v_2}{\partial x^2}+\frac{\partial^2 \Imz\,v_2}{\partial y^2}\\
&=&\Delta\,\Imz\,v_2 .
\end{array}
\end{equation}
Puisque le noyau de $\Delta$ est constitué des fonctions constantes, la partie imaginaire de $v_2$ est constante égale à $c$. On pose\index{$\psi$} :
\begin{equation}
\psi=\Rez\,v_2
\end{equation}
et on a donc $v_2=\psi+ic$.\\
{\bf $8^{eme}$ étape :} {\it On montre que $\phi$ s'annule en $z_0$.}\\
Les équations (\ref{expression-pour-a}) se réduisent donc à:
\begin{equation}\label{expression-pour-a-Dietmaer-salamon}
\begin{array}{rcl}
a_y&=&-\pi h_{2,x}-\frac{\partial \psi}{\partial x}\\
a_x&=&\pi h_{2,y}+\frac{\partial \psi}{\partial y} .
\end{array}
\end{equation}
Puisque le couple $(\phi,\overrightarrow{a})$ appartient à ${\cal A}$, l'intégrale du potentiel vecteur $\overrightarrow{a}$ sur $\Omega$ est nulle. Puisque $\psi$ est une fonction $C^{\infty}$, ${\cal L}$-périodique sur $\R^2$, l'intégrale sur $\Omega$ des fonctions $\frac{\partial \psi}{\partial x}$ et $\frac{\partial \psi}{\partial y}$ est aussi nulle.\\
Cela ne peut se faire que si $( h_{2,x},h_{2,y})=(0,0)$. On a donc $\phi_{h_2}=\phi_0$ et $z_{h_2}=z_0$.\\
{\bf $9^{eme}$ étape :} {\it On précise la forme de $\psi$.}\\
Exprimons $\rot\,\overrightarrow{a}$ avec la formule (\ref{expression-pour-a-Dietmaer-salamon}):
\begin{equation}\label{calcul-rotationnel-a}
\begin{array}{rcl}
\rot\,\overrightarrow{a}&=&\frac{\partial a_y}{\partial x}-\frac{\partial a_x}{\partial y}\\
&=&-\frac{\partial^2 \psi}{\partial x^2}-\frac{\partial^2 \psi}{\partial y^2}\\
&=&-\Delta\,\psi .
\end{array}
\end{equation}
La fonction $\psi$ est alors solution de l'équation 
\begin{equation}\label{equation-bogolomnyi-plaitpas}
0=2\pi-\lambda+|\phi_0|^2e^{2\psi}-\Delta\,\psi .
\end{equation}
La fonction $\eta=2\psi$ est alors solution de l'équation
\begin{equation}\label{equation-bogolomnyi-2}
0=2(2\pi-\lambda)+2|\phi_0|^2e^{\eta}-\Delta\,\eta .
\end{equation}
On sait par le théorème \ref{Kazdan-Warner-Salamon-Auroux} que l'équation (\ref{equation-bogolomnyi-2}) possède une unique solution. L'équation (\ref{equation-bogolomnyi-plaitpas}) possède donc aussi une unique solution que l'on note \index{$f_s$}$f_s$.\\
On donc bien les solutions annoncées. $\hfill{\bf CQFD}$\\
\\
Pour la suite, on note\index{$\phi_s$}\index{$\overrightarrow{a_s}$}
\begin{equation}\label{definition-solution-dirac}
\begin{array}{rcl}
\phi_s&=&\phi_0e^{f_s}\\
\overrightarrow{a_s}&=&(\frac{\partial f_s}{\partial y},-\frac{\partial f_s}{\partial y}),
\end{array}
\end{equation}
o\`u $f_s$ est l'unique solution de 
\begin{equation}\label{equation-bogolomnyi}
\lambda-2\pi=|\phi_0|^2e^{2\psi}-\Delta\,\psi .
\end{equation}

\begin{corollaire}.\label{calculs-energie}
La fonction $m_F(\lambda,\frac{1}{\sqrt{2}})=\inf_{(\phi, \overrightarrow{a})\in{\cal A}}F_{\lambda,k}(\phi, \overrightarrow{a})$ est égale à:
\begin{equation}
\left\lbrace\begin{array}{lcl}
\frac{\lambda^2}{4}\mbox{~si~}\lambda\leq 2\pi\\
\lambda\pi-\pi^2\mbox{~si~}\lambda>2\pi .
\end{array}\right.
\end{equation}
\end{corollaire}
{\it Preuve.} Le théorème~\ref{merveille-praguoise} nous dit précisément que $m_F(\lambda,k)=\frac{\lambda^2}{4}$ si $\lambda\leq 2\pi$.\\
La formule~(\ref{BKN}) nous indique que $m_F(\lambda,k)\geq \lambda\pi-\pi^2$ car $A_+\geq 0$.\\
Si $\lambda>2\pi$, le théorème~\ref{piratage-salamon} nous indique que:
\begin{equation}
F_{\lambda,k}(\phi_s,\overrightarrow{a_s})=\lambda\pi-\pi^2+A_+(\phi_s,\overrightarrow{a_s})=\lambda\pi-\pi^2 .
\end{equation}
Donc si $\lambda>2\pi$, alors $m_F(\lambda,k)=\lambda\pi-\pi^2$.$\hfill{\bf CQFD}$

\begin{proposition}.\label{sol-BKN-sont-bifurques}
Si $k=\frac{1}{\sqrt{2}}$, il existe $\epsilon_0>0$ tel que, pour $\lambda\in]2\pi,2\pi+\epsilon_0]$, le couple bifurqué $(\phi_+,\overrightarrow{a}_+)$ soit de la forme (\ref{forme-speciale-sol-BKN}).
\end{proposition}
{\it Preuve.} Il nous suffit de démontrer que le couple 
\begin{equation}
(\phi_0e^{\psi},(\frac{\partial \psi}{\partial y},-\frac{\partial \psi}{\partial x}))
\end{equation}
o\`u $\psi$ est la solution régulière de l'équation 
\begin{equation}\label{equation-psi}
\lambda-2\pi=|\phi_0|^2e^{2\psi}-\Delta\,\psi
\end{equation}
possède une norme $H^1$ qui tend vers $0$ quand $\lambda$ tend vers $2\pi$.\\
En effet le théorème \ref{existence-de-la-bifurcation} nous dit que toutes les solutions non triviales dans un voisinage de $(0,0)$ dans ${\cal A}$ des équations (\ref{Odeh-equations}) sont pour $\lambda>2\pi$ proche de $2\pi$ des solutions bifurquées.\\
Le théorème \ref{Kazdan-Warner-Salamon-Auroux} nous indique que la solution $\psi$ de l'équation (\ref{equation-psi}) est $C^{\infty}$ car $|\phi_0|^2\in C^{\infty}(\Tore{\cal L},\R)$ et $C^{\infty}(\Tore{\cal L},\R)=\cap_{n\geq 0} H^n(\Tore{\cal L},\R)$.\\
On pose $\psi=\frac{1}{2}u+\frac{\ln\,(\lambda-2\pi)}{2}$, la fonction $u$ est $C^{\infty}$ et vérifie l'équation 
\begin{equation}
-\Delta\,u+2(\lambda-2\pi)|\phi_0|^2e^{u}=2(\lambda-2\pi) .
\end{equation}
La fonction $u$ est donc solution de l'équation (\ref{equation-Kazdan-Warner}) avec $A=2(\lambda-2\pi)$ et $h=2(\lambda-2\pi)|\phi_0|^2$.\\
La proposition \ref{lemme-majoration-a-priori} nous indique que 
\begin{equation}
\Vert u\Vert_{L^{\infty}}\leq \phi_A(t,B)
\end{equation}
avec $t=\Vert h\Vert_{L^2}$, $B=\int_{\Omega} h$ et $\phi_A$ est défini à l'équation (\ref{definition-phiA}).\\
Dans notre cas, cela donne 
\begin{equation}
\left\lbrace\begin{array}{rcl}
t&=&2(\lambda-2\pi)\Vert |\phi_0|^2\Vert_{L^2}=2(\lambda-2\pi)\sqrt{I}\\
B&=&2(\lambda-2\pi)\\
\phi_A(t,B)&=&4c_0(\lambda-2\pi)+4c_0(\lambda-2\pi)\sqrt{I}\exp\,[8c_0(\lambda-2\pi)\sqrt{I}]\\
&=&O(\lambda-2\pi) .
\end{array}\right.
\end{equation}
La constante $I$ est définie à l'équation (\ref{defcon-I}) et on utilise que $\phi_0$ est par définition un vecteur unitaire.\\
L'expression $\phi_A$ tend vers $0$ quand $\lambda$ tend vers $2\pi$, il en est donc de m\^eme de $\Vert u\Vert_{L^{\infty}}$.\\
La fonction $u$ étant $C^{\infty}$ on peut intégrer librement et on obtient
\begin{equation}
\begin{array}{rcl}
\int_{\Omega}|-\Delta\,u|^2
&=&4(\lambda-2\pi)^2\int_{\Omega}[1-|\phi_0|^2e^{u}]^2\\
&=&2(\lambda-2\pi)^2[1+e^{2O(\Vert u\Vert_{L^{\infty}})}]\\
&=&(\lambda-2\pi)^2O(1) .
\end{array}
\end{equation}
On en conclut que $\Vert \Delta\,u\Vert_{L^{2}}=O(\lambda-2\pi)$. Vu que l'on a aussi un contr\^ole sur la norme $L^{\infty}$, on obtient facilement 
\begin{equation}
\Vert u\Vert_{H^2}=O(\lambda-2\pi)
\end{equation}
La définition (\ref{definition-solution-dirac}) nous permet de dire que
\begin{equation}
\Vert \overrightarrow{a}\Vert_{H^1}=
\Vert(\frac{1}{2}\frac{\partial u}{\partial y},-\frac{1}{2}\frac{\partial u}{\partial x})\Vert_{H^1}
=O(\lambda-2\pi)
\end{equation}
On a donc le contr\^ole sur le potentiel vecteur.\\
L'expression de $\phi$ (cf~(\ref{definition-solution-dirac})) est 
\begin{equation}
\phi_0\sqrt{\lambda-2\pi}e^{\frac{u}{2}}
\end{equation}
et sa norme $H^1$ est trivialement estimée par $O(\sqrt{\lambda-2\pi})$.$\hfill{\bf CQFD}$

\begin{lemme}.\label{lemme-CDI-yohann}
Supposons $k\geq\frac{1}{\sqrt{2}}$; si le couple $(\phi,\overrightarrow{a})$ minimise la fonctionnelle $F_{\lambda,k}$, alors:
\begin{equation}\label{inegalite-merveilleuse-CDI}
\int_{\Omega}|\rot\,\overrightarrow{a}|^2\leq 2(\lambda\pi-2\pi^2) .
\end{equation}
\end{lemme}
{\it Preuve.}\\
{\bf $1^{ere}$ étape :}\\
Montrons le résultat pour $k=\frac{1}{\sqrt{2}}$. Dans ce cas, le couple minimisant $(\phi,\overrightarrow{a})$ est de la forme $(e^{ic}\phi_s,\overrightarrow{a_s})$ avec $c\in\R$. Le potentiel vecteur est donc forcément égal à $\overrightarrow{a_s}$.\\
Si on intègre sur $\Omega$ l'équation~(\ref{equation-bogolomnyi}), on obtient:
\begin{equation}
\int_{\Omega}|\phi_{s}|^2=\lambda-2\pi .
\end{equation}
L'équation~(\ref{equation-bogolomnyi}) nous donne également:
\begin{equation}
|\phi_{s}|^4=(\lambda-2\pi)^2+2(\lambda-2\pi)\Delta\,f_s+|\Delta\,f_s|^2 .
\end{equation}
Par conséquent, on a la relation :
\begin{equation}
\int_{\Omega}|\phi_{s}|^4=(\lambda-2\pi)^2+\int_{\Omega}|\Delta\,f_s|^2 .
\end{equation}
Or on sait par~(\ref{calcul-rotationnel-a}) que: $\rot\,\overrightarrow{a_s}=-\Delta\,f_s$; par conséquent, on a la relation:
\begin{equation}
\begin{array}{rcl}
\int_{\Omega}(\lambda-|\phi_s|^2)^2&=&\lambda^2-2\lambda\int_{\Omega}|\phi_s|^2+\int_{\Omega}|\phi_s|^4\\
&=&\lambda^2-2\lambda(\lambda-2\pi)+(\lambda-2\pi)^2+\int_{\Omega}|\Delta\,f_s|^2\\
&=&(\lambda-(\lambda-2\pi))^2+\int_{\Omega}|\Delta\,f_s|^2\\
&=&(2\pi)^2+\int_{\Omega}|\Delta\,f_s|^2\\
&=&(2\pi)^2+\int_{\Omega}|\rot\,\overrightarrow{a_s}|^2 .
\end{array}
\end{equation}
L'état $(\phi_s,\overrightarrow{a_s})$ défini au théorème~\ref{piratage-salamon} vérifie :
\begin{equation}
\int_{\Omega}\Vert i\overrightarrow{\nabla}\phi_s+(\overrightarrow{A}_0+\overrightarrow{a_s})\phi_s\Vert^2+\frac{1}{4}(\lambda-|\phi_s|^2)^2+\frac{1}{4}\int_{\Omega}|\rot\,\overrightarrow{a_s}|^2=\lambda\pi-\pi^2 .
\end{equation}
Par conséquent, on a l'inégalité:
\begin{equation}
\pi^2+\frac{1}{2}\int_{\Omega}|\rot\,\overrightarrow{a_s}|^2\leq \lambda\pi-\pi^2 ,
\end{equation}
qui implique l'inégalité:
\begin{equation}
\int_{\Omega}|\rot\,\overrightarrow{a_s}|^2\leq 2[\lambda\pi-2\pi^2] .
\end{equation}
{\bf $2^{eme}$ étape :}\\
Il suffit d'appliquer la proposition \ref{majoration-concert-Divine-comedy} pour obtenir le résultat voulu. $\hfill{\bf CQFD}$\\
\\
Si $k\geq \frac{1}{\sqrt{2}}$, la formule~(\ref{BKN}) nous donne des résultats moins précis mais qui permettront néanmoins de décrire la structure qualitative du diagramme des phases.

\begin{theorem}.\label{encadrement-des-minimums}
On suppose $k\geq\frac{1}{\sqrt{2}}$.\\
La fonction $m_F(\lambda,k)=\inf_{(\phi, \overrightarrow{a})\in{\cal A}}F_{\lambda,k}(\phi, \overrightarrow{a})$ vérifie 
\begin{equation}
\left\lbrace\begin{array}{lcl}
m_F(\lambda,k)=\frac{\lambda^2}{4}&\mbox{~si~}&\lambda\leq 2\pi\\
\lambda\pi-\pi^2\leq m_F(\lambda,k)\leq\pi^2+(\lambda\pi-2\pi^2)(k^2+\frac{1}{2})&\mbox{~si~}&\lambda>2\pi .
\end{array}\right.
\end{equation}
\end{theorem}
{\it Preuve.} L'égalité $m_F(\lambda,k)=\frac{\lambda^2}{4}$ si $\lambda\leq 2\pi$ provient du théorème~\ref{merveille-praguoise}.\\
La fonctionnelle $F_{\lambda,k}$ vérifie l'inégalité:
\begin{equation}\label{inegalite-sur-F}
\begin{array}{rcl}
F_{\lambda,k}(\phi, \overrightarrow{a})&=&\lambda\pi-\pi^2+A_+(\phi, \overrightarrow{a})+\frac{1}{2}(k^2-\frac{1}{2})\int_{\Omega}|\rot\,\overrightarrow{a}|^2\\
&\geq&\lambda\pi-\pi^2+\frac{1}{2}(k^2-\frac{1}{2})\int_{\Omega}|\rot\,\overrightarrow{a}|^2\\
&\geq&\lambda\pi-\pi^2 .
\end{array}
\end{equation}
On en déduit l'inégalité: $\lambda\pi-\pi^2\leq m_F(\lambda,k)$.\\
%Le lemme~\ref{lemme-CDI-yohann} implique que 
L'égalité~(\ref{inegalite-sur-F}), combinée avec le lemme~\ref{lemme-CDI-yohann}, nous donne:
\begin{equation}
\begin{array}{rcl}
F_{\lambda,k}(\phi_s,\overrightarrow{a_s})&=&\lambda\pi-\pi^2+\frac{1}{2}(k^2-\frac{1}{2})\int_{\Omega}|\rot\,\overrightarrow{a_s}|^2\\
&\leq&\lambda\pi-\pi^2+(k^2-\frac{1}{2})(\lambda\pi-2\pi^2)\\
&\leq&\pi^2+(\lambda\pi-2\pi^2)[1+(k^2-\frac{1}{2})]\\
&\leq&\pi^2+(\lambda\pi-2\pi^2)(k^2+\frac{1}{2}) .
\end{array}
\end{equation}
Et donc on a l'inégalité: $m_F(\lambda,k)\leq \pi^2+(\lambda\pi-2\pi^2)(k^2+\frac{1}{2})$.$\hfill{\bf CQFD}$\\
\\
La formule suivante va nous permettre d'écrire les développements asymptotiques de l'énergie pour $k\geq \frac{1}{\sqrt{2}}$.

\begin{theorem}.\label{theorem-egalite-somme-clef}
On a la relation suivante entre $K$ et $I$:
\begin{equation}
I-4K=1
\end{equation}
o\`u $K$ et $I$ sont les constantes définies aux équations (\ref{defcon-I}) et (\ref{defcon-K}).
\end{theorem}
{\it Preuve.} Au corollaire~\ref{calculs-energie}, on a calculé explicitement $m_F(\lambda,\frac{1}{\sqrt{2}})$ pour tout $\lambda$. On a obtenu: 
\begin{equation}
m_F(\lambda,\frac{1}{\sqrt{2}})=\left\lbrace\begin{array}{ll}
\frac{\lambda^2}{4}&\mbox{~si~}\lambda<2\pi, \\
\lambda\pi-\pi^2&\mbox{~si~}\lambda\geq 2\pi .
\end{array}\right.
\end{equation}
Par la proposition \ref{sol-BKN-sont-bifurques} les états du théorème \ref{piratage-salamon} sont des états bifurqués pour $\lambda$ proche de $2\pi$. On peut donc utiliser le chapitre III.\\
Ces résultats s'écrivent aussi en utilisant la variable $\epsilon$ définie à l'équation~(\ref{approx-premier-niveau}) et les fonction $F_S$ et $F_N$ défini à l'équation (\ref{definition-FN-FS})
\begin{equation}
\left\lbrace\begin{array}{l}
F_{N}(\epsilon,\frac{1}{\sqrt{2}})=\frac{(2\pi+\epsilon)^2}{4}, \\
F_{S}(\epsilon,\frac{1}{\sqrt{2}})=\epsilon\pi+\pi^2 .
\end{array}\right.
\end{equation}
On obtient alors:
\begin{equation}
(F_{S}-F_{N})(\epsilon,\frac{1}{\sqrt{2}})=\frac{-\epsilon^2}{4} .
\end{equation}
Mais on sait par le théorème~\ref{evolution-de-F} que
\begin{equation}
\lim_{\epsilon\rightarrow 0}\frac{(F_{S}-F_{N})(\epsilon,\frac{1}{\sqrt{2}})}{\epsilon^2}=-\frac{1}{4(I-\frac{2}{k^2}K)}=-\frac{1}{4(I-4K)} .
\end{equation}
En égalisant, on obtient la relation: $I-4K=1$.$\hfill{\bf CQFD}$

\begin{corollaire}.\label{corollaire-condition-hard}
La condition $I-\frac{2}{k^2}K>0$ des théorèmes~\ref{existence-de-la-bifurcation},~\ref{evolution-de-F} et~\ref{stabilite} est vérifiée, si $k\geq\frac{1}{\sqrt{2}}$.
\end{corollaire}
{\it Preuve.} La fonction $k\rightarrow I-\frac{2}{k^2}K$ est une fonction croissante sur $[\frac{1}{\sqrt{2}},\infty[$ et prend la valeur $1$ en $\frac{1}{\sqrt{2}}$; par conséquent, elle est strictement positive sur cet intervalle de définition.$\hfill{\bf CQFD}$

\saction{Le diagramme des phases, l'état normal}\label{sec:diagramme-des-phases}
\noindent Dans cette section, on trouve les valeurs de $(k,H_{ext})$ pour lesquelles ${\cal E}^V_{k,H_{ext}}=\frac{1}{4}$. Physiquement cela correspond à une énergie qui est atteinte par un état normal. On aura alors résolu le problème \ref{quelles-peine-pour-ca}.\\
On définit les ensembles\index{$F_k$}\index{$G_k$}:
\begin{equation}\label{definition-Fk-Gk}
\begin{array}{l}
F_{k}=\lbrace H_{ext}\mbox{~tel~que~}\forall H_{int}\in\R_+^{*}, \frac{1}{\lambda^2}m_F(\lambda,k)+\frac{1}{2}(H_{ext}-H_{int})^2\geq \frac{H_{ext}^2}{2}\rbrace ,\\
G_{k}=\lbrace H_{ext}\mbox{~tel~que~}\forall H_{int}\in\R_+^{*}, \frac{1}{\lambda^2}m_F(\lambda,k)+\frac{1}{2}(H_{ext}-H_{int})^2\geq \frac{1}{4}\rbrace .
\end{array}
\end{equation}
Par rapport à la définition \ref{caracterisation-etat-possibles}, $(k,H_{ext})$ correspond à un état pur si et seulement si $H_{ext}\in F_k$. $(k,H_{ext})$ correspond à un état normal si et seulement si $H_{ext}\in G_k$.\\
\\
Dans la section~\ref{sec:analyse-asymptotique}, on a montré que les couples vérifiant les équations de Ginzburg-Landau sont de norme assez petite si $k$ est assez grand. On montrera ici un résultat un peu différent pour les couples minimisants la fonctionnelle mais dans un domaine de valeur de $k$ plus large.\\
On peut aussi voir ce théorème comme une extension du théorème~\ref{merveille-praguoise}.

\begin{theorem}.\label{mur-1-racine2-franchi}
$\forall \epsilon>0$, $\exists \delta >0$ tel que $\forall (\phi,\overrightarrow{a})$ minimisant la fonctionnelle $F_{\lambda,k}$ avec $\lambda \leq 2\pi+\delta$ et $k\geq\frac{1}{\sqrt{2}}-\delta$, on a :
\begin{equation}
\Vert \phi\Vert _{H^1}+\Vert \overrightarrow{a}\Vert_{H^1} \leq \epsilon .
\end{equation}
\end{theorem}
{\it Preuve.}\\
{\bf $1^{ere}$ étape :}\\
On sait que $F_{\lambda,k}(0,0)=\frac{\lambda^2}{4}$; donc, si $(\phi, \overrightarrow{a})$ minimise $F_{\lambda,k}$, alors $F_{\lambda,k}(\phi, \overrightarrow{a})\leq \frac{\lambda^2}{4}$.\\
On sait par le théorème~\ref{regularite} que le couple $(\phi, \overrightarrow{a})$ est $C^{\infty}$ et par le théorème~\ref{principe-maximum} que $|\phi|\leq\sqrt{\lambda}$.\\
On a alors l'inégalité
\begin{equation}
\frac{k^2}{2}\int_{\Omega}|\rot\,\overrightarrow{a}|^2\leq F_{\lambda,k}(\phi,\overrightarrow{a})\leq \frac{\lambda^2}{4},
\end{equation}
qui nous donne:
\begin{equation}
\int_{\Omega}|\rot\,\overrightarrow{a}|^2\leq \frac{\lambda^2}{2k^2} .
\end{equation}
On obtient donc dans notre cas:
\begin{equation}
\int_{\Omega}|\rot\,\overrightarrow{a}|^2\leq\frac{(2\pi+\delta)^2}{(1-\sqrt{2}\delta)^2} .
\end{equation}
Par conséquent, la norme $H^1$ de $\overrightarrow{a}$ est bornée par une constante ne dépendant que de $\delta$.\\
{\bf $2^{eme}$ étape :} {\it On majore la fonctionnelle $A_+$.}\\
Par le théorème~\ref{Bochner-Kodaira-Nakano}, on a la relation:
\begin{equation}
\lambda\pi-\pi^2+A_+(\phi, \overrightarrow{a})+\frac{1}{2}(k^2-\frac{1}{2})\int_{\Omega}|\rot\,\overrightarrow{a}|^2\leq\frac{\lambda^2}{4}
\end{equation}
que l'on réécrit sous la forme:
\begin{equation}
A_+(\phi, \overrightarrow{a})+\frac{1}{2}(k^2-\frac{1}{2})\int_{\Omega}|\rot\,\overrightarrow{a}|^2\leq\frac{(\lambda-2\pi)^2}{4} .
\end{equation}
Puisque le couple $(\phi,\overrightarrow{a})$ minimise $F_{\lambda,k}$, on a la majoration, si $k\geq\frac{1}{\sqrt{2}}-\delta$:
\begin{equation}
\begin{array}{rcl}
\frac{1}{2}(\frac{1}{2}-k^2)\int_{\Omega}|\rot\,\overrightarrow{a}|^2&\leq&\D\frac{1}{2}(\frac{1}{2}-[\frac{1}{\sqrt{2}}-\delta]^2)\int_{\Omega}|\rot\,\overrightarrow{a}|^2\\
&\leq&\D\frac{\delta}{\sqrt{2}}\frac{(2\pi+\delta)^2}{(1-\sqrt{2}\delta)^2} .
\end{array}
\end{equation}
On obtient donc la majoration:
\begin{equation}
A_+(\phi, \overrightarrow{a})\leq\frac{(\lambda-2\pi)^2}{4}+\frac{\delta}{\sqrt{2}}\frac{(2\pi+\delta)^2}{(1-\sqrt{2}\delta)^2} .
\end{equation}
{\bf $3^{eme}$ étape :} {\it On majore le terme $\rot\,\overrightarrow{a}+|\phi|^2$.}\\
Compte tenu de la définition de $A_+$ (cf~(\ref{fonctionnelle-critique-2-ecriture-2})), on a trivialement 
\begin{equation}
\int_{\Omega}|\rot\overrightarrow{a}+|\phi|^2+2\pi-\lambda|^2\leq(\lambda-2\pi)^2+\frac{4\delta}{\sqrt{2}}\frac{(2\pi+\delta)^2}{(1-\sqrt{2}\delta)^2}.
\end{equation}
Maintenant on développe la somme à l'intérieur de l'intégrale et on obtient
\begin{equation}\label{etape3}
2(2\pi-\lambda)\int_{\Omega}|\phi|^2+\int_{\Omega}|\rot\overrightarrow{a}+|\phi|^2|^2\leq\frac{4\delta}{\sqrt{2}}\frac{(2\pi+\delta)^2}{(1-\sqrt{2}\delta)^2}.
\end{equation}
Puisque le couple $(\phi,\overrightarrow{a})$ minimise $F_{\lambda,k}$ on a $|\phi|\leq \sqrt{\lambda}$. On en déduit, si $\lambda\leq 2\pi+\delta$:
\begin{equation}\label{maj-triviale}
(\lambda-2\pi)\int_{\Omega}|\phi|^2\leq \delta\int_{\Omega}\lambda\leq \delta(2\pi+\delta) .
\end{equation}
Les équations (\ref{maj-triviale}) et (\ref{etape3}) nous donnent la majoration:
\begin{equation}
\int_{\Omega}|\rot\overrightarrow{a}+|\phi|^2|^2\leq\frac{4\delta}{\sqrt{2}}\frac{(2\pi+\delta)^2}{(1-\sqrt{2}\delta)^2} +2\delta(2\pi+\delta).
\end{equation}
On pose\index{$\eta(\delta)$} :
\begin{equation}
\eta(\delta)=\frac{4\delta}{\sqrt{2}}\frac{(2\pi+\delta)^2}{(1-\sqrt{2}\delta)^2} +2\delta(2\pi+\delta).
\end{equation}
{\bf $4^{eme}$ étape :} {\it Obtention du contr\^ole $H^1$ sur $\overrightarrow{a}$:}\\
Le couple $(\phi, \overrightarrow{a})$ est $C^{\infty}$ donc on peut intégrer librement.
\begin{equation}\label{controle-L2-phi}
\begin{array}{rcl}
\int_{\Omega}|\phi|^2&=&\int_{\Omega}\rot\,\overrightarrow{a}+|\phi|^2\\
&\leq&\sqrt{\int_{\Omega}|\rot\,\overrightarrow{a}+|\phi|^2|^2}\\
&\leq&\sqrt{\eta(\delta)} .
\end{array}
\end{equation}
Par conséquent, on a un contrôle sur la norme $L^2$ de $\phi$:
\begin{equation}\label{controle-H1-a}
\begin{array}{rcl}
\int_{\Omega}|\rot\,\overrightarrow{a}|^2&\leq&\Vert \rot\,\overrightarrow{a}\Vert ^2_{L^2}\\
&\leq&(\Vert \rot\,\overrightarrow{a}+|\phi|^2\Vert_{L^2}+\Vert \phi\Vert_{L^4}^2)^2\\
&\leq&(\sqrt{\eta(\delta)}+\sqrt{\int_{\Omega}|\phi|^4})^2\\
&\leq&(\sqrt{\eta(\delta)}+\sqrt{\lambda}\sqrt{\int_{\Omega}|\phi|^2})^2\\
&\leq&(\sqrt{\eta(\delta)}+\sqrt{2\pi+\delta}\sqrt[4]{\eta(\delta)})^2 .
\end{array}
\end{equation}
{\bf $5^{eme}$ étape :} {\it Obtention du contr\^ole $H^1$ sur $\phi$.}\\
L'inéquation (\ref{majorante-best}) que l'on réécrit ci-dessous
\begin{equation}
\begin{array}{rcl}
\Vert (i\overrightarrow{\nabla}+\overrightarrow{A}_{0})\phi\Vert _{L^2}
&\leq&\Vert \overrightarrow{a}\Vert_{L^4}\Vert\phi\Vert_{L^4}+\sqrt{2\Vert \overrightarrow{a}\Vert^2_{L^4}\Vert\phi\Vert^2_{L^4}+\lambda \Vert \phi\Vert ^2_{L^2}}
\end{array}
\end{equation}
est vraie pour notre couple $(\phi,\overrightarrow{a})$ puisqu'il est solution des équations (\ref{Odeh-equations}).\\
La norme $L^2$ de $\phi$ est contr\^olée par l'équation (\ref{controle-L2-phi}). La norme $L^4$ de $\overrightarrow{a}$ est contr\^olée par l'équation (\ref{controle-H1-a}) et le théorème (\ref{sobolev-imbedding}). La norme $L^4$ de $\phi$ est contr\^olée par l'inégalité
\begin{equation}
\begin{array}{rcl}
\Vert\phi\Vert_{L^4}\leq \sqrt[4]{2\pi+\delta}\sqrt{\Vert\phi\Vert_{L^2}}
\end{array}
\end{equation}
et l'équation (\ref{controle-L2-phi}).\\
On a donc bien montré le résultat voulu; il suffit de prendre $\delta$ assez petit. $\hfill{\bf CQFD}$\\
\\
%Le théorème précédent nous permet de montrer que le minimum est atteint par le couple bifurqué dans un certain domaine de $k$. Il faut remarquer que ce théorème peut sans doute \^etre étendu jusqu'a $k=\frac{1}{\sqrt{2}}$
Le théorème suivant dit que si $H_{ext}$ est proche de $k$ et si $(H_{int},\phi,\overrightarrow{a})$ minimise la fonctionnelle $E^V_{k,H_{ext}}$ alors le champ $H_{int}$ est proche de $k$.\\
On rappelle que $\lambda=\frac{2\pi k}{H_{int}}$.

\begin{theorem}.\label{limitation-voyage-Hint}
Soit $k_0>\frac{1}{\sqrt{2}}$ fixé. Alors pour tout $\epsilon>0$, il existe $\eta>0$ tel que, si $H_{ext}\in [k-\eta,k]$ et $k\geq k_0$ et $H_{int}$ est un champ vérifiant ${\cal E}^V_{k,H_{ext}}=G^V_{k,H_{ext}}(H_{int})$ alors $|H_{int}-k|\leq \epsilon$ et $\lambda<2\pi+\epsilon$.
%$H_{ext}\in[k-,k]$ et $k\geq k_0$ alors le minimum de la fonctionnelle $E^V_{k,H_{ext}}$ est atteint par le couple bifurqué $(\phi_+,\overrightarrow{a_+})$.
\end{theorem}
{\it Preuve.}\\
{\bf $1^{ere}$ étape :}\\
On écrit $H_{ext}=k-\delta$; puisque $H_{ext}\leq k$ on a $\delta\geq 0$. Soit $H_{int}$ un champ minimisant la fonction
\begin{equation}
\left\lbrace\begin{array}{rcl}
\R_+&\mapsto&\R_+\\
H_{int}&\mapsto&G^V_{k,H_{ext}}(H_{int})
\end{array}\right.
\end{equation}
qui existe par le théorème \ref{infimum_minimum}.\\
On a alors, puisque $E^V_{k,H_{ext}}(H_{ext},0,0)=\frac{1}{4}$, l'inégalité:
\begin{equation}
G^V_{k,H_{ext}}(H_{int})=\frac{1}{\lambda^2}m_{F}(\lambda,k)+\frac{1}{2}(H_{int}-H_{ext})^2\leq \frac{1}{4} .
\end{equation}
Le théorème \ref{ecrantement} et l'inégalité: $H_{ext}\leq k$ nous donnent: 
\begin{equation}
H_{int}\leq k.
\end{equation}
On écrit maintenant $H_{int}=k-\omega$ avec $\omega\geq 0$.\\
L'inégalité $H_{int}\leq H_{ext}$ nous donne $\omega\geq\delta$.\\
{\bf $2^{eme}$ étape :}\\
L'inégalité $m_F(\lambda,k)\geq \lambda\pi-\pi^2$ du théorème \ref{encadrement-des-minimums} implique :
\begin{equation}
\frac{1}{\lambda^2}[\lambda\pi-\pi^2]+\frac{1}{2}(H_{int}-H_{ext})^2\leq \frac{1}{4} .
\end{equation}
Cela se simplifie en
\begin{equation}\label{equation-Haint-Hext}
\begin{array}{rcl}
\frac{1}{2}(H_{int}-H_{ext})^2
&\leq&\D \frac{(\lambda-2\pi)^2}{4\lambda^2}\\
&\leq&\D \frac{1}{4}(1-\frac{2\pi}{\lambda})^2\\[2mm]
&\leq&\D \frac{1}{4}(1-\frac{H_{int}}{k})^2\mbox{~puisque~}\lambda=\frac{2\pi k}{H_{int}}\\[1mm]
&\leq&\D \frac{1}{4k^2}(k-H_{int})^2 .
\end{array}
\end{equation}
L'inégalité (\ref{equation-Haint-Hext}) se réécrit
\begin{equation}
\frac{(\omega-\delta)^2}{2}\leq \frac{1}{4k^2}\omega^2 .
\end{equation}
En prenant la racine carré et en utilisant le fait que $\omega-\delta\geq 0$ on obtient 
\begin{equation}
\omega-\delta\leq \frac{1}{\sqrt{2}k}\omega
\end{equation}
qui implique alors
\begin{equation}\label{inegalite-cle-de-voute}
[1-\frac{1}{\sqrt{2}k}]\omega\leq \delta .
\end{equation}
{\bf $3^{eme}$ étape :}\\
%Par le théorème \ref{mur-1-racine2-franchi} et le théorème \ref{existence-de-la-bifurcation}, il existe $\epsilon>0$ tel que, si $\lambda<2\pi+\epsilon$, alors le minimum de la fonctionnelle $F_{\lambda,k}$ est atteint par le couple bifurqué.\\
Par l'inégalité (\ref{inegalite-cle-de-voute}) on a
\begin{equation}
\begin{array}{rcl}
\omega
&\leq&\frac{\eta}{[\frac{1}{2}-\frac{1}{4k_0^2}]} .
\end{array}
\end{equation}
Donc on a l'inégalité:
\begin{equation}
\begin{array}{rcl}
\lambda
&=&\D\frac{2\pi k}{H_{int}}\\
&=&\D\frac{2\pi k}{k-\omega}\\
&=&\D\frac{2\pi}{1-\frac{\omega}{k}}\\
&\leq&\D \frac{2\pi}{1-\frac{\eta}{k[\frac{1}{2}-\frac{1}{4k_0^2}]}}\\
&\leq&\D \frac{2\pi}{1-\frac{\eta}{k_0[\frac{1}{2}-\frac{1}{4k_0^2}]}} .
\end{array}
\end{equation}
Par conséquent, en prenant $\eta$ assez petit, on a: $\lambda\leq 2\pi+\epsilon$.  $\hfill{\bf CQFD}$
\begin{remarque}
Cette localisation du problème n'a pas à notre connaissance été effectuée par Chapman dans \cite{pirate-I}.\\
Le résultat suivant est énoncé dans Chapman (\cite{pirate-I}, p.~463) mais la démonstration reste formelle.
\end{remarque}
\begin{theorem}.\label{Bifurcation-Abrikosov-real-nature}(Bifurcation d'Abrikosov)
Pour tout  $k>\frac{1}{\sqrt{2}}$, il existe $u>0$ tel que, si $H_{ext}\in[k-u,k]$, alors le minimum ${\cal E}^{V}_{k,H_{ext}}$ est atteint sur le couple bifurqué. On a, pour $H_{ext}<k$, l'estimation:
\begin{equation}
{\cal E}^{V}_{k,H_{ext}}=\frac{1}{4}-(k-H_{ext})^2\frac{1}{2I(2k^2-1)}+o(k-H_{ext})^2
\end{equation}
o\`u le reste $o(k-H_{ext})^2$ est une fonction analytique.
\end{theorem}
{\it Preuve.} On écrit $H_{ext}=k-\delta$, $H_{int}=k-\omega$ et on utilise les notations $\lambda=2\pi+\epsilon$ et $\lambda=\frac{2\pi k}{H_{int}}$ définies aux équations~(\ref{approx-premier-niveau}) et (\ref{quelques-definitions}).\\
Si $H_{ext}$ est suffisamment proche de $k$, la quantité $\lambda$ devient proche de $2\pi$ par le théorème \ref{limitation-voyage-Hint} et le théorème \ref{mur-1-racine2-franchi} nous donne une localisation des couples minimisants la fonctionnelle $E^V_{k,H_{ext}}$.\\
Par conséquent, si $H_{ext}$ est suffisamment proche de $k$ alors les couples minimisants la fonctionnelle $E^V_{k,H_{ext}}(H_{int},\phi,\overrightarrow{a})$ sont les couples bifurqués définis au théorème~\ref{existence-de-la-bifurcation}.\\
Le couple bifurqué $(\phi_+(\epsilon,k), \overrightarrow{a_+}(\epsilon,k))$ défini au théorème~\ref{existence-de-la-bifurcation} vérifie :
\begin{equation}
{\cal E}^V_{k,H_{ext}}=E^V_{k,H_{ext}}(H_{int},\phi_+(\epsilon,k), \overrightarrow{a_+}(\epsilon,k))
\end{equation}
pour un certain champ $H_{int}$ proche de $k$.\\
La preuve du théorème~\ref{evolution-de-F} nous indique alors que:
\begin{equation}
E^{V}_{k,H_{ext}}(H_{int}, \phi_+,\overrightarrow{a_+})=\frac{1}{\lambda^2}[F_N(\lambda)+\epsilon^2(\frac{-1}{4(I-\frac{2}{k^2}K)})+o(\epsilon^2)]+\frac{1}{2}(H_{ext}-H_{int})^2 .
\end{equation}
On simplifie cette expression en utilisant le théorème \ref{limitation-voyage-Hint}. On a:
\begin{equation}\label{expression-Hint}
\begin{array}{rcl}
E^{V}_{k,H_{ext}}(H_{int}, \phi_+,\overrightarrow{a_+})
&=&\frac{1}{4}+(\frac{\lambda-2\pi}{\lambda})^2(\frac{-1}{4(I-\frac{2}{k^2}K)})\\
&+&o(H_{int}-k)^2+\frac{1}{2}(H_{ext}-H_{int})^2\\[2mm]
&=&\frac{1}{4}+(1-\frac{H_{int}}{k})^2(\frac{-1}{4(I-\frac{2}{k^2}K)})\\
&+&o(H_{int}-k)^2+\frac{1}{2}(H_{ext}-H_{int})^2\\[2mm]
&=&\frac{1}{4}+\frac{\omega^2}{k^2}(\frac{-1}{4(I-\frac{2}{k^2}K)})\\
&+&o(\omega^2)+\frac{1}{2}(\omega-\delta)^2 .
\end{array}
\end{equation}
La fonction $F_S(\epsilon,k)$ est analytique par rapport à $\epsilon$ donc $E^{V}_{k,H_{ext}}(H_{int}, \phi_+,\overrightarrow{a_+})$ est analytique par rapport à $\omega$. On cherche le minimum ${\cal E}^V_{k,H_{ext}}$. Par conséquent, on dérive par rapport à $\omega$:
\begin{equation}\label{expression-Hint-derivee}
\begin{array}{rcl}
\frac{\partial}{\partial \omega}E^{V}_{k,H_{ext}}(H_{int}, \phi_+,\overrightarrow{a_+})
&=&\frac{2\omega}{k^2}(\frac{-1}{4(I-\frac{2}{k^2}K)})\\
&+&o(\omega)+(\omega-\delta)
\end{array}
\end{equation}
o\`u $o(\omega)$ est une quantité qui est analytique par rapport à $\omega$.\\
On utilise la formule $I-4K=1$ du théorème \ref{theorem-egalite-somme-clef}. L'équation à résoudre est alors 
\begin{equation}\label{expression-Hint-eq-a-resoudre}
\begin{array}{rcl}
\delta&=&\omega[1+\frac{2}{k^2}(\frac{-1}{4(I-\frac{2}{k^2}K)})]+o(\omega)\\
&=&\omega[\frac{I(2k^2-1)}{2(k^2I-2K)}]+o(\omega) .
\end{array}
\end{equation}
Les quantités $2k^2-1$, $I$ et $k^2I-2K$ sont strictement positives par 
l'hypothèse $k\geq \frac{1}{\sqrt{2}}$, le lemme \ref{positivite-K-et-I} et l'hypothèse \ref{hypothese-suite} (qui est vérifiée par le corollaire \ref{corollaire-condition-hard}).\\
On peut appliquer le théorème des fonctions implicites (version analytique) et on obtient
\begin{equation}
\omega(\delta)=\frac{[2k^2I-4K]\delta}{I(2k^2-1)}+o(\delta), 
\end{equation}
le $o(\delta)$ étant analytique.\\
On calcule l'énergie avec cette valeur
\begin{equation}\label{calcul-energie-bifurcation}
\begin{array}{rcl}
{\cal E}^{V}_{k,H_{ext}}
&=&\D E^{V}_{k,H_{ext}}(H_{int}, \phi_+,\overrightarrow{a_+})\\
&=&\D\frac{1}{4}+\frac{\omega^2}{k^2}(\frac{-1}{4(I-\frac{2}{k^2}K)})+\frac{1}{2}(\omega-\delta)^2+o(\omega^2)\\
&=&\D\frac{1}{4}+(\frac{[2k^2I-4K]\delta}{I(2k^2-1)})^2(\frac{-1}{4k^2(I-\frac{2}{k^2}K)})+\frac{1}{2}(\frac{[I-4K]\delta}{I(2k^2-1)})^2+o(\delta^2)\\[2mm]
&=&\D\frac{1}{4}+\frac{4k^4[I-\frac{2}{k^2}K]^2\delta^2}{I^2(2k^2-1)^2}(\frac{-1}{4k^2(I-\frac{2}{k^2}K)})+\frac{1}{2}\frac{\delta^2}{I^2(2k^2-1)^2}+o(\delta^2)\\[2mm]
&=&\D\frac{1}{4}+\delta^2[-k^2I+2K+\frac{1}{2}]\frac{1}{I^2(2k^2-1)^2}+o(\delta^2)\\
&=&\D\frac{1}{4}+\delta^2[-k^2I+\frac{I}{2}]\frac{1}{I^2(2k^2-1)^2}+o(\delta^2)\\
&=&\D\frac{1}{4}-\delta^2\frac{1}{2I(2k^2-1)}+o(\delta^2) .
\end{array}
\end{equation}
On a bien l'estimation voulue sur l'énergie.$\hfill{\bf CQFD}$

%L'énergie de l'état normal est $E_N=\frac{1}{4}$ l'énergie de l'etat supraconducteur pur est
%\begin{equation}
%\begin{array}{rcl}
%E_S&=&\frac{H_{ext}^2}{2}\\
%&=&\frac{(k-\delta)^2}{2}\\
%&=&\frac{k^2}{2}+o(1)
%\end{array}
%\end{equation}
%Par conséquent si $\delta$ est assez petit, l'énergie du couple bifurqué est inférieure à l'énergie du couple $(0,0)$. Puisque $E_N<E_S$ on en déduit que le minimum ${\cal E}^{V}_{k,H_{ext}}$ est forcément atteint pour un état mixte.

%Décrivons tout d'abord le comportement quand $H_{ext}\rightarrow\infty$
\begin{theorem}.\label{apparition-intervalle2}
Soit $H_{c2}$\index{$H_{c2}$} l'application: 
\begin{equation}
\begin{array}{rcl}
\R_+^*&\mapsto&\R_+\\
k&\mapsto&\left\lbrace\begin{array}{ll}
\frac{1}{\sqrt{2}}&\mbox{~si~}k\leq\frac{1}{\sqrt{2}}\\
k&\mbox{~si~}k\geq\frac{1}{\sqrt{2}}
\end{array}\right.
\end{array}
\end{equation}
L'ensemble $G_{k}$ est de la forme $[H_{c2}(k),\infty[$ et pour tout $H_{ext}\geq H_{c2}(k)$ le minimum de la fonctionnelle $E^V_{k,H_{ext}}(H_{int},\phi,\overrightarrow{a})$ est atteint par l'état normal.
\end{theorem}
{\it Preuve.} Le théorème \ref{usage-theoreme-monotonie} nous donne le résultat si $k\leq \frac{1}{\sqrt{2}}$.\\
Si $k\geq \frac{1}{\sqrt{2}}$ et $H_{ext}\in[k,+\infty[$, les théorèmes \ref{normal-H-inegalite} et \ref{normal-k-inegalite} nous indique que le minimum est atteint par l'état normal.\\
Supposons par l'absurde que le minimum de la fonctionnelle $E^V_{k,H_{ext}}$ en $(k,H_{ext})$ soit atteint par l'état normal avec $H_{ext}<k$ alors, par le théorème de monotonie \ref{normal-H-inegalite} on sait que, si $H'_{ext}\in[H_{ext},k]$ le minimum de la fonctionnelle $E^V_{k,H_{ext}}$ est atteint par l'état normal.\\
Or le théorème \ref{Bifurcation-Abrikosov-real-nature} nous indique que le minimum de la fonctionnelle n'est pas atteint sur l'état normal si $H'_{ext}$ est assez proche de $k$; c'est absurde. Donc $G_{k}$ est de la forme indiquée.  $\hfill{\bf CQFD}$

\begin{theorem}.
Soit $k>0$ fixé. Si $H_{ext}>H_{c2}(k)$ alors
\begin{equation}
\left\lbrace\begin{array}{l}
\forall H_{int}>0, \forall (\phi, \overrightarrow{a})\in{\cal A}-\{(0,0)\}\\
E^{V}_{k,H_{ext}}(H_{int},\phi, \overrightarrow{a})>\frac{1}{4} .
\end{array}\right.
\end{equation}
Autrement dit l'état supraconducteur normal est l'unique état d'équilibre de la fonctionnelle.
\end{theorem}
{\it Preuve.} Soit $H_{ext}>H_{c2}(k)$ un champ magnétique. L'énergie de l'état normal est égale à $\frac{1}{4}$.\\
On a donc, puisque $H_{c2}(k)\in G_k$, l'inégalité:
\begin{equation}\label{normal-phase-pour-B-critique}
\forall H_{int}>0,\,\,\frac{1}{\lambda^2}m_F(\lambda,k)+\frac{1}{2}(H_{int}-H_{c2}(k))^2\geq \frac{1}{4} .
\end{equation}
En particulier, si $H_{int}=H_{c2}(k)$, cette inégalité implique :
\begin{equation}\label{inegalite-particuliere-usage-numero2}
(\frac{H_{c2}(k)}{2\pi k})^2m_F(\frac{2\pi k}{H_{c2}(k)},k)\geq \frac{1}{4} .
\end{equation}
Il faut démontrer que $\forall H_{int}>0, \forall (\phi, \overrightarrow{a})\in{\cal A}-\{(0,0)\}$, l'inégalité
\begin{equation}\label{inegalite-a-demontrer-2}
\begin{array}{l}
\frac{1}{\lambda^2}F_{\lambda,k}(\phi, \overrightarrow{a})+\frac{1}{2}(H_{ext}-H_{int})^2>\frac{1}{4} ,
\end{array}
\end{equation}
est vérifiée.\\
Si $H_{int}\leq H_{c2}(k)$, alors on a trivialement:
\begin{equation}
(H_{int}-H_{c2}(k))^2<(H_{int}-H_{ext})^2.
\end{equation}
Par conséquent, l'inégalité~(\ref{inegalite-a-demontrer-2}) est démontrée pour $H_{int}\leq H_{c2}(k)$.\\
Supposons maintenant $H_{int}>H_{c2}(k)$.\\
Si $(\phi, \overrightarrow{a})\in {\cal A}-\{(0,0)\}$, alors la contraposée du lemme~\ref{annulation-deux-integrale} s'écrit
\begin{equation}
\begin{array}{l}
\int_{\Omega}\Vert i\overrightarrow{\nabla}\phi+(\overrightarrow{A}_0+\overrightarrow{a})\phi\Vert^2+\int_{\Omega}|\rot\,\overrightarrow{a}|^2\not=0 .
\end{array}
\end{equation}
On a alors, par l'expression de $D$ (cf~(\ref{definition-Dklambdaphia})) et si $H_{int}>H_{c2}(k)$,
\begin{equation}\label{inegalite-stricte}
D_{\frac{2\pi k}{H_{c2}(k)},k}(\phi, \overrightarrow{a})<D_{\frac{2\pi k}{H_{int}},k}(\phi, \overrightarrow{a}) .
\end{equation}
En combinant les équations~(\ref{inegalite-stricte}) et~(\ref{inegalite-particuliere-usage-numero2}), on obtient la relation:
\begin{equation}
\frac{1}{4}<D_{\frac{2\pi k}{H_{int}},k}(\phi, \overrightarrow{a}) .
\end{equation}
%par conséquent l'inégalité~(\ref{inegalite-a-demontrer-2}) est démontrée.
On a donc bien montré que l'état normal est le seul état minimisant la fonctionnelle 
\begin{equation}
\begin{array}{rcl}
E^{V}_{k,H_{ext}}:\R_+^*\times {\cal A}&\mapsto&\R\\
(H_{int},\phi,\overrightarrow{a})&\mapsto&E^{V}_{k,H_{ext}}(H_{int},\phi,\overrightarrow{a})
\end{array}
\end{equation}
$\hfill{\bf CQFD}$

\saction{Diagramme de phase, l'état supraconducteur pur}\label{sec:diagramme-de-phase-2}
\noindent Passons à l'étude plus complexe de l'ensemble $F_{k}$.
\begin{theorem}.\label{apparition-intervalle}
L'ensemble $F_{k}$ est de la forme $]0,H_{c1}(k)]$\index{$H_{c1}$} avec 
\begin{equation}
\left\lbrace\begin{array}{ll}
H_{c1}(k)=\frac{1}{\sqrt{2}}&\mbox{~si~}k\leq \frac{1}{\sqrt{2}}\\
\frac{1}{\sqrt{2}}\geq H_{c1}(k)\geq \frac{1}{2k}&\mbox{~si~}k > \frac{1}{\sqrt{2}}.
\end{array}\right.
\end{equation}
De plus la fonction $H_{c1}(k)$ est une fonction décroissante de $k$.
\end{theorem}
{\it Preuve.}\\
{\bf $1^{ere}$ étape :}\\
On sait par le théorème~\ref{usage-theoreme-monotonie} que, si $k\leq\frac{1}{\sqrt{2}}$, alors $]0,\frac{1}{\sqrt{2}}]=F_{k}$. On a donc le résultat dans ce cas.\\
{\bf $2^{eme}$ étape :}\\
On suppose $k\geq \frac{1}{\sqrt{2}}$. Pour que le minimum de la fonctionnelle $E^{V}_{k,H_{ext}}(H_{int},\phi,\overrightarrow{a})$ soit atteint par l'état supraconducteur pur, il faut que:
\begin{equation}\label{inegalite_champ-critique}
\forall H_{int}>0, G^{V}_{k,H_{ext}}(H_{int})\geq \frac{H_{ext}^2}{2} .
\end{equation}
En vertu du théorème~\ref{encadrement-des-minimums}, l'inégalité~(\ref{inegalite_champ-critique}) est impliquée par l'inégalité:
\begin{equation}
\frac{1}{\lambda^2}(\lambda\pi-\pi^2)+\frac{1}{2}(H_{ext}-H_{int})^2\geq \frac{H_{ext}^2}{2} .
\end{equation}
On utilise l'expression $\lambda=\frac{2\pi k}{H_{int}}$; cette inégalité se réécrit en :
\begin{equation}
(\frac{H_{int}}{2 k}-(\frac{H_{int}}{2k})^2)+\frac{1}{2}(H_{int}^2-2H_{int}H_{ext})\geq 0 .
\end{equation}
Cette dernière est elle même équivalente à :
\begin{equation}
H_{int}[\frac{1}{2k}-H_{ext}]+H_{int}^2[\frac{1}{2}-\frac{1}{4k^2}]\geq 0 .
\end{equation}
Cette inégalité est vérifiée si $H_{ext}\leq \frac{1}{2k}$ puisque on a toujours $\frac{1}{2}-\frac{1}{4k^2}\geq 0$; on a donc l'inclusion $]0,\frac{1}{2k}]\subset F_k$.\\
{\bf $3^{eme}$ étape :}\\
Si $H_{ext}>\frac{1}{\sqrt{2}}$, alors l'énergie de l'état normal est strictement inférieure à l'énergie de l'état supraconducteur pur; en effet:
\begin{equation}
E_{N}=\frac{1}{4}<\frac{H_{ext}^2}{2}=E_S .
\end{equation}
Cela signifie qu'aucun point de l'intervalle $]\frac{1}{\sqrt{2}},\infty[$ n'appartient à $F_k$. Par conséquent $F_k\subset ]0,\frac{1}{\sqrt{2}}]$.\\
{\bf $4^{eme}$ étape :}\\
On sait par le théorème~\ref{monotonie-etat-pur} que, si $H_{ext}\in F_k$ alors pour tout $H'_{ext}\leq H_{ext}$, on a $H'_{ext}\in F_k$. Par conséquent, l'ensemble $F_{k}$ est un intervalle qui peut être de la forme $]0,H_{c1}(k)[$ ou $]0,H_{c1}(k)]$ avec $\frac{1}{\sqrt{2}}\geq H_{c1}(k)>\frac{1}{2k}$.\\
De par la definition de $H_{c1}(k)$ on a 
\begin{equation}
F_k=\cap_{(\phi,\overrightarrow{a})\in{\cal A}}  U_{\lambda,k,H_{int},\phi,\overrightarrow{a}}^{-1}([0,\infty])
\end{equation}
avec $U_{\lambda,k,H_{int},\phi,\overrightarrow{a}}(H_{ext})=\frac{1}{\lambda^2}F_{\lambda,k}(\phi,\overrightarrow{a})+\frac{(H_{int}-H_{ext})^2}{2}-\frac{H_{ext}^2}{2}$.\\
L'ensemble $F_k$ est un fermé dans $\R_+^*$ comme intersection de fermés donc on a $F_k=]0,H_{c1}(k)]$.\\
%
%Montrons que c'est le second cas qui est réalisé. Soit la suite $H_{ext}^{\nu}=H_{c1}(k)(1-\frac{1}{\nu+2})$ cette suite appartient à $F_k$ et tend vers $H_{c1}(k)$. On a donc $\forall \nu\geq 0$ les relations
%\begin{equation}
%\forall H_{int}>0, \forall (\phi, \overrightarrow{a})\in{\cal A}, \frac{1}{\lambda^2}F_{\lambda,k}(\phi, \overrightarrow{a})+\frac{1}{2}(H_{ext}^{\nu}-H_{int})^2\geq \frac{(H_{ext}^{\nu})^2}{2} .
%\end{equation}
%Si on fait tendre $\nu$ vers l'infini, ces inégalités se transforment par passage à la limite en 
%\begin{equation}
%\frac{1}{\lambda^2}F_{\lambda,k}(\phi, \overrightarrow{a})+\frac{1}{2}(H_{c1}(k)-H_{int})^2\geq \frac{H_{c1}(k)^2}{2}
%\end{equation}
%ce qui signifie que $H_{c1}(k)\in F_k$.\\
Si $k'\leq k$  alors $F_k\subset F_{k'}$ (cf~\ref{monotonie-etat-pur}). On a alors trivialement $H_{c1}(k)\leq H_{c1}(k')$. $\hfill{\bf CQFD}$\\
\\
On a un résultat plus précis:
\begin{proposition}.\label{pur-unique-etat-equilibre}
Si $H_{ext}<H_{c1}(k)$, alors pour tout $H_{int}>0$, pour tout $(\phi, \overrightarrow{a})\in{\cal A}$:
\begin{equation}
\frac{1}{\lambda^2}F_{\lambda,k}(\phi, \overrightarrow{a})+\frac{1}{2}(H_{ext}-H_{int})^2>\frac{H_{ext}^2}{2} .
\end{equation}
Autrement dit l'état supraconducteur pur est l'unique état d'équilibre de la fonctionnelle.
\end{proposition}
{\it Preuve.} On sait par le théorème~\ref{apparition-intervalle} que:\\
Pour tout $H_{int}>0$, pour tout $(\phi, \overrightarrow{a})\in{\cal A}$, on a:
\begin{equation}
\frac{1}{\lambda^2}F_{\lambda,k}(\phi, \overrightarrow{a})+\frac{1}{2}(H_{c1}(k)-H_{int})^2\geq\frac{H_{c1}(k)^2}{2} .
\end{equation}
Cette inégalité est équivalente à 
\begin{equation}
\frac{1}{\lambda^2}F_{\lambda,k}(\phi, \overrightarrow{a})+\frac{1}{2}[-2H_{int}H_{c1}(k)+H_{int}^2]\geq 0 .
\end{equation}
Si $H_{ext}<H_{c1}(k)$, alors:
\begin{equation}
-H_{int}(H_{ext}-H_{c1}(k))>0 .
\end{equation}
On a donc pour tout $H_{int}>0$, pour tout $(\phi, \overrightarrow{a})\in{\cal A}$:
\begin{equation}
\frac{1}{\lambda^2}F_{\lambda,k}(\phi, \overrightarrow{a})+\frac{1}{2}[-2H_{int}H_{ext}+H_{int}^2]> 0 .
\end{equation}
Cette inégalité équivaut à :
\begin{equation}
\frac{1}{\lambda^2}F_{\lambda,k}(\phi, \overrightarrow{a})+\frac{1}{2}(H_{ext}-H_{int})^2>\frac{H_{ext}^2}{2} .
\end{equation}
On a donc bien montré que l'état supraconducteur pur est le seul état qui minimise la fonctionnelle $E^{V}_{k,H_{ext}}(H_{int},\phi,\overrightarrow{a})$. $\hfill{\bf CQFD}$

\begin{theorem}.\label{expression-en-inf-assez-utile}
On a l'expression suivante pour le champ $H_{c1}(k)$:
\begin{equation}\label{expression-en-sup-du-champ-Hc1}
\begin{array}{r}
H_{c1}(k)=\inf_{(\phi,\overrightarrow{a})\in{\cal A}}\lbrace\frac{1}{4\pi k}\int_{\Omega}\Vert i\overrightarrow{\nabla}\phi+(\overrightarrow{A}_0+\overrightarrow{a})\phi\Vert^2\\
+\sqrt{[\frac{1}{2}+\frac{1}{2(2\pi)^2}\int_{\Omega}|\rot\,\overrightarrow{a}|^2][\int_{\Omega}(1-|\phi|^2)^2]}\rbrace .
\end{array}
\end{equation}
\end{theorem}
{\it Preuve.} Par les définitions (\ref{definition-Fk-Gk}), on a:
\begin{equation}
\begin{array}{rcl}
H_{ext}\in F_k
&\Leftrightarrow&\forall H_{int},\phi,\overrightarrow{a},\frac{\D F_{\lambda,k}(\phi,\overrightarrow{a})}{\lambda^2}+\frac{1}{2}(H_{int}-H_{ext})^2\geq \frac{H_{ext}^2}{2}\\
&\Leftrightarrow&\forall H_{int},\phi,\overrightarrow{a},D_{\lambda,k}(\phi,\overrightarrow{a})+\frac{1}{2}(H_{int}-H_{ext})^2\geq \frac{H_{ext}^2}{2}\\
&\Leftrightarrow&\forall H_{int},\phi,\overrightarrow{a},\frac{1}{H_{int}}D_{\lambda,k}(\phi,\overrightarrow{a})+\frac{H_{int}}{2}\geq H_{ext}\\
&\Leftrightarrow&\left\lbrace\begin{array}{l}
\forall H_{int}>0, \forall(\phi, \overrightarrow{a})\in{\cal A}\\
\frac{H_{int}}{2}+\frac{1}{4\pi k}\int_{\Omega}\Vert i\overrightarrow{\nabla}\phi+(\overrightarrow{A}_0+\overrightarrow{a})\phi\Vert^2\\
+\frac{1}{4H_{int}}\int_{\Omega}(1-|\phi|^2)^2+\frac{H_{int}}{2(2\pi)^2}\int_{\Omega}|\rot\,\overrightarrow{a}|^2\geq H_{ext} .
\end{array}\right. 
\end{array}
\end{equation}
Or le minimum de la fonction
\begin{equation}
\begin{array}{rcl}
\R_+^*&\mapsto&\R\\
t&\mapsto&at+\frac{b}{t}
\end{array}
\end{equation}
est égal à $2\sqrt{ab}$ et est atteint pour $t=\sqrt{\frac{b}{a}}$. On obtient donc:
\begin{equation}
\begin{array}{rcl}
H_{ext}\in F_k
&\Leftrightarrow&
\left\lbrace\begin{array}{rcl}
H_{ext}&\leq& \inf_{(\phi,\overrightarrow{a})\in{\cal A}}\lbrace\frac{1}{4\pi k}\int_{\Omega}\Vert i\overrightarrow{\nabla}\phi+(\overrightarrow{A}_0+\overrightarrow{a})\phi\Vert^2\\
&&+\sqrt{[\frac{1}{2}+\frac{1}{2(2\pi)^2}\int_{\Omega}|\rot\,\overrightarrow{a}|^2][\int_{\Omega}(1-|\phi|^2)^2]}\rbrace .
\end{array}\right. 
\end{array}
\end{equation}
On a donc l'égalité voulue, car $F_k=]0,H_{c1}(k)]$. $\hfill{\bf CQFD}$

\begin{theorem}.\label{estimee-asymptotique-weak-sylvia-saint-serfaty}
Le champ $H_{c1}(k)$ vérifie l'estimation asymptotique lorsque  $k\rightarrow\infty$
\begin{equation}
H_{c1}(k)=O(\frac{\ln\,k}{k}) .
\end{equation}
\end{theorem}
{\it Preuve.}\\
{\bf $1^{ere}$ étape :}\\
On sait par la proposition~\ref{unique-zero} que les zéros de $\phi_0$ sont simples et que l'ensemble de ces zéros est de la forme $z_0+{\cal L}$. On peut donc trouver un domaine fondamental $\Omega'$ tel que $z_0$ soit le seul zéro appartenant à l'intérieur de $\Omega'$ et qu'il n'y ait pas d'autres zéros dans $\overline{\Omega'}$.\\
Soit $\delta>0$ tel que les boules $B(z_0+l,\delta)$ soient distinctes dans $\C$ avec $l\in{\cal L}$.\\
Soit la fonction $f$ définie par
\begin{equation}
\begin{array}{rcl}
f:\C&\mapsto&\R\\
z&\mapsto&\left\lbrace\begin{array}{rl}
\frac{1}{|\phi_0|(z)}&\mbox{~si~} z\notin \cup_{l\in{\cal L}}B(z_0+l,\frac{\delta}{2})\\
\frac{2|z-(z_0+l)|}{\delta}\frac{1}{|\phi_0|(z)}&\mbox{~si~}z\in B(z_0+l,\frac{\delta}{2})
\end{array}\right.
\end{array}
\end{equation}
La fonction $f\phi_0$ est de module $1$ en dehors de $\cup_{l\in{\cal L}}B(z_0+l,\frac{\delta}{2})$, est de module $\frac{2|z-(z_0+l)|}{\delta}$ dans la boule $B(z_0+l,\frac{\delta}{2})$. Elle est également de classe $H^1_{loc}$.\\
La fonction $\frac{\phi_0(z)}{z-z_0}$ ne s'annule pas sur $B(z_0,\delta)$; par conséquent il existe une section $\theta$ définie sur $B(z_0,\delta)$ pour l'argument de cette fonction complexe puisque $B(z_0,\delta)$ est simplement connexe. La section $\theta$ est $C^{\infty}$.\\
Soit $\theta_1$ la fonction $C^{\infty}$ définie sur $B(z_0,\delta)$ par
\begin{equation}
\begin{array}{l}
\theta_1=\theta\mbox{~dans~}B(z_0,\frac{\delta}{2})\\
\theta_1=0\mbox{~dans~}B(z_0,\delta)-B(z_0,\frac{3\delta}{4})\\
\end{array}
\end{equation}
Une telle fonction existe. On la prolonge à $\C$ en imposant que $\theta_1$ est ${\cal L}$-périodique et que la fonction $\theta_1$ s'annule hors de $\cup_{l\in{\cal L}}B(z_0+l,\frac{\delta}{2})$.\\
La fonction $f(z)\phi_0(z)e^{i\theta_1(z)}$ est donc de module $1$ hors de $\cup_{l\in{\cal L}}B(z_0+l,\frac{\delta}{2})$ et est égale à $\frac{2(z-z_0)}{\delta}$ dans $B(z_0,\frac{\delta}{2})$.\\
On appelle \index{$\phi'$}cette fonction $\phi'$. Elle est de classe $H^1_{loc}$.\\
Il existe $\delta>0$ tel que, $B(z_0+l,\frac{\delta}{2}) \cap \Omega'=\emptyset$ si $l\not= 0$ et $B(z_0,\frac{\delta}{2})\subset \Omega'$.\\
C'est cette fonction que l'on va modifier maintenant pour arriver à nos fins.\\
Soit $\frac{\delta}{2}>\epsilon>0$. On définit la fonction \index{$\phi_{\epsilon}$}$\phi_{\epsilon}$ par
\begin{equation}
\begin{array}{rcl}
\phi_{\epsilon}:\C&\mapsto&\C\\
z&\mapsto&\left\lbrace\begin{array}{rl}
\phi'&\mbox{~si~} z\notin \cup_{l\in{\cal L}}B(z_0+l,\frac{\delta}{2})\\
\frac{z-z_0}{\epsilon}&\mbox{~si~} z\in B(z_0,\epsilon)\\
\frac{z-z_0}{|z-z_0|}&\mbox{~si~} z\in B(z_0,\frac{\delta}{2})-B(z_0,\epsilon) .
\end{array}\right.
\end{array}
\end{equation}
cette fonction est continue et peut \^etre considérée comme une section du fibré $E_1$, toute la question est d'estimer les quantités $\int_{\Omega}\Vert i\overrightarrow{\nabla}\phi_{\epsilon}+(\overrightarrow{A}_0+\overrightarrow{a})\phi_{\epsilon}\Vert^2$ et $\int_{\Omega}(1-|\phi_{\epsilon}|^2)^2$.\\
{\bf $2^{eme}$ étape :}\\
La fonction $\phi_{\epsilon}$ est de module inférieur à $1$. On calcule sa norme $H^1$.
\begin{equation}
\nabla\frac{z}{|z|}=(\frac{y^2-ixy}{(x^2+y^2)^{\frac{3}{2}}},\frac{ix^2-xy}{(x^2+y^2)^{\frac{3}{2}}})
\end{equation}
On obtient aussi:
\begin{equation}
\begin{array}{rcl}
\int_{B(z_0,\frac{\delta}{2})-B(z_0,\epsilon)}\Vert \nabla\phi_{\epsilon}\Vert^2&=&\int_{B(z_0,\frac{\delta}{2})-B(z_0,\epsilon)}\frac{1}{x^2+y^2}dxdy\\
&=&\int_{\epsilon}^{\frac{\delta}{2}}\frac{1}{r^2}2\pi rdr\\
&=&\int_{\epsilon}^{\frac{\delta}{2}}\frac{1}{r}2\pi dr\\
&=&2\pi[\ln\,\frac{\delta}{2}-\ln\,\epsilon]\\
&=&O(-\ln\,\epsilon) .
\end{array}
\end{equation}
L'expression $\int_{\Omega'-B(z_0,\frac{\delta}{2})}\Vert\nabla\phi_{\epsilon}\Vert^2$ ne dépend pas de $\epsilon$ et
\begin{equation}
\begin{array}{rcl}
\int_{B(z_0,\epsilon)}\Vert\nabla\phi_{\epsilon}\Vert^2&=&\int_{B(z_0,\epsilon)}\frac{2}{\epsilon^2}dxdy\\
&=&\frac{2}{\epsilon^2}\pi \epsilon^2\\
&=&2\pi\\
&=&O(1) .
\end{array}
\end{equation}
On a donc l'estimée:
\begin{equation}
\begin{array}{rcl}
\int_{\Omega}\Vert i\overrightarrow{\nabla}\phi_{\epsilon}+\overrightarrow{A}_0\phi_{\epsilon}\Vert^2&=&O(-\ln\,\epsilon) .
\end{array}
\end{equation}
Enfin, vu la définition de $\phi_{\epsilon}$, on obtient:
\begin{equation}
\int_{\Omega'}(1-|\phi_{\epsilon}|^2)^2\leq \int_{B(z_0,\epsilon)}dxdy=\pi\epsilon^2=O(\epsilon^2) .
\end{equation}
On introduit le couple $(\phi_{\epsilon},0)$ dans l'expression à l'intérieur de l'infimum de l'expression~(\ref{expression-en-sup-du-champ-Hc1}). On obtient la valeur:
\begin{equation}
\frac{O(-\ln\,\epsilon)}{k}+O(\epsilon) .
\end{equation}
Si on prend $\epsilon=\frac{1}{k}$, on obtient l'estimée:
\begin{equation}
\frac{O(\ln\,k)}{k}+\frac{O(1)}{k} .
\end{equation}
On a donc trouvé une valeur pour l'expression entre accolade de l'infimum.\\
Comme le champ $H_{c1}(k)$ est l'infimum sur les couples possibles, on en déduit l'estimée annoncée. $\hfill{\bf CQFD}$

\begin{remarque}.
La thèse de Sylvia Serfaty contient des estimées de ce type pour le cas de la fonctionnelle de Ginzburg-Landau sur un domaine du plan.\\
A priori notre cadre (minimisation sur des fonctions périodiques) est plus simple et ces résultats doivent également \^etre vrais.\\
Dans \cite{serfaty1} elle montre l'existence de solution minimisant localement la fonctionnelle. Il est montré dans \cite{serfaty2} que cette solution minimisante est en fait un minimum global
\end{remarque}

\begin{proposition}.\label{proposition-merci-marie-monier}
La fonction $k\mapsto kH_{c1}(k)$ est croissante. La fonction $H_{c1}(k)$ est continue.
\end{proposition}
{\it Preuve.} On sait que le champ $H_{c1}(k)$ a pour expression 
\begin{equation}
\begin{array}{c}
H_{c1}(k)=\inf_{(\phi,\overrightarrow{a})\in{\cal A}}\lbrace\frac{1}{4\pi k}\int_{\Omega}\Vert i\overrightarrow{\nabla}\phi+(\overrightarrow{A}_0+\overrightarrow{a})\phi\Vert^2\\
+\sqrt{[\frac{1}{2}+\frac{1}{2(2\pi)^2}\int_{\Omega}|\rot\,\overrightarrow{a}|^2][\int_{\Omega}(1-|\phi|^2)^2]}\rbrace .
\end{array}
\end{equation}
La fonction $k\mapsto\frac{1}{k}$ est décroissante donc la fonction $k\mapsto H_{c1}(k)$ est décroissante. On a aussi
\begin{equation}
\begin{array}{c}
kH_{c1}(k)=\inf_{(\phi,\overrightarrow{a})\in{\cal A}}\lbrace\frac{1}{4\pi}\int_{\Omega}\Vert i\overrightarrow{\nabla}\phi+(\overrightarrow{A}_0+\overrightarrow{a})\phi\Vert^2\\
+k\sqrt{[\frac{1}{2}+\frac{1}{2(2\pi)^2}\int_{\Omega}|\rot\,\overrightarrow{a}|^2][\int_{\Omega}(1-|\phi|^2)^2]}\rbrace .
\end{array}
\end{equation}
La croissance de $k\mapsto kH_{c1}(k)$ est également claire.\\
Si $0<k\leq k'$ on a $H_{c1}(k')\in [\frac{k}{k'}H_{c1}(k),H_{c1}(k)]$ donc 
\begin{equation}
\lim_{k'\rightarrow k, k'\geq k}H_{c1}(k')=H_{c1}(k)
\end{equation}
de m\^eme si $0<k'\leq k$ on a $H_{c1}(k')\in [H_{c1}(k),\frac{k}{k'}H_{c1}(k)]$ donc on a aussi une limite à gauche. La fonction $k\mapsto H_{c1}(k)$ est continue. $\hfill{\bf CQFD}$

\saction{Conclusion et références}\label{sec:conclusion}
\noindent La première étude sur l'état d'Abrikosov a été faite par Abrikosov dans l'article \cite{Abrikosov}. Elle montrait à partir des équations de Ginzburg-Landau qu'il existe un état intermédiaire entre la supraconductivité pure et l'état normal.\\
Cette étude a été reprise par Lasher \cite{lasher} et Odeh \cite{odehII} qui décrivent la transition de phase vers l'état mixte.\\
Dans les années $90$, Bethuel, Hélein et Brézis ont analysé la limite $k\rightarrow\infty$, sans champ magnétique dans \cite{helein}, puis avec champ magnétique dans \cite{riviere}.\\
L'article \cite{pirate-II} est une étude de l'existence de solutions du problème variationnel sur le tore.\\
L'étude de la bifurcation d'Abrikosov est faite dans \cite{Barany-Golubitsky}. Les articles \cite{almogI} et \cite{pirate-I} étudient l'énergie des états bifurqués et leur stabilité. L'article \cite{Takac} se consacre aux zéros des états bifurqués. L'article \cite{matamo-qi} montre que l'on ne peut pas avoir de lignes de zéros dans le cas d'un ouvert simplement connexe.\\
Le cas $k=\frac{1}{\sqrt{2}}$ a été étudié dans \cite{almogII} o\`u une condition nécessaire et suffisante pour l'existence de solution à $N$ vortex est donnée. Les références \cite{caffarelli}, \cite{Garcia-prada}, \cite{Taubes}, \cite{wang-yang} sont consacrés entre autre à l'existence et l'unicité de couples minimisants.\\
L'analyse asymptotique quand $k$ tend vers l'infini du champ critique $H_{c1}$ est faite dans \cite{serfaty1} et \cite{serfaty2} en dimension $2$. La dimension $1$ est traitée dans \cite{Mes-chers-chefsI} et \cite{Mes-chers-chefsII}.\\
Les références \cite{adams}, \cite{Hormander-ALPDE-I}, \cite{taylor} et \cite{brezis} sont des références relativement standards sur les espaces de Sobolev, les opérateurs pseudodifférentiels et l'analyse fonctionnelle qui nous ont été utiles dans notre travail.\\
%Les références \cite{Chabat}, \cite{spe-M'-dolbeault}, \cite{Griffiths-Harris}, \cite{spin-geom}, \cite{Malliavin}, \cite{Nakahara} et \cite{Siegel} sont plus personnelles.
\printindex


\begin{thebibliography}{plain}
\baselineskip=12truept
\bibitem[Ab]{Abrikosov}A.A. Abrikosov: Magnetic properties of group II Superconductor.\\
(J. Exptl. Theoret. Phys. (USSR){\bf 32}, 1174-1182, Volume 5, 1957).

\bibitem[Ad]{adams}R. D. Adams: Sobolev Spaces.\\
Birkhäuser, 1975.

\bibitem[Al1]{almogI}Y. Almog: On the bifurcation and stability of periodic solutions of the Ginzburg Landau equations in the plane. Preprint

\bibitem[Al2]{almogII}Y. Almog: Periodic solutions to the first order Ginzburg-Landau equations. Preprint.

\bibitem[BGT]{Barany-Golubitsky}E. Barany, M. Golubitsky and J. Turski: Bifurcations with local gauge symmetries in the Ginzburg-Landau equations.\\
Physica-D 56 (1992), p.~36-56 .

\bibitem[BBH]{helein}F. Bethuel, H. Brézis, F. Hélein: Ginzburg-Landau vortices.\\
Progress in Nonlinear Differential Equations and their Applications, 13. Birkhäuser, .

\bibitem[BR]{riviere}F. Bethuel, T. Rivière: Vortices for a variational problem related to superconductivity.\\
Annales de l'Institut Henri Poincaré (Analyse Non Linéaire), 12 (1995), No. 3, p.~243--303.

\bibitem[Bo]{bogolom}E. B. Bogmol'nyi: The stability of classical solutions. Sov. J. Nucl. Phys. 24, 449 (1977).

\bibitem[BH1]{Mes-chers-chefsI}C. Bolley, B. Helffer: Asymptotics for critical fields in an approximate Ginzburg-Landau model in the large size limit.\\
Math. Models Methods Appl. Sci. 8 (1998), no. 5, 821--849.

\bibitem[BH2]{Mes-chers-chefsII}C. Bolley, B. Helffer: The Ginzburg-Landau equations in a semi-infinite superconducting film in the large $\kappa$ limit.\\
European J. Appl. Math. 8 (1997), no. 4, 347--367.

\bibitem[Br]{brezis}H. R. Brézis: Analyse fonctionnelle, théorie et applications.\\
Collection Mathématique Appliquée pour la maîtrise. Masson, 1983.

\bibitem[CY]{caffarelli}L.A. Caffarelli, Y. Yang: Vortex condensation in the Chern-Simons Higgs model: An existence theorem.\\
Communications in Mathematical Physics, 168, 321-336 (1995)

\bibitem[Chab]{Chabat}B. Chabat: Introduction à l'analyse complexe, Tome
2: Fonctions de plusieurs variables. MIR, 1990.

\bibitem[Chap]{pirate-I}S. J. Chapman: Nucleation of superconductivity in decreasing fields. I, II.\\
European J of Appl Math. 5 (1994), No. 4, p.~449--468 et p.~469--494. 

\bibitem[Cr]{Crandall}M. G. Crandall, P. H. Rabinowitz: Bifurcation from simple eigenvalues.\\
Journal of Functional analysis, 8, 321-340 (1971)

\bibitem[De]{Demally}J. P. Demailly: Sur l'identité de Bochner Kodaira Nakano en géométrie hermitienne. Séminaire d'analyse P. Lelong, P. Dolbeault, H. Skoda année 83/84 p.~88-97, Lecture Notes in maths 1198.

%\bibitem[DM]{top-defect}M.J.W. Dodgson, M.A. Moore: Topological defects in the Abrikosov Lattice of vortices in Type II superconductors.\\
%(preprint)

\bibitem[Do]{spe-M'-dolbeault}P. Dolbeault: Analyse complexe. Masson, Collection ma\^itrise de mathématique pure, 1990.

%Lediteur reste a trouver, une recherche bibliographique s'impose

%\bibitem[FeMa]{fedoriuk}M. V. Fedoriuk, M. P. Maslov: Semi-classical approximation in quantum mechanics

\bibitem[EMQ]{matamo-qi}C. Elliot, H. Matamo, T. Qi: Zeros of complex Ginzburg-Landau order parameter with applications to superconductivity.\\
European J of Appl Math. 5 (1994), No. 4, p.~431--448.

\bibitem[Ga]{Garcia-prada}O. Garcia Prada: A direct existence proof for the vortex equations over a compact Riemannian manifold. Bull London Math Soc 26 (1994) 86-96

\bibitem[Ge]{PG-de-Gennes}P. G. de Gennes: Superconducting properties of metals and alloys.\\
Addison Wesley, 1966.

\bibitem[GS]{golubitsky}M. Golubitsky, D.G. Schaeffer: Singularities and groups in bifurcation theory I,II.\\
Springer Verlag, 1985-1988.

\bibitem[GH]{Griffiths-Harris}P. Griffiths, J. Harris: Principles of algebraic Geometry.\\
Wiley Classics Library Edition Published, 1994.

\bibitem[DGP]{pirate-II}Q. Du, M. D. Gunzburger, J. S. Peterson: Modeling and analysis of a periodic Ginzburg-Landau model for Type-II superconductors.\\
SIAM Journal on applied mathematics 53 (1993), No 3, p.~689-717.

\bibitem[Ho]{Hormander-ALPDE-I} L. Hörmander, The Analysis of Linear Partial Differential Operators I,II.\\
Springer Verlag, 1983.

\bibitem[JaTa]{jaffe-taubes}A. Jaffe, C.H. Taubes: Vortices and Monopoles.\\
Birkhäuser, 1980.

\bibitem[JT]{Acker}T. J\"urgen, A. Thomas: the generalized Lichnerowicz formula and analysis of Dirac operators. J Reine Angew Math 471 (1996) p23-42.

\bibitem[Ki]{Kittel} C. Kittel, Physique de l'état solide.\\
Dunod Collection Université, 1983.

\bibitem[La]{lasher} G. Lasher: Series solutions of the Ginzburg-Landau equations for the Abrikosov mixed state.\\
Physical review, Volume 140, No 2A, p. 523-528. 1965

\bibitem[LM]{spin-geom}H.B. Lawson and M.-L. Michelsohn: Spin geometry.\\
Princeton University press, 1989.

\bibitem[Li]{Lichne}A. Lichnerowicz: Spineurs harmonique. CRAS(A) 257 (1963).

%\begin{theorem}
%Si la fonction f vérifie
%\begin{equation}
%f(\frac{x+y}{2})\leq \frac{f(x)+f(y)}{2}
%\end{equation}
%alors si $f$ est mesurable $f$ est convexe
%\end{theorem}

%\bibitem[Si]{func-int}B. Simon: Functional integration and quantum physics.\\
%Academic Press.

\bibitem[MaAi]{Malliavin}P. Malliavin et H. Airault: Intégration, analyse de Fourier, analyse gaussienne.\\
Masson, 1994.

\bibitem[Na]{Nakahara}M. Nakahara: Geometry, topology and physics.\\
Institute of Physics Publishing, 1990.

\bibitem[Od1]{odehI} F. Odeh: Existence and bifurcation theorems for the Ginzburg-Landau equation.\\
Journal of Mathematical Physics, Volume 8, No 12, December 1967.

\bibitem[Od2]{odehII} F.Odeh: A bifurcation problem in superconductivity, in Bifurcation Theory and nonlinear eigenvalue problems.\\
Edited by J.B. Keller, S. Antman, W.A. Benjamin, Inc (1969), p. 99-112.

\bibitem[RS]{reed-simon-IV}M. Reed, B. Simon: Methods of modern mathematical physics, IV Analysis of operators.\\
Academic Press, 1978.

\bibitem[RV]{convex-plait-helffer}A. Roberto, D. Varberg: Convex functions, Pure and Applied Mathematics, Vol 57.\\
Academic Press, New York-London, 1973.

\bibitem[SST]{sarma} D. Saint-James, G. Sarma, E.J. Thomas: Type-II-Superconductivity.\\
Pergamon Press, 1969.

\bibitem[Sa]{merci-MXK}D. Salamon: Spin geometry and Seiberg-Witten invariants.\\
(preprint communiqué par Denis Auroux).

\bibitem[Se]{serfaty1} S. Serfaty: Local minimizers for the Ginzburg-Landau Energy near Critical Magnetic Field, part I.\\
à para\^itre dans Communications in Contemporary Mathematics.

\bibitem[SeSa]{serfaty2} S. Serfaty, E. Sandier: Global minimizers for the Ginzburg-Landau Energy below the First Critical Magnetic Field.\\
à para\^itre dans Annales IHP, Analyse non linéaire.

\bibitem[Sie]{Siegel} C. L. Siegel: Topics in complex function theory, Volumes I,II,III.\\
Wiley Classics Library, 1988.

\bibitem[Tak]{Takac}P. Takac: Bifurcation and vortex formation in the Ginzburg-Landay-equations. Preprint.

\bibitem[Tau]{Taubes}C. H. Taubes: Arbitrary N-Vortex solutions to the first order Ginzburg-Landau equations.
Communications in Mathematical Physics 72, 277-292, 1980.

\bibitem[Tay]{taylor}M.E. Taylor: Pseudodifferential operators.\\
Princeton University Press, 1981.

\bibitem[T]{tink} M. Tinkham: Introduction to superconductivity, 2nd edition.\\
McGraw-Hill, 1996.
%\bibitem[TrDo]{troy}R. Troy, A T Dorsey: A self consistent microscopic theory of surface superconductivity.\\
%(http://xxx.lpthe.jussieu.fr/abs/cond-mat/9411099)
%
%\bibitem[myst]{myst}Analysis of convex functions

\bibitem[WY]{wang-yang}S. Wang, Y.Yang: Abrikosov's vortices in the critical coupling.\\
SIAM J. Math. Anal. 23 (1992), no. 5, 1125--1140.


\end{thebibliography}
\end{document}